\documentclass[12pt]{ecsthesis}

\usepackage{wasysym}
\usepackage{amsmath}
\usepackage{amssymb,amsfonts,amsxtra,mathrsfs,makeidx,graphics,graphicx,amsthm,bbm}
\usepackage{epsfig}
\usepackage[all]{xy}
\usepackage{slashed}
\usepackage{caption}
\usepackage{bm}
\usepackage{multirow}
\usepackage[utf8]{inputenc}
\usepackage{fancyhdr}

\newenvironment{myenumerate}{
\begin{enumerate}
   \setlength{\itemsep}{1pt}
   \setlength{\parskip}{0pt}
   \setlength{\parsep}{0pt}}{\end{enumerate}}

\newenvironment{myitemize}{
\begin{itemize}
   \setlength{\itemsep}{1pt}
   \setlength{\parskip}{0pt}
   \setlength{\parsep}{0pt}}{\end{itemize}}
   
\renewcommand{\arraystretch}{1.3}

\newcommand{\CC}{\mathcal{C}}
\newcommand{\CF}{\mathcal{F}}

\newcommand{\CK}{\mathcal{K}}
\newcommand{\CL}{\mathcal{L}}
\newcommand{\CM}{\mathcal{M}}
\newcommand{\CN}{\mathcal{N}}
\newcommand{\CO}{\mathcal{O}}
\newcommand{\CH}{\mathcal{H}}
\newcommand{\CV}{\mathcal{V}}

\newcommand{\Tr}{\mbox{Tr}}

\newcommand{\IR}{\mathbb{R}}
\newcommand{\IZ}{\mathbb{Z}}

\newcommand{\SU}{\mathrm{SU}}

\def\p{\partial}

\newcommand{\half}{\frac{1}{2}}

\newcommand{\nn}{\nonumber}

\newcommand{\ft}[2]{{\textstyle\frac{#1}{#2}}}

\def\Slash#1{\rlap{\hbox{$\mskip 3 mu /$}}#1}      
\def\oneone{\rlap 1\mkern4mu{\rm l}}

\def\bea{\begin{eqnarray}}
\def\eea{\end{eqnarray}}
\def\be{\begin{equation}}
\def\ee{\end{equation}}

\makeatletter
\renewcommand*\env@matrix[1][\arraystretch]{%
  \edef\arraystretch{#1}%
  \hskip -\arraycolsep
  \let\@ifnextchar\new@ifnextchar
  \array{*\c@MaxMatrixCols c}}
\makeatother

\def\={\; = \;}
\def\+{\, + \,}

\def\wt{\widetilde}
\def\wh{\widehat}

\def\t{\tau}
\def\a{\alpha}
\def\z{\zeta}
\def\m{\mu}
\def\s{\sigma}
\def\r{\rho}             
\def\l{{\lambda}}

\def\nv{n_\text{v}}
\def\nh{n_\text{h}}
\def\vth{\vartheta}

\begin{document}


\thispagestyle{empty} 
\vspace*{-3cm} \vskip10.truecm
\begin{center}
	{\LARGE Quantum Black Hole Entropy and Localization \\ in Supergravity}
\end{center}


\clearpage
\thispagestyle{empty}
\vspace*{-3cm}
\vskip23.truecm
Ph.D. Thesis Utrecht University, June 2016\\
ISBN: 978-90-393-6583-0\\
Printed by: Ipskamp Drukkers, Enschede, The Netherlands\\
Cover: Diagonal Records, used with permission\\


\cleardoublepage
\thispagestyle{empty}
\begin{center}

{\LARGE Quantum Black Hole Entropy and Localization \\ in Supergravity} \\

\vskip2.truecm

{\large Quantum Entropie voor Zwarte Gaten en Localisatie \\ in Supergravitatie} \\

\vskip1.truecm

(met een samenvatting in het Nederlands) \\

\vskip7.truecm

{\large Proefschrift} \\

\vskip1.truecm

ter verkrijging van de graad van doctor aan de Universiteit Utrecht op gezag van de rector magnificus, prof. dr. G.~J.~van der Zwaan, ingevolge het besluit van het college voor promoties in het openbaar te verdedigen op maandag 13 juni 2016 des middags te 12.45 uur \\

\vskip1.truecm

{\large door} \\

\vskip2.truecm

{\Large Valentin Reys} \\

\vskip1.truecm

geboren op 28 oktober 1988 te Strasbourg, Frankrijk

\end{center}

\newpage
\thispagestyle{empty}

\begin{flushleft}
Promotor: Prof. dr. B.Q.P.J. de Wit \\
Copromotor: Dr. S. Murthy 
\end{flushleft}

\vskip19.truecm
This work is part of the research programme of the Foundation for Fundamental Research on Matter (FOM), which is part of the Netherlands Organization for Scientific Research (NWO). This work was supported by the ERC Advanced Grant no. 246974, ``Supersymmetry: a window to non-perturbative physics''. The research was conducted at Nikhef, Amsterdam. 


\input{Thesis_My_Papers.tex}
\cleardoublepage
\thispagestyle{empty}


\begin{flushright}
\vskip35.truecm
\textit{Hold to the now, the here, \\ through which all future \\ plunges to the past.} \\ \vspace{.4cm}
-- James Joyce, \textit{Ulysses}
\end{flushright}
\newpage
\thispagestyle{empty}


\pagestyle{plain}
\pagenumbering{roman}
\mbox{}
\newpage
\setcounter{page}{1}
\setcounter{tocdepth}{1}
\makeatletter \renewcommand{\@dotsep}{10000} \makeatother
\tableofcontents
\cleardoublepage\newpage
\mainmatter
\pagenumbering{arabic}
\setcounter{page}{1}
\pagestyle{fancy}
\cleardoublepage


\chapter{Introduction}
\label{chap:intro}
Black holes are fascinating objects: from an observational or experimental standpoint, numerous candidates have been identified in Nature. Nowadays, there are roughly 20 observed stellar binaries in our galaxy alone which are believed to contain black holes of some solar masses, and super-massive black holes provide the only explanation, as of yet, for the phenomena observed in the centers of active galaxies~\cite{Frolov:1998wf}. Gravitational wave detectors such as LIGO~\cite{Ligo} and VIRGO~\cite{Virgo} also aim to directly observe processes involving black holes in our cosmic neighborhood. Interestingly enough, a black hole merger has been confirmed experimentally very recently~\cite{Abbott:2016blz}. From a theoretical standpoint, black holes allow scientists to test and refine a variety of novel ideas having appeared in the vast arena of gravitational physics. Black holes bridge gaps between various areas of research and make highly non-trivial connections among different tentative descriptions of our reality. This is because they possess a number of interesting properties which make their study a rich field of research.\\
Classically, black holes are solutions of Einstein's theory of General Relativity. They are intrinsically gravitational objects and describe a region of space-time where a large mass (typically a few solar masses or more) is concentrated, giving rise to a curvature singularity surrounded by an event horizon. This classical horizon is the boundary of a region in space-time from which ``nothing can escape'', not even light, due to the extreme gravitational pull exerted by the black hole. The interior of the horizon, and the singularity itself, are therefore hidden from view and causally disconnected from the exterior. It is believed that whenever a curvature singularity forms in Nature, it is always accompanied by an event horizon, so that there are no ``naked'' singularities observable. This is the Cosmic Censorship hypothesis~\cite{Wald:1998de}.\\
Semi-classically, perhaps the most important property of black holes is the possibility to identify within their description a quantity behaving exactly as a thermodynamical entropy, according to a proposal made by Bekenstein and Hawking. This entropy acts as a measure of disorder or randomness in the internal constituents of the black holes. Furthermore, one can also identify a corresponding temperature, in agreement with the standard laws of thermodynamic. This led to the crucial realization that black holes are, in fact, not entirely ``black'' but that they must radiate in order to reach thermodynamical equilibrium with their environment. One should stress again that this is a semi-classical effect, which arises upon considering a classical black hole interacting with a quantum field. Arbitrarily close to the black hole, pairs of particles and anti-particles are created from the vacuum due to the quantum nature of the field. A member of one such pair can then fall into the interior of the black hole, while its companion escapes to infinity. This process may be interpreted as the emission of a thermal radiation by the black hole. The associated thermodynamical entropy defined by Bekenstein and Hawking is a function of the parameters of the black hole solutions as measured by an observer far away (at infinity). These parameters therefore play the role of state variables. For the simplest black holes, which are solutions of general relativity in a vacuum, the only parameter entering their description is their mass~$M$. There exist other interesting cases where the black holes are also electrically and/or magnetically charged. These black holes are solutions of Einstein-Maxwell theory, a combination of general relativity and Maxwell's electro-magnetism. Their entropy depends on their mass and on their electric and/or magnetic charges~$(M,Q)$. Further, stationary rotating black holes also exist theoretically and their entropy is parametrized by an additional angular momentum~$(M,Q,J)$.\\
Since the discovery of black hole entropy by Bekenstein and Hawking, considerable effort has been deployed to understand precisely how this property arises at a fundamental level. But to do so comes with obvious difficulties, as there is at present no way to efficiently probe the interior of a black hole past its event horizon and observe what makes up its internal constituents. This failure is due to the posited extreme environment of the interior, where quantum and gravitational effects are expected to be comparable, and one needs to take both into account to arrive at a correct description. To learn more about the internal structure of black holes would thus require a quantum description of gravity.\\
Unfortunately there exists no consistent Quantum Field Theory (QFT) of gravity based on general relativity. A special extension of general relativity exists, which combines Einstein's theory with supersymmetry to give rise to Supergravity. A theory of supergravity possesses all the usual space-time symmetries of general relativity, and in addition is made up of a specific type of matter and gauge fields symmetric under the exchange of bosons and fermions. In such theories, it is also possible to describe black hole solutions. These black holes are the supersymmetric analogues of the classical black holes of general relativity, and their quantum behavior is under much better theoretical and computational control due to the additional constraints imposed by supersymmetry. This symmetry should therefore be viewed as a convenient and controlled theoretical framework to begin gathering clues as to how black holes behave in the quantum regime, even if supersymmetry itself has not yet been observed experimentally in Nature~\cite{Schorner-Sadenius:2015cga}.\\
A number of other theories have been put forward to try and describe the quantum regime of gravity. Among these, String Theory seems to be the most promising to many.
At its core, this theory is a somewhat radical departure from the fundamental tools of description available to the high-energy physicist using QFT. In string theory, the fundamental objects are not fields defined at every point in space-time, but extended objects: extremely small (typically of size close to the Planck length,~$10^{-33}$ cm) vibrating strings of energy, the spectrum of which generates what we observe in our macroscopic world as particle manifestations. This includes all the known particles of the Standard Model, but also gravitons (the fundamental quanta of gravity) and other more exotic particles. Another important difference as compared to the usual QFTs is that the theory is consistently formulated in ten space-time dimensions. Upon ``curling up'', or compactifying, six of these extra dimensions on an internal space, it is possible to make contact with our four-dimensional world and understand how its properties arise from a higher-dimensional perspective. In this framework, it is also possible to describe black holes (and more generally black objects, which may have a different topology than the classical black holes of general relativity): they are realized as stacks of D-branes, which are extended objects in the spectrum of the theory endowed with special properties. These D-branes interact quantum mechanically in the internal six-dimensional space and have a well-defined (and computable) number of energy states. Using this description, string theory therefore offers a microscopic view of the degrees of freedom available to the black holes and provides a statistical interpretation of their thermodynamical entropy, in a way entirely similar to Boltzmann's description of the macroscopic entropy of a gas based on the number of microstates available to the atoms making up the system.\\
In some simple cases, it is possible to compute scattering amplitudes in string theory, and these results have been found to agree with the ones obtained in an effective field theory of supergravity. This suggests that, when considering the low-energy limit of string theory, one obtains theories of supergravity in certain specific situations. When such a connection is available, the quantum description of a black hole is readily available in both formulations, and it is therefore of great interest to compare the predictions made by the high-energy, microscopic description of string theory and the low-energy, macroscopic description offered by supergravity. This has been the topic of many years of research, and fascinating advances and insights have been gained from this connection. We could list here for example the holographic principle and the AdS/CFT conjecture. It will be the purpose of this work to study such a connection between string theory and supergravity by analyzing in great detail the entropy of specific black holes whose description is available in both frameworks.

The outline of the present thesis is as follows. In the remainder of this Chapter, we discuss examples of classical black hole solutions in general relativity and Einstein-Maxwell theory. We then review the proposal to define their thermodynamical entropy, and we introduce concepts of string theory and supersymmetry necessary to interpret this entropy statistically. In Chapter~\ref{chap:modern-BH-S}, we present a more refined and complete analysis of the entropy of black holes using string theory, and put forward a concrete proposal to define their quantum entropy in the macroscopic, low-energy theory. We then introduce the mathematical framework which will be required for an exact calculation of said quantum entropy. This ultimately leads to a precise program for computing the exact quantum entropy of specific black holes in supergravity and to compare it to string theory predictions. In Chapter~\ref{chap:sugra}, we derive the supergravity theory which will be used in the rest of this work in order to lay a solid foundation for explicit calculations. In Chapter~\ref{chap:n8-loc}, we review the first example where the quantum entropy of a maximally supersymmetric black hole was computed at all orders in supergravity and successfully matched with the string-theoretic, microscopic description. We discuss two major assumptions which entered this calculation, and we then justify the first of these assumptions in a rigorous manner. The second assumption is examined and found to be correct in Chapter~\ref{chap:n2-loc}, where we develop a general framework to compute one of the central ingredients in the recipe for the quantum entropy of black holes. Putting these new ingredients together, we push the quantum entropy program forward to less supersymmetric black holes and compute their quantum entropy in Chapter~\ref{chap:n4-loc}. Finally, we close with some conclusions and important open questions in Chapter~\ref{chap:conclusion}. Three Appendices are used to gather the conventions chosen throughout this work, some facts regarding the mathematical theory of modular, Jacobi and Siegel forms, and the technical details underlying the construction of the supergravity theory which is used in the main text.

\section{Classical black holes}
\label{sec:classical-BH}

General relativity is based on the Einstein-Hilbert action and describes a theory invariant under local coordinate transformations of the space-time manifold, which is taken to be Riemannian. The field encoding the dynamics of space-time is the metric tensor~$g_{\mu\nu}$, which acts as a gauge field for the local coordinate transformations. For definiteness, we will restrict ourselves to four space-time dimensions and Minkowski signature in this Section, so we take~$\mu,\nu = 0,\ldots,3$. The conventions used in this work are summarized in Appendix~\ref{app:conv}. The Einstein-Hilbert action describing general relativity takes the form
\be
\label{eq:EH-action}
S_{EH} = \int\,d^4x\,\sqrt{-g}\Bigl[-\frac{1}{16\pi G}\bigl(R - \Lambda\bigr) + L_{\textnormal{mat.}}\Bigr] \, ,
\ee
where~$G$ is Newton's constant,~$g$ is the determinant of the metric,~$R$ is the Ricci scalar of the manifold which measures the curvature of space-time,~$\Lambda$ is the cosmological constant and~$L_{\textnormal{mat.}}$ describes the matter content of the theory, minimally coupled so that all derivatives are covariant with respect to the space-time symmetries. The equations of motion associated to this action are Einstein's equations,
\be
\label{eq:GR-EOM}
R_{\mu\nu} - \frac12\bigl(R - \Lambda\bigr)g_{\mu\nu} = 16\pi G\,T_{\mu\nu} \, ,
\ee
with~$R_{\mu\nu}$ the Ricci tensor and~$T_{\mu\nu}$ the stress-energy tensor derived from~$L_{\textnormal{mat.}}$. Throughout this work, we will set the cosmological constant to zero.\footnote{Modern experimental observations indicate that the cosmological constant is in fact non-zero and very small~\cite{Riess:1998cb}, but this has no bearing on the study of black hole entropy which is carried on in this work.} In most of this work, natural units where~$\hbar = c = G = 1$ are used.

Shortly after the discovery of general relativity, Schwarzschild (and independently Droste) obtained one of the first solution to Einstein's equations in an asymptotically flat vacuum (\textit{i.e.} with $L_{\textnormal{mat.}} = 0$)~\cite{Schwarzschild:1916uq,Droste:1917aa}. This is the gravitational analogue of the Coulomb charge in Maxwell's theory of electro-magnetism. The Schwarzschild line-element written in spherical coordinates~$(t,r,\theta,\phi)$ is
\be
\label{eq:Schwarz-BH}
ds^2 = g_{\mu\nu}dx^\mu dx^\nu = -\left(1 - \frac{2M}{r}\right)dt^2 + \left(1 - \frac{2M}{r}\right)^{-1}dr^2 + r^2\left(d\theta^2 + \sin^2\theta\,d\phi^2\right) \, .
\ee
This solution describes a spherically symmetric black hole of mass~$M$ located at the origin of the space-time~$r=0$. At this point lies a \emph{curvature singularity} where the Ricci scalar diverges. This is a ``true'' singularity which cannot be eliminated by a change of coordinate system. On the other hand, the surface specified by~$r = 2M$ is a coordinate singularity and signals the presence of an \emph{event horizon}.

One can also find analytic solutions to Einstein's equations when matter is present in the space-time. In the present work, we will be interested in electro-magnetically charged black holes, also known as \emph{dyonic} black holes, and we therefore consider the case where~$L_{\textnormal{mat.}} = -\tfrac14 F_{\mu\nu}F^{\mu\nu}$ is the Maxwell Lagrangian. The line-element describing a charged black hole under this Maxwell field is now given by the Reissner-Nordstr{\"o}m solution~\cite{Reissner:1916aa, Nordstrom:1918aa}
\begin{align}
\label{eq:RN-BH}
ds^2 =&\, -\left(1-\frac{2M}{r}+\frac{q^2+p^2}{r^2}\right)dt^2 + \left(1-\frac{2M}{r}+\frac{q^2+p^2}{r^2}\right)^{-1}dr^2 + r^2\,d\Omega_2{}^2 \, , \nonumber \\
F_{rt} =&\, \frac{q}{r^2} \, , \quad F_{\theta\phi} = p\sin\theta \, .
\end{align}
Here,~$d\Omega_2{}^2$ is the line-element of the 2-sphere and~$(q,p)$ are the electric and magnetic charges of the black hole. As in the Schwarzschild solution, a curvature singularity sits at the origin of space-time~$r=0$, and the coordinate singularities are now located at~$r_{\pm} = M \pm \sqrt{M^2 - (q^2 + p^2)}$. This indicates that the charged black hole in fact possesses two horizons, an inner and an outer one. In the limiting case where~$M^2=(q^2 + p^2)$, these two horizons coalesce at~$r_+ = r_- = M$, and the black hole is called \textit{extremal}. Dyonic extremal black holes are especially interesting as they exhibit a symmetry enhancement close to their horizon. This can be shown, for instance, by making the following change of coordinates with an arbitrary constant~$\alpha$:
\be
\tau = \frac{\alpha}{r_+^2}\,t \, , \quad \rho = \alpha^{-1}(r-r_+) \, .
\ee
Taking~$\alpha \rightarrow 0$ while keeping~$\rho$ fixed, the original radial coordinate~$r$ approaches the horizon located at~$r_+$. In this near-horizon region, the line-element~\eqref{eq:RN-BH} then becomes:
\begin{align} 
\label{eq:RN-BH-Extr}
ds^2 =&\, r_+^2\left(-\rho^2d\tau^2 + \frac{d\rho^2}{\rho^2}\right) + r_+^2\,d\Omega_2{}^2 \, , \nonumber \\
F_{\rho\tau} =&\, q \, , \quad F_{\theta\phi} = p\sin\theta \, .
\end{align}
This metric is a direct product of~$AdS_2$ (the two-dimensional anti-de Sitter space) and the 2-sphere~$S^2$. This product space is invariant under the~$\mathrm{SO}(3)$ group of rotations of the 2-sphere, just like the non-extremal solution, but also possesses an additional~$\mathrm{SO}(2,1)$ symmetry inherited from the~$AdS_2$ factor which was not present in the non-extremal case. As we will see in Section~\ref{sec:higherdim-S}, a similar and in fact stronger kind of symmetry enhancement also occurs for supersymmetric black holes. Generically, a symmetry enhancement leads to what is known as an \textit{attractor mechanism}, where one obtains a stronger set of constraints which the field configuration has to satisfy, arising from the larger group of symmetries acting on the system. This phenomenon will be especially relevant in our study of extremal, supersymmetric black holes, as will be explained in Chapter~\ref{chap:modern-BH-S}. The Schwarzschild and (extremal) Reissner-Nordstr{\"o}m black holes discussed above will provide examples for various concepts introduced in the rest of this Chapter.
 
\section{Semi-classical black holes}
\label{sec:semiclassical-BH}

Having presented two classical black hole solutions to Einstein's equations in four dimensions, we now introduce the fundamental discovery made by Bekenstein and Hawking~\cite{Bekenstein:1973ur, Hawking:1971tu}. In the early 1970s, Penrose, Floyd and Christodoulou realized that black holes exhibit a remarkable tendency to increase their horizon's surface area when undergoing perturbations~\cite{Penrose:1971uk, Christodoulou:1970wf}. This led Bekenstein and Hawking to formulate an analogue of the second law of thermodynamics for black holes, since this second law states that changes in a closed thermodynamic system always take place in the direction of increasing entropy. Therefore, they posited that one could formally define a \emph{thermodynamical entropy} for black holes as 
\be
\label{eq:BH-entropy}
\mathcal{S}_{BH} = \frac{k_B\,c^3}{G\,\hbar}\,\frac{A}{4} \, ,
\ee
where~$A$ is the area of the black hole horizon and~$k_B$ is the Boltzmann constant. Here we have reinstated all fundamental constants to point to the presence of~$\hbar$, which indicates that such a quantity is defined in a semi-classical theory, where quantum effects are expected to start playing a part in the story. Equation~\eqref{eq:BH-entropy} is known as the Bekenstein-Hawking \emph{area-law}, and it aims to identify the amount of disorder within black holes, or our lack of information about them, with the surface area of their horizon. Since the latter depends on the macroscopic parameters associated to the black holes as measured by an observer at infinity (like their mass or electro-magnetic charges), these parameters should be formally understood as coarse-grained thermodynamical variables specifying the state in which the black holes are.

Based on this formal analogy, Bardeen, Carter and Hawking went on to establish the general \emph{laws of black hole mechanics}~\cite{Bardeen:1973gs}.
These laws apply to the static electro-magnetically neutral Schwarzschild black holes~\eqref{eq:Schwarz-BH} parametrized by their mass~$M$, to the static electro-magnetically charged Reissner-Nordstr{\"o}m black holes~\eqref{eq:RN-BH} parametrized by their mass and charges~$(M,Q)$, and to the stationary axisymmetric rotating black holes (known as Kerr solutions~\cite{Kerr:1963aa}) parametrized by their mass, charges and angular momentum~$(M,Q,J)$.
\begin{myitemize}
\item $0^{\textnormal{th}}$ law: The surface gravity~$\kappa$ of a stationary black hole, defined as the force required to be applied by an observer infinitely far away to maintain a mass at a fixed location on the black hole horizon, is constant over the event horizon.\\
\item $1^{\textnormal{st}}$ law: Any two neighboring stationary axisymmetric black hole solutions are related by
\be
\label{eq:1stlawBH}
\delta M = \frac{\kappa}{8\pi}\,\delta A + \omega\,\delta J + \mu\,\delta Q \, ,
\ee
where~$\delta M$,~$\delta A$,~$\delta J$ and~$\delta Q$ denote the change in mass, area, angular momentum and electro-magnetic charge, respectively, when going from one solution to the other,~$\omega$ is the angular velocity measured at infinity, and~$\mu$ the chemical potential conjugate to the electro-magnetic charge of the black hole, also measured at infinity. \\
\item $2^{\textnormal{nd}}$ law: The area~$A$ of a black hole never decreases in any process,~$\delta A \geq 0$. For example, if and when two black holes collide, they will coalesce and form a single black hole whose area is necessarily greater or equal to the sum of the areas of the initial black holes.\footnote{This has been observed very recently by the LIGO and VIRGO collaborations~\cite{Abbott:2016blz}.}\\
\item $3^{\textnormal{rd}}$ law: It is impossible by any procedure, no matter how idealized, to reduce the surface gravity~$\kappa$ to zero by a finite sequence of operations.
\end{myitemize}

A comparison to the usual laws of thermodynamics is indeed suggestive of a \emph{thermodynamical interpretation} of black hole dynamics. Historically however, a central ingredient was still missing to take this formal analogy to the level of a true correspondence: while at this stage it seems tempting to interpret the surface gravity~$\kappa$ as the analogue of the temperature for a black hole, can a first-principle derivation of such a relation be obtained?

This piece of the puzzle was provided by Hawking in 1975~\cite{Hawking:1974sw}. Upon considering a classical black hole interacting with a quantum field, he came to the realization that black holes are almost perfect black bodies which can absorb and emit \emph{radiation}, at a temperature proportional to their surface gravity. Such radiation may \textit{a priori} seem in contradiction with the naive picture that nothing can escape from a black hole, but it is precisely the quantum character of the field with which the black hole interacts that makes it possible. Pictorially, a quantum field close to the horizon of the black hole undergoes quantum fluctuations, which result in particle/anti-particle pair creations from the vacuum. Arbitrarily close to the horizon, it is possible for the anti-particle to fall into the black hole while the particle escapes, or vice-versa. The net effect is then a radiation emission from the black hole horizon. It is important to stress again here that this is a semi-classical picture, where the black hole is still thought of as a classical solution to Einstein's equations but the field it interacts with is intrinsically quantum to allow for vacuum fluctuations. The radiation occurs at a temperature formally defined as the \emph{black hole temperature}, which is given by
\be
\label{eq:BH-temperature}
T_{BH} = \frac{\kappa}{2\pi} \, ,
\ee
where fundamental constants have been set to unity. Armed with this missing piece of the puzzle, it is easy to see that the first law of black hole mechanics~\eqref{eq:1stlawBH} can be stated precisely as the first law of thermodynamics:
\be
\label{eq:first-law-BH}
\delta M = T_{BH}\,\delta\Bigl(\frac{A}{4}\Bigr) + \omega\,\delta J + \mu\,\delta Q \, .
\ee
This cements the interpretation that the entropy of a black hole is to be identified with its area according to the area-law~\eqref{eq:BH-entropy}, and also takes the formal analogy of the second law of black hole mechanics and the second law of thermodynamics into a true equivalence. It is capital to emphasize here that~\eqref{eq:BH-entropy} is, at first sight, an extremely puzzling formula. In usual thermodynamical objects, the entropy behaves as an extensive quantity, which means it grows proportionally to the volume of the system. But for black holes, it is instead the area of the system which controls the entropy. This realization has led to a variety of interpretations, the most famous of which is probably the holographic principle and its concrete realization, the AdS/CFT correspondence~\cite{'tHooft:1993gx, Susskind:1994vu, Maldacena:1997re, Aharony:1999ti}.

There is also an interesting consequence of the third law of black hole mechanics for Reissner-Nordstr{\"o}m black holes. Their surface gravity is given by
\be
\kappa = \frac{r_+ - r_-}{2\,r_+{}^2} \, ,
\ee
so that, for an extremal Reissner-Nordstr{\"o}m black hole, the surface gravity becomes zero due to the coalescence of the inner and outer horizons~$r_+ = r_-$. First, this shows that for an extremal black hole and from the point of view of an observer at infinity, a massive object located precisely at the horizon will remain there indefinitely, since no force is required from the observer to keep it there. Second, the third law of black hole mechanics  makes the extremal Reissner-Nordstr{\"o}m black hole quite peculiar: starting from a non-extremal solution, it is impossible to obtain an extremal solution in any finite amount of time. Extremal black holes therefore stand on their own as limiting cases which cannot be obtained from more realistic non-extremal black holes by any physical process. In view of the symmetry enhancement in their near-horizon region alluded to in Section~\ref{sec:classical-BH}, this makes extremal black holes idealized objects very useful to study gravity, albeit disconnected from the black holes we expect to observe in our universe.

As an illustration of the concepts introduced above, let us compute the Bekenstein-Hawking entropy of the Schwarzschild and extremal Reissner-Nordstr{\"o}m black holes. Both these black holes are spherically symmetric with an horizon sitting at~$r_h=2M$ and $r_h=\sqrt{(q^2 + p^2)}$, respectively, and the area-law~\eqref{eq:BH-entropy} gives
\be
\label{eq:BH-Entropy-S-ERN}
\mathcal{S}_{BH} = 4\pi M^2 \, , \quad \textnormal{and} \quad \mathcal{S}_{BH} = \pi(q^2 + p^2) \, ,
\ee
respectively (we have set all fundamental constants to unity). As previously stated, both entropies only depend on the macroscopic parameters of the black hole solutions, namely the mass~$M$ and the electric and magnetic charges~$(q,p)$. This type of dependence will be central to the tentative interpretation of the Bekenstein-Hawking entropy which will be proposed later in this Chapter in terms of a \emph{microscopic} description of the black holes.

At this stage, it will be instructive to derive the Hawking temperature~\eqref{eq:BH-temperature} for a Schwarzschild black hole in a way which makes use of the notion of Euclidean time, following~\cite{Gibbons:1976ue}. Here, we intend to show that the Hawking temperature can be recovered using standard methods of statistical physics in Euclidean signature, thus showing that even though the notion of temperature (and entropy) associated to a black hole may at first sight appear counter-intuitive, it fits within the ``standard'' derivation of these quantities known from statistical quantum mechanics.\\ 
In quantum mechanics, the time-evolution operator of a given system is defined as~$e^{-itH}$, where~$H$ is the Hamiltonian of the system. If we now consider a single scalar field~$\Phi$ and a Euclidean continuation~$t \rightarrow -i\tau$, the trace over the quantum Hilbert space of the time-evolution operator is given by
\be
\label{eq:time-evol-op}
\Tr_\mathcal{H}\,e^{-\tau H} = \int d\phi \, \langle\phi|e^{-\tau H}|\phi\rangle \, .
\ee
Using the path-integral representation of the Euclidean time-evolution operator, this can be written as
\be
\Tr_\mathcal{H}\,e^{-\tau H} = \int d\phi \int \mathcal{D}\Phi\,e^{-S_E[\Phi]} \, ,
\ee
where~$S_E[\Phi]$ is the Euclidean action over periodic field configurations satisfying boundary conditions~$\Phi(\tau) = \Phi(0) = \phi$. Let us now examine the Euclidean line-element of the Schwarzschild black hole~\eqref{eq:Schwarz-BH}. By a change of the radial coordinate~$(r - 2M) =\rho^2/(8M)$, it is possible to zoom-in on the near-horizon region when taking~$\rho \rightarrow 0$. In this near-horizon region, the Euclidean line-element takes the form
\be
\label{eq:Schwarz-NHG}
ds^2 = \rho^2\frac{d\tau^2}{16M^2} + d\rho^2 + 4M^2d\Omega_2{}^2 \, .
\ee
If we now make an additional change of coordinate,
\be
\frac{\tau}{4M} = \theta \, ,
\ee
the metric~\eqref{eq:Schwarz-NHG} is simply the metric of a two-dimensional flat Euclidean space times a 2-sphere, 
\be
ds^2 = \rho^2d\theta^2 + d\rho^2 + 4M^2d\Omega_2{}^2 \, ,
\ee
provided the variable~$\theta$ has the periodicity~$0\leq\theta<2\pi$ (otherwise there would be a conical singularity at the origin of the two-dimensional plane). This identification implies that the Euclidean time coordinate~$\tau$ of the near-horizon Schwarzschild metric must have periodicity~$8\pi M$.\\
Now, in quantum mechanics, the \emph{thermal partition function} is given by
\be
Z(\beta) = \Tr_\mathcal{H}\,e^{-\beta H} \, ,
\ee
where~$\beta$ is the inverse temperature,~$H$ is the Hamiltonian, and the trace is again taken over the Hilbert space of the theory. This partition function is related to the trace of the Euclidean time-evolution operator~\eqref{eq:time-evol-op} upon identifying~$\beta = \tau$. For a Euclidean Schwarschild black hole,~$\tau$ must have periodicity~$8 \pi M$, so we deduce
\be
\label{eq:Schwarz-BH-Temp}
T = \frac{1}{\beta} = \frac{1}{8\pi M} = \frac{\kappa}{2\pi} \, ,
\ee
where we have used that the surface gravity of the Schwarzschild black hole is given by~$\kappa = 1/(4M)$. This is precisely the Hawking temperature~\eqref{eq:BH-temperature}. This simple calculation shows that the familiar interplay between the periodicity of Euclidean time and the temperature of a physical system learned from statistical quantum mechanics can be successfully exploited for black holes to reproduce the findings of Hawking. 

The close analogy between the laws of black hole mechanics and the laws of thermodynamics opens the way for a natural, and ultimately profound, question. We have learned since the work of Boltzmann that there exists a fundamental link between the thermodynamics of a system, describing its \emph{macroscopic} behavior, and the \emph{microscopic} state configurations accessible to this system when described in terms of its internal constituents. This is summarized by the Boltzmann equation
\be
\label{eq:Boltzmann}
\mathcal{S}_B = k_B\,\log\Omega \, .
\ee
Here~$\Omega$ denotes the number of microstates available to the internal constituents of the system under consideration,~$\mathcal{S}_B$ is the statistical entropy of the system and~$k_B$ is the Boltzmann constant (which we will set to one hereafter). This relation explains, for example, how the entropy of a gas can be obtained from the microscopic kinetic theory describing the motion of~$N$ atoms or molecules making up the gas, where~$N$ is very large. This microscopic description uses methods of statistical physics, and the macroscopic, thermodynamical quantities are seen as averaged or coarse-grained properties of this complicated mechanical system. Both the macroscopic and microscopic descriptions match when we take the \textit{thermodynamic limit}~$N\rightarrow\infty$ along with the volume of the system~$V\rightarrow\infty$ while keeping~$V/N$ fixed and finite.

Semi-classically, we have identified the macroscopic entropy of a black hole with the area of its horizon. Thus we should ask what is the analogue of the~$\Omega$ quantity for the black hole. This turns out to be a deep and difficult question, since it requires us to identify the internal constituents of the black hole and compute the number of microstates available to them. Black holes being intrinsically gravitational objects, our search for a statistical interpretation of their entropy takes us into the realm of \textit{quantum gravity}, where a description of the black hole in terms of its internal ``gravitational constituents'' or ``atoms'' should be available. Within this description, a limit akin to the thermodynamic limit should allow us to recover the Bekenstein-Hawking area-law. Were this to be achieved, it would then truly warrant a thermodynamical interpretation of black hole mechanics.

In this context, we can see the need for a theory of quantum gravity as arising from the search for a statistical interpretation of the area-law~\eqref{eq:BH-entropy}. A straightforward way to build such a theory and study the corresponding microscopic description of black holes would be to quantize the Einstein-Hilbert action within the standard framework of QFT. However, general relativity is notoriously difficult to quantize: at the perturbative level, it is known to be non-renormalizable in four dimensions~\cite{'tHooft:1973us, 'tHooft:1974bx, Deser:1974nb}.
There are however other theories of quantum gravity which are not directly based on quantizing the Einstein-Hilbert action. The most prominent of these to this day, and the one that we will make use of in the rest of this work, is string theory. As we will explain in Section~\ref{sec:higherdim-S}, this theory indeed provides a microscopic picture of the internal constituents of specific black holes, and it is possible to evaluate the degeneracies~$\Omega$ of this system in great detail. In fact, it turns out to be even more powerful: not only is it possible to recover the Bekenstein-Hawking entropy in a certain limit, but one can also probe the sub-leading \emph{corrections} to~\eqref{eq:BH-entropy}. Such corrections are expected to be present since, as we stressed in the beginning of this Section, the area-law was derived for a \emph{classical} black hole solution of Einstein's equations. Within the context of quantum gravity, it will therefore naturally receive quantum corrections, and it will be the main focus of this work to study such corrections.

It has been known for some time how to incorporate a specific kind of corrections to~\eqref{eq:BH-entropy}, namely perturbative corrections arising in a low-energy \emph{effective} theory of quantum gravity. This effective theory is obtained by focusing on the low-energy degrees of freedom (the massless modes) and neglecting the heavier degrees of freedom by integrating them out. Its action describes the dynamics of a classical background metric field for sufficiently weak curvatures at sufficiently large distances. Any quantum field theory can be described as such by focusing on its low-energy degrees of freedom. The dynamics of the massive degrees of freedom are then encoded in corrections to the action describing the massless degrees of freedom. When these corrections are suppressed, it is possible to conduct a perturbative expansion in the effective theory. For example, in string theory, it is possible to show that the low-energy effective action generically contains higher-curvature terms~\cite{Green:1987sp}, which are generated both through quantum loop corrections and stringy~$\alpha'$-corrections, where~$\alpha'$ is the dimensionful parameter of the theory. The string length is defined as~$l_{\textnormal{string}} = \sqrt{\alpha'}$ and the string mass is given by~$m_{\textnormal{string}} = (2\,\alpha')^{-1/2}$. When analyzing the action for the massless modes of string theory at tree-level, one obtains the Einstein-Hilbert action along with an infinite tower of higher-derivative couplings suppressed by increasing powers of~$\alpha'$, which shows that these higher-derivative interactions are sub-leading in the low-energy limit. If the black hole entropy picture is to be consistent, we should like to know how these sub-leading corrections affect the Bekenstein-Hawking area-law.

To examine this question, let~$L$ be a covariant Lagrangian built out of dynamical fields, including the metric and collectively denoted as~$\Phi$. Within such a theory, there exists a prescription due to Wald to describe the entropy of black hole solutions based on the Noether current associated with diffeomorphisms~\cite{Wald:1993nt, Iyer:1994ys, Jacobson:1994qe}. It is therefore instructive to first review the notions of Noether currents and their associated charges.\\
Generically, under any field variation~$\delta\Phi$, the Lagrangian~$L$ always varies into
\be
\delta(\sqrt{-g}L) = \sqrt{-g}\,E\cdot\delta\Phi + \sqrt{-g}\,\nabla_\mu\Theta^\mu(\delta\Phi) \, ,
\ee
where~$E=0$ are the equations of motion. If we now consider field variations which leave the action invariant up to boundary terms, or in other words when considering a \emph{symmetry} of the action functional (denoted by~$\delta_\textrm{S}$ to distinguish from generic variations), the Lagrangian must be invariant up to a total derivative:~$\delta_\textrm{S}(\sqrt{-g}L) = \sqrt{-g}\,\nabla_\mu N^\mu$. General relativity is an example of a theory in which~$N^\mu$ is always non-vanishing, while gauge theories usually have~$N^\mu = 0$ unless Chern-Simons terms are present. The \emph{Noether current} associated with symmetries of the theory is defined as
\be
J^\mu = \Theta^\mu(\delta_\textrm{S}\Phi) - N^\mu \, ,
\ee
and it satisfies~$\nabla_\mu J^\mu = 0$ when~$E=0$. Associated to this Noether current is the \emph{Noether potential}~$\mathcal{Q}^{\mu\nu}$, defined as~$J^\mu = \nabla_\nu\mathcal{Q}^{\mu\nu}$. The total \emph{Noether charge} contained in a spatial volume~$\Sigma$ can be expressed as a boundary integral of this potential
\be
\mathcal{Q} = \oint_{\partial\Sigma} d^2 x\,\sqrt{h}\,\epsilon_{\mu\nu}\mathcal{Q}^{\mu\nu} \, ,
\ee
where~$h_{\mu\nu}$ and~$\epsilon_{\mu\nu}$ are the induced measure and binormal on the boundary~$\partial\Sigma$.

We now specialize the discussion to theories which are invariant under diffeomorphisms of the space-time manifold (the local coordinate transformations). In such theories, one can define a Noether charge associated to these transformations, and it will be expressed as a boundary integral of the corresponding Noether potential. A crucial observation is that, when a black hole is present in the space-time, there are two boundaries to take into account: one is the boundary at asymptotic infinity, where the macroscopic parameters of the solution~$(M,Q,J)$ are measured, and the other is the horizon of the black hole itself, since the inaccessible interior should not be thought of as part of the space-time manifold. In this situation, there exists a relation between surface integrals defined at asymptotic infinity and surface integrals at the horizon. Wald showed that this relation takes precisely the form of the first law of black hole mechanics~\eqref{eq:1stlawBH}, which led him to define the entropy as the surface integral of the Noether potential associated with diffeomorphisms over the horizon of the black hole\footnote{It should be noted that Wald's derivation requires the existence of a so-called bifurcation point on the horizon, and thus applies to non-extremal black holes but \textit{a priori} not to extremal ones. However, it has also been argued, \textit{e.g.} in~\cite{Jacobson:1994qe}, that such a bifurcate horizon is not necessary to define the entropy as in~\eqref{eq:Wald-entropy}. We will adopt the latter point of view and assume that Wald's definition of the entropy also applies to extremal black holes.}
\be
\label{eq:Wald-entropy}
\mathcal{S}_{\textnormal{Wald}} = 2\pi \oint_{H} d^2 x\,\sqrt{h}\,\epsilon_{\mu\nu}\mathcal{Q}^{\mu\nu} \, ,
\ee
with~$h_{\mu\nu}$ and~$\epsilon_{\mu\nu}$ the induced measure and binormal on the horizon~$H$. The formula~\eqref{eq:Wald-entropy} gives a beautiful \emph{local} geometric definition of the thermodynamical entropy for black holes in any theory invariant under diffeomorphisms. 

As a simple application, one can derive corrections to the Bekenstein-Hawking area-law in the presence of higher-derivative terms using Wald's formula. Suppose the theory of gravity under consideration is described by the following higher-curvature deformation of the Einstein-Hilbert action~\eqref{eq:EH-action}:
\be
\label{eq:S-EH-def}
S_{EH\;\text{def.}} = -\frac{1}{16\pi}\int d^4 x \sqrt{-g} \, \bigl(R + \alpha R^2 \bigr) \, ,
\ee
where we have included in the Lagrangian a term proportional to the square of the Ricci scalar with a dimensionful constant~$\alpha$. In this case, the Noether potential associated with diffeomorphism invariance is given by~\cite{Iyer:1994ys, Jacobson:1994qe}:
\be
\mathcal{Q}^{\mu\nu\rho\sigma} = \frac{\delta L_{EH\;\text{def.}}}{\delta R_{\mu\nu\rho\sigma}} = \frac{1}{32\pi}\bigl(1 + 2\alpha R\bigr)\bigl(g^{\mu\sigma}g^{\nu\rho} - g^{\mu\rho}g^{\nu\sigma}\bigr) \, .
\ee
Using~\eqref{eq:Wald-entropy}, one obtains the following entropy 
\be
\mathcal{S}_{\text{Wald}} = \frac{1}{4}\oint_H d^2 x \sqrt{h} \, \bigl(1 + 2\alpha R\bigr) = \frac{A}{4} + \frac{\alpha}{2}\oint_H d^2 x \sqrt{h} \, R \, .
\ee
The first term in the above expression is the Bekenstein-Hawking entropy~\eqref{eq:BH-entropy}. The second term captures the \emph{sub-leading corrections} to the area-law coming from the higher-derivative term in the action~\eqref{eq:S-EH-def}.
 
The clear advantage of Wald's approach is that it allows for the incorporation of higher-derivative terms, which naturally arise in effective field theories, in the entropy of black hole solutions. This formalism thus goes beyond the semi-classical approximation of the Bekenstein-Hawking area-law and can take into account quantum corrections, as long as they are encoded in a \emph{local} effective action invariant under diffeomorphisms. However, this is not the end of the story. This is because the full effective action of quantum gravity is expected to also contain \emph{non-local} terms which arise when integrating out massless degrees of freedom, as well as~\emph{non-perturbative} effects originating from the full UV theory, which are invisible in a local effective theory. Because Wald's proposal is based on a local action functional, it does not provide a framework to deal with these non-local and non-perturbative terms. So while the Wald entropy is indeed a generalization of the Bekenstein-Hawking area-law, it may fail to provide the full quantum answer for the black hole entropy in a theory of quantum gravity.

One should, however, not lose hope that this complete answer might be within reach. In the next Section, we explain how string theory provides a powerful higher-dimensional picture to understand the origin of the fundamental parameters of black holes (such as their mass and electro-magnetic charges), which in turn provides a string-theoretic origin of their thermodynamical entropy. In this description, which relies on quantum mechanically interacting D-branes, it is also possible to give an estimate for the number of microstates available to the system, which leads to a beautiful statistical interpretation of the entropy. This will serve as the template upon which we will build a formalism to define and compute the quantum entropy of black holes in the next Chapter. 

\section{The higher-dimensional origin of charges and supersymmetric black holes}
\label{sec:higherdim-S}

At the perturbative level, superstring theory is defined by quantizing the relativistic supersymmetric string in a fixed background geometry. It is now known that five consistent perturbative formulations of string theory exist, and they are all based on a ten-dimensional supersymmetric description (they admit the ten-dimensional Minkwoski vacuum as their ground state). In order to make contact with the observed four-dimensional world, one must compactify the six extra dimensions on an internal manifold, the shape and nature of which determine the properties of the theory in the remaining four non-compact directions. In perturbation theory, string theories provide a consistent description of quantum gravity, in the sense that one can compute loop corrections involving gravitons. At the non-perturbative level however, no background-independent formulation is known. This is why string-theoretical calculations are conducted in a perturbative expansion (in the parameter~$\alpha'$). Let us note that in the past, tremendous progress has been made in understanding non-perturbative properties of string theories by studying solitons, instantons, and string dualities. See~\cite{Polchinski:1998rr} for an overview and references.

There are objects of fundamental interest in all string theories, which are called branes~\cite{Polchinski:1995mt, Polchinski:1996na}. Here we simply recall that branes are supersymmetric objects in the theory on which open strings can end, and they source the various~$p$-form gauge fields of the theory. When compactifying the theory from ten down to four dimensions, the branes are taken to wrap the internal six dimensions, so that brane configurations in string theory are \emph{point-like} from the four-dimensional perspective. We summarize the field content and D-branes (branes with Dirichlet boundary conditions) of so-called Type II string theories in Tables~\ref{table:string-bosonic-field-content},~\ref{table:string-fermionic-field-content} and~\ref{table:string-brane-sources}. We will mainly discuss Type IIB string theory in what follows. In this theory, D-branes provide the ten-dimensional starting point for describing black holes, as we now explain.

The microscopic quantum description of black holes in string theory typically starts with a ten-dimensional brane configuration of given charges and mass at weak coupling. To describe the influence this system has on the four-dimensional world we observe, six dimensions must be compactified on some internal space, and the branes are taken to wrap various cycles in this internal space. One then computes an appropriate partition function in the QFT living on the world-volume of the branes. By ``appropriate'', we mean here a partition function which is \emph{topologically protected}, in the sense that it is invariant under changes in the string coupling constant. These type of quantities are often very useful to extract information about the strong coupling behavior of a system by first going to the weak coupling regime (where computations are generally expected to be technically easier thanks to perturbation theory) and then extrapolate the result to strong coupling.\footnote{In technical terms, we are thinking here of the so-called \emph{elliptic genus}~\cite{Witten:1986bf}, or some generalization thereof.} At strong coupling, the branes under consideration gravitate and form a black hole. The partition function computed at weak coupling is therefore expected to count the microstates of the corresponding macroscopic gravitational configurations. We illustrate these concepts here by presenting the first evidence, discovered by Strominger and Vafa~\cite{Strominger:1996sh}, that stacks of branes do indeed capture the microscopic degeneracies of certain black holes.
\begin{table}
\begin{center}
\begin{tabular}{|c||c|c|}
	\hline 
	Theory & NS-NS Bosons & R-R Bosons \\
	\hline \hline
	Type IIA & $g_{\mu\nu}$, $B_2$, $\phi$ & $C_1$, $C_3$ \\
	\hline
	Type IIB & $g_{\mu\nu}$, $B_2$, $\phi$ & $C_0$, $C_2$, $C_4$ \\
	\hline
\end{tabular}
\caption{Bosonic field content of Type II string theories. The Ramond-Ramond~$p$-form field strengths are denoted by~$C_p$. The NS-NS sector always contains the graviton, the Kalb-Ramond 2-form, and the dilaton.} \label{table:string-bosonic-field-content}
\end{center}
\end{table}
\begin{table}
\begin{center}
\begin{tabular}{|c||c|c|}
	\hline
	Theory & Chiral fermions (MW) & Non-chiral fermions (MW) \\
	\hline \hline
	Type IIA & - & $(\tilde{\psi}_\mu^L,\tilde{\psi}_\mu^R)$, $(\tilde{\lambda}^L,\tilde{\lambda}^R)$ \\
	\hline
	Type IIB & $(\psi_\mu^L,\psi_\mu^L)$, $(\lambda^R,\lambda^R)$ & - \\
	\hline
\end{tabular}
\caption{Fermionic field content of Type II string theories. The fermions are always Majorana-Weyl (MW) in ten dimensions.} \label{table:string-fermionic-field-content}
\end{center}
\end{table}
\begin{table}
\begin{center}
\begin{tabular}{|c||c|c|c|}
	\hline 
	Theory & R-R Form & D$p$-brane source & Dual D$(6-p)$-brane source \\
	\hline \hline
	\multirow{2}{*}{Type IIA} & $C_1$ & $D0$ & $D6$ \\ \cline{2-4}
	 & $C_3$ & $D2$ & $D4$ \\
	\hline
	 & $C_0$ & - & $D7$ \\ \cline{2-4}
	Type IIB & $C_2$ & $D1$ & $D5$ \\ \cline{2-4}
	 & $C_4$ & $D3$ & $D3$ \\ 
	\hline
\end{tabular}
\caption{D-brane sources of the various gauge fields in Type II string theories. In ten dimensions, the D$p$ branes are dual to D$(6-p)$ branes.} \label{table:string-brane-sources}
\end{center}
\end{table}

Starting from Type IIB string theory in ten dimensions, Strominger and Vafa considered a compactification on~$\mathrm{K}3\times S^1$ to obtain a five-dimensional theory. Here~$\mathrm{K}3$ is a four-dimensional space, which is of standard use in string theory compactifications because it is endowed with special properties. One of the most important of these properties encodes the behavior of spinor fields living on the manifold under parallel transport. This is refered to as the \emph{holonomy group} of the manifold, and it specifies the number of unbroken supersymmetries after the compactification. Here, the original supersymmetric ten-dimensional theory has 32 real supercharges (the dimension of a fundamental spinor in ten dimensions). The internal space~$\mathrm{K}3$ has~$\mathrm{SU}(2)$ holonomy and the circle~$S^1$ has trivial holonomy, which means that the resulting five-dimensional theory preserves 16 real supercharges~\cite{Duff:1986hr}. Moreover, since string theory includes a graviton field in its spectrum, the theory obtained after compactification is a theory of gravity invariant under 16 real supersymmetries,~\textit{i.e.} a supergravity theory. The minimal amount of real supersymmetries which can be preserved in five dimensions is 8, since a fundamental spinor in five dimensions has complex dimension four. Thus, we are dealing with an~$\mathcal{N}=2$ supergravity theory after compactification, where~$\CN$ refers to the number of ``copies'' of minimal supersymmetry.\footnote{Note that we could also adopt a nomenclature based on a four-dimensional perspective: there, a fundamental spinor has real dimension 4, twice as less as the five-dimensional spinors owing to the possibility of imposing a Majorana condition in four dimensions~\cite{VanProeyen:1999ni}. From this perspective, a supergravity theory preserving 16 real supercharges is naturally denoted as an $\CN=4$ supergravity theory.} The low-energy effective action of this supergravity theory (in the Einstein frame) contains the following terms:
\be
\label{eq:5d-eff-action-IIB}
-\frac{1}{16\pi}\int d^5x\, \sqrt{-g}\Bigl(R - \frac43(\nabla \phi)^2 - \frac{e^{-4\phi/3}}{4} H^2 - \frac{e^{2\phi/3}}{4}F^2\Bigl) \, ,
\ee
where~$H$ is a 2-form field strength arising from the NS-NS 3-form of Type IIB with one component along the~$S^1$,~$F$ is the R-R 2-form field strength and~$\phi$ is the dilaton. In this theory, a black hole solution can carry charges with respect to both~$H$ and~$F$:
\be
Q_H = -\frac{1}{4\pi^2} \int_{S^3} *e^{-4\phi/3}H \, , \qquad Q_F = -\frac{1}{16\pi} \int_{S^3} *e^{2\phi/3}F \, .
\ee
An extremal dyonic black hole solution to the equations of motion associated to~\eqref{eq:5d-eff-action-IIB} is given by~\cite{Strominger:1996sh}:
\be
\label{eq:5d-BH-Sol}
ds^2 = -\left(1 - \frac{r_h^2}{r^2}\right)^2 dt^2 + \left(1 - \frac{r_h^2}{r^2}\right)^{-2} dr^2 + r^2 d\Omega_3{}^2 \, ,
\ee
where~$d\Omega_3{}^2$ is the line-element of the 3-sphere and the horizon is located at
\be
r_h = \Bigl(\frac{8\, Q_H Q_F^2}{\pi^2}\Bigr)^{1/6} \, .
\ee
This is simply a five-dimensional dyonic extremal Reissner-Nordstr{\"o}m black hole with charges~$(Q_F,\,Q_H)$ and a near-horizon geometry~$AdS_2 \times S^3$. This solution preserves 4 of the 16 real supercharges of the theory, and this is usually denoted by saying that the black hole is 1/4-BPS. The Bekenstein-Hawking entropy~\eqref{eq:BH-entropy} for this black hole is given by the area of the 3-sphere of radius~$r_h$ divided by four,
\be
\label{eq:5d-BH}
\mathcal{S}_{BH} = 2\pi\sqrt{\frac{1}{2}Q_F^2\,Q_H} \, .
\ee 
Note that the action~\eqref{eq:5d-eff-action-IIB} (as well as the entropy~\eqref{eq:5d-BH}) receives corrections from both string loop and sigma model perturbation theory~\cite{Strominger:1996sh}. Type IIB string loop corrections are suppressed by powers of the string coupling constant~$g_s$ which is proportional to~$Q_F / Q_H$. Sigma model corrections are suppressed by inverse powers of the Schwarzschild radius which is proportional to~$\sqrt{Q_F^2 / Q_H}$. Therefore, validity of~\eqref{eq:5d-BH} requires that both charges~$Q_F$ and~$Q_H$ be large. 

How can one recover the thermodynamical entropy~\eqref{eq:5d-BH} by counting the microstates available to a D-brane system? Strominger and Vafa gave us the answer by analyzing the dynamics of a system composed of
\begin{myitemize}
\item one D5-brane wrapping~$C\times S^1$, where~$C$ is the holomorphic 4-cycle in~$\mathrm{K}3$,
\item $\bigl(\tfrac12 Q_F^2 + 1\bigr)$ D1-branes wrapping~$S^1$. 
\end{myitemize}
The R-R 2-form of Type IIB string theory is sourced by both the D1- and D5-branes (see Table~\ref{table:string-brane-sources}), and since D5-branes carry a negative D1-brane charge, the total charge under this 2-form is~$\tfrac12 Q_F^2$. Microscopically, the other charge~$Q_H$ arises from momentum along the~$S^1$ of the internal space.\\
Since we wish to describe the black hole~\eqref{eq:5d-BH-Sol} with this microscopic set-up, we should count states which preserve a quarter of the space-time supersymmetries (1/4-BPS states). To do so, one can count the states which preserve half of the supersymmetries of the D-brane worldvolume theory. This is because 1/2-BPS states in space-time correspond to supersymmetric ground states of the D-brane worldvolume theory. In the limit where the internal~$\mathrm{K}3$ is small compared to the size of the circle~$S^1$, this worldvolume theory is a supersymmetric sigma model with target space~$Sym^{\bigl(\tfrac12 Q_F^2 + 1\bigr)}\bigl[\mathrm{K}3\bigr]$, the symmetric product of~$\bigl(\tfrac12 Q_F^2 + 1\bigr)$ copies of~$\mathrm{K}3$~\cite{Bershadsky:1995qy}. This target space can be intuitively understood as the moduli space of~$\bigl(\tfrac12 Q_F^2 + 1\bigr)$ un-ordered points on~$\mathrm{K}3$, corresponding to the D1-branes moving on the single D5-brane.\\
For large~$Q_H$, the degeneracy of 1/2-BPS states in this sigma model can be evaluated using the Cardy formula~\cite{Cardy:1986ie}:
\be
\label{eq:Cardy}
d(Q_H,c) \sim \exp\Bigl(2\pi\sqrt{\frac{Q_H\,c}{6}}\Bigr) \, ,
\ee
where~$c$ is the central charge of the sigma model.
For the case at hand, the central charge is~$c=6\bigl(\tfrac12 Q_F^2 + 1\bigr)$, which leads to the following statistical entropy for the D-brane system
\be
\label{eq:5d-stat}
\mathcal{S}_B = \log d(Q_H,Q_F) = 2\pi\sqrt{Q_H\Bigl(\frac{1}{2}Q_F^2 + 1\Bigr)} \, .
\ee
In the limit where~$Q_F$ is also large, the statistical entropy of the microscopic D-brane configurations agrees with the Bekenstein-Hawking entropy~\eqref{eq:5d-BH}. Note that this limit of large charges is the analogue of the thermodynamic limit discussed in the context of Boltzmann's equation: large~$Q_F$ means a large number of D1-brane configurations making up the internal constituents of the black hole (\textit{i.e.} a large number of ``particles''~$N$) and large~$Q_H$ means a large circle~$S^1$ in the internal manifold.
Scaling both the charges uniformly keeps~$Q_H/Q_F \sim 1/g_s$ fixed and finite. In this limit, the statistical and thermodynamical entropies do indeed agree.

This result was the first tantalizing hint that the microscopic degrees of freedom accessible to the interior of a supersymmetric black hole could be successfully described by the dynamics of a D-brane system in string theory. Shortly after, other compactifications of string theory were considered, mostly down to five- and four-dimensional theories of supergravity admitting BPS black hole solutions (see~\textit{e.g.}~\cite{Maldacena:1996ky}). Again in these cases, it was found that the Bekenstein-Hawking entropy of black holes could be reproduced from a stack of interacting branes in the microscopic picture. This correspondence highlights the higher-dimensional origin of the charges in the black hole solutions, which is always central in the derivation. In the Strominger-Vafa case, the black hole charges arose as the number of D1-branes and the momentum along a compactified direction in the microscopic string theory. It is also important to stress that the success of this benchmark example and the ones that followed relies on the fact that one can compute (or at least estimate in some limit) the degeneracies of BPS states in the microscopic string theory, in part thanks to the constraints imposed by supersymmetry. In fact, we will see in the next chapter how supersymmetry and further mathematical tools allow us to go beyond the Cardy formula and obtain also sub-leading corrections to the microscopic degeneracies of BPS states. But for now, we discuss how supersymmetry also constrains the macroscopic black hole solutions.

In the rest of this work, we will mainly be interested in so-called 1/2-BPS black hole solutions of four-dimensional~$\mathcal{N}=2$ supergravity coupled to vector and scalar fields. These black hole solutions preserve four out of the eight real supercharges present in the theory and interpolate between a flat Minkowski vacuum at spatial infinity and their near-horizon region. Much like in the bosonic case of the extremal Reissner-Nordstr{\"o}m solution~\eqref{eq:RN-BH-Extr}, a symmetry enhancement takes place in this region: the 1/2-BPS solution is in fact \emph{full-BPS} near the horizon, which means that it preserves the full set of eight supercharges present in the theory. This is known as the \textit{BPS attractor mechanism}, and was first exhibited in~\cite{Ferrara:1995h}. Full supersymmetry of the near-horizon region has a wealth of consequences. Among them, imposing the vanishing of all fermionic variations under supersymmetry shows that four-dimensional extremal 1/2-BPS black hole solutions have an~$AdS_2 \times S^2$ near-horizon geometry. It also constrains the scalar fields interacting with the black hole to take definite values in the near-horizon region, and these values are fixed entirely by the electric and magnetic charges~$(q_I,\,p^I)$ of the black hole (here~$I$ is an index running over all the gauge fields in the supergravity theory). In particular, this near-horizon field configuration is \emph{independent} of the values the fields take at space-time infinity: by the time the scalar fields reach the horizon, they have lost all information about their initial conditions. 

As was pointed out earlier, this attractor behavior does not rely on supersymmetry specifically: it will occur whenever any symmetry gets enhanced. Recalling once more the case of~\eqref{eq:RN-BH-Extr}, we have seen that extremality enhances the bosonic symmetries of the near-horizon region in a black hole. Therefore, there also exists a formulation of the attractor mechanism for extremal black holes which does not rely on supersymmetry. It is usually referred to as the \emph{AdS attractor mechanism} and, in the black hole context, was proposed by Sen in~\cite{Sen:2005wa}. For supersymmetric black holes, this mechanism coincides with the BPS attractor mechanism~\cite{Cardoso:2006xz}, and it amounts to asking what are the consequences of imposing a certain symmetry on the black hole horizon (an AdS symmetry in the former case, supersymmetry in the latter). We now present this mechanism in some detail for four-dimensional extremal black holes interacting with scalar and vector fields.

Starting from such a black hole solution, we impose~$\mathrm{SO}(2,1) \times \mathrm{SO}(3)$ symmetry in the near-horizon region. This fixes the value of all the fields in the theory up to undetermined constants -- the near-horizon geometry is~$AdS_{2} \times S^{2}$ with sizes~$v_1$ and~$v_2$ for the two factors, respectively, the gauge fields under which the black hole is charged have a constant electric field strength~$e^{I}_*$ on the $AdS_{2}$ factor and a constant magnetic flux on the 2-sphere with charge~$p^{I}$,
and the scalar fields take constant values~$u_s$:
\begin{align}
\label{eq:Sen-NHG}
ds^2 =&\ v_1\Bigl(-r^2 dt^2 + \frac{dr^2}{r^2}\Bigr) + v_2\bigl(d\theta^2+\sin^2\theta \, d\phi^2\bigr) \, , \nonumber \\
\phi_s =&\ u_s \, , \qquad F^I_{rt} = e^I_* \, , \qquad F^I_{\theta\phi} = p^I\sin\theta \, .
\end{align}
In this setting, let
\be
\label{eq:Sen-Entr-Funct}
\mathcal{E}(v_1,v_2,u_s,e^I_*,p^I) := 2\pi\Bigl(-\frac12 q_I e_*^I - \int_{S^2} d\theta \, d\phi \, \sqrt{-g} \, L\Bigr) \, ,
\ee
denote the \emph{entropy function}, which is built out of the charges and the Lagrangian~$L$ of the theory (possibly including higher-derivative interactions) evaluated on the near-horizon geometry~\eqref{eq:Sen-NHG} and integrated over the~$S^2$. In terms of this function, the classical equations of motion and Bianchi identities for the various fields correspond to extremizing~$\mathcal{E}$ with respect to the parameters,
\be
\label{eq:Sen-Extrem}
\frac{\p \mathcal{E}}{\p v_i} = 0 \, , \quad \frac{\p \mathcal{E}}{\p u_s} = 0 \, , \quad \frac{\p \mathcal{E}}{\p e^I_*} = 0 \, .
\ee
The Bekenstein-Hawking-Wald entropy of the black hole is then equal to the entropy function taken at the attractor values of the fields determined by~\eqref{eq:Sen-Extrem}:
\be 
\label{eq:S-SenBHW}
\mathcal{S}_{BHW} = \mathcal{E}\vert_{\text{attr.}} \, . 
\ee
The equations~\eqref{eq:Sen-Extrem} and~\eqref{eq:S-SenBHW} are a concise and elegant way to cast the entropy of black holes as a \emph{variational principle} in the near-horizon region (such a formulation also exists for the BPS attractor mechanism and is based on a BPS entropy function defined in~\cite{Cardoso:2006xz}). We stress again here that this derivation of the entropy is not based on the specific Einstein-Hilbert action, but relies solely on the existence of the~$\mathrm{SO}(2,1) \times \mathrm{SO}(3)$ symmetry in the near-horizon region. This means that the Lagrangian from which the function~$\mathcal{E}$ is built can include for example higher-derivative terms, in which case the entropy computed with the method outlined above is the Wald entropy introduced in Section~\ref{sec:semiclassical-BH}. Moreover, this variational procedure admits straightforward generalizations to dimensions other than four, and the possibility to include higher-rank gauge symmetries~\cite{Sen:2005wa}.

As an example, we can use the AdS attractor mechanism to derive the entropy of the extremal, four-dimensional Reissner-Nordstr{\"o}m black hole introduced in Section~\ref{sec:classical-BH}. In this example, there are no scalar fields present but the black hole is indeed charged under a single~$\mathrm{U}(1)$ gauge field, so the near-horizon geometry takes the form:
\begin{align}
\label{eq:RN-NHG}
ds^2 =&\ v_1\Bigl(-r^2 dt^2 + \frac{dr^2}{r^2}\Bigr) + v_2\bigl(d\theta^2+\sin^2\theta \, d\phi^2\bigr) \, , \nonumber \\
F_{rt} =&\ e_* \, , \qquad F_{\theta\phi} = p\sin\theta \, .
\end{align}
The Lagrangian of the theory is the sum of the Einstein-Hilbert and Maxwell Lagrangians
\be
L = -\frac{1}{16\pi}R - \frac{1}{4}F_{\mu\nu}F^{\mu\nu} \, .
\ee
Note that we do not include higher-derivative interactions in this example, so that the Wald entropy is equal to the Bekenstein-Hawking entropy. We may evaluate~\eqref{eq:Sen-Entr-Funct} on the field configuration~\eqref{eq:RN-NHG} to find
\be
\mathcal{E}(v_1,v_2,e_*,p) = -\pi q\,e_* - 8\pi^2 v_1 v_2\Bigl[-\frac{1}{16\pi}\Bigl(\frac{2}{v_1} - \frac{2}{v_2}\Bigr) + \frac{1}{2}v_1^{-2}e_*^2 - \frac{1}{2}v_2^{-2}p^2\Bigr] \, .
\ee
The first equation of~\eqref{eq:Sen-Extrem} yields
\be
v_1 v_2(v_1-v_2) = 0 \, ,
\ee
which sets the~$AdS_2$ and~$S^2$ factors to have the same overall size~$v_1 = v_2 := v$, with~$v = 4\pi(e_*^2 + p^2)$. The last extremization equation of~\eqref{eq:Sen-Extrem} yields~$q = -8\pi e_*$. Finally,~\eqref{eq:S-SenBHW} gives the entropy of the black hole
\be
\mathcal{S}_{BHW} = \pi(q^2 + p^2) \, .
\ee
Comparing to~\eqref{eq:BH-Entropy-S-ERN}, this is precisely the Bekenstein-Hawking entropy of the extremal Reissner-Nordstr{\"o}m black hole.

The entropy function~$\mathcal{E}$ can be thought of as an effective action in the near-horizon~$AdS_2$ factor of extremal four-dimensional black holes, since the equations of motion and the Bianchi identities correspond to the extremization equations~\eqref{eq:Sen-Extrem}. The definition~\eqref{eq:S-SenBHW} correctly reproduces the Bekenstein-Hawking-Wald entropy for these black holes, although it only provides us with the semi-classical answer: we have not yet reached a complete answer to the quantum black hole entropy problem in this Section. However, the next Chapter will show how the notions introduced above can be nicely generalized to finally allow us to go beyond the Bekenstein-Hawking-Wald entropy of black hole solutions in supergravity. In parallel, string-theoretic generalizations of the Strominger-Vafa picture will also provide a more complete and accurate description of the microscopic degrees of freedom in black holes. This program will rely on the effectiveness of computing the \emph{sub-leading} corrections to the Bekenstein-Hawking-Wald entropy in both string theory and supergravity, and comparing these corrections to investigate the statistical interpretation of black hole entropy in the quantum regime.

\chapter{Quantum black hole entropy}
\label{chap:modern-BH-S}
Over the past 15 years or so, a more extensive treatment of black hole entropy has been put forward, relying on string theory results generalizing the Strominger-Vafa analysis presented in the previous Chapter, as well as improvements made to the Bekenstein-Hawking-Wald entropy formula in theories of quantum gravity. These advances have made it possible to go beyond the semi-classical limit and explore \emph{quantum corrections} to the Bekenstein-Hawking-Wald formula.

The Strominger-Vafa result~\eqref{eq:5d-stat} was obtained using the approximate Cardy formula~\eqref{eq:Cardy} for the statistical entropy of the brane system. There exists standard D-brane methods to evaluate this statistical entropy with much more accuracy, eventually leading to an \emph{exact} result for the degeneracies of a black hole predicted by string theory. In the following, we present an example that will be especially relevant in this work, and stress the connection between these results and the mathematical theory of \emph{modular and Jacobi forms}. We will then introduce new concepts in supergravity theories which have allowed for a refined definition of the thermodynamical entropy of a certain class of extremal dyonic black holes. Computing this quantum entropy and comparing to the results predicted by string theory will be the focus of the rest of this thesis. 

To begin this investigation, it shall be useful to recall how modular forms naturally appear in the context of microstate counting in string theory. Suppose we are interested in computing the degeneracy of 1/2-BPS states in Type II string theory compactified on an internal manifold~$\mathrm{K}3\times T^2$. This theory is dual to the heterotic string compactified on a six-dimensional torus~$T^6$. The resulting four-dimensional low-energy theory is an~$\mathcal{N}=4$ supergravity theory since the~$\mathrm{K}3$ breaks half of the original supersymmetries. The 1/2-BPS states in this theory have zero magnetic charge but non-zero electric charge~$Q$. They are known as Dabholkar-Harvey states~\cite{Dabholkar:1989jt, Dabholkar:1990yf}. They are purely electric and perturbative in the heterotic frame. The partition function for such states is given by the partition function of the chiral conformal field theory of 24 left-moving transverse bosons of the heterotic string~\cite{Dabholkar:2012zz}. The Hilbert space~$\mathcal{H}$ of this theory is the Fock space representation of the commutator algebra of 24 harmonic oscillators representing the transverse oscillation modes of the string:
\be
\left[a_{i\,n},\,a^\dagger_{j\,m}\right] = \delta_{ij}\delta_{n+m,0} \, ,
\ee
where~$i,j=1,\ldots,24$ and~$n,m=1,2,\ldots,\infty$. The Hamiltonian is simply
\be
H = \sum_{i=1}^{24} n a^\dagger_{i\,n}a_{i\,n} - 1 \, ,
\ee
and the partition function is given by~$Z(\tau) = \Tr_{\mathcal{H}}\left(q^H\right)$, where we denote~$q := e^{2\pi i \tau}$. Each oscillator mode of energy~$n$ contributes to the trace, and using the sum of a geometric series, we immediately find
\be
\label{eq:1/2BPS-n4-partition}
Z(\tau) = \frac{1}{q}\prod_{n=1}^{\infty}\frac{1}{(1-q^n)^{24}} \, .
\ee
This is the inverse of the product representation of the \emph{discriminant function}~$\Delta(\tau)$ which is a modular form of weight 12. The modularity of the partition function is naturally inherited from the modularity of the torus used in the heterotic string compactification. This modular symmetry is extremely convenient to evaluate the Fourier coefficients~$d(n)$ of the partition function. By an inverse Fourier transform, we have that
\be
d(n) = \int d\tau \, Z(\tau) e^{-2\pi i n \tau} = \int d\tau \, \frac{e^{-2\pi i n \tau} }{\Delta(\tau)} \, .
\ee
What is the behavior of this quantity as~$n$ becomes very large? Most of the contributions to the integral for large~$n$ will arise from the small~$\tau$ region, so the large~$n$ asymptotics for the degeneracies can be extracted from the knowledge of the partition function at small~$\tau$. As~$\tau \rightarrow 0$ (or equivalently~$q\rightarrow 1$), the asymptotics of~$Z(\tau)$ are very difficult to read off from~\eqref{eq:1/2BPS-n4-partition} since it is an infinite product of divergent quantities. But here, we can make use of the fact that the partition function is the inverse of the discriminant function. Since~$\Delta$ is a modular form of weight 12, we have the identity\footnote{We refer the reader to Appendix~\ref{app:modular} for a collection of detailed facts regarding modular forms and their generalizations.}~$\Delta(e^{2\pi i \tau}) = \tau^{-12}\Delta(e^{-2\pi i/\tau})$, which yields for the partition function
\be
\label{eq:mod-partition}
Z(\tau) = \tau^{12}Z\Bigl(-\frac{1}{\tau}\Bigr) \, .
\ee
As~$\tau\rightarrow 0$,~$-1/\tau \rightarrow -\infty$ or equivalently~$\tilde{q} := e^{-2\pi i/\tau} \rightarrow 0$. It is now straightforward to obtain the~$\tilde{q}\rightarrow 0$ asymptotics of the partition function,
\be
Z\Bigl(-\frac{1}{\tau}\Bigr) = \frac{1}{\tilde{q}}\prod_{n=1}^{\infty}\frac{1}{(1-\tilde{q}^n)^{24}} \sim \frac{1}{\tilde{q}}\, .
\ee
Using~\eqref{eq:mod-partition}, this allows us to write the degeneracies of the 1/2-BPS states in Type II string theory compactified on~$\mathrm{K}3 \times T^2$ for large~$n$ as
\be
d(n) \sim \int d\tau \, \tau^{12} \, e^{-2\pi i n \tau + \frac{2\pi i}{\tau}} = 2\pi\,n^{-13/2}\,I_{13}\bigl(4\pi\sqrt{n}\bigr) \, ,
\ee
which is a Bessel function of the first kind of weight 13. In obtaining this result, the modular properties of the partition function were of crucial importance. Various generalizations of the model just presented exist for other types of BPS states and in different string compactifications, using the more general Jacobi and Siegel counterparts of modular forms. Before presenting in more detail how this happens in a specific example, we make some comments about the degeneracies of Dabholkar-Harvey states just derived.

One may use the asymptotic expansion of the Bessel function of the first kind for large values of~$n$~\eqref{eq:Bessel-Exp} to find the statistical entropy of Dabholkar-Harvey states in the thermodynamic limit:
\be
\label{eq:DH-stat-S}
\mathcal{S}_B = \log d(n) \sim 4\pi\sqrt{n} \, .
\ee
Here,~$n$ is given in terms of the electric charge of the Dabholkar-Harvey state as~$n = Q^2/2$~\cite{Dabholkar:2012zz}. Therefore, the statistical entropy~\eqref{eq:DH-stat-S} scales  \emph{linearly} in the charges,~$\mathcal{S}_B \sim Q$. One can construct extremal BPS black hole solutions carrying the same charge quantum numbers as the string states considered here, and it is reasonable to expect that their Bekenstein-Hawking entropy will reproduce the leading order statistical entropy~\eqref{eq:DH-stat-S}. Unfortunately, the corresponding black holes (often referred to as \emph{small black holes}, as their size is comparable to the string size in the string frame) are mildly singular and have a vanishing classical horizon~\cite{Sen:1994eb}, and therefore vanishing thermodynamical entropy! The solution to this apparent discrepancy emphasizes the importance of higher-derivative corrections to the Bekenstein-Hawking entropy: the black hole solution with vanishing entropy was constructed using only the tree-level low-energy effective action of the heterotic string and neglecting higher-derivative terms. It was shown in~\cite{Sen:1995in} that after taking into account the effects due to these higher-derivative terms, the geometry of the black hole is modified and the area of their horizon becomes non-zero. Within this effective theory, one can use Wald's formalism and show that the Bekenstein-Hawking-Wald entropy now precisely matches the statistical entropy~\eqref{eq:DH-stat-S}~\cite{Dabholkar:2004yr}.

We now present how the modularity of the partition function of BPS states in certain string theory compactifications can be used to extract the exact statistical entropy of a brane system following the original derivation of Maldacena, Moore and Strominger~\cite{Maldacena:1999bp}. This will serve as the basic string-theoretical prediction for the quantum entropy of a black hole, which we will strive to reproduce using a low-energy supergravity description. 

\section{1/8-BPS black holes in $\CN=8$ string theory}
\label{sec:1/8stringBH}

We begin by considering type IIB string theory compactified on~$T^{6}$. The internal manifold has trivial holonomy and therefore does not break any of the 32 supersymmetries present in the original ten-dimensional theory. Thus, at low energies, the effective description of the theory is given by~$\CN=8$ supergravity in four dimensions. This theory has a macroscopic 1/8-BPS black hole solution carrying electric and magnetic charges under the various gauge fields in the theory. The~$\CN=8$ string theory has an~$E_{7,7}(\IZ)$ duality group\footnote{The discrete nature of this group originates from the fact that the charges in string theory are quantized and take their values on a discrete lattice.} with a duality invariant~$\Delta$ which is quartic in the charges. In order to compute the microscopic degeneracies, one goes to a particular duality frame in which there is an explicit description of the charges of the black hole as charges in the microscopic string theory. 

A simple description consists of at least four charges which can be represented as follows. Writing~$T^{6} = T^{4} \times S^{1} \times \wt S^{1}$, one has:
\begin{myitemize}
\item a D1-brane wrapped on~$S^1$, 
\item a D5-brane wrapped on~$T^4 \times S^{1}$, 
\item $n$ units of momentum along~$S^{1}$,
\item $\ell$ units of momentum along~$\wt S^1$,
\item one unit of Kaluza-Klein monopole charge on~$\wt S^{1}$~\cite{Gross:1983hb}.
\end{myitemize}
In the following, we will refer to this brane system as the D1-D5-P-KK$\vert_{\CN=8}$ system. The electric and magnetic charge vectors of the black hole are given in terms of the microscopic charges of the system as
\be
\label{eq:charges-N8}
Q_e^2/2 = n \, , \quad Q_e\cdot Q_m = \ell \, , \quad Q_m^2/2 = 1 \, .
\ee
The U-duality invariant in this configuration is~$\Delta := Q_e^2 Q_m^2 - (Q_e \cdot Q_m)^2 = 4n-\ell^{2}$. This invariant is quartic in the charges. We have already seen in Section~\ref{sec:semiclassical-BH} that the area of a macroscopic dyonic extremal black hole scales quadratically in the charges, so we should already expect the area and the related Bekenstein-Hawking entropy to scale as~$\sqrt{\Delta}$. 

Using this brane description, one can compute the BPS partition function which is the generating function of the microscopic index of 1/8-BPS states in the theory:\footnote{Here and in the following, we use a notation which is common in number theory and the discussion of modular and Jacobi forms,~$q := \exp(2\pi\mathrm{i}\tau)$ and $y := \exp(2\pi\mathrm{i}z)$.} 
\be
Z^\text{BPS}(\t,z) \= \sum_{n, \ell \, \in \, \IZ} c(n, \ell)\,q^n\, y^\ell \, . 
\ee
This quantity was shown to have a simple explicit form in terms of known theta and eta functions~\cite{Maldacena:1999bp}:
\be 
\label{eq:phi21}
Z^\text{BPS}(\t,z) \=  \varphi_{-2,1}(\t,z) \, := \, \frac{\vartheta_1(\t,z)^2}{\eta(\t)^{6}} \, , 
\ee
where
\begin{align}
\vartheta_1(\t,z) =&\, q^{\tfrac18}\bigl(y^{\tfrac12} - y^{-\tfrac12}\bigr)\prod_{n=1}^{\infty}\bigl(1-q^n\bigr)\bigl(1-yq^n\bigr)\bigl(1-y^{-1}q^n\bigr) \, , \nonumber \\
\eta(\t) =&\, q^{\tfrac1{24}}\prod_{n=1}^{\infty}\bigl(1-q^n\bigr) \, .
\end{align}
The black hole degeneracies are related to the index of 1/8-BPS states in the theory as~\cite{Sen:2009vz,Dabholkar:2010rm}:
\be
d(n,\ell) \= (-1)^{\ell} \, c(n,\ell) \, . 
\ee 
The function~$\varphi_{-2,1}$ is an example of a \emph{Jacobi form} of weight $-2$ and index 1. We have collected a number of technical facts regarding Jacobi forms and their generalizations in Appendix~\ref{app:modular}. At this stage, we simply want to point out that the transformation properties obeyed by Jacobi forms (see~\eqref{eq:modtransform} and~\eqref{eq:elliptic}) are extremely constraining and give us great control over their Fourier coefficients~$c(n, \ell)$. As a simple example, the elliptic transformation property~\eqref{eq:elliptic} implies that the Fourier coefficients of a Jacobi form of index~$m$ obey
\be
c(n,\ell) = C_\ell(4nm-\ell^2) \, , \; \textnormal{where~$\,C_\ell(4nm-\ell^2)$ depends only on~$\ell$ mod~$2m$} \, .
\ee 
Therefore, for the specific 1/8-BPS black hole in~$\CN=8$ supergravity corresponding to the brane system introduced above (where the Jacobi form has index 1), the degeneracies~\eqref{eq:degen} are a function of $4n-\ell^2 = \Delta$ only,
\be
\label{eq:degen}
d(\Delta) =  (-1)^{\Delta +1} c(n,\ell) \, , \quad \text{with~$\ell = \Delta$ mod 2} \, ,
\ee
which is a manifestation of the physical~$U$-duality symmetry. It is also consistent with the expectation borne out of the attractor mechanism, which guarantees that the entropy of the black hole must be a function of its electric and magnetic charges only. The latter are indeed given in terms of the microscopic momenta~$n$ and~$\ell$ of the brane description according to~\eqref{eq:charges-N8}.

The modular transformation property~\eqref{eq:modtransform} is so powerful that one has an analytic formula for all the coefficients of a Jacobi form in terms of its \emph{polar coefficients}, which are the Fourier coefficients associated to terms with a negative power of~$q$ in the Fourier expansion. This formula, called the \emph{Hardy-Ramanujan-Rademacher} expansion and displayed in~\eqref{eq:radi}, takes the form of an infinite convergent sum of Bessel functions (see~\cite{Dijkgraaf:2000fq} for a nice exposition). 

For the 1/8-BPS states' partition function~\eqref{eq:phi21}, which is a weak Jacobi form of weight~$-2$ and index 1, the Rademacher expansion~\eqref{eq:radi} yields:
\be
\label{eq:rademsp} 
c(n,\ell) = 2{\pi} \, \Bigl(\frac{\pi}{2}\Bigr)^{7/2} \, \sum_{c=1}^\infty c^{-9/2} \, K_{c}(\Delta) \; \wt I_{7/2} \Bigl(\frac{\pi \sqrt{\Delta}}{c} \Bigr)  \, , \quad \textnormal{with} \quad \Delta = 4n-\ell^2 \, .
\ee
Here~$K_{c}$ is a particular combination of so-called Kloosterman sums with the property~$K_{1}(\Delta)=1$, and~$\wt I_\r(z)$ denotes the modified Bessel function of order~$\r$ (see~\eqref{eq:intrep} for definitions). Equation~\eqref{eq:rademsp} (together with~\eqref{eq:degen}) can be interpreted as an \emph{exact} formula for the degeneracies of the D1-D5-P-KK$\vert_{\CN=8}$ system. 

In the limit of large charges (\textit{i.e.} large~$\Delta$), we may use the asymptotic series of the modified Bessel function~\eqref{eq:Bessel-Exp} to estimate the leading contribution to the black hole entropy. Evidently, this is given by the~$c=1$ terms in the sum~\eqref{eq:rademsp}, and for~$\Delta \rightarrow \infty$, we have
\be
\wt I_{7/2} \bigl(\pi\sqrt{\Delta}\bigr) \sim \exp\bigl(\pi\sqrt{\Delta}\bigr) \, ,
\ee
thus showing that for large~$\Delta$ (that is, in the thermodynamic limit), the statistical entropy of the D1-D5-P-KK$\vert_{\CN=8}$ system computed in~\cite{Maldacena:1999bp} is given by
\be
\mathcal{S}_B \sim \pi\sqrt{\Delta} \, .
\ee
This agrees with the Bekenstein-Hawking entropy since the latter scales as~$\sqrt{\Delta}$ in the limit of large charges.

In Chapter~\ref{chap:n4-loc}, we will investigate 1/4-BPS black hole solutions of~$\CN=4$ supergravity obtained by compactifying Type IIB on~$\mathrm{K}3 \times T^{2}$. We have already seen at the beginning of this Chapter how to obtain the degeneracies of 1/2-BPS states in this theory, but the 1/4-BPS states come with additional subtleties. A similar, albeit more technical, analysis of their degeneracies than the one presented for 1/8-BPS states in~$\CN = 8 $ string theory can still be conducted, as we will review later. 

The case discussed in this Section shows that the microscopic string theory can compute the exact degeneracies of certain D-brane systems very efficiently, owing to the modular or Jacobi symmetries of the BPS states' partition function (naturally inherited from the properties of the internal space used in the compactification down to four dimensions). We now would like to ask the following question: can these results be reproduced in the low-energy effective description of string theory? This question amounts to asking whether there exist a recipe in supergravity to compute the quantum entropy of black holes \emph{exactly}, that is by re-summing all quantum corrections to the Bekenstein-Hawking-Wald entropy fomula. Remarkably, the answer to this question is in the positive for extremal supersymmetric black holes. To present these results, we should first and foremost \emph{define} what we mean by the quantum entropy of these black holes in supergravity theories. To this end, we now introduce the \emph{Quantum Entropy Function} (QEF). Subsequently, we discuss the method used to carry out the computation of this quantum entropy.

\section{Sen's Quantum Entropy Function}
\label{sec:QEF}

We have seen in Section~\ref{sec:higherdim-S} that, for extremal black holes, it is possible to cast the attractor mechanism as a variational principle for the function~$\mathcal{E}$ introduced in~\eqref{eq:Sen-Entr-Funct}. To include the effect of quantum fluctuations of the fields on the black hole entropy, Sen promotes this variational principle to a functional integral, called the \textit{quantum entropy}, over all the fields of the theory which asymptote to the attractor configuration specified by~\eqref{eq:Sen-Extrem}~\cite{Sen:2008vm}. This is a very natural extension of the notion of entropy into the quantum realm, analogous to the Feynman path-integral extension of the classical motion of a physical system obtained by extremizing the action functional. 

The functional integral for the partition function is weighted by the exponential of the Wilsonian effective action at some fundamental scale defining the theory, such as the string scale. To make the classical variational problem well-defined, it is necessary to add a boundary term $-i q_{I} \int A^{I}$ to the action. With this boundary term, the quantum partition function can be naturally interpreted as the expectation value of a Wilson line inserted at the boundary
\be 
\label{eq:QEF}
\exp\left[\mathcal{S}_{Q} (q, p)\right] := W(q, p) = \left\langle \exp[-i \, q_I \oint_{\tau}  A^I]  \right\rangle_{\rm{AdS}_2}^\text{finite}\ . 
\ee 
The angular brackets indicate an integration (with an appropriate measure) over all the field fluctuations with appropriate~$AdS_{2}$ boundary conditions~\cite{Sen:2008vm, Sen:2008yk, Castro:2008ms}. Note that these boundary conditions fix all the electric and magnetic charges in the theory, and naturally lead to a microcanonical ensemble for the statistical interpretation of this quantum entropy. The superscript ``finite'' in the expression~\eqref{eq:QEF} refers to the fact that the action of the theory is divergent due to the infinite volume of $AdS_{2}$, and one therefore needs to regularize it. This is done by putting a cutoff $r_{0}$ on the $AdS_{2}$ geometry so that the proper length of the boundary scales as $2 \pi \sqrt{v} r_{0}$, where~$v$ is the size of~$AdS_2$. Since the classical action is an integral of a \textit{local} Lagrangian, it scales as~${S_1 r_0 + S_0 + \CO(r_0^{-1})}$. The linearly divergent part can be subtracted by an appropriate boundary counter-term, and this procedure sets the origin of energy in the boundary theory. After this renormalization we can take the cutoff~$r_0$ to infinity to obtain a finite functional integral weighted by the exponential of the finite piece $S_0$. This finite piece is a functional of all fields and contains arbitrary higher-derivative terms. It is referred to as the \emph{renormalized action} $S_{\rm ren}$. 

The main interest of the above definition for the quantum entropy is that it should correctly reproduce the entropy obtained from a microscopic description of the black holes provided by string theory. A one-loop evaluation of the functional integral~\eqref{eq:QEF} for supersymmetric black holes was conducted in~\cite{Sen:2011ba, Banerjee:2011jp}, and the leading logarithmic corrections were successfully matched to the microscopic predictions. Even a preliminary reading of these papers allows us to appreciate the technical power used in computing these one-loop corrections. This direct method of computing logarithmic corrections is applicable in a wide variety of black holes, including non-supersymmetric ones. On the other hand, for supersymmetric solutions, the method of \textit{supersymmetric localization} allows us to sum up the contributions from all orders of perturbation theory at one shot. We now present this method in some generality, in view of applying it to specific four-dimensional supersymmetric black holes in Chapters~\ref{chap:n8-loc},~\ref{chap:n2-loc} and~\ref{chap:n4-loc}.

\section{Supersymmetric localization}
\label{sec:susy-loc}

Supersymmetric localization relies on a number of mathematical theorems derived in the 1980s~\cite{Duistermaat1982, berline1983, Atiyah:1984px}. It was first suggested that it could be applied to physical situations to obtain highly non-trivial results in~\cite{Witten:1988ze}. The work of Pestun~\cite{Pestun:2007rz} provided definitive evidence that localization in supersymmetric QFTs could be used to extract meaningful results from \textit{a priori} very complicated situations.

The basic principle underlying supersymmetric localization can be stated as follows.\footnote{The rigor of localization is based on the mathematical work quoted above. We refer the reader interested in a more formal presentation of the localization arguments to these references, along with the excellent review~\cite{Szabo:1996md}. For the purpose of the present work, it will suffice to give a more physical approach to the localization argument.} Suppose we are interested in computing the partition function of a quantum system, which is given by the path-integral
\be
Z = \int \mathcal{D}\Phi\,e^{\,S[\Phi]} \, ,
\ee
where~$S[\Phi]$ is the action functional for the system and~$\Phi$ denotes the collection of quantum fields. Although the computation of this quantity looks at first sight near impossible (it requires us to perform an infinite-dimensional integral over the entire field configuration space of the system!), localization shows that in the presence of a specific symmetry, it is in fact exactly computable. To understand how this happens, we introduce the following:
\begin{myitemize}
\item Let~$Q$ be a fermionic symmetry of the theory, and~$Q^2$ be such that it is compact and generates isometries of the space-time on which the QFT lives.\\
\item Let~$S$ be a~$Q$-invariant action functional, \textit{i.e.}~$QS = 0$.\\
\item Let~$\mathcal{V}$ be a fermionic functional of fields such that~$Q^2\mathcal{V} = 0$.
\end{myitemize}
We can then \emph{deform} the partition function and define
\be
Z(\lambda) := \int \mathcal{D}\Phi\,e^{\,S[\Phi] + \lambda\,Q\mathcal{V}} \, .
\ee
It can now be shown~\cite{Witten:1988ze} that this deformed partition function is in fact independent of the parameter~$\lambda$,
\begin{equation}
\label{eq:loc-principle}
\frac{d}{d\lambda} Z(\lambda) = 0 \, ,
\end{equation}
provided the integration measure of the path-integral is itself invariant under~$Q$ (which we will assume). Equation~\eqref{eq:loc-principle} shows that one can deform the initial action by the bosonic functional~$Q\mathcal{V}$, hereafter referred to as the \emph{localizing action}, without changing the value of the path-integral under consideration. This is extremely convenient: being interested in the original path-integral~$Z(0)$, we can compute it for any value of the~$\lambda$ parameter, and especially for~$\lambda \rightarrow \infty$. In this regime, the path-integral is entirely dominated by the saddle-point field configuration~$Q\mathcal{V}(\Phi) = 0$. The solution(s) to this equation specify a submanifold of the full field configuration space, called the \textit{localizing manifold}~$\mathcal{M}_Q$, and the path-integral can be evaluated using the sole knowledge of this submanifold. More precisely, we have the following \textit{exact} equation:
\begin{equation}
\label{eq:localized-int}
\int \mathcal{D}\Phi \,e^{\,S[\Phi]} = Z(\lambda = 0) = Z(\lambda \rightarrow \infty) = \int_{\mathcal{M}_Q} [d\phi] \, e^{\,S[\phi]} \, Z_{\textnormal{1-loop}}(\phi) \, ,
\end{equation}
where~$\phi$ denote the coordinates on~$\mathcal{M}_Q$,~$[d\phi]$ a measure taking into account the curvature of~$\mathcal{M}_Q$, and~$Z_{\textnormal{1-loop}}$ is a one-loop functional determinant factor arising from the quadratic fluctuations of the fields orthogonal to~$\mathcal{M}_Q$. 

Supersymmetric localization shows that the exact evaluation of a complicated path-integral can be reduced to a much simpler one-loop computation involving only finite-dimensional, regular integrals. This drastic simplification entirely stems from the constraining powers of the fermionic symmetry generated by~$Q$. Using supersymmetry as the fermionic symmetry and applying this formalism to the quantum entropy function introduced in Section~\ref{sec:QEF}, we will see in subsequent Chapters how this general principle allows for an exact computation of the path-integral~\eqref{eq:QEF}. But before doing so, let us discuss a few key aspects required (or simply desirable on technical grounds) for supersymmetric localization in general.

Evidently, the most important ingredient of the localization recipe is the supercharge~$Q$ used to build the deformation functional~$Q\mathcal{V}(\Phi)$. This supercharge specifies the localizing manifold and indirectly, the one-loop determinant factor. It will therefore be extremely convenient to work in a formalism in which the action of this supercharge on all fields of the theory is known and fixed once and for all. This is possible when one works in an \textit{off-shell} supersymmetric theory, since in this case the algebra of supercharges closes on all the fields without the need for imposing equations of motion. In such an off-shell setting, any modifications to the original action one wishes to localize (for example upon including higher-derivative terms) will have no bearing on the~$Q$-transformations of the fields and therefore on the characterization of the localizing manifold. Also, note that for localization to work, it is only necessary to use a single supercharge. This will be relevant when dealing with off-shell hypermultiplets in Chapter~\ref{chap:n2-loc}.\\
Another key aspect in the supersymmetric localization technique is that the path-integral~\eqref{eq:QEF} is defined in a \textit{Euclidean} theory (as evidenced for example by the periodic integral of the gauge field over the Euclidean time circle parametrized by~$\tau$). Hence, we will have to work with a Euclidean supergravity theory. Such theories can be obtained using a Wick-rotation and analytic continuation starting from their Minkowskian counterparts, but this procedure usually relies on a number of prescriptions which may be convention-dependent. In an effort to unambiguously define the Euclidean theory we will make use of in the calculation of~\eqref{eq:QEF}, we will describe in Chapter~\ref{chap:sugra} how to obtain a fully off-shell Euclidean theory of supergravity by the method of time-like dimensional reduction.

In the localization procedure, the choice of the fermionic functional~$\mathcal{V}$ is free. Choosing two different~$\mathcal{V}$'s will give different-looking intermediate steps in the localization (for instance different localizing manifolds), but it is a mathematical theorem that at the very end of the calculation, the two choices should yield the same final answer. We can therefore exploit this freedom to choose a particularly convenient fermionic functional:
\be
\label{eq:specificV}
\mathcal{V} = \sum_\a \bigl(Q\psi_\a\,,\,\psi_\a\bigr) \, ,
\ee
where~$(.\,,\,.)$ is an appropriate inner product, and~$\psi_\a$ denote the fermions of the theory (labelled by the index~$\a$). With this choice (and in a bosonic background), the localizing equations specifying the manifold~$\mathcal{M}_Q$ reduce to \emph{BPS equations}
\be
\label{eq:QV-BPS}
Q\mathcal{V} = 0 \, \Longleftrightarrow \, Q\psi_\a = 0 \, ,
\ee
for all the fermions. This is particularly convenient in the supergravity context, where BPS equations are extensively studied and already encode much of the information regarding the geometry of space-time.

We can now summarize our strategy for studying the quantum entropy of supersymmetric black hole solutions and exploring the connection to the statistical entropy of string theory:
\begin{myenumerate}
\item We pick a four-dimensional, Euclidean, off-shell, supergravity theory. In this theory, we focus on an extremal supersymmetric dyonic black hole solution preserving at least one supercharge~$Q$.\\
\item We \textit{define} the macroscopic quantum entropy of the black hole using~\eqref{eq:QEF}.\\
\item We apply the localization method by finding the localizing manifold~$\mathcal{M}_Q$ for all the fields present in the theory, and we compute the one-loop functional determinant arising from quadratic fluctuations orthogonal to~$\mathcal{M}_Q$.\\
\item We evaluate the resulting finite-dimensional integral to obtain an exact answer for the quantum entropy of the black hole under consideration.\\
\item We compare the result obtained for this macroscopic entropy against the microscopic predictions of string theory for the same black hole.
\end{myenumerate}
If the last step is conclusive, so that there is an agreement between the macroscopic and microscopic descriptions of the black hole, it provides a non-trivial test that supergravity is indeed an appropriate low-energy description of string theory and sheds light on the statistical interpretation of the black hole's thermodynamical entropy, including all possible quantum corrections to the area-law of Bekenstein and Hawking.

To initiate the localization program in supergravity, it will be useful to formally study the Euclidean supergravity theory in which we will work for step 1. This theory is built using the method of \emph{off-shell time-like dimensional reduction}, as is explained in detail in the next Chapter. Once this off-shell Euclidean supergravity theory is constructed, we will focus on evaluating the QEF for specific black hole solutions.

\chapter{Supergravity}
\label{chap:sugra}
Quantum Field Theories (QFTs) are in general invariant under certain space-time and internal symmetries. 
The familiar space-time symmetries are generated by the energy-momentum operator~$P_\mu$ and the Lorentz operators~$M_{[\mu\nu]}$ which make up the Poincar\'{e} group. Internal symmetries constitute flavor (global) and gauge (local) symmetries acting on the fields themselves. One can also consider a fermionic symmetry relating bosons and fermions. This is the notion of \emph{supersymmetry}.\footnote{According to the Haag-\L{}opusza\'{n}ski-Sohnius theorem, this fermionic symmetry is compatible with the generic group of symmetries of the S-matrix in a local and unitary QFT~\cite{Haag:1974qh}.} This symmetry is generated by spinor charges~$Q_\alpha^i$, where~$\alpha$ is a space-time spinor index and~$i=1,\,\ldots,\,\mathcal{N}$ labels the distinct supercharges. For the simplest~$\mathcal{N}=1$ case, the supersymmetry algebra is given by the standard Poincar\'{e} algebra of bosonic charges supplemented by the following commutation relations involving the spinor charges:
\begin{align}
	\left\{Q_\alpha,\,\bar{Q}^\beta\right\} =&\, 2\,\left(\gamma^\mu\right)_\alpha{}^\beta\,P_\mu \, , \nonumber \\
	\left[M_{\mu\nu},\,Q_\alpha\right] =&\, 2\,\left(\gamma_{\mu\nu}\right)_\alpha{}^\beta\,Q_\beta \, , \\
	\left[P_\mu,\,Q_\alpha\right] =&\, 0 \, , \nonumber
\end{align}
other (anti-)commutators being zero.

The standard construction of gauge theories starting from the symmetry algebra and gauging it by making the invariance hold locally can naturally be applied to supersymmetric theories. Doing so, one obtains \emph{supergravity} theories. Local invariance under supersymmetry has a wealth of interesting consequences for these theories, one of the most important of which is that they necessarily must contain a spin-2 field associated with diffeomorphism invariance of the space-time manifold. In other words, local supersymmetry implies the presence of a metric tensor in the spectrum, and thus implies gravity. The spin-2 graviton field has a superpartner called the gravitino, which is a spin-$3/2$ field, along with possible other lower-spin fields which furnish the irreducible gravity multiplet. In addition to the graviton multiplet, one can also couple various interacting matter multiplets to it.

In the vast majority of this work, we will be concerned with specific supergravity theories which exist in four space-time dimensions, along with their black hole solutions. However, as was emphasized in Section~\ref{sec:susy-loc}, we will need to work in Euclidean signature in order to apply localization to the computation of black hole entropy. The four-dimensional Minkowski supergravity theories are well established in the literature, but their Euclidean counterparts have so far not been given the same treatment, so we will derive the theory we need by the method of time-like dimensional reduction from a five-dimensional Minkowski supergravity theory. As was also alluded to in Section~\ref{sec:susy-loc}, it will be convenient for the purposes of localization to use an \textit{off-shell} formulation of supergravity. This can be conveniently implemented using the method of superconformal multiplet calculus~\cite{deWit:1979ug, deWit:1980tn}.
  
We now proceed to describe the five-dimensional conformal supergravity theory which we will dimensionally reduce down to four dimensions.

\section{Conformal supergravity}
 
The conformal group is the group of symmetries which leave the light-cone invariant. It contains the Poincar\'{e} group, along with additional symmetry generators: the dilatations $D$ and the conformal boosts, or special conformal transformations, $K^A$. In five dimensions, it is given by the group $\mathrm{SO}(5,2)$. To each of these operators, we associate a gauge field and a transformation parameter according to Table~\ref{table:conformal-alg}.
\begin{table}
\begin{center}
\begin{tabular}{|c|c|c|c|c|}
	\hline
	Generator & $P^A$ & $M^{AB}$ & $D$ & $K^A$ \\
	Gauge field & $e_M{}^A$ & $\omega_M{}^{AB}$ & $b_M$ & $f_M{}^A$ \\
	Parameter & $\xi^A$ & $\epsilon^{AB}$ & $\Lambda_D$ & $\Lambda_K^A$ \\
	\hline
\end{tabular}
\caption{The generators of the conformal algebra, along with their associated gauge fields and parameters.} \label{table:conformal-alg}
\end{center}
\end{table}  
In this Table, the indices~$A,B=0,\ldots,4$ label the coordinates of a flat manifold of Minkowski signature. At this point, this is still an internal manifold, and we will shortly see how this manifold can be identified with the tangent space associated to the space-time manifold. Using the~$\mathrm{SO}(5,2)$ Lie algebra, one obtains the transformation rules of the gauge fields under conformal transformations:
\begin{align}
	\label{eq:bosonic-conf-transfos}
	\delta e_M{}^A =&\, \mathcal{D}_M\xi^A - \Lambda_D\,e_M{}^A + \epsilon^{AB}\,e_{MB} \, , \nonumber \\
	\delta \omega_M{}^{AB} =&\, \mathcal{D}_M\epsilon^{AB} + 4\,\Lambda_K{}^{[A}e_M{}^{B]} + 2\,\xi^{[A}f_M{}^{B]} \, , \nonumber \\
	\delta b_M =&\, \partial_M\Lambda_D + 2\,\Lambda_K{}^A e_{MA} - \xi^A f_{MA} \, , \\
	\delta f_M{}^A =&\, \mathcal{D}_M\Lambda_K{}^A + \Lambda_D f_M{}^A + \epsilon^{AB} f_{MB} \, . \nonumber
\end{align}
Here, the derivative~$\mathcal{D}_M$ is covariant with respect to Lorentz and dilatation transformations. From these transformation rules, one builds the following curvature tensors:
\begin{align}
	R(P)_{MN}{}^A =&\, 2\,\partial_{[M}e_{N]}{}^A + 2\,b_{[M}e_{N]}{}^A - 2\,\omega_{[M}{}^{AB}e_{N]B} \, , \nonumber \\
	R(M)_{MN}{}^{AB} =&\, 2\,\partial_{[M}\omega_{N]}{}^{AB} - 2\,\omega_{[M}{}^{AC}\omega_{N]C}{}^B - 8\,e_{[M}{}^{[A}f_{N]}{}^{B]} \, .
\end{align}
Upon imposing algebraic constraints on the curvature, we can relate the internal transformations~\eqref{eq:bosonic-conf-transfos} to space-time transformations. Imposing~$R(P)_{MN}{}^A = 0$ shows that the P-transformation of the vielbein reduces to a covariant general coordinate transformation of the space-time manifold. This constraint can also be solved for the gauge field~$\omega_M{}^{AB}$, which is then identified with the natural spin-connection of the space-time manifold. Note that because of the dilatations, this spin-connection contains a term proportional to the gauge field~$b_M$ and so differs from the spin-connection one may be familiar with from general relativity. A second constraint~$e_A{}^M\,R(M)_{MN}{}^{AB} = 0$ can be used to solve for the gauge field of special conformal transformations:
\begin{equation}
	f_M{}^A = \tfrac16\,R(\omega,e)_M{}^A - \tfrac1{48}\,R(\omega,e)e_M{}^A \, ,
\end{equation}
where~$R(\omega,e)_M{}^A = R(\omega)_{MN}{}^{AB}e_B{}^N$ is the Ricci tensor and~$R(\omega,e)$ the corresponding Ricci scalar. As mentioned above, the curvature~$ R(\omega)_{MN}{}^{AB}$ reduces to the usual Riemann curvature of general relativity upon setting~$b_M = 0$.

We now combine the conformal algebra with supersymmetry. We will work with extended~$\mathcal{N}=2$ supersymmetry. The~$\mathcal{N}=2$ superconformal group in five dimensions is given by the supergroup~$\mathrm{F}^2(4)$~\cite{Freedman:2012zz}. In addition to the symmetry generators presented above, it contains two distinct type of supersymmetry generators, denoted by~$Q^i$ and~$S^i$ (where~$i = 1,2$ for~$\mathcal{N}=2$). For these generators, we have (suppressing space-time spinor indices) the usual anti-commutator for the~$Q$'s,
\begin{equation}
	\left\{Q^i,\,\bar{Q}^j\right\} = 2\,\gamma^A\,P_A\,\delta^{ij}\, ,
\end{equation}
and the S-supersymmetries close into the generator of conformal boosts:
\begin{equation}
	\left\{S^i,\,\bar{S}^j\right\} = -\gamma^A\,K_A\,\delta^{ij} \, .
\end{equation}
All commutators and anti-commutators of the~$\mathrm{F}^2(4)$ superalgebra are invariant under a $\mathrm{USp}(2) \simeq \mathrm{SU}(2)$ group, called the automorphism or~\emph{$R$-symmetry} group of the superalgebra, and one can associate a gauge field and parameter to the generator of this transformation to gauge it like all the other symmetries. This extends Table~\ref{table:conformal-alg} to Table~\ref{table:superconformal-alg}.
\begin{table}
\begin{center}
\begin{tabular}{|c|c|c|c|c|c||c|c|}
	\hline
	Generator & $P^A$ & $M^{AB}$ & $D$ & $K^A$ & $V_i{}^j$ & $Q^i$ & $S^i$ \\
	Gauge field & $e_M{}^A$ & $\omega_M{}^{AB}$ & $b_M$ & $f_M{}^A$ & $V_{M\,i}{}^j$ & $\psi_M{}^i$ & $\phi_M{}^i$  \\
	Parameter & $\xi^A$ & $\epsilon^{AB}$ & $\Lambda_D$ & $\Lambda_K^A$ & $\Lambda_{SU(2)}$ & $\epsilon^i$ & $\eta^i$ \\
	\hline
\end{tabular}
\caption{The generators of the superconformal algebra, along with their associated gauge fields and parameters. Bosonic and fermionic generators are separated by a double line.} \label{table:superconformal-alg}
\end{center}
\end{table}

The five-dimensional fields of this conformal supergravity theory organize themselves into multiplets of the superconformal algebra~$\mathrm{F}^2(4)$. We distinguish between the \textit{Weyl multiplet}, which contains the graviton and its superpartner, and the \textit{matter multiplets}, which consist of additional fields living on the curved space-time whose dynamic is encoded by the Weyl multiplet. One can conformally couple these matter multiplets to the graviton multiplet to describe the full dynamics of space-time and matter. Since the theory is completely off-shell, we will also incorporate auxiliary fields into each multiplets, so that the superconformal algebra closes without the need to impose equations of motion.

Starting from the five-dimensional Weyl multiplet, we explain in the next Section how to effect the time-like dimensional reduction to four dimensions in order to obtain a Euclidean version of four-dimensional~$\mathcal{N}=2$ supergravity. We then repeat the analysis for the matter multiplets, where we consider vector and hyper multiplets.
These results will be used in subsequent Chapters to conduct localization computations related to the quantum entropy of four-dimensional supersymmetric black holes in Euclidean signature.

\section{Conformal~$\CN = 2$ supergravity in four Euclidean dimensions}
\label{sec:euclid-superg-from-dim-red}
    
The independent fields of the $5D$ Weyl multiplet consist of the
f\"unfbein $e_M{}^A$, the gravitino fields $\psi_M{}^i$, the
dilatational gauge field $b_M$, the R-symmetry gauge fields $V_{M
  i}{}^j$ (which is an anti-hermitian, traceless matrix in the
$\mathrm{SU}(2)$ indices $i,j$),  a tensor $T_{AB}$, a scalar
$D$ and a spinor $\chi^i$. All spinor fields are
symplectic Majorana spinors. Our conventions here are as in
\cite{Banerjee:2011ts}. The three gauge fields $\omega_M{}^{AB}$,
$f_M{}^A$ and $\phi_M{}^i$, associated with local Lorentz
transformations, conformal boosts and S-supersymmetry, respectively,
are not independent as will be discussed later. The infinitesimal Q, S
and K transformations of the independent fields, parametrized by
spinors $\epsilon^i$ and $\eta^i$ and a vector
$\Lambda_\mathrm{K}{}^A$, respectively, are as
follows,\footnote{
  In five dimensions we consistently use world indices $M,N,\ldots$
  and tangent space indices $A,B,\ldots$ For fields that do not carry
  such indices the distinction between $5D$ and $4D$ fields may not
  always be manifest, but it will be specified in the text whenever
  necessary.} 
\begin{align}
  \label{eq:Weyl-susy-var}
  \delta e_M{}^A =&\,  \bar\epsilon_i \gamma^A \psi_M{}^i\,,
  \nonumber\\ 
  \delta \psi_{M}{}^i  =&\, 2\,
  {\cal  D}_M \epsilon^i + \tfrac1{2}\mathrm{i}\,
  T_{AB}( 3\,\gamma^{AB}\gamma_M-\gamma_M\gamma^{AB}) \epsilon^i
  -\mathrm{i}  \gamma_M\eta^i \,, \nonumber\\ 
  \delta V_{M i}{}^j =&\, 6 \mathrm{i}\,
  \bar\epsilon_{i} \phi_{M}{}^{j}
  -16\, \bar\epsilon_{i}\gamma_M\chi^{j} -3 \mathrm{i}\,
  \bar\eta_{i}\psi_M{}^{j} + 
  \delta^i{}_j\,[-{3}\mathrm{i}\,\bar\epsilon_{k}\phi_{M}{}^{k}
  +8\,\bar\epsilon_{k}\gamma_M\chi^{k}+\ft{3}{2}\mathrm{i}\,
  \bar\eta_{k}\psi_M{}^{k}] \,, \nonumber \\
  \delta b_M =&\,
  \mathrm{i} \bar\epsilon_i\phi_M{}^i -4 \bar\epsilon_i\gamma_M
  \chi^i + \ft12\mathrm{i} \bar\eta_i\psi_M{}^i +2\,\Lambda _\mathrm{K}{\!}^A\,
  e_{MA} \,, \\
  \delta T_{AB} =&\,  \ft43 \mathrm{i}\, \bar\epsilon_i \gamma_{AB}
  \chi^i -\ft{1}{4} \mathrm{i}\, \bar\epsilon_i R_{AB}{}^i(Q)\,,
  \nonumber\\  
  \delta \chi^i =&\,  
  \ft 12 \epsilon^i D +\ft{1}{64} 
  R_{MN j}{}^{i}(V) \,\gamma^{MN} \epsilon^j 
  + \tfrac3{64}\mathrm{i}(3\, \gamma^{AB} \Slash{D}
  +\Slash{D}\gamma^{AB})T_{AB} \, \epsilon^i \nonumber\\
  &\,
  -\tfrac 3{16} T_{AB}T_{CD}\gamma^{ABCD}\epsilon^i 
  +\tfrac3{16} T_{AB}\gamma^{AB} \eta^i  \,, \nonumber\\
  \delta D =&\,
  2\, \bar\epsilon_i \Slash{D} \chi^i - 2\mathrm{i}\,
  \bar\epsilon_i  T_{AB}\,\gamma^{AB} \chi^i - \mathrm{i}
  \bar\eta_i\chi^i \,. \nonumber
\end{align}
Under local scale transformations the fields and transformation
parameters transform as indicated in Table
\ref{tab:weyl-multiplet}. The derivatives $\mathcal{D}_M$ are
covariant with respect to all the bosonic gauge symmetries with the
exception of the conformal boosts. In particular we note
\begin{equation}
  \label{eq:D-epsilon}
\mathcal{D}_{M} \epsilon^i = \big( \partial_M - \tfrac{1}{4}
\omega_M{}^{CD} \, \gamma_{CD} + \tfrac1{2} \, b_M\big)
\epsilon^i + \tfrac1{2} \,{V}_{M j}{}^i \, \epsilon^j  \,, 
\end{equation}
where the gauge fields transform under their respective gauge
transformations according to
$\delta\omega_M{}^{AB}=\mathcal{D}_M\epsilon^{AB}$, $\delta b_M=
\mathcal{D}_M\Lambda_D$ and $\delta V_{M i}{}^j= -2\mathcal{D}_M
\Lambda_i{}^j$, with $(\Lambda_i{}^j)^\ast\equiv \Lambda^i{}_j=
- \Lambda_j{}^i$. The derivatives $D_M$ are covariant with
respect to all the superconformal symmetries. 
\begin{table}
\begin{center}
\begin{tabular}{|c||ccccccc|ccc||ccc|} 
\hline 
 & &\multicolumn{8}{c}{Weyl multiplet} & &
 \multicolumn{2}{c}{parameters} & \\  \hline \hline
 field & $e_M{}^{A}$ & $\psi_M{}^i$ & $b_M$ &
 ${V}_{M\,i}{}^j$ & $T_{AB} $ & 
 $ \chi^i $ & $D$ & $\omega_{M}{}^{AB}$ & $f_M{}^A$ &$\phi_M{}^i$&
 $\epsilon^i$ & $\eta^i$  &
  \\ \hline
$w$  & $-1$ & $-\tfrac12 $ & 0 &  0 & 1 & $\tfrac{3}{2}$ & 2 & 0 &
1 & $\tfrac12 $ & $-\tfrac12$  & $\tfrac12$ & \\ \hline 
\end{tabular}
\caption{Weyl weights $w$ of the Weyl multiplet component fields and the supersymmetry transformation parameters in five space-time dimensions. \label{tab:weyl-multiplet}}
\end{center}
\end{table}

The above supersymmetry variations and the conventional
constraints involve a number of supercovariant curvature tensors,
denoted by $R(P)_{MN}{}^A$, $R(M)_{MN}{}^{AB}$, $R(D)_{MN}$,
$R(K)_{AB}{}^A$ $R(V)_{MN i}{}^j$, $R(Q)_{MN}{}^i$ and $R(S)_{MN}{}^i$
whose explicit form can be found in \cite{Banerjee:2011ts}. The conventional constraints,
\begin{align}
  \label{eq:conv-constraints-5}
  R(P)_{MN}{}^A =&\, 0\,,\nonumber \\
  \gamma^M R(Q)_{MN}{}^i =&\, 0\,,\\
  e_A{}^M\, R(M)_{MN}{}^{AB} =&\, 0 \,, \nonumber
\end{align}
determine the gauge fields $\omega_M{}^{AB}$, $f_M{}^A$ and
$\phi_M{}^i$. These constraints lead to additional
conditions on the curvatures when combined with the Bianchi
identities. In this way one can derive $R(M)_{[ABC]D} =0= R(D)_{AB}$
and the pair-exchange property $R(M)_{ABCD}=R(M)_{CDAB}$ from the
first and the third constraint.  The second constraint, which implies
also that $\gamma_{[MN} R(Q)_{PQ]}{}^i =0$, determines the curvature
$R(S)_{MN}{}^i$. We are not primarily interested in exhibiting the detailed
relation between the $5D$ and $4D$ fields, although this is an obvious
by-product of the dimensional reduction. 

The reduction to four space-time dimensions is effected by first
carrying out the standard Kaluza-Klein decompositions on the various
fields that will ensure that the resulting $4D$ fields will transform
consistently under four-dimensional diffeomorphisms. The $5D$
space-time coordinates $x^M$ are decomposed into four coordinates
$x^\mu$ and a fifth coordinate $x^{\hat 5}$. The dependence on this
fifth coordinate will be suppressed in the reduction. Likewise the
tangent-space indices $A$ decompose into the four indices $a=1,2,3,4$
and a fifth index $A=5$. In Pauli-K\"all\'en notation one of the
coordinates is imaginary so that the $5D$ space-time signature will be
a permutation of $(-++++)$. In \cite{Banerjee:2011ts} the fifth
coordinate $x^{\hat 5}$ was real, so that the reduced theory was based
on a four-dimensional Minkowskian space-time. Here, we
consider the time-like reduction where the fifth coordinate is purely
imaginary. Upon the reduction, where the dependence on the fifth
coordinate is suppressed, the resulting theory will then be based on a
four-dimensional Euclidean space. An important observation is that the
results of \cite{Banerjee:2011ts} were obtained using Pauli-K\"all\'en
conventions, which enables a direct conversion into the Euclidean theory
by an appropriate change of the reality conditions on the fields. One
simply has to include factors $\pm \mathrm{i}$ whenever dealing with
the fifth world or tangent-space component. For instance, the fifth
coordinate of $x^M$ takes the form $x^{\hat5}= \mathrm{i} x^0$, so
that the fifth component of a contravariant vector field $V^{\hat5}$
will be imaginary and can be written as $\mathrm{i} V^0$, where $V_0$
is real. For a covariant vector the fifth component $W_{\hat 5}$ will
instead be equal to $-\mathrm{i} W_0$, where $W_0$ is real. A
corresponding rule applies to tangent-space vectors.

After this general introduction we will exhibit the consequences of
the above strategy. As is standard, the vielbein field and the
dilatational gauge field are first written in special form, by means
of an appropriate local Lorentz transformation and a conformal boost
in the time direction, respectively. In obvious notation,
\begin{equation}
  \label{eq:kk-ansatz}
  e_M{}^A= \begin{pmatrix} e_\mu{}^a & \mathrm{i} B_\mu\phi^{-1} \\[4mm]
    0 & \phi^{-1}
    \end{pmatrix} \;,\qquad
    e_A{}^M= \begin{pmatrix} e_a{}^\mu & -\mathrm{i} e_a{}^\nu B_\nu \\[4mm]
    0 & \phi
    \end{pmatrix}\;,\qquad
    b_M = \begin{pmatrix} b_\mu \\[4mm]  0
  \end{pmatrix} \,.
\end{equation}
Note that the vielbein field is not real because we will keep using
the tangent-space indices $A=1,\ldots,5$.  As compared to the
space-like reduction \cite{Banerjee:2011ts}, the field $\phi$ has remained unchanged while the
Kaluza-Klein gauge field $B_\mu$ requires a factor $\mathrm{i}$ so
that it remains real.  All the fields on the right-hand side of
\eqref{eq:kk-ansatz} are now real and possible sign factors depend on
whether we have suppressed an upper coordinate $A=5$ and/or a lower
coordinate $M=\hat5$. The fields now carry only four-dimensional world
and tangent-space indices, $\mu,\nu,\ldots$ and $a,b,\ldots$, taking
four values while the components referring to the fifth direction will be
suppressed. Observe that the scaling weights for $e_M{}^A$ and
$e_\mu{}^a$ are equal to $w=-1$, while for $\phi$ we have $w=1$. The
fields $b_M$, $b_\mu$ and $B_\mu$ carry weight $w=0$. 

For the fermions there is no need to introduce a new notation,
because the spinors have an equal number of components in five and
four space-time dimensions. In five dimensions, we employ symplectic Majorana spinors~$\chi^i$ with~$i=1,2$ subject to the reality constraint,\footnote{\label{foot:charge-conj-5D} 
  The charge conjugation matrix $C$ has the properties $C\gamma_A
  C^{-1} = \gamma_A{}^{\rm T}$, with $C^{\rm T} = -C$ and $ C^\dagger
  = C^{-1}$. The $5D$ gamma matrices in Pauli-K\"all\'en notation
  satisfy $\gamma_{ABCDE} = {\bf 1}\,\varepsilon_{ABCDE}$.
} 
\be
\label{eq:sympl-Majo-5D}
C^{-1}\bar{\chi}_i{}^\mathrm{T} = \varepsilon_{ij}\chi^j \, ,
\ee
where the Dirac conjugate is defined as~$\bar{\chi} = \chi^\dagger\gamma^5$. Observe that we adhere to the convention according to which raising or
lowering of $\mathrm{SU}(2)$ indices is effected by complex
conjugation. For fermionic bilinears, with five-dimensional spinor fields $\psi^i$ and
$\varphi^i$ and a spinor matrix $\Gamma$ constructed from products of
gamma matrices, we note the following result,
\begin{equation}
  \label{eq:bilinear}
  (\bar\varphi_j\gamma^5\Gamma^{\dagger}\gamma^5\psi^i)^\dagger = \bar\psi_i\,\Gamma\,\varphi^j= - \delta_i{}^j \,\bar\varphi_k\, C^{-1}\, \Gamma^{\rm T}\,C\, \psi^k   + \bar\varphi_i\,C^{-1}\, \Gamma^{\rm T}\,C\, \psi^j\,. 
\end{equation}
Hence the bilinears $O_i{}^j$ equal to $\mathrm{i}\,\bar \psi_i\,
\varphi^j$, $\bar \psi_i\gamma_A \varphi^j$ and $\mathrm{i}\,\bar
\psi_i\gamma_{AB} \varphi^j$ are pseudo-hermitean: $O^i{}_j=
\varepsilon^{ik}\varepsilon_{jl} \,O_k{}^l$ 
(provided $A,B,\ldots=
1,\ldots,4$; in Pauli-K\"all\'en convention $\gamma^5$ acquires an
additional minus sign which is related to the definition of the Dirac
conjugate). 
In the context of the spinors special care is required in
converting from Minkoswki to Euclidean signature, because (Fierz)
reordering of the spinors depends sensitively on whether the spinor is
a Majorana or an anti-Majorana field. Observe that the gravitino field
$\psi_{\hat 5}$ with its world index in the fifth directions will be
an anti-Majorana field. This will be properly accounted for in the
Kaluza-Klein ans\"atze, which will include the proper factors of the
imaginary unit, as can be seen in Appendix~\ref{app:sugra}.


\subsection{Off-shell dimensional reduction: the Weyl multiplet}
\label{sec:euclidean-sg-4D}
Here we summarize the results for the superconformal transformation rules in $4D$ Euclidean supergravity which are obtained in Appendix \ref{app:sugra}. 
%
%
We present the transformation rules of the superconformal fields, as well as the covariant curvatures which are needed. We refrain from presenting the full list of superconformal curvatures and the identities one can derive for them, as these will not be needed later on. We do refer the interested reader to the original publication~\cite{BdW-Me} where more details are given. 
Note that in contrast to the Minkowski case where four-dimensional~$\CN=2$ conformal supergravity has a~$\mathrm{SU}(2)\times\mathrm{U}(1)$ R-symmetry, the Euclidean theory has a non-compact~$\mathrm{SU}(2)\times\mathrm{SO}(1,1)$ R-symmetry as exhibited in Appendix~\ref{app:sugra}. The conventions used for four-dimensional spinors are given in Appendix~\ref{app:sugra} as well, and in particular they satisfy the symplectic Majorana reality condition~\eqref{eq:sympl-Majo-4D}.
The Weyl and chiral weights of the independent fields of the Weyl multiplet have been collected in Table \ref{table:weyl}. 
%
\begin{table}
\begin{center}
\begin{tabular}
    {|c||cccccccc|ccc||ccc|}
\hline
 & &\multicolumn{9}{c}{Weyl multiplet} & &
 \multicolumn{2}{c}{parameters} & \\ \hline \hline
 field & $e_\mu{}^{a}$ & $\psi_\mu{}^i$ & $b_\mu$ & $A_\mu$ &
 $\mathcal{V}_\mu{}^i{}_j$ & $T_{ab}^\pm $ &
 $ \chi^i $ & $D$ & $\omega_\mu^{ab}$ & $f_\mu{}^a$ & $\phi_\mu{}^i$ &
 $\epsilon^i$ & $\eta^i$
 & \\[.5mm] \hline
$w$  & $-1$ & $-\tfrac12 $ & 0 &  0 & 0 & 1 & $\tfrac{3}{2}$ & 2 & 0 &
1 & $\tfrac12 $ & $ -\tfrac12 $  & $ \tfrac12  $ & \\[.5mm] \hline
$c$  & $0$ & $-\tfrac12 $ & 0 &  0 & 0 & $\pm 1$ & $-\tfrac{1}{2}$ & 0 &
0 & 0 & $-\tfrac12 $ & $ -\tfrac12 $  & $ -\tfrac12  $ & \\[.5mm] \hline
 $\gamma_5$   &  & + &   &    &   &   & + &  &  &  & $-$ & $ + $  & $
 -  $ & \\ \hline
\end{tabular}

\caption{
Weyl and chiral weights ($w$ and $c$) and fermion
chirality $(\gamma_5)$ of the Weyl multiplet fields and the transformation parameters in four space-time dimensions.\label{table:weyl}}
\end{center}
\end{table}

Under Q-supersymmetry, S-supersymmetry and special conformal transformations the independent fields of the Weyl multiplet transform as follows,
\begin{align}
  \label{eq:weyl-multiplet}
  \delta e_\mu{}^a =&\, \bar{\epsilon}_i\gamma^5\gamma^a\psi_\mu{}^i \, , \nonumber\\
  \delta\psi_\mu{}^i =&\, 2\,\mathcal{D}_\mu\epsilon^i + \tfrac1{16}\mathrm{i}\,(T^+_{ab} + T^-_{ab})\gamma^{ab}\gamma_\mu\epsilon^i - \mathrm{i}\,\gamma_\mu\eta^i \, , \nonumber\\
  \delta b_\mu =&\, \tfrac12\mathrm{i}\,\bar{\epsilon}_i\gamma^5\phi_\mu{}^i - \tfrac34\,\bar{\epsilon}_i\gamma^5\gamma_\mu\chi^i + \tfrac12\mathrm{i}\,\bar{\eta}_i\gamma^5\psi_\mu{}^i + \Lambda_K{}^a e_{\mu a}\, , \nonumber \\
  \delta A_\mu =&\, -\tfrac12\mathrm{i}\,\bar{\epsilon}_i\phi_\mu{}^i - \tfrac34\,\bar{\epsilon}_i\gamma_\mu\chi^i - \tfrac12\mathrm{i}\,\bar{\eta}_i\psi_\mu{}^i \, , \nonumber \\
  \delta \mathcal{V}_\mu{}^i{}_j =&\, 2\mathrm{i}\,\bar{\epsilon}_j\gamma^5\phi_\mu{}^i - 3\,\bar{\epsilon}_j\gamma^5\gamma_\mu\chi^i - 2\mathrm{i}\,\bar{\eta}_j\gamma^5\psi_\mu{}^i \\
  &\,-\tfrac12\delta^i{}_j\bigl(2\mathrm{i}\,\bar{\epsilon}_k\gamma^5\phi_\mu{}^k - 3\,\bar{\epsilon}_k\gamma^5\gamma_\mu\chi^k - 2\mathrm{i}\,\bar{\eta}_k\gamma^5\psi_\mu{}^k \bigr) \, , \nonumber \\
  \delta T^\pm_{ab} =&\, -8\mathrm{i}\,\bar{\epsilon}_i \gamma^5 R(Q)^\pm_{ab}{}^i \, , \nonumber \\
  \delta\chi^i =&\, \tfrac1{24}\mathrm{i}\gamma^{ab}\Slash{D}(T^+_{ab} + T^-_{ab})\epsilon^i + \tfrac16 R(\mathcal{V})_{ab}{}^i{}_j\gamma^{ab}\epsilon^j - \tfrac13 R(A)_{ab}\gamma^{ab}\gamma^5\epsilon^i + D\epsilon^i \nonumber \\
  &\,+ \tfrac1{24}(T^+_{ab} + T^-_{ab})\gamma^{ab}\eta^i \, , \nonumber \\
  \delta D =&\, \bar{\epsilon}_i\gamma^5\Slash{D}\chi^i \, . \nonumber
\end{align}
Here $\epsilon^i$ denotes the symplectic Majorana parameter of Q-supersymmetry, $\eta^i$ the symplectic Majorana parameter of S-supersymmetry, and $\Lambda_K{}^a$ is the transformation parameter for special conformal boosts. 
The full superconformal covariant derivative is denoted by $D_\mu$, while $\mathcal{D}_\mu$ denotes a covariant derivative with respect to Lorentz, dilatation, chiral $\mathrm{SO}(1,1)$, and $\mathrm{SU}(2)$ transformations. In particular,
\begin{align}
  \label{eq:D-epslon}
  \mathcal{D}_{\mu} \epsilon^i = \big(\partial_\mu - \tfrac{1}{4}\omega_\mu{}^{ab} \, \gamma_{ab} + \tfrac1{2} \, b_\mu + \tfrac{1}{2}\, A_\mu\gamma^5\big) \epsilon^i + \tfrac1{2} \, \mathcal{V}_{\mu}{}^i{}_j \, \epsilon^j \, .
\end{align}
We will also need the covariant curvatures of certain gauge symmetries, which take the following form,
\begin{align}
  \label{eq:curvatures}
  R(P)_{\mu \nu}{}^a  =&\, 2 \, \partial_{[\mu} \, e_{\nu]}{}^a + 2 \, b_{[\mu} \, e_{\nu]}{}^a -2 \, \omega_{[\mu}{}^{ab} \, e_{\nu]b} - \tfrac1{2}\,\bar{\psi}_{i[\mu}\gamma^5\gamma^a\psi_{\nu]}{}^i \, , \nonumber\\[.2ex]
  R(Q)_{\mu \nu}{}^i = & \, 2 \, \mathcal{D}_{[\mu} \psi_{\nu]}{}^i - \mathrm{i}\,\gamma_{[\mu} \phi_{\nu]}{}^i + \tfrac{1}{16}\mathrm{i}\,(T_{ab}^+ + T_{ab}^-) \, \gamma^{ab}\gamma_{[\mu} \psi_{\nu]}{}^i \, , \nonumber\\[.2ex]
  R(D)_{\mu \nu} = & \,2\,\partial_{[\mu} b_{\nu]} - 2\,f_{[\mu}{}^a e_{\nu]a} - \tfrac{1}{2}\mathrm{i}\,\bar{\psi}_{i[\mu}\gamma^5\phi_{\nu]}{}^i + \tfrac{3}{4}\,\bar{\psi}_{i[\mu}\gamma^5\gamma_{\nu]} \chi^i \, ,  \nonumber\\[.2ex]
  R(A)_{\mu \nu} = &\, 2 \, \partial_{[\mu} A_{\nu]} + \tfrac12\mathrm{i}\,\bar{\psi}_{i[\mu}{}\phi_{\nu]}{}^i + \tfrac34\,\bar{\psi}_{i[\mu}\gamma_{\nu]}\chi^i \, , \nonumber\\[.2ex]
  R(\mathcal{V})_{\mu \nu}{}^i{}_j =& \, 2\, \partial_{[\mu}\mathcal{V}_{\nu]}{}^i{}_j + \mathcal{V}_{[\mu}{}^i{}_k \, \mathcal{V}_{\nu]}{}^k{}_j  \\
  &\,-  2\mathrm{i}\,\bar{\psi}_{j[\mu}\gamma^5\phi_{\nu]}{}^i + 3\,\bar{\psi}_{j[\mu}\gamma^5\gamma_{\nu]} \chi^i + \tfrac12\delta^i{}_j\bigl(2\mathrm{i}\,\bar{\psi}_{k[\mu}\gamma^5\phi_{\nu]}{}^k -3\,\bar{\psi}_{k[\mu}\gamma^5\gamma_{\nu]} \chi^k \bigr) \, , \nonumber \\[.2ex]
  R(M)_{\mu \nu}{}^{ab} =& \, 2 \,\partial_{[\mu} \omega_{\nu]}{}^{ab} - 2\, \omega_{[\mu}{}^{ac}\omega_{\nu]c}{}^b - 4 f_{[\mu}{}^{[a} e_{\nu]}{}^{b]} + \tfrac12\mathrm{i}\,\bar{\psi}_{i[\mu}\gamma^5\phi_\nu{}^i \nonumber\\[.2ex]
  &\, -\tfrac18\mathrm{i}\,\bar{\psi}_{\mu\,i}\gamma^5\psi_\nu{}^i\,(T^{ab+} + T^{ab-}) - \tfrac34\bar{\psi}_{i[\mu}\gamma^5\gamma_{\nu]}\gamma^{ab}\chi^i - \bar{\psi}_{i[\mu}\gamma^5\gamma_{\nu]}R(Q)^{ab\,i} \, . \nonumber
\end{align}
The theory includes three conventional constraints (which have already been incorporated in \eqref{eq:curvatures}), given by
\begin{align}
  \label{eq:conv-constraints}
  &R(P)_{\mu \nu}{}^a =  0 \, , \nonumber \\[1mm]
  &\gamma^\mu R(Q)_{\mu \nu}{}^i + \tfrac32 \gamma_{\nu}
  \chi^i = 0 \, , \\[1mm]
  &
  e^{\nu}{}_b \,R(M)_{\mu \nu a}{}^b - \wt{R}(A)_{\mu a} + \tfrac1{16} T_{ab}^+\,T_\mu^{-\;b} - \tfrac{3}{2}\, D \, e_{\mu a} = 0
  \, . \nonumber
\end{align}
These constraints are S-invariant, and they determine the fields $\omega_{\mu}{}^{ab}$, $\phi_{\mu}{}^i$ and $f_{\mu}{}^a$ in terms of the independent fields of the Weyl multiplet. For instance,
\begin{align}
  \label{eq:dependent}
  \omega_\mu{}^{ab} =&\, -2\,e^{\nu[a}\partial_{[\mu}e_{\nu]}{}^{b]}
     -e^{\nu[a}e^{b]\sigma}e_{\mu c}\,\partial_\sigma e_\nu{}^c
     -2\,e_\mu{}^{[a}e^{b]\nu}b_\nu  \nonumber \\
     &\,-\ft{1}{4}\bigl(2\,\bar{\psi}_{\mu\,i}\gamma^5\gamma^{[a}\psi^{b]i} + \bar{\psi}^a{}_i\gamma^5\gamma_\mu\psi^{b\,i}\bigr) \, , \\
     \phi_{\mu}{}^i =& \, -\tfrac12\mathrm{i}\left( \gamma^{\rho \sigma} \gamma_\mu - \tfrac{1}{3} \gamma_\mu \gamma^{\rho \sigma} \right) \left(\mathcal{D}_\rho \psi_{\sigma}{}^i + \tfrac{1}{32}\mathrm{i}\,(T_{ab}^+ + T_{ab}^-) \gamma^{ab} \gamma_\rho\psi_{\sigma}{}^i + \tfrac{1}{4} \gamma_{\rho \sigma} \chi^i \right) \, . \nonumber
\end{align}
We will also need the bosonic part of the expression for the
uncontracted connection $f_\mu{}^a$,
\begin{equation}
  \label{eq:f-bos-uncon}
  f_\mu{}^a = \tfrac12\,R(\omega,e)_\mu{}^a - \tfrac14\,\bigl(D+\tfrac13 R(\omega,e)\bigr) e_\mu{}^a - \tfrac12\,\wt{R}(A)_\mu{}^a - \tfrac1{32}\,T_{\mu b}^-\,T^{+\,ba} \,,
\end{equation}
where $R(\omega,e)_\mu{}^a= R(\omega)_{\mu\nu}{}^{ab} e_b{}^\nu$ is
the non-symmetric Ricci tensor, and $R(\omega,e)$ the corresponding
Ricci scalar. The curvature $R(\omega)_{\mu\nu}{}^{ab}$ is associated
with the spin connection field $\omega_\mu{}^{ab}$, given in
(\ref{eq:dependent}).

The transformations of $\omega_{\mu}{}^{ab}$, $\phi_{\mu}{}^i$ and
$f_{\mu}{}^a$ are induced by the constraints \eqref{eq:conv-constraints}. We refrain from presenting their explicit expressions, as they will not be needed in what follows. They can be found in~\cite{BdW-Me}.
\subsection{Electric-magnetic duality}
\label{sec:electr-magn-dual}

We now briefly discuss how electric-magnetic duality in the Euclidean theory is the same as in the Minkowski theory~\cite{Gaillard:1981rj,deWit:2001pz}.

Start from a Lagrangian $L(F)$ which depends on $n$ abelian selfdual and
anti-selfdual field strengths $F_{\mu\nu}{\!}^{I\pm}$ (but not on
their derivatives) and possibly on other fields. The field equations
are defined in terms of the tensors 
\begin{equation}
  \label{eq:dual-G}
  G^{\mu\nu}{\!}_I{}^\pm = 2\, \frac{\partial L}{\partial F_{\mu\nu}{\!}^{I\pm}} \;. 
\end{equation}
The corresponding Bianchi identities and equations of motion then take
a functionally similar form, 
\begin{equation}
  \label{eq:bianch-eom}
 \partial^\mu\big( F_{\mu\nu}{\!}^{I+} - F_{\mu\nu}{\!}^{I-}\big)  = 0
 =  \partial^\mu\big( G_{\mu\nu  I}{\!}^+ + G_{\mu\nu  I}{\!}^-\big)
 \,. 
\end{equation}
Obviously these equations are invariant under the electric-magnetic
duality transformations,
\begin{equation}
  \label{eq:emdual-rot}
  \begin{pmatrix} 
     F_{\mu\nu}{\!}^{I\pm}\\[2mm]
     G_{\mu\nu  J}{\!}^\pm 
    \end{pmatrix}
    \longrightarrow  
    \begin{pmatrix} 
     \tilde F_{\mu\nu}{\!}^{I\pm}\\[2mm]
    \tilde G_{\mu\nu  J}{\!}^\pm 
    \end{pmatrix}
    = 
  \begin{pmatrix}
   U^I{}_K&\pm Z^{IL}\\[2mm]
   \pm W_{JK}& V_J{}^L
   \end{pmatrix}
   \begin{pmatrix} 
     F_{\mu\nu}{\!}^{K\pm}\\[2mm]
     G_{\mu\nu  L}{\!}^\pm 
    \end{pmatrix}\,,
\end{equation}
where $\tilde F_{\mu\nu}{\!}^{I\pm}$, and $\tilde G_{\mu\nu
  J}{\!}^\pm$ denote the transformed field strengths (and not the
Hodge dual). Here the $n\times n$ submatrices $U^I{}_K$, $Z^{IL}$, $
W_{JK}$ and $V_J{}^L$ are real. The question is now whether the rotated
tensors $\tilde G_{\mu\nu J}{\!}^\pm$ can again follow from a new
Lagrangian $\tilde L(\tilde F)$ in analogy with \eqref{eq:dual-G}. In
that case there may be a different Lagrangian that leads to an
equivalent set of Bianchi identities and equations of motion. 
As it turns out, this imposes the following restriction on the matrices in \eqref{eq:emdual-rot}, 
\begin{align}
  \label{eq:sympl-matrix}
  &U^\mathrm{T} V - W^\mathrm{T} Z = VU^\mathrm{T} - W Z^\mathrm{T} =
  \oneone\,, \nonumber\\
  & U^\mathrm{T} W = W^\mathrm{T} U\,,\qquad Z^\mathrm{T} V=
  V^\mathrm{T} Z\,. 
\end{align}
which are equivalent to
\begin{equation}
  \label{eq:sympl-prop-O}
  \begin{pmatrix} U& \pm Z\\[2mm] \pm W &V\end{pmatrix} 
\begin{pmatrix} 0&-\oneone\\[2mm] \oneone &0 \end{pmatrix}  
  \begin{pmatrix} U& \pm Z\\[2mm] \pm W
    &V\end{pmatrix}^{\!\!\mathrm{T}}  =  
\begin{pmatrix} 0&-\oneone\\[2mm] \oneone &0 \end{pmatrix} \\,
\end{equation}
Hence the electric-magnetic dualities form a group of equivalence
transformations that connect different Lagrangians which describe the
same physics since the equations of motion and the Bianchies identities are the same. As is clear from~\eqref{eq:sympl-matrix}, this group is
$\mathrm{Sp}(2n;\mathbb{R})$, the same group as in the Minkowski theory \cite{Gaillard:1981rj,deWit:2001pz}.

Note that the above results follow from the observation that the two Lagrangians
$\tilde L$ and $L$ are related by
\be
  \label{eq:F-versus-Ftilde}
  \tilde L(\tilde F) - \tfrac14 \tilde F_{\mu\nu}{\!}^{I+}\, \tilde
  G^{\mu\nu}{\!}_I {\!}^+ - \tfrac14 \tilde F_{\mu\nu}{\!}^{I-}\, \tilde
  G^{\mu\nu}{\!}_I {\!}^-
  = L(F) - \tfrac14  F_{\mu\nu}{\!}^{I+}\, 
  G^{\mu\nu}{\!}_I {\!}^+ - \tfrac14  F_{\mu\nu}{\!}^{I-}\, 
  G^{\mu\nu}{\!}_I {\!}^-\,,
\ee
up to terms that are independent of $ F_{\mu\nu}{\!}^{I\pm}$

\subsection{Off-shell dimensional reduction; matter multiplets}
\label{sec:shell-dimens-reduct-matter}
In this Section we repeat the same analysis as in Section~\ref{sec:euclidean-sg-4D}, but now applied to the vector multiplet and the
hypermultiplet. We refrain from presenting similar results for tensor
multiplets. They can be derived by the same method, or, alternatively,
they can be found by considering a composite tensor multiplet
constructed from the square of a vector multiplet.

In five space-time dimensions the vector supermultiplet consists of a
real scalar $\sigma$, a gauge field $W_\mu$, a triplet of (auxiliary)
fields $Y^{ij}$, and a fermion field $\Omega^i$. Under Q- and
S-supersymmetry these fields transform as follows~\cite{Banerjee:2011ts},
\begin{align}
  \label{eq:sc-vector-multiplet}
  \delta \sigma =&\,
  \mathrm{i}\,\bar{\epsilon}_i\Omega^i \,,
  \nonumber \\ 
  \delta\Omega^i =&\,
  - \ft12 (\hat{F}_{AB}- 4\,\sigma T_{AB}) \gamma^{AB} \epsilon^i
  -\mathrm{i} \Slash{D} \sigma\epsilon^i -2\varepsilon_{jk}\,
  Y^{ij} \epsilon^k 
  + \sigma\,\eta^i \,,   \nonumber\\ 
  \delta W_M =&\,
  \bar{\epsilon}_i\gamma_M\Omega^i - \mathrm{i}
  \sigma \,\bar\epsilon_i \psi_M{}^i  \,, \nonumber\\ 
  \delta Y^{ij}=&\,  
   \varepsilon^{k(i}\, \bar{\epsilon}_k \Slash{D} \Omega^{j)} 
  + 2{\mathrm{i}}\varepsilon^{k(i}\, \bar\epsilon_k (-\ft1{4}
  T_{AB}\gamma^{AB}\Omega^{j)}+ 4 \sigma \chi^{j)})
  -\ft1{2}{\mathrm{i}}  \varepsilon^{k(i}\, \bar{\eta}_k
  \Omega^{j)} \,.  
\end{align}
where $(Y^{ij})^\ast\equiv Y_{ij}= \varepsilon_{ik}\varepsilon_{jl}
Y^{kl}$, and the supercovariant field strength is defined as, 
\begin{equation}
  \label{eq:W-field-strength}
 \hat F_{MN}(W) = 2\, \partial_{[M} W_{N]}  -
 \bar\Omega_i\gamma_{[M} \psi_{N]}{}^i +\ft12
 \mathrm{i}\sigma\,\bar\psi_{[M i} \psi_{N]}{}^i \,.
\end{equation}
The fields behave under local scale transformations according to the
weights shown in Table~\ref{tab:w-weights-matter-5D}. 

%
\begin{table}
\begin{center}
\begin{tabular}{|c||cccc||cc|} 
\hline 
 &  \multicolumn{4}{c|}{vector multiplet}&\multicolumn{2}{c|} {hypermultiplet} \\ \hline \hline 
 field & $\sigma$ & $W_\mu$ & $\Omega_i$  & $Y_{ij}$ &$A_i{}^\alpha$&
 $\zeta^\alpha$  \\[.4mm] \hline 
$w$  & $1$ & $0$ &$\tfrac{3}{2}$ & 2 & $\tfrac{3}{2}$ & $2$ \\[.4mm] \hline

\end{tabular}
\caption{Weyl weights $w$ of the
    vector multiplet and the hypermultiplet component fields in five
    space-time dimensions. \label{tab:w-weights-matter-5D}}
\end{center}   
\end{table}

The reduction proceeds in the same way as before, except that we now have the advantage that some of the $4D$
fields belonging to the $4D$ Weyl multiplet have already been identified. We decompose the $5D$
gauge field $W_M$ into a four-dimensional gauge field $W_\mu$ and a
scalar $-\mathrm{i}\,W= W_{\hat{5}}$ by using the standard Kaluza-Klein ansatz, and write
the Q- and S-transformation rules, including the compensating Lorentz
transformation \eqref{eq:comp-Lor}. Just as in
\eqref{eq:compensating-chiral-tr} we introduce an R-covariant spinor
field field,
\begin{equation}
  \label{eq:new-Omega}
  (\Omega^i + W\,\hat{\psi}^i)\big|^\mathrm{Rcov}= \exp[-\tfrac12
  \varphi\,\gamma^5]\, (\Omega^i +  W\,\hat{\psi}^i)\,,
\end{equation}
which transforms under $\mathrm{SO}(1,1)$. In terms of the chiral R-covariant spinor
fields, we derive the following transformation rules,
\begin{align}
  \label{eq:vector}
  \delta\big[\mathrm{e}^{\mp\varphi} (\sigma \pm \phi W)\big] =&\, \pm\mathrm{i}\,\bar\epsilon_i(1\pm\gamma^5) \big(\Omega^i + W\,\hat{\psi}^i\big)\, , \nonumber \\  
  \delta W_\mu=&\, \tfrac12\,\bar\epsilon_i\big[\gamma_\mu(1-\gamma^5)(\Omega^i+W\,\hat{\psi}^i) -\mathrm{i} (\sigma-\phi W)\mathrm{e}^{\varphi}(1+\gamma^5) \psi_{\mu}{}^i\big] \nonumber\\ 
  &- \tfrac12\,\bar\epsilon_i\big[\gamma_\mu(1+\gamma^5)(\Omega^i + W\,\hat{\psi}^i) - \mathrm{i} (\sigma + \phi W)\mathrm{e}^{-\varphi}(1-\gamma^5)\psi_{\mu}{}^i\big] \, , \nonumber \\ 
  \delta(1\pm\gamma^5)\big(\Omega^i + W\,\hat{\psi}^i\big) =&\, -\tfrac12\big[\hat F(W)_{ab}-\tfrac18(\sigma\mp\phi W) \,\hat T_{ab} \big] \gamma^{ab}(1\pm\gamma^5)\epsilon^i \\ 
  &- \mathrm{i}\,\Slash{D}\big[(\sigma \pm\phi W)\mathrm{e}^{\mp\varphi}\big](1\mp\gamma^5)\epsilon^i - 2 \hat Y^{ik}\varepsilon_{kj}(1\pm\gamma^5)\epsilon^j \nonumber \\
  &+ (\sigma\pm\phi W)\mathrm{e}^{\mp\varphi}(1\pm\gamma^5)\eta^i  \,, \nonumber
\end{align}
where $\hat Y^{ij}$ is defined by:
\begin{equation}
  \label{eq:hat-Y}
  \hat Y^{ij} = Y^{ij} - \tfrac12  W\,\hat{V}_k{}^{i} \, \varepsilon^{jk} +
  \tfrac12\mathrm{i}\,\phi^{-1}\,(\bar\Omega_k\gamma^5-\tfrac{1}{2}\,\sigma
  \phi^{-1}\, \bar{\hat{\psi}}_k)\hat\psi^{(i} \, \varepsilon^{j)k} \, .
\end{equation}
Note that in \eqref{eq:vector}, we have again
suppressed the field-dependent S-supersymmetry and $\mathrm{SU}(2)$
R-symmetry transformations given in~\eqref{eq:Q-susy decom-par}.

At this stage, we can cast the transformation rules~\eqref{eq:vector} in a form which we will use in the following. This is done by first making the following field redefinitions:
\begin{align}
(\sigma + \phi W)e^{-\varphi} :=&\, 2\,X \, , \qquad (\sigma - \phi W)e^{\varphi} := 2\,\bar{X} \, , \nonumber \\
(\Omega^i + W\hat{\psi}^i) :=&\, \lambda^i \, , \qquad \qquad \qquad \;\;\; \hat{Y}_{ij} := \tfrac12 Y_{ij} \, .
\end{align}
In terms of the newly introduced fields,~\eqref{eq:vector} takes the form
\begin{align}
&\delta X = \mathrm{i}\,\bar{\epsilon}_i P_+ \lambda^i \, , \cr 
&\delta \bar{X} = -\mathrm{i}\,\bar{\epsilon}_i P_- \lambda^i \, ,\cr
&\delta W_\mu = \bar{\epsilon}_i\gamma_\mu P_- \lambda^i - \bar{\epsilon}_i \gamma_\mu P_+ \lambda^i + 2\mathrm{i}\bigl[X\,\bar{\epsilon}_i P_- \psi_{\mu}{}^i - \bar{X}\,\bar{\epsilon}_i P_+ \psi_{\mu}{}^i\bigr] \, , \\
&\delta(P_+\lambda^i) = -\tfrac12\bigl[\hat{F}(W)_{ab} - \tfrac14\bar{X}T_{ab}\bigr]\gamma^{ab}P_+\epsilon^i - 2\mathrm{i}\,\Slash{D}X\,P_-\epsilon^i - Y^i{}_j P_+ \epsilon^j + 2\,X P_+ \eta^i \, , \cr
&\delta(P_-\lambda^i) = -\tfrac12\bigl[\hat{F}(W)_{ab} - \tfrac14 X T_{ab}\bigr]\gamma^{ab} P_-\epsilon^i - 2\mathrm{i}\,\Slash{D}\bar{X}\,P_+\epsilon^i - Y^i{}_j P_- \epsilon^j + 2\,\bar{X} P_-\eta^i \, , \nonumber \\
&\delta Y^i{}_j = -2\,\bar{\epsilon}_j \gamma^5 \Slash{D}\lambda^i + \delta_j^i\,\bar{\epsilon}_k\gamma^5\Slash{D}\lambda^k \, , \nonumber
\end{align}
where we have introduced the short-hand notation for the Weyl projectors
\be
\label{eq:weyl-proj}
P_\pm = \tfrac12(1\pm\gamma^5) \, .
\ee
We now make additional field redefinitions. This is done so as to match the conventions which were chosen in the original works~\cite{Murthy:2013xpa, Murthy:2015yfa} forming the basis of the next Chapters, since the analysis conducted there (based on previous works like~\cite{Dabholkar:2010uh}) specifically uses an imaginary~$T$ tensor, a complex vector field~$W_\mu$ and complex scalar fields~$X,\,\bar{X}$. This is not the most natural choice from the point of view of the dimensional reduction, where the reality conditions of fields are inherited from five dimensions, but it is still a consistent set of conventions. We should point out that this choice is an artefact of using a Wick-rotated and analytically continued Minkowski theory as the basis for the Euclidean theory, a procedure which relies on a number of prescriptions and convention choices. Thanks to the results of this Chapter we can now avoid using such a procedure and work directly in the Euclidean theory that was derived from five dimensions, albeit after making simple field redefinitions to match the conventions and reality conditions on fields used in the Wick-rotated theory. To implement this, we make the following replacements:
\be
\label{eq:field-redefs}
T_{ab}^- \rightarrow -\mathrm{i} T^-_{ab} \, , \quad T_{ab}^+ \rightarrow \mathrm{i} T_{ab}^+ \, , \quad X \rightarrow \mathrm{i}X \, , \quad \bar{X} \rightarrow -\mathrm{i}\bar{X} \, , \quad W_\mu \rightarrow -W_\mu \, .
\ee
With these changes, the transformations under Q- and S-supersymmetry of the fermions in the Weyl multiplet and of the vector multiplet fields now take the form
\begin{align}
\label{eq:weyl-fermions-good-vars}
\delta(P_\pm\psi_\mu{}^i) =&\, 2\,\mathcal{D}_\mu(P_\pm\epsilon^i) \pm \tfrac1{16}(T^+_{ab} + T^-_{ab})\gamma^{ab}\gamma_\mu P_\mp\epsilon^i - \mathrm{i}\,\gamma_\mu P_\mp \eta^i \, , \\
\delta(P_\pm\chi^i) =&\, \pm\tfrac1{24}\,\gamma^{ab}\Slash{D}(T^+_{ab} + T^-_{ab}) P_\mp \epsilon^i + \tfrac16 R(\mathcal{V})_{ab}{}^i{}_j\gamma^{ab}P_\pm\epsilon^j \mp \tfrac13 R(A)_{ab}\gamma^{ab}P_\pm\epsilon^i \cr
&\,+ D P_\pm\epsilon^i \mp \tfrac1{24}\mathrm{i}\,(T^+_{ab} + T^-_{ab})\gamma^{ab}P_\pm\eta^i \, , \nonumber
\end{align}
and
\begin{align}
\label{eq:vector-good-vars}
\delta X =&\, \bar{\epsilon}_i P_+ \lambda^i \, , \cr
\delta\bar{X} =&\, \bar{\epsilon}_i P_- \lambda^i \, , \cr
\delta W_\mu =&\, \bar{\epsilon}_i \gamma_\mu P_+ \lambda^i  - \bar{\epsilon}_i\gamma_\mu P_- \lambda^i + 2\bigl[X\,\bar{\epsilon}_i P_- \psi_{\mu}{}^i + \bar{X}\,\bar{\epsilon}_i P_+ \psi_{\mu}{}^i\bigr] \, , \\
\delta(P_+\lambda^i) =&\, \tfrac12\bigl[\hat{F}(W)_{ab} - \tfrac14\bar{X}T_{ab}\bigr]\gamma^{ab}P_+\epsilon^i + 2\Slash{D}X P_-\epsilon^i - Y^i{}_j P_+ \epsilon^j + 2\mathrm{i}\,X P_+\eta^i \, , \cr
\delta(P_-\lambda^i) =&\, \tfrac12\bigl[\hat{F}(W)_{ab} - \tfrac14 X T_{ab}\bigr]\gamma^{ab}P_-\epsilon^i - 2\Slash{D}\bar{X}P_+\epsilon^i - Y^i{}_j P_- \epsilon^j - 2\mathrm{i}\,\bar{X}P_-\eta^i \, , \cr
\delta Y^i{}_j =&\, -2\,\bar{\epsilon}_j \gamma^5 \Slash{D}\lambda^i + \delta_j^i\,\bar{\epsilon}_k\gamma^5\Slash{D}\lambda^k \, . \nonumber
\end{align}
These two sets of transformation rules will be extensively used in the following Chapters. We do not present the details of the other transformation rules after the redefinitions~\eqref{eq:field-redefs}, as they will not be needed in what follows.\\

Hypermultiplets are associated with target spaces of dimension $4r$
that are hyperk\"ahler cones \cite{deWit:1999fp}. The five-dimensional supersymmetry
transformations are most conveniently written in terms of the sections
$A_i{}^\alpha(\phi)$, where $\alpha= 1,2,\ldots,2r$,
\begin{align} 
  \label{eq:hypertransf}
  \delta A_i{}^\alpha=&\, 2\mathrm{i}\,\bar\epsilon_i\zeta^\alpha\,,
  \nonumber\\ 
  \delta\zeta^\alpha =&\, - \mathrm{i}\Slash{D}
  A_i{}^\alpha\epsilon^i 
  + \tfrac3{2} A_i{}^\alpha\eta^i \,.
\end{align}
The $A_i{}^\alpha$ are the local sections of an
$\mathrm{Sp}(r)\times\mathrm{Sp}(1)$ bundle. We also note the
existence of a covariantly constant skew-symmetric tensor
$\Omega_{\alpha\beta}$ (and its complex conjugate
$\Omega^{\alpha\beta}$ satisfying
$\Omega^{\beta\gamma}\Omega_{\gamma\alpha}= -\delta_\alpha{}^\beta$),
and the symplectic Majorana condition for the spinors reads as
$C^{-1}\bar\zeta_\alpha{}^\mathrm{T} = \Omega_{\alpha\beta}
\,\zeta^\beta$. Covariant derivatives contain the $\mathrm{Sp}(r)$
connection $\Gamma_A{}^\alpha{}_\beta$, associated with rotations of
the fermions. The sections $A_i{}^\alpha$ are pseudo-real, \textit{i.e.} they
are subject to the constraint, $A_i{}^\alpha \varepsilon^{ij}
\Omega_{\alpha\beta} = A^j{}_\beta\equiv (A_j{}^\beta)^\ast$. The
information on the target-space metric is contained in the so-called
hyperk\"ahler potential. For our purpose the geometry of the
hyperk\"ahler cone is not relevant. Hence we assume that the cone is
flat, so that the target-space connections and curvatures will
vanish. The extension to non-trivial hyperk\"ahler cone geometries is
straightforward. For the local scale transformations we refer again to the weights
shown in Table~\ref{tab:w-weights-matter-5D}. 

The hypermultiplet is
not realized as an off-shell supermultiplet. Closure of the
superconformal transformations is only realized upon using fermionic
field equations, but this fact does not represent a serious problem in
what follows. The $4D$ fields have, however, different Weyl weights as
is shown in Table \ref{table:w-weights-matter-4D}. This has been taken
into account in the reduction, by scaling $A_i{}^\alpha$ by a factor
$\phi^{-1/2}$, as can be seen below. Furthermore we define an
R-covariant spinor combination,
\begin{equation}
  \label{eq:zeta-new}
  (\phi^{-1/2} \zeta^\alpha-\tfrac12\phi^{-3/2} A_j{}^\alpha \gamma^5
  \hat{\psi}^j)\big|^\mathrm{Rcov}= \exp[\tfrac12
  \varphi\,\gamma^5]\, (\phi^{-1/2} \zeta^\alpha-\tfrac12\phi^{-3/2}
  A_j{}^\alpha \gamma^5 \hat{\psi}^j)\,. 
\end{equation}
The $4D$ Q- and S-supersymmetry variations take the following
form, again in terms of R-covariant chiral spinors, 
\begin{align}
  \label{eq:hyper}
  &\delta( \phi^{-1/2} A_i{}^\alpha) =
  2 \mathrm{i}\,\bar\epsilon_i P_+\big(\phi^{-1/2} \zeta^\alpha
  -\tfrac{1}2\, \phi^{-3/2} A_j{}^\alpha \gamma^5
  \hat{\psi}^j\big) \nonumber \\
  &\qquad\qquad\qquad-2\mathrm{i}\,\bar\epsilon_i P_-\big(\phi^{-1/2} \zeta^\alpha
  -\tfrac{1}2\, \phi^{-3/2} A_j{}^\alpha \gamma^5
  \hat{\psi}^j\big)\, , \\
  &\delta\bigl(P_\pm\big(\phi^{-1/2} \zeta^\alpha -\tfrac{1}2\,
  \phi^{-3/2} A_j{}^\alpha \gamma^5 \hat{\psi}^j\big)\bigr) =
  - \mathrm{i}\Slash{D}( \phi^{-1/2}
  A_i{}^\alpha)\,P_\mp\epsilon^i  +\,\phi^{-1/2} A_i{}^\alpha P_\pm\eta^i  \, , \nonumber
\end{align}
where, as before, we suppressed the S-supersymmetry and R-symmetry
transformations with field-dependent parameters. Note that the proportionality factor in
front of the $4D$ S-supersymmetry variation has changed as compared to the $5D$ result \eqref{eq:hypertransf}. 
%
\begin{table}
\begin{center}
\begin{tabular}{|c||cccc|cc|}
\hline
 & \multicolumn{4}{c|}{vector multiplet} & 
 \multicolumn{2}{c|}{hypermultiplet} \\
 \hline \hline
 field & $X$ & $W_\mu$  & $\Omega_i$ & $Y^{ij}$& $A_i{}^\alpha$ & $\zeta^\alpha$ \\[.5mm] \hline
$w$  & $1$ & $0$ & $\tfrac32$ & $2$ & $1$ &$\tfrac32$
\\[.5mm] \hline
$c$  & $-1$ & $0$ & $-\tfrac12$ & $0$ & $0$ &$-\tfrac12$
\\[.5mm] \hline
$\gamma_5$   & && $+$  &   &  &$-$ \\ \hline
\end{tabular}
\caption{Weyl
      and chiral weights ($w$ and $c$) and fermion chirality
      $(\gamma_5)$ of the vector multiplet and the hypermultiplet
      component fields in four space-time
      dimensions. \label{table:w-weights-matter-4D}}
\end{center}
\end{table}
We can again make a simple field redefinition to bring these transformation rules into a convenient form. Let
\be
\phi^{-1/2}A_i{}^\alpha := \wt{A}_i{}^\alpha \, , \quad \phi^{-1/2} \zeta^\alpha -\tfrac{1}2\,\phi^{-3/2} A_j{}^\alpha \gamma^5 \hat{\psi}^j := \wt{\zeta}^\alpha \, .
\ee
In terms of these fields,~\eqref{eq:hyper} takes the form
\begin{align}
\label{eq:hyper-good-vars}
\delta\wt{A}_i{}^\alpha =&\, 2\mathrm{i}\,\bar{\epsilon}_i P_+ \wt{\zeta}^\alpha - 2\mathrm{i}\,\bar{\epsilon}_i P_- \wt{\zeta}^\alpha \, , \cr
\delta\bigl(P_\pm\wt{\zeta}^\alpha\bigr) =&\, -\mathrm{i}\,\Slash{D}\wt{A}_i{}^\alpha P_\mp\epsilon^i + \wt{A}_i{}^\alpha P_\pm\eta^i \, ,
\end{align}
These transformation rules will be used in Chater~\ref{chap:n2-loc}, where for clarity we will drop the tilde on the scalar sections and the hyperini.

\subsection{Supergravity algebra}
\label{sec:sugra-alg}

With the help of the transformation rules for the various fields just derived, we can write down the algebra of Q- and S-supersymmetries for the Euclidean~$\CN=2$ supergravity theory. This algebra closes \emph{off-shell} on all the fields of the Weyl and vector multiplets, and closes on-shell on the hypermultiplet fields (for which one needs to use fermionic equations of motion). As expected from the general discussion at the beginning of this Chapter, two successive Q-transformations generate a general coordinate transformation, along with additional symmetries: a Lorentz transformation, a K- and an S-transformation and in addition a gauge transformation acting on the vector fields present in the vector multiplets, since they come with their own gauge symmetries:
\begin{equation}
\bigl[\delta_Q(\epsilon_1),\delta_Q(\epsilon_2)\bigr] = \delta_{\mathrm{cgct}}(\xi) + \delta_M(\epsilon_{ab}) + \delta_K(\hat{\Lambda}_K{}^a) + \delta_S(\hat{\eta}^i) + \delta_{\mathrm{gauge}} \, .
\end{equation}
The parameters of the various transformations are given by
\begin{align}
\xi^\mu =&\, 2\,\bar{\epsilon}_{2\,i}\gamma^\mu P_-\epsilon_1{}^i - 2\,\bar{\epsilon}_{2\,i}\gamma^\mu P_+\epsilon_1{}^i \, , \nonumber \\
\epsilon_{ab} =&\, \tfrac12\,\bar{\epsilon}_{2\,i}P_-\epsilon_1{}^i\,T^-_{ab} + \tfrac12\,\bar{\epsilon}_{2\,i}P_+\epsilon_1{}^i\,T^+_{ab} \, , \\
\hat{\Lambda}_K{}^a =&\, -\tfrac12\,\bar{\epsilon}_{2\,i}P_-\epsilon_1{}^i\,D_b T^{-ba} - \tfrac12\,\bar{\epsilon}_{2\,i}P_+\epsilon_1{}^i\,D_b T^{+ba} + \tfrac32\,\bar{\epsilon}_{2\,i}\gamma^a\gamma^5\epsilon_1{}^i\,D \, , \nonumber \\
\hat{\eta}^i =&\, -6\mathrm{i}\,\bar{\epsilon}_{j[1}P_-\epsilon_{2]}{}^i\,\chi^j \, . \nonumber
\end{align}
The remaining gauge transformation acting on vector fields is parametrized by
\begin{equation}
\delta_{\mathrm{gauge}} W_\mu = \partial_\mu\bigl(4\,\bar{\epsilon}_{2\,i}P_+\epsilon_1{}^i\,\bar{X} + 4\,\bar{\epsilon}_{2\,i}P_-\epsilon_1{}^i\,X\bigr) \, .
\end{equation}
The additional commutators of a Q-transformation with an S-transformation and of two successive S-transformations will not be needed in what follows.\\

This concludes the analysis of the Weyl, vector and hyper multiplets in Euclidean signature. All the transformation rules have been obtained using the method of time-like dimensional reduction from the Minkowski $5D$ theory, as explained in Appendix~\ref{app:sugra}. Moreover, the four-dimensional superconformal algebra on the Weyl multiplet and the vector multiplets is realized \emph{off-shell} (without imposing equations of motion). These off-shell Euclidean transformation rules will serve as a basis for all the following Chapters.

Obviously, many more details regarding the Euclidean theory can be given, and they are interesting in their own right. Most of these details are discussed in the publication~\cite{BdW-Me}, but they are not needed for the analysis which is presented in the following Chapters, so we do not elaborate on them here. We should however point out that, as in the Minkwoski case, the superconformal gravity theory obtained here is \emph{gauge-equivalent} to the~$\mathcal{N}=2$ Poincar\'{e} supergravity theory. That is, upon gauge-fixing the extra conformal (super)symmetries (the dilatations, special conformal transformations and S-supersymmetry), one obtains the Poincar\'{e} theory. 

Recall also from the Minkowski case that gauge-fixing the dilatations requires an additional vector multiplet~\cite{deWit:1980tn}, which is called a \emph{compensating} multiplet (it is required so that the Einstein-Hilbert term in the gauge-fixed supergravity action has the canonical form). Thus, the coupling of $\nv+1$ vector multiplets to the Weyl multiplet is gauge-equivalent to a Poincar\'{e} supergravity theory of one graviton multiplet coupled to~$\nv$ vector multiplets. In the conformal setting, we will always consider the coupling of~$\nv+1$ vector multiplets to the Weyl multiplet to take into account this extra compensating vector multiplet.

In the Euclidean theory just derived, we will be interested in 1/2-BPS black hole solutions and their quantum entropy defined as the path-integral~\eqref{eq:QEF}. In the previous Chapter, we presented an exact result for the degeneracies of 1/8-BPS states in~$\CN=8$ string theory, which are given by~\eqref{eq:rademsp}. With some anticipation, we also mentioned that a similar result can be derived for 1/4-BPS states in $\CN=4$ string theory (this will be the topic of Chapter~\ref{chap:n4-loc}). An inquisitive reader might then wonder: why focus on~$\CN=2$ supergravity theories if we want to compare the results of the localizing program applied to the QEF to results obtained in $\CN=4$ or~$\CN=8$ string theories? The reason is that, at present, no covariant off-shell formulations of~$\CN=4$ and~$\CN=8$ supergravity theories is known. This hampers the localization program laid out in the previous Chapter since, as we explained, it is desirable to use an off-shell theory to characterize the localizing manifold~$\mathcal{M}_Q$. Nevertheless, it is still possible to make contact with the string theory degeneracies when we consider \emph{truncated} supergravity theories, as we now briefly explain following the original idea of~\cite{Shih:2005he}.

Starting from Type II string theory compactified on the torus~$T^6$, we obtain a four-dimensional~$\CN=8$ theory with 28 massless~$\mathrm{U}(1)$ gauge fields. A charged state in this theory is characterized by 28 electric and 28 magnetic charges, which form the {\bf 56} representation of the U-duality group~$E_{7,7}(\mathbb{Z})$. Under the~$\mathrm{SO}(6,6;\mathbb{Z})$ T-duality group, the 28 gauge fields decompose as {\bf 28} = {\bf 12} + {\bf 16}. We can now truncate the theory by dropping four~$\CN=4$ gravitini multiplets, each containing four gauge fields. This amounts to dropping 16 gauge fields, which we take to be the {\bf 16} in the decomposition of {\bf 28}. The resulting theory is a reduced~$\CN=4$ theory with a U-duality group~$\mathrm{SO}(6,6;\mathbb{Z}) \times \mathrm{SL}(2;\mathbb{Z})$, with~$\mathrm{SL}(2;\mathbb{Z})$ the S-duality group. We now drop two more~$\CN=2$ gravitini multiplets (containing two gauge fields each) and the hypermultiplets to obtain a truncated~$\CN=2$ theory with 8 gauge fields and a U-duality group~$\mathrm{SO}(6,2;\mathbb{Z}) \times \mathrm{SL}(2;\mathbb{Z})$. The 1/8-BPS states in the original~$\CN=8$ theory correspond to 1/2-BPS states in the truncated~$\CN=2$ theory, and we can now use~$\CN=2$ conformal supergravity coupled to~$\nv+1=8$ vector multiplets as the low-energy limit of the latter. This effective theory being off-shell, we can apply localization to the QEF for the 1/2-BPS black hole solutions, and compare to the degeneracies of 1/8-BPS states in~$\CN=8$ string theory.

Similarly, one can start from Type II string theory compactified on~$\mathrm{K}3 \times T^2$ to obtain an~$\CN=4$ theory with 28 massless gauge fields. The U-duality group of this theory is~$\mathrm{SO}(6,22;\mathbb{Z})\times\mathrm{SL}(2;\mathbb{Z})$. Upon dropping two~$\CN=2$ gravitini multiplets and the hypermultiplets, the resulting truncated theory has~$\CN=2$ supersymmetry and contains 24 massless gauge fields, with a U-duality group~$\mathrm{SO}(2,22;\mathbb{Z})\times\mathrm{SL}(2;\mathbb{Z})$. 1/4-BPS states in the original theory now correspond to 1/2-BPS in the truncated~$\CN=2$ theory, whose low-energy description is that of~$\CN=2$ conformal supergravity coupled to~$\nv+1=24$ vector multiplets. The degeneracies computed using the QEF for 1/2-BPS black holes in this theory can then be compared with the ones for 1/4-BPS states in the~$\CN=4$ string theory.

Of course, any calculation conducted in a truncated theory will be valid in the full theory \emph{provided} the truncation scheme is consistent. For the black hole degeneracies, this amounts to proving that the fields being dropped during the truncation effectively carry no entropy. While we do not attempt to prove such statements on general grounds in this work, we will see in the coming Chapters that we can provide an \textit{a posteriori} justification of this fact for the~$\CN=8$ and~$\CN=4$ truncations outlined above. A first-principle understanding of this seems highly desirable at the moment and is left for future work.  

\chapter{Quantum entropy of 1/2-BPS black holes and localization}
\label{chap:n8-loc}
In Chapter~\ref{chap:sugra}, we outlined the construction of the four-dimensional~$\CN=2$ conformal supergravity in Euclidean signature. This theory admits black hole solutions, and in the present Chapter we will review its 1/2-BPS black hole solutions and their Bekenstein-Hawking-Wald entropy. We then summarize the computation of their exact quantum entropy using the localization formalism applied to the QEF presented in Chapter~\ref{chap:modern-BH-S}. This will lead to a Master Formula~\eqref{eq:master-integral} for the quantum entropy of 1/2-BPS black holes in~$\CN=2$ conformal supergravity, which will play a central role in what follows. As will be explained, this formula relies on an assumption which we will prove in the remainder of this Chapter by examining the influence of a large class of full-superspace supergravity actions on the quantum entropy using the formalism of the kinetic multiplet.

\section{Semi-classical 1/2-BPS black hole entropy}
\label{sec:semi-class-1/2bps-S}

We use the Euclidean theory of~$\CN=2$ conformal supergravity developed in Chapter~\ref{chap:sugra}. This theory describes the Weyl multiplet coupled to $\nv+1$ vector multiplets and~$n_h$ hypermultiplets. Recall that the Weyl multiplet of the theory contains the following independent fields:
\be
\label{eq:Weylfields}
{\bf W} = \left( e_{\mu}^{a}, \,  P_\pm\psi_{\mu}{}^i, \, b_{\mu}, \, A_{\mu}, \CV_\mu{}^i{}_j,  \, T_{\mu\nu}^\pm, \, P_\pm\chi^i, \, D \right) \, ,
\ee
along with composite fields built out of the above fields. In the two-derivative gauge-fixed Poincar\'e theory, the field $e_{\mu}^{a}$ is the vielbein, and the $T_{\mu\nu}$ tensor is an auxiliary field without kinetic term. These two fields will play an important role in this Chapter and the following ones.

The independent fields of the Euclidean vector multiplet are
\be
\label{eq:Vectorfields}
{\bf X}^{I} = \left( X^{I}, \, \bar{X}^I, \, P_\pm\lambda^{I\,i}, \, W_{\mu}^{I}, \, Y^{I}_{ij}  \right) \, , 
\ee
where $X^{I}$ and~$\bar{X}^I$ are two scalars, the gaugini $\lambda^I$ form an $\SU(2)$ doublet of symplectic Majorana-Weyl fermions, $W^{I}_{\mu}$ is a vector field, and the $Y^{I}_{ij}$ form an $\SU(2)$ triplet of auxiliary scalars. The index~$I$ runs over the~$\nv+1$ vector multiplets.

The action of the theory under consideration is specified by the \emph{holomorphic prepotential}~$F(X^I,\wh{A})$, describing the coupling of the vector multiplets to the background Weyl multiplet through chiral-superspace integrals (see \textit{e.g.}~\cite{Mohaupt:2000mj}). Here,~$\wh{A} := (T_{\mu\nu}^-)^2$ is the lowest component of the~${\bf W}^2$-multiplet. The latter dependence encodes higher-derivative terms in the supergravity action proportional to the square of the Weyl tensor. Supersymmetry requires that this prepotential be holomorphic and homogeneous of degree two,
\be 
\label{eq:prepXA}
F(\Lambda\,X^I,\Lambda^2\,\wh{A}) = \Lambda^2\,F(X^I,\wh{A}) \, .
\ee
As explained in Section~\ref{sec:electr-magn-dual}, electric-magnetic duality of the theory is realized as symplectic transformations, under which the pair~$(X^{I}, F_{I})$ with~$F_{I} := \p F(X^I,\wh{A})/\p X^{I}$ transforms linearly. 

Conformal~$\CN=2$ supergravity admits 1/2-BPS black hole solutions. They carry electric and magnetic charges~$(q_{I},p^{I})$, where $I$ runs over the ~$\nv +1$ vector multiplets, and interpolate between fully supersymmetric asymptotically flat space and the near-horizon Euclidean~$AdS_{2} \times S^{2}$ region. The near-horizon region is itself fully BPS due to the supersymmetry enhancement granted by the attractor mechanism (recall the discussion in Section~\ref{sec:higherdim-S}), and in the low-energy limit it can be decoupled from the environment and studied on its own. We parametrize it as follows (with all other fields not related by symmetries set to zero): 
\begin{align}
\label{eq:full-BPS-metric}
ds^2 =&\, v\Bigl[(r^2-1)d\t^2 + \frac{dr^2}{r^2-1}\Bigr] + v\bigl[d\psi^2 + \sin^2(\psi)d\phi^2\bigr] \, , \\
F^I_{r\t} =&\, -ie^I_*, \;\; F^I_{\psi\phi} = p^I\sin\psi, \;\; X^I = X^I_*, \;\; \bar{X}^I = \bar{X}^I_*, \;\; T^-_{r\t} = -ivw, \;\; T^+_{r\t} = -ivw \, . \nonumber
\end{align}
Here~$F^I_{\mu\nu}$ is the field strength of the vector field~$W_\mu^I$ in the vector multiplet~$I$ and $(v,\,w,\,X^I_*,\,\bar{X}^I_*\,e^I_*,\,p^I)$ are constants. The choice of a complex vector field strength is an artifact of using conventions which have been derived in a Wick-rotated version of the Minkowski theory. Note also that, in contrast to the extremal Reissner-Nordstr\"{o}m line-element~\eqref{eq:RN-BH-Extr}, we have chosen the position of the horizon to be at a fixed distance~$r=1$ when taking the near-horizon limit.

The full-BPS solution~\eqref{eq:full-BPS-metric} has an~$\mathrm{SL}(2) \times \mathrm{SU}(2)$ bosonic symmetry, the two factors acting on the~$AdS_{2}$ and~$S^{2}$ parts respectively. It also admits eight real supersymmetries, which together with the bosonic symmetries form the~$\mathrm{SU}(1,1 | 2)$ superalgebra. One of the supercharges of this algebra, which we shall call~$Q$, will play an important role in the following. It obeys the algebra
\be 
\label{eq:specificQ}
Q^{2} = L_{0} - J_{0} \, , 
\ee
where~$L_{0}$ and~$J_{0}$ are the Cartan generators of the~$\mathrm{SL}(2)$ and the~$\mathrm{SU}(2)$ algebras respectively. 

The two scalar fields~$X^{I}$ and~$\bar{X}^I$ of the vector multiplets are determined completely in terms of the fluxes by the BPS attractor mechanism~\cite{LopesCardoso:1998wt, LopesCardoso:2000qm}, or equivalently using the AdS attractor mechanism introduced in Section~\ref{sec:higherdim-S}:
\be
F_{ab}^{+ I} = \frac{1}{4} X^I_* \, T^+_{ab} \, , \qquad F_{ab}^{- I} = \frac{1}{4}\bar{X}^I_* \, T^-_{ab} \, .
\ee
For our solution~\eqref{eq:full-BPS-metric}, we have:
\be
\label{eq:attractor-scalars-metric}
X_*^I = \frac{2}{w}(e_*^I + ip^I) \, , \quad \bar{X}_*^I = \frac{2}{w}(e_*^I - ip^I) \, , \quad v= \frac{16}{w^2}\, .
\ee
At this stage, one can choose a dilatation gauge~$w = 4$ which brings the determinant of the metric~\eqref{eq:full-BPS-metric} to unity, but we will keep this scale factor manifest in this Chapter. The electric fields~$e_{*}^{I}$ are determined in term of the charges~$q_{I}$ by the extremization equation~\eqref{eq:Sen-Extrem},
\be
\label{eq:attractor-charges}
F_I\Bigl(\frac{2}{w}(e_*^I + ip^I)\Bigr) - \bar{F}_I\Bigl(\frac{2}{w}(e_*^I - ip^I)\Bigr)\Big|_{\wh{A}=-4w^2} = i\frac{w}{4}\,q_I \, .
\ee
Finally, the Bekenstein-Hawking-Wald entropy of this black hole is given by~\eqref{eq:S-SenBHW}:
\be
\label{eq:semi-class-1/2bps}
\mathcal{S}_{BHW} = -\pi\,q_I e^I_* - 2\pi i\Bigl[F\Bigl(\frac{2}{w}(e_*^I + ip^I)\Bigr) - \bar{F}\Bigl(\frac{2}{w}(e_*^I - ip^I)\Bigr)\Bigr]\Big|_{\wh{A}=-4w^2} \, .
\ee
This entropy is completely determined by the prepotential of the theory, and is a function of the electric and magnetic charges of the black hole only. 
The apparent dependence on the constant~$w$ reflects the fact that we haven't gauge-fixed the dilatations here, so that there is still a scaling symmetry. 
As we mentioned in the beginning of this Section, the prepotential can have a dependence on the lowest component of the~${\bf W}^2$-multiplet which encode higher-derivative terms in the supergravity action.~$F(X,\wh{A})$ can thus, for instance, take the form of an infinite series in powers of~$\wh{A}$ corresponding to a higher-derivative expansion of the supergravity action, and the entropy inherits this expansion structure through~\eqref{eq:semi-class-1/2bps}. A consequence of this fact will be examined in Chapter~\ref{chap:n2-loc}.

This concludes the semi-classical analysis of 1/2-BPS black holes in~$\CN=2$ conformal supergravity. The main players in this story were the vector multiplets,\footnote{In particular, note that the hypermultiplets played essentially no role. This will change in Chapter~\ref{chap:n2-loc} when analyzing the quantum entropy in more detail.} and the Bekenstein-Hawking-Wald entropy of the black hole is entirely determined using the attractor mechanism (BPS or AdS equivalently, since the black hole solution is supersymmetric). We now turn to the quantum entropy of these black holes using the QEF formalism of Section~\ref{sec:QEF}.

\section{Quantum 1/2-BPS black hole entropy}
\label{sec:quantum-BH-S}

According to the discussion in Section~\ref{sec:QEF}, the quantum entropy of the 1/2-BPS black hole solutions presented in the previous Section is defined as a functional integral over all the fields of the supergravity theory:
\be
\label{eq:QEF2}
\exp[\mathcal{S}_Q(q,p)] = W (q, p) = \left\langle \exp[-i \, q_I \oint_{\t}  A^I]  \right\rangle_{\rm{AdS}_2}^{\textnormal{finite}}\ . 
\ee 
To apply localization, we pick the specific supercharge~$Q$ with algebra~\eqref{eq:specificQ}~\cite{Dabholkar:2010uh}. With this choice, the first step in the localization program of Section~\ref{sec:susy-loc} is to find all solutions to the localizing equations~\eqref{eq:QV-BPS} 
\be
\label{eq:qpsi}
Q \, \psi_{\a} =0 \, , 
\ee
where $\psi_{\a}$ runs over all the fermions of the theory. The space of solutions to these equations is the localization manifold~$\CM_{Q}$. For four-dimensional~$\CN=2$ conformal supergravity, the complete localization manifold was found in~\cite{Gupta:2012cy} and is described as follows.
 
When the field strength of the $\mathrm{SU}(2)$ R-symmetry gauge field $\CV_{\mu \;\, j}^{\;\; i}$ is set to zero, the full set of bosonic solutions to the localization equations in $\CN=2$ off-shell supergravity coupled to vector and hyper multiplets is labeled by $\nv+1$ real parameters. These parameters label the size of fluctuations of a certain shape (fixed by supersymmetry) of the conformal mode of the metric and of the scalars in the vector multiplets, and can be taken to be the values of these $\nv+1$ fields at the center of $AdS_{2}$. Using the dilatation gauge symmetry of the theory, one can trade the conformal mode of the metric for the scalar of the compensating vector multiplet. We set the metric of Euclidean $AdS_{2} \times S^{2}$ to have unit determinant, and the scalar fields of the vector multiplet have the solution:
\be 
\label{eq:scalars}
X^I = X^I_* + \frac{w}{4}\frac{C^{I}}{r}, \quad\quad \bar{X}^I = \bar{X}^I_* + \frac{w}{4}\frac{C^{I}}{r} \, , \qquad I = 0\,,\ldots , n_v\, .
\ee
These fluctuations are off-shell 1/2-BPS solutions, and they are supported by the auxiliary fields in the vector multiplets:
\be 
\label{eq:auxfields}
Y_1^{I\,1} = -Y_2^{I\,2} = \frac{w^2}{8}\frac{C^{I}}{r^{2}} \, .
\ee
The rest of the fields in the solution remain unchanged with respect to the full-BPS $AdS_{2} \times S^{2}$ solution \eqref{eq:full-BPS-metric}. In particular, the hypermultiplets have a trivial bosonic profile~$A_i{}^\alpha =0$.
An important point to note at the end of the first step is that the localization manifold~$\CM_{Q}$ is universal insofar that it is independent of the choice of the action, since the supersymmetry variations~\eqref{eq:qpsi} are defined entirely in the off-shell theory. 

The next step is to evaluate the effective action of the supergravity theory on the localizing solutions and correctly integrate over the localizing manifold. As explained in Section~\ref{sec:susy-loc}, the integral over~$\mathcal{M}_Q$ includes the measure~$[d\phi]$ induced from the supergravity field space, as well as the one-loop determinant of the deformation action~$Q\mathcal{V}$ coming from integration over the (non-supersymmetric) directions orthogonal to the localizing manifold in field space:\footnote{The hat on~$W$ refers to the fact that we will only consider smooth supergravity configurations in the functional integral. There can be other configurations which are only smooth in the full string theory, such as orbifold configurations. Their contribution is highly subleading as explained in~\cite{Banerjee:2008ky, Murthy:2009dq, Dabholkar:2014ema}.\label{footnote:WHat}}
\be
\wh W (q, p) = \int_{\mathcal{M}_{Q}} [d\phi] \exp\bigl[S_{\rm ren}(\phi, q, p) \bigr] \, Z_{\text{1-loop}}(\phi) \, ,
\ee
where~$S_{\rm ren}$ is the renormalized action defined below~\eqref{eq:QEF}.
In~\cite{Dabholkar:2010uh}, this integral was computed \emph{assuming} that the effective renormalized action $S_{\rm ren}$ only contains chiral-superspace integral terms which are governed by the holomorphic prepotential~$F$ of the theory. With this assumption, and defining the new variables
\be
\label{eq:ephi}
\phi^I := e_{*}^I+2 C^I \ ,
\ee
the renormalized action is given by
\be
\label{eq:Sren}
S_{\rm ren}(\phi, q, p) = - \pi  q_I   \phi^I - 2\pi i\Bigl[ F\Bigr(\frac{2}{w}(\phi^I+ ip^I)\Bigl) - \bar{F} \Bigl(\frac{2}{w}(\phi^I- ip^I)\Bigr) \Bigr]\Big|_{\wh{A}=-4w^2} \, ,
\ee
where the prepotential~$F(X,\wh{A})$ is evaluated at the attractor value~$\wh{A} = -4w^2$. The quantum entropy of 1/2-BPS black holes in~$\CN=2$ conformal supergravity coupled to~$\nv +1$ vector multiplets and hypermultiplets therefore takes the form
\be
\label{eq:master-integral}
\wh{W} (q, p) = \int_{\mathcal{M}_{Q}}  \, \prod_{I=0}^{\nv} [d\phi^{I}] \, \exp\Bigl[- \pi  \, q_I  \, \phi^I +4\pi\,\textnormal{Im}\,F\bigl(\tfrac{2}{w}(\phi^I + ip^I)\bigr) \Bigr] \, Z_\text{1-loop}(\phi^{I})\, ,
\ee
This is the Master Formula which will be used in the remainder of this work. It has a number of important features:
\begin{myenumerate}
\item It is \emph{universal}, in the sense that it only depends on the prepotential~$F$ of the supergravity theory one wishes to consider and not on the details of the action itself. \\
\item It is a \emph{finite,~$(\nv+1)$-dimensional integral}, making its evaluation infinitely easier than the partition function~\eqref{eq:QEF2} defined as a path-integral. \\
\item By definition, it encodes all quantum corrections to the Bekenstein-Hawking-Wald entropy formula~\eqref{eq:semi-class-1/2bps}. Therefore, it should match string theory predictions for a given compactification specifying the prepotential~$F$ of the corresponding low-energy supergravity theory.\\
\item It is expected to be correct as long as the assumption of~\cite{Dabholkar:2010uh} is valid, namely that only chiral-superspace integrals contribute to the quantum entropy.
\end{myenumerate}

Note also that~\eqref{eq:master-integral} shares a number of interesting features with the OSV proposal of~\cite{Ooguri:2004zv}, and it can be seen as part of an attempt to derive and refine this conjecture from the gravitational theory. Details of the comparison with this proposal are given in~\cite{Dabholkar:2010uh, Dabholkar:2011ec}. We shall make a comparison with the related proposal of~\cite{Denef:2007vg} in Chapter~\ref{chap:n2-loc}.

The Master Formula~\eqref{eq:master-integral} was applied in~\cite{Dabholkar:2010uh, Dabholkar:2011ec} to the $\CN=2$ truncation of Type IIB string theory compactified on~$T^6=T^4\times S^1\times \widetilde{S}^1$ outlined at the end of the previous Chapter. In the full~$\CN=8$ string theory, the microscopic degeneracy of 1/8-BPS states (preserving 4 real supercharges) is given by~\eqref{eq:rademsp}. In the~$\CN=2$ truncation, the prepotential~$F(X)$ of the low-energy supergravity theory entering the integral formula is given by the cubic prepotential
\be
\label{eq:prepot-n8}
F(X) = -\frac{X^1 X^a C_{ab} X^b}{X^0} \, , \qquad a,b = 2,\,\ldots,7
\ee
where~$C_{ab}$ is the intersection matrix of the six 2-cycles of~$T^4$. This prepotential describes the classical two-derivative action of~$\CN=2$ conformal supergravity. Note that it does not depend on~$\widehat{A}$ because there are no higher-derivative quantum corrections in the case of toroidal compactification. It was further assumed in~\cite{Dabholkar:2010uh, Dabholkar:2011ec} that the functional determinant~$Z_{\text{1-loop}}$ is trivial and equal to unity in this specific truncation. We will come back to this in Chapter~\ref{chap:n2-loc} and verify this assumption.

In this setting, the quantum entropy for the 1/2-BPS~$\CN=2$ black holes computed using~\eqref{eq:master-integral} is in agreement with the microscopic degeneracy of 1/8-BPS states in $\CN=8$ string theory~\eqref{eq:rademsp} to \emph{exponential accuracy}, as evidenced by Table 2 in~\cite{Dabholkar:2011ec}. This was the first successful application of the  localization program in a theory of supergravity. We note here that the successful matching to~$\CN=8$ string theory predictions hints at the fact that the truncated~$\CN=2$ supergravity theory considered in~\cite{Dabholkar:2010uh, Dabholkar:2011ec} is in fact consistent and encodes all the relevant information about the complete string theory. In our current understanding of the truncation, this seems quite non-trivial and we do not know how to justify this from first principles. Nevertheless, we will take this result as a sign that such truncations down to an~$\CN=2$ conformal supergravity theory coupled to vector and hyper multiplets can be used in such situations. We will make use of this assumption later in Chapter~\ref{chap:n4-loc} for 1/4-BPS states in~$\CN=4$ string theory.

Furthermore, the success of this analysis points to a non-renormalization theorem of the quantum entropy computed using the prepotential. Namely, it seems like full-superspace integrals in the effective action do not contribute to the quantum entropy of supersymmetric black holes. In the rest of this Chapter, we shall provide evidence in support of this non-renormalization theorem and effectively prove the assumption of point 4. in the list below~\eqref{eq:master-integral}, thereby ensuring that one can use the Master Formula for the computation of the quantum entropy of generic 1/2-BPS black holes in~$\CN=2$ supergravity coupled to vector and hyper multiplets.

\section{Full-superspace integrals and classical entropy}
\label{sec:D-terms-S}

In this section, we review the construction of a large class of full-superspace integrals that can be built in a theory of~$\CN=2$ supergravity coupled to~$\CN=2$ vector multiplets. This is done using the technology of the so-called \emph{kinetic multiplet}~\cite{deWit:1980tn}. We then review the fact that the semi-classical black hole entropy does not change on adding these full-superspace terms to the effective action. These results were first reported in~\cite{deWit:2010za} which we follow. We will suppress fermionic terms in what follows since we are interested in purely bosonic configurations (the black hole).

\subsection{A large class of full-superspace integral Lagrangians}
\label{subsec:kinetic-mult}

Constructing the $\CN=2$ supersymmetric Lagrangians of various matter fields coupled to supergravity is quite an intricate technical task. The coupling of a chiral multiplet $\Phi$ to supergravity through a chiral-superspace integral was worked out in the early days~\cite{deWit:1980tn}: 
\be 
\label{eq:chiralcoup}
S = \int d^{4}x \, \CL  \= \int d^{4}x \, d^{4} \theta \, \varepsilon \, \Phi \, , 
\ee
where~$\varepsilon$ is an appropriately defined chiral superspace measure and~$\theta$ are superspace Grassmanian coordinates. This result was then adapted and modified to construct the coupling of vector multiplets (by writing the vector multiplet as a reduced chiral multiplet), and to construct higher-derivative terms (by considering a holomorphic function $F$ of chiral multiplets as a chiral multiplet itself). Since~$\theta$ has a Weyl weight~$1/2$, the coupling~\eqref{eq:chiralcoup} is consistent only if the superfield~$\Phi$ has weight 2 (so that the action has weight zero). 

The same technique can be further modified to construct full-superspace integrals. The idea is to construct a kinetic multiplet out of an anti-chiral multiplet, which involves four covariant~$\bar{\theta}$-derivatives, \textit{i.e.}~$\mathbb{T}(\bar{\Phi}) \propto \bar{D}^4\bar{\Phi}$. This means that~$\mathbb{T}(\bar{\Phi})$ contains up to four space-time derivatives, so that the expression 
\be 
\label{eq:kinetic_int}
\int d^4\theta \, d^4\bar{\theta}\;\Phi \, \bar{\Phi} \sim \int d^4\theta\;\Phi \, \mathbb{T}(\bar{\Phi})
\ee
corresponds to a usual higher-derivative coupling Lagrangian. Here we are being slightly schematic and we have not shown the superspace measure.

The field~$\Phi$ and $\bar \Phi$ entering the expression~\eqref{eq:kinetic_int} can be composite fields built out of the basic field content of the theory, and can very well be two independent fields. We use this fact later in Section~\ref{sec:Dterms-quantum}. A more subtle point concerns the nature of the composite field~$\bar \Phi$ entering this expression~\cite{Butter:2013lta}. We will assume that~$\bar \Phi$ is a physical field that is a local functional of the fluctuating fields of the theory.

From the above expression, one sees that the operator~$\mathbb{T}$ increases the Weyl weight by 2, and so the superfield~$\Phi$ should have Weyl weight 0 for the coupling to be consistent. For a chiral multiplet~$\Phi$ with components~$(A,\Psi_i,B_{ij},F^-_{ab},\Lambda_i,C)$, the Lagrangian~\eqref{eq:kinetic_int} is~\cite{deWit:2010za}:
\begin{align} 
\label{KinLag}
e^{-1}L =&\,
  4\,\mathcal{D}^2 A\,\mathcal{D}^2\bar A + 8\,\mathcal{D}^\mu A\, \big[R_\mu{}^a(\omega,e) -\tfrac13 R(\omega,e)\,e_\mu{}^a \big]\mathcal{D}_a\bar A + C\,\bar C \nonumber \\[.1ex]
  &\, - \mathcal{D}^\mu B_{ij} \,\mathcal{D}_\mu B^{ij} + (\tfrac16 R(\omega,e) +2\,D) \, B_{ij} B^{ij} \nonumber \\[.1ex]
  &\, - \big[\varepsilon^{ik}\,B_{ij} \,F^{+\mu\nu} \, R(\mathcal{V})_{\mu\nu}{}^{j}{}_{k} +\varepsilon_{ik}\,B^{ij} \,F^{-\mu\nu} R(\mathcal{V})_{\mu\nu j}{}^k \big] \nonumber \\[.1ex]
  &\, - 8\, D\, \mathcal{D}^\mu A\, \mathcal{D}_\mu\bar A + \big(8\, \mathrm{i} R(A)_{\mu\nu} + 2\, T_\mu{}^{cij} \, T_{\nu cij}\big) \mathcal{D}^\mu A \,\mathcal{D}^\nu\bar A \\[.1ex]
  &\, - \big[\varepsilon^{ij} \mathcal{D}^\mu T_{bc ij}\mathcal{D}_\mu A\,F^{+bc} - \varepsilon_{ij} \mathcal{D}^\mu T_{bc}{}^{ij}\mathcal{D}_\mu \bar A\,F^{-bc}\big] \nonumber \\[.1ex]
  &\, - 4 \big[\varepsilon^{ij}T^{\mu b}{}_{ij}\,\mathcal{D}_\mu A\,\mathcal{D}^c F^{+}_{cb}-\varepsilon_{ij}T^{\mu bij}\,\mathcal{D}_\mu\bar A\,\mathcal{D}^c F^{-}_{cb}\big] \nonumber \\[.1ex]
  &\, + 8\,\mathcal{D}_a F^{-ab}\,\mathcal{D}^c F^+{}_{cb} + 4\,F^{-ac}\, F^+{}_{bc}\, R(\omega,e)_a{}^b +\tfrac1{4} T_{ab}{}^{ij} \,T_{cdij} F^{-ab} F^{+cd} \,. \nonumber
\end{align}
By making various choices for the chiral multiplet~$\Phi$ which enters this formula, we can construct a large class of full-superspace Lagrangians. In our theory, we have a Weyl multiplet of weight 1 and $n_{\rm v}+1$ vector multiplets~${\bf X}^{I}$ of weight 1. Associated to each vector multiplet~${\bf X}^{I}$ is a 
reduced chiral multiplet~$\CC^{I}$~\cite{deRoo:1980mm}. 
We can build a class of Lagrangians by choosing the chiral multiplet~$\Phi$ above to be equal to an arbitrary holomorphic function~$f(\CC^{I})$ and similarly~$\bar \Phi$ to be equal to an anti-holomorphic function~$\bar g(\bar \CC^{I})$. The weight zero conditions on~$\Phi$, $\bar \Phi$ translate to the condition that the functions~$f$, $\bar g$ are homogeneous functions of degree zero. More generally, we can consider a sum of products of such functions
\be 
\label{eq:genH}
\CH(\CC^{I}, \bar \CC^{I}) = \sum_{n, \bar n} f^{(n)}(\CC^{I}) \; \bar g^{(\bar n)} (\bar \CC^{I}) \, . 
\ee
The full-superspace integral 
\be 
e^{-1}L \= \int d^4\theta \, d^4\bar{\theta}\; \CH(\CC^{I}, \bar \CC^{I}) \, ,
\ee
written in components is as follows~\cite{deWit:2010za}
\begin{align} 
\label{eq:HLagcomps}
  e^{-1}L =&\, \mathcal{H}_{IJ\bar K \bar L}\Big[\tfrac14
    \big( F_{ab}^-{}^I\, F^{-ab\,J}
                -\tfrac12 B_{ij}{}^I\, B^{ijJ} \big)
                \big( F_{ab}^+{}^K \, F^{+ab\,L} -\tfrac12 B^{ijK}\,
                  B_{ij}{}^L  \big)
              \nonumber\\
              & \qquad\quad +4\,\mathcal{D}_a A^I\, \mathcal{D}_b \bar A^K
                \big(\mathcal{D}^a A^J \,\mathcal{D}^b \bar A^L
                  - 2\, F^{-\,ac\,J}\,F^{+\,b}{}_c{}^L -
                  \tfrac14 \delta^{ab}\, B^J_{ij}\,B^{L\,ij}\big)
              \Big]\nonumber\\[.5ex]
   +\Big\{ \mathcal{H}_{IJ\bar K}&\Big[4\,\mathcal{D}_a A^I\,
     \mathcal{D}^a A^J\, \mathcal{D}^2\bar A^K
      + \big(F^{-ab\,I}F_{ab}^{-\,J} +\tfrac12 B^I_{ij}\, B^{Jij})
      \big( \Box_\mathrm{c} A^K - \tfrac18 F^{-\,K}_{ab} T^{ab ij}
          \varepsilon_{ij}\big)  \nonumber\\
  & \; -8 \,\mathcal{D}^a A^I F^{-\,J}_{ab}
  \big( \mathcal{D}_cF^{+\,cb\,K} - \tfrac12 \mathcal{D}_c\bar A^K
              T^{ij\,cb} \varepsilon_{ij}\big) - \mathcal{D}_a
            A^I\, B^J_{ij}\,\mathcal{D}^aB^{K\,ij}\Big]
            +\mathrm{h.c.}\Big\} \nonumber
            \\[.5ex]
     +\mathcal{H}_{I\bar J}&\Big[ 4\big( \Box_\mathrm{c} \bar A^I - \tfrac18
         F_{ab}^{+\,I}\, T^{ab}{}_{ij} \varepsilon^{ij}\big)
     \big( \Box_\mathrm{c}  A^J - \tfrac18 F_{ab}^{-\,J}\, T^{abij}
       \varepsilon_{ij}\big) + 4\,\mathcal{D}^2 A^I \,\mathcal{D}^2
       \bar A^J \nonumber\\
       & \; - 8\,\mathcal{D}_{a}F^{-\,abI\,}\,
       \mathcal{D}_cF^{+c}{}_{b}{}^J   - \mathcal{D}_a B_{ij}{}^I\,
            \mathcal{D}^a B^{ij\,J}
            +\tfrac1{4} T_{ab}{}^{ij} \,T_{cdij}
            \,F^{-ab\,I}F^{+cd\,J}
     \nonumber\\
     &\;
     +\big(\tfrac16 R(\omega,e) +2\,\hat{D}\big) B_{ij}{}^I B^{ij\,J} - 4\,
     F^{-ac\,I}\, F^{+}{}_{bc}{}^J \, R(\omega,e)_a{}^b  \nonumber\\
     &\; + 8\big(R^{\mu\nu}(\omega,e)-\tfrac13 g^{\mu\nu}
     R(\omega,e) -\tfrac1{4} T^\mu{}_{b}{}^{ij}T^{\nu b}{}_{ij}
     +\mathrm{i} R(A)^{\mu\nu} - g^{\mu\nu}D\big) \mathcal{D}_\mu A^I
     \mathcal{D}_\nu \bar A^J  \nonumber\\
     &\;
     + \big[\mathcal{D}_c \bar A^J \big(\mathcal{D}^c
     T_{ab}{}^{ij}\,F^{-\,I\,ab} +4
       \,T^{ij\,cb} \,\mathcal{D}^aF^{-\,I}_{ab} \big)\varepsilon_{ij}
       +[\mathrm{h.c.}; I\leftrightarrow J]  \big]\nonumber\\
     &\; -\big[\varepsilon^{ik}\, B_{ij}{}^I\, F^{+ab\,J}\,
      R(\mathcal{V})_{ab}{}^j{}_k +[\mathrm{h.c.}; I\leftrightarrow J]
      \big]  \Big] \,.
\end{align}
This can be further generalized by including the Weyl multiplet in the construction of the weight-zero superfields~$\Phi$, $\bar \Phi$. In this case, the corresponding function~$\CH$ is homogeneous of degree zero with~$\CC^{I}$ having scaling weight 1 and ${\bf W}^2$ having scaling weight 2. The resulting Lagrangian generalizes~\eqref{eq:HLagcomps} with additional terms (see~(4.10), (4.11) in~\cite{deWit:2010za}). When the~${\bf W}^{2}$ multiplet is a non-zero constant, the additional terms drop out, and in this case the Lagrangian is proportional to~\eqref{eq:HLagcomps}. 

\subsection{Non-renormalization of the semi-classical entropy}

As reviewed in Section~\ref{sec:semi-class-1/2bps-S}, the semi-classical entropy is computed by evaluating the local effective Lagrangian of the theory on the full-BPS solutions~\eqref{eq:full-BPS-metric}. In addition, the first derivative of the Lagrangian enters the answer through the definition of the charges~\eqref{eq:Sen-Extrem}. As we now review, all the full superspace integrals discussed in the previous subsection, as well as their first derivatives, vanish when evaluated on the full-BPS configuration~\cite{deWit:2010za}.

The Euclidean $AdS_2 \times S^2$ form of the metric~\eqref{eq:full-BPS-metric} implies
\be
\label{eq:full-BPS-gauge-aux}
R(A)_{\mu\nu} = R(\mathcal{V})_{\mu\nu}{}^i{}_j = D = R(\omega,e) = 0 \, .
\ee
The components of ${\bf W}^2$ then take the simple form:
\be 
\label{eq:weylclass}
A|_{W^2} = (T_{ab}^-)^2 = -4w^2 \, , \qquad B_{ij}|_{W^2} = F^{-ab}|_{W^2} = C|_{W^2} = 0 \, .
\ee
In the gauge-fixed theory, when $w$ is constant, the full Weyl-squared multiplet is a constant multiplet (the lowest component is a constant and the higher components vanish). 
It is convenient to write down the explicit values of the (anti)self-dual component of the $T$-tensor:
\be
\label{eq:TTensor}
T^-_{ab} = \begin{pmatrix}[1.0] 0 & iw & 0 & 0 \\ -iw & 0 & 0 & 0 \\ 0 & 0 & 0 & iw \\ 0 & 0 & -iw & 0 \end{pmatrix} \, , \quad T^+_{ab} = \begin{pmatrix}[1.0] 0 & iw & 0 & 0 \\ -iw & 0 & 0 & 0 \\ 0 & 0 & 0 & -iw \\ 0 & 0 & iw & 0 \end{pmatrix} \, .
\ee

Similarly, the reduced chiral multiplet in the full-BPS configuration~\eqref{eq:full-BPS-metric} is also a constant. 
\be 
\label{eq:constmult}
A|_{\CC^{I}} = X^I_* \, , \quad  B_{ij}|_{\CC^{I}}  = F^-_{ab}|_{\CC^{I}}  = C|_{\CC^{I}} = 0 \, . 
\ee
The Lagrangian~\eqref{eq:HLagcomps} involves only derivatives of~$A|_{\CC^{I}}$, and therefore vanishes on this constant solution. The generalized Lagrangian including the contribution from the Weyl multiplet also vanishes for the full-BPS solution with the Weyl and vector multiplets being constant. Using similar arguments,~ \cite{deWit:2010za} also shows that the first derivative of the Lagrangian with respect to all the fields vanish. We thus deduce that the charges,
and therefore the entropy, are not modified by the addition of the full-superspace integrals. 

In the next section, we shall consider half-BPS solutions wherein the scalar fields are not constant and have a non-trivial profile in the bulk of~$AdS_{2}$ given by~\eqref{eq:scalars}.

\section{Full-superspace integrals and quantum entropy}
\label{sec:Dterms-quantum}

Our goal is to examine the effect of the full-superspace integrals described in the previous section on the functional integral~\eqref{eq:QEF2} for the quantum entropy of 1/2-BPS black holes. We will show now that the quantum entropy is completely insensitive to any of these full-superspace integrals.

Our method of proof is conceptually very simple. As stressed in Chapter~\ref{chap:modern-BH-S} and elsewhere, the localizing manifold is defined using the off-shell supersymmetry variations and does not depend on the action. This means that a full-superspace integral added to the effective action can potentially affect the quantum entropy in the following three ways:
\begin{enumerate}
\item It can change the value of the effective action evaluated on the localizing solutions and therefore change the value of $S_{\rm ren}$.
\item It can change the measure on the localizing manifold either through the classical induced measure $[d \phi]$ or the value of the one-loop determinant $Z_{\text{1-loop}}$.
\item It can change the functional dependence of the electric charges $q_{I}$ on the fluctuating fields\footnote{The actual charges~$q_{I}$ take integer values and are fixed once and for all.}. (The magnetic charges $p^{I}$ are topological quantities and do not depend on the action.) 
\end{enumerate}
In the following, we will discuss point~1 and we will show that all known full-superspace integrals which can be constructed in~$\CN=2$ supergravity at any level in the derivative expansion do not contribute to the renormalized action~$S_{\rm ren}$. Before doing so, we examine the effect on the measure, the one-loop determinant, and the electric charges, assuming that point~1 holds.
 
The classical induced measure arises from considering the localizing manifold as an embedded submanifold of the full field space of supergravity. It is a function of the action evaluated on the submanifold and of the determinant of the embedding matrix. The localizing solutions are solutions of the BPS equations which, in our off-shell supergravity formalism, do not change under any modification of the action. This means that the embedding matrix is also independent of the action. Since, by assumption, the action evaluated on the localizing manifold does not change, the induced measure does not change\footnote{Note here that the determinant coming from the modes orthogonal to the embedding surface will change in general, but this fact is irrelevant to our computation here.} on addition of the full-superspace integrals. The one-loop determinant, by definition, is evaluated using the deformation action that is fixed once and for all in our first step of localization, and manifestly does not depend on the higher-derivative terms that we add to the effective action of supergravity. 

The electric charges~$q_{I}$ enter the functional integral in two different places, each time as a boundary term in the effective action. The first occurrence is the explicit coefficient of the Wilson line~\eqref{eq:QEF2} which clearly does not depend on the higher-derivative action. The other occurrence is through the boundary conditions of the gauge fields and scalar fields in the functional integral~\eqref{eq:QEF2}. Since the boundary conditions are completely fixed by the full-BPS solutions~\eqref{eq:full-BPS-metric}, the charge is completely determined by the semi-classical theory, and the off-shell deformation inside the~$AdS_{2}$ does not affect it. We have already seen in Section~\ref{sec:D-terms-S} that the functional form of the charges in the semi-classical theory are not modified by the addition of full-superspace terms.

We now turn to the first point in the list above. As described in Section~\ref{sec:D-terms-S}, the Weyl-squared multiplet of the localizing solutions is fixed to its classical full-BPS value that was displayed explicitly in~\eqref{eq:weylclass}. We now turn to the vector multiplet. For clarity, we parametrize the fluctuation away from the attractor solution by an arbitrary real field~$\varphi (r)$, and we will plug back the half-BPS localizing value $\varphi (r) = \frac{C }{r}$ at the end of the computation. The scalars are given by:
\be
X  = X _* + \frac{w}{4}\varphi , \quad\quad \bar{X}  = \bar{X} _* + \frac{w}{4}\varphi  \, ,
\ee
and the auxiliary fields are determined by supersymmetry in terms of $\varphi$:
\be
Y_1^{I,1} = -Y_2^{I,2} = \frac{w^2}{8}\left((r^2-1)\partial_r\varphi  + r\varphi \right) \, . 
\ee
This localizing solution is extended to all the components of a reduced chiral multiplet~$\mathcal{C}$ following~\cite{deRoo:1980mm}:
\begin{align}
A|_{\mathcal{C} } =&\, X  = X _* + \tfrac{w}{4}\varphi( r ) \cr
B_{ij}|_{\mathcal{C} } =&\, Y_{ij} \cr
F_{ab}^-|_{\mathcal{C} } =&\, -\tfrac{w}{16} T^-_{ab} \, \varphi (r) \\
C|_{\mathcal{C} } =&\, -\tfrac{w}{2}\,\mathcal{D}^2\varphi ( r ) + \tfrac{w}{64}(T^+_{ab})^2 \varphi (r ) \nn \, .
\end{align}
We also remind the reader that in Euclidean signature, the anti-chiral multiplet $\bar{\mathcal{C}} $ is not the complex conjugate of ${\mathcal{C}}$. Note that when $\varphi  = 0$, the half-BPS localizing configuration reduces to the full-BPS attractor solution, and we recover the constant multiplet~\eqref{eq:constmult}.

We now need to build weight zero chiral multiplets to use the full-superspace formula~\eqref{KinLag} built out of kinetic multiplets. As a simple example, using the reduced chiral multiplet~$\CC$ associated with one vector multiplet~$\bf X$ and the Weyl-squared multiplet, we can build a chiral multiplet $\Phi$ of weight $w=0$ by taking the combination
\be 
\label{XtoPhi}
\Phi = \mathcal{C} \otimes \left({\bf W}^2\right)^{-\tfrac{1}{2}} \, .
\ee
This composite chiral superfield has the following components:
\begin{align}
\label{phimult}
A|_{\Phi } =&\, \tfrac{1}{2iw}\,X _* + \tfrac{1}{8i}\,\varphi ( r ) \, ,\cr
B_{ij}|_{\Phi } =&\, \tfrac{1}{2iw}\,Y_{ij}  \, , \cr
F^-_{ab}|_{\Phi } =&\,  \tfrac{i}{32} T^-_{ab} \, \varphi ( r ) \, , \\
C|_{\Phi } =&\, \tfrac{i}{4}\,\mathcal{D}^2\varphi ( r ) - \tfrac{i}{128}(T^+_{ab})^2 \varphi ( r )\, . \nn
\end{align}
The corresponding composite superfield built out of an anti-chiral multiplet and the Weyl-squared multiplet is given by
\begin{align}
\label{phimultbar}
\bar{A}|_{\bar{\Phi}} =&\, -\tfrac{1}{2iw}\,\bar{X}_* - \tfrac{1}{8i}\,\varphi ( r ) \, ,\cr
B_{ij}|_{\bar{\Phi}} =&\, -\tfrac{1}{2iw}\,Y_{ij}  \, , \cr
F^+_{ab}|_{\bar{\Phi}} =&\, \tfrac{i}{32}T^+_{ab} \, \varphi ( r ) \, , \\
\bar{C}|_{\bar{\Phi}} =&\, -\tfrac{i}{4}\,\mathcal{D}^2\varphi ( r ) + \tfrac{i}{128}(T^-_{ab})^2 \varphi ( r )\, . \nn
\end{align}
The Lagrangian~\eqref{KinLag} evaluated on the field configuration~\eqref{phimult},~\eqref{phimultbar} is:
\begin{align}
\label{Lagrangianphi}
e^{-1}L =&\, \tfrac{1}{16}\mathcal{D}^2\varphi \mathcal{D}^2\varphi  + \tfrac{1}{8}\mathcal{D}^\mu \varphi  R(\omega,e)_\mu^{\:\:\:a} \mathcal{D}_a\varphi  + \tfrac{1}{16}\mathcal{D}^2\varphi \mathcal{D}^2\varphi  \cr
&\,-\tfrac{1}{512}\varphi \mathcal{D}^2\varphi \left[\left(T^+_{ab}\right)^2 + \left(T^-_{cd}\right)^2\right] + \tfrac{1}{16384}\left(T^+_{ab}\right)^2\left(T^-_{cd}\right)^2\left(\varphi \right)^2 \cr
&\,+\tfrac{w^2}{128}\,\partial^\mu\left[(r^2-1)\partial_r\varphi  + r\varphi \right]\partial_\mu\left[(r^2-1)\partial_r\varphi  + r\varphi \right] \\
&\,+ \tfrac{1}{64}T^{-c}_\mu T^+_{\nu c}\mathcal{D}^\mu\varphi \mathcal{D}^\nu\varphi - \tfrac{1}{64}\left[T^{+\;\mu b}T^+_{cb} - T^{-\;\mu b}T^-_{cb}\right]\mathcal{D}_\mu\varphi \mathcal{D}^c\varphi  \cr
&\,-\tfrac{1}{128}\mathcal{D}_a\varphi  \mathcal{D}^c\varphi  T^{-ab}T^+_{cb} - \tfrac{1}{256}T^{-\;ac}R(\omega,e)_a^{\:\:\:b}T^+_{bc}\left(\varphi \right)^2 \cr
&\,- \tfrac{1}{8192}\left(T^-_{ab}\right)^2\left(T^+_{cd}\right)^2\left(\varphi \right)^2 \nn \, .
\end{align}
The Riemann tensor of the near-horizon solution is determined completely by supersymmetry in terms of the $T^+$, $T^-$ components 
\be
R_a^{\;\;b} = \tfrac{1}{16}T_{ac}^-T^{+cb} \, .
\ee
Using this relation, and the explicit values of the~$T$-tensor \eqref{eq:TTensor}, the Lagrangian \eqref{Lagrangianphi} reduces to 
\be
e^{-1}L = \frac{1}{8}\mathcal{D}^2\varphi \mathcal{D}^2\varphi  + \frac{1}{64}\varphi \mathcal{D}^2\varphi 
+\frac{w^2}{128}\partial^\mu\left[(r^2-1)\partial_r\varphi  + r\varphi \right]\partial_\mu\left[(r^2-1)\partial_r\varphi  + r\varphi \right] \, .
\ee
Here we have used the fact that the covariant derivative on the scalar fields reduces to the ordinary partial derivative. This Lagrangian can be rewritten as follows:
\be \label{Lagfin}
e^{-1}L = \frac{1}{8}\mathcal{D}^2\varphi \left[r^2\mathcal{D}^2\varphi  + \frac{w^2}{8}\varphi \right] + \frac{w^2}{64}(r^2-1)\,\partial_r\left(r\varphi \right)\left[\mathcal{D}^2\varphi  + \frac{w^2}{32}\,\partial_r\left(r\varphi \right)\right] \, . 
\ee
Finally, plugging in the value $\varphi (r) = \tfrac{C^{I}}{r}$ shows that each of the two terms in the above Lagrangian vanishes, and we obtain:
\be \label{Liszero}
e^{-1}L = 0 \, .
\ee
We thus have that the simplest full-superspace Lagrangian 
\be
\label{kineticintagain}
\int d^4\theta \, d^4\bar{\theta}\;\Phi \, \bar{\Phi} \, ,
\ee
for the field~$\Phi$ of~\eqref{XtoPhi} vanishes when evaluated on our localizing solutions. It is easy to check that this result also holds for a chiral field multiplied by an anti-chiral field built out of different vector multiplets:
\be
\label{kineticintIJ}
\int d^4\theta \, d^4\bar{\theta}\;\Phi^{I} \, \bar{\Phi}^{J}  \, . 
\ee
The reason is that such a Lagrangian is quadratic in the fluctuation~$\varphi$ and, when evaluated on the localizing solutions labeled by the real parameters~$C^{I}$, is proportional to $C^{I} C^{J}$. The $r$-dependent part of the Lagrangian is exactly the same as in~\eqref{Lagfin} and vanishes for the same reason.

To discuss more general functions,  it is convenient to go to a gauge-fixed frame where $w$ and therefore the Weyl-squared multiplet is a constant. This means that the formula~\eqref{eq:HLagcomps} for the vector multiplets that was written down for functions of only vector multiplets can be used for functions of the vector multiplets and the Weyl-squared multiplet by simply replacing the weight one field~${\bf X}^{I}$ by the weight zero field~$\Phi^{I} = \mathcal{C}^{I} \otimes \left({\bf W}^2\right)^{-\tfrac{1}{2}}$. In this case, the function~$\CH$ can be an arbitrary real function~$\CH(\Phi^{I}, \bar \Phi^{I})$. As noted below~\eqref{eq:weylclass}, there are additional terms in the full Lagrangian, but these drop out for a constant Weyl multiplet, and the Lagrangian~\eqref{eq:HLagcomps} is thus the most general Lagrangian of this type.

Our task is now clear -- we need to evaluate the Lagrangian \eqref{eq:HLagcomps} on our localizing solutions~\eqref{phimult}. The Lagrangian splits into quadratic, cubic, and quartic terms in~$\Phi^{I}$ (and~$\bar \Phi^{I}$). The Lagrangian~\eqref{KinLag} follows from taking~$\CH = \Phi \, \bar \Phi$, in which case~\eqref{eq:HLagcomps} reduces to its quadratic piece, which vanishes on the localizing solutions as we've already seen in~\eqref{Liszero}. We note that the first term in the quadratic piece of~\eqref{eq:HLagcomps}  is equal to the term $C \bar C$ in~\eqref{KinLag}

We have already seen above that the Lagrangian~\eqref{eq:HLagcomps} vanishes when the chiral or anti-chiral superfield is a constant (namely of the type~\eqref{eq:constmult} with only the lowest component being non-zero and constant). This means that the Lagrangian evaluated on our localizing solutions is proportional to the fluctuations~$\varphi^{I}( r )$. Therefore, the quadratic, cubic, and quartic pieces in the Lagrangian are proportional to $\CH_{I\bar J} \, C^{I} \, C^{J}$, $\CH_{IJ \bar K} \, C^{I} \, C^{J} \, C^{K}$, and $\CH_{IJ \bar K \bar L} \, C^{I} \, C^{J} \, C^{K} \, C^{L}$ (recall that $C^{I}$ is real). The $r$-dependent part of the Lagrangian~\eqref{eq:HLagcomps} can therefore be extracted using a single superfield~$\Phi$ and its conjugate~$\bar \Phi$.

From our computation above, it is manifest that the quadratic piece vanishes on the full localizing solutions.  We find that the cubic and quadratic part of the Lagrangian~\eqref{eq:HLagcomps} also vanish identically. Therefore, the full Lagrangian~\eqref{eq:HLagcomps} vanishes on the localizing solutions. 

\section{Summary of results and assumptions}

The conclusion of the analysis presented in the previous Section is that the full-superspace integrals whose contribution were discarded in the originial localization computation of~\cite{Dabholkar:2010uh, Dabholkar:2011ec} indeed do not contribute to the final result for the quantum entropy of the 1/2-BPS black hole solutions of Euclidean~$\CN=2$ supergravity. Note that we have only considered a large class of such full-superspace integrals,but have not exhausted all the possibilities yet: there are, for instance, full superspace integrals built out of \emph{nested} kinetic multiplets,~\textit{i.e.} coming from kinetic multiplets built out of other kinetic multiplets, and so on and so forth. Nevertheless, we will consider in what follows that the same methods developed here can be applied to such ``nested'' full-superspace integrals. A preliminary investigation using manifest superspace methods was also conducted and seemed to corroborate the result, although it has not been published. 

Effectively, we will consider having proven point 4. in the list of features of the Master Formula, and we will therefore be able to safely use~\eqref{eq:master-integral} in the following Chapters to compute the quantum entropy of 1/2-BPS black holes. We repeat it here for convenience:
\be
\wh{W} (q, p) = \int_{\mathcal{M}_{Q}}  \, \prod_{I=0}^{\nv} [d\phi^{I}] \, \exp\Bigl[- \pi  \, q_I  \, \phi^I +4\pi\,\textnormal{Im}\,F\bigl(\tfrac{2}{w}(\phi^I + ip^I)\bigr) \Bigr] \, Z_\text{1-loop}(\phi^{I})\, .
\ee
Recall that the hat on~$W$ indicates that this formula still receives corrections from orbifold configurations, see footnote~\ref{footnote:WHat}. Since these contributions are exponentially suppressed, we will discard them in what follows.

Coming back to the impressive agreement the authors of~\cite{Dabholkar:2010uh, Dabholkar:2011ec} found between the quantum entropy of 1/8-BPS black holes in~$\CN = 8$ supergravity and the Maldacena-Moore-Strominger microscopic degeneracies of the corresponding D-brane system~\eqref{eq:rademsp}, recall that there was another assumption which entered the computation: the one-loop determinant factor~$Z_\textnormal{1-loop}$ in~\eqref{eq:master-integral} was taken to be trivial in the~$\CN=2$ truncation they considered. In the next Chapter, we will justify this assumption by computing this factor in a general theory of Euclidean~$\CN=2$ supergravity coupled to an arbitrary number of vector and hyper multiplets, and subsequently applying the result to the specific~$\CN=2$ truncation of~$\CN=8$ Type IIB string theory compactified on~$T^6$ which~\cite{Dabholkar:2010uh, Dabholkar:2011ec} examined. Anticipating the results, we will show that the functional determinant is indeed trivial in this particular case. Later in Chapter~\ref{chap:n4-loc}, we will make use of the general formula for this determinant and apply it to another~$\CN=2$ truncation, this time of Type IIB string theory compactified on~$\mathrm{K}3 \times T^2$  which gives rise to a low-energy four-dimensional~$\CN=4$ theory where 1/4-BPS black hole solutions exist.

\chapter{One-loop functional determinants in localization}
\label{chap:n2-loc}
In this Chapter, we examine another important aspect of supersymmetric localization of the QEF~\eqref{eq:QEF} and its application to the quantum entropy of black holes, namely the one-loop functional determinants entering the Master Formula~\eqref{eq:master-integral}. The measure along the localizing manifold itself~$[d\phi]$ has been discussed (in a slightly different context) in~\cite{Cardoso:2008fr}, and we will also comment on these results in due course. 

The task that we set ourselves here is to compute the one-loop fluctuation determinant of the localizing action~$Q\CV$~\eqref{eq:specificV} for vector multiplets and hypermultiplets. We compute the determinant of the fluctuations of the fields in the theory normal to the localization manifold at an arbitrary point~$\phi^{I}$ on~$\mathcal{M}_Q$, focusing on the dependence of this determinant on the charges and on the fields~$\phi^{I}$ and ignoring overall numerical constants. A non-trivial dependence on~$\phi^{I}$ means that the non-zero modes (under $Q$) of bosons and fermions do not cancel in the functional integral~\eqref{eq:QEF}. As we will see, the dependence of the determinant on the fields~$\phi^{I}$ appears only through the scale of the fluctuating geometry. 

In the vector multiplet sector, fixing the gauge symmetry associated to the vector fields~$A_\mu^I$  does not commute with the off-shell supersymmetry, and to address this problem, we develop a formalism to treat BRST symmetries for vector multiplets consistent with the off-shell closure of the supersymmetry algebra. We do so using the standard rules of quantization for theories with multiple gauge invariances~\cite{Batalin:1981jr, deWit:1978cd}. Our results are applicable to four-dimensional~$\CN=2$ supergravity coupled to vector multiplets in any background that preserves some supersymmetry. In the case of the (deformed) 4-sphere, it agrees with the treatment of~\cite{Pestun:2007rz, Hama:2012bg}. In the Euclidean~$AdS_{2} \times S^{2}$ background, our formalism leads to a different algebra. Note that we will only consider Abelian vector multiplets in what follows, although the generalization of these results to non-Abelian vector multiplets should be straightforward.

In the Weyl multiplet sector, we will make a particular choice of gauge in order to perform explicit calculations. The physical observables are, of course, gauge invariant. Recall that the superconformal algebra includes a local dilatation invariance under which the vierbein has scaling weight~$w=-1$ and the vector multiplet scalars~$X^{I}$ have~$w=+1$. The associated gauge field is~$b_{\mu}$. There is also a local invariance under special conformal transformations with gauge field~$f_\mu{}^{a}$. To gauge-fix the latter, we impose the K-gauge condition~$b_\mu=0$. To gauge-fix the former, it is convenient to introduce the symplectically invariant scalar~$\CK$ via:
\be 
\label{eq:EminK}
e^{-\CK} \, := \,  -i(X^I \bar{F}_I   - \bar{X}^I F_I) \, . 
\ee 
The field~$e^{-\CK}$ has scaling weight $w=2$, and it appears in the supergravity action as a conformal compensator, with the kinetic term for the graviton appearing through the combination:
\be 
\label{eq:OmLag}
\sqrt{g} \,  e^{-\CK} \, R   \, .
\ee
The physical, dilatation-invariant metric is given by~$G_{\mu\nu} := e^{-\CK} \, g_{\mu\nu}$.

The local scale invariance is generically gauge-fixed by setting a field with non-zero scaling weight to a constant value. A common choice of gauge is the condition~$e^{-\CK} =1$ in which we have only~$\nv$ fluctuating vector multiplets. In this gauge the original metric~$g_{\mu\nu}$ has the standard Einstein-Hilbert Lagrangian for the graviton, as seen easily from~\eqref{eq:OmLag}. In this Chapter however, we shall use the gauge condition~$\sqrt{g}=1$, which is also very convenient to analyze the solutions to the localizing equations~\cite{Dabholkar:2010uh}. In this gauge, the fluctuations of the graviton~$g_{\mu\nu}$ are constrained to have fixed volume, but we gain a linearly acting symplectic symmetry on the $(\nv+1)$ freely fluctuating fields~$X^{I}$ and~$\bar{X}^I$. 

As was briefly mentioned at the end of Chapter~\ref{chap:sugra}, we see from this discussion that one of the~($\nv+1$) vector multiplet plays the role of a~\emph{compensating} multiplet. In addition, we need another compensating multiplet to gauge-fix the extra gauge symmetries of the conformal supergravity theory, and we choose this to be a hypermultiplet as in~\cite{deWit:1980tn}. With these two compensators, the conformal supergravity theory is gauge-equivalent to Poincar\'{e} supergravity. Unlike the case for vector multiplets however, a formalism to treat off-shell~$\CN=2$ supersymmetry transformations on hypermultiplets with a finite number of auxiliary fields is not known. The compensating hypermultiplet is therefore treated using its equations of motion. We will briefly comment on the consequences of this in the following.

The gauge-fixed version of the full-BPS attractor solution of Section~\ref{sec:semi-class-1/2bps-S} is found by setting~$w=4$ in~\eqref{eq:full-BPS-metric},~\eqref{eq:attractor-scalars-metric} and~\eqref{eq:attractor-charges}. In particular, the R-symmetry gauge fields~$\mathcal{V}_\mu{}^i{}_j$ and~$A_\mu$ are pure gauge according to~\eqref{eq:full-BPS-gauge-aux} and we set them to zero in what follows. Note that, at the two-derivative level in the supergravity action (that is, when the prepotential does not depend on~$\wh{A}$), one may recast the semi-classical entropy formula~\eqref{eq:semi-class-1/2bps} in terms of the field~$\CK$ introduced in~\eqref{eq:EminK} as follows~\cite{Ferrara:1995ih}:
\be
\label{eq:semi-class-S-2deriv}
\mathcal{S}_{\text{BHW}} = \pi\,e^{-\CK} \, .
\ee
In this form, it is clear that if we uniformly scale all charges as~$(q_{I},p^{I}) \rightarrow \Lambda (q_{I},p^{I})$ with~$\Lambda \rightarrow \infty$, the semi-classical entropy of the 1/2-BPS black hole solution scales as~$\Lambda^2$. We will refer to this scaling behavior later in this Chapter.

Supersymmetric localization of the QEF yielding the Master Formula~\eqref{eq:master-integral} is performed using the supercharge~\eqref{eq:specificQ} contained in the superconformal algebra $\mathrm{SU}(1,1|2)$. Since we will discuss this supercharge and the localizing action~$Q\mathcal{V}$ built from it in great detail, we begin by finding an explicit expression for the associated spinor parameters on the Euclidean~$AdS_2 \times S^2$ background. These are found by analyzing the conformal Killing spinor equations obtained from requiring that the variations of the Weyl multiplet fermions~\eqref{eq:weyl-fermions-good-vars} vanish:
\be
\label{eq:CKS1}
2\,\mathcal{D}_\mu(P_\pm\epsilon^i) \pm \tfrac1{16}(T^+_{ab} + T^-_{ab})\gamma^{ab}\gamma_\mu P_\mp\epsilon^i - \mathrm{i}\,\gamma_\mu P_\mp\eta^i = 0\, ,
\ee
\vspace{-1cm}
\be
\label{eq:CKS2}
\pm\tfrac1{24}\gamma^{ab}\Slash{\mathcal{D}}(T_{ab}^+ + T_{ab}^-)P_\mp\epsilon^i + D\,P_\pm\epsilon^i \mp \tfrac1{24}\mathrm{i}\,(T_{ab}^+ + T_{ab}^-)\gamma^{ab}P_\pm\eta^i = 0 \, ,
\ee
with~$P_\pm = \tfrac12(1\pm\gamma^5)$. In contrast to~\eqref{eq:CKS1}, which determines the Killing spinors of the space-time and thus contains geometrical information, Equation~\eqref{eq:CKS2} does not impose any additional constraints on the geometry and is used to fix the value of the background auxiliary fields~$T_{ab}$ and~$D$ compatible with the conformal Killing spinors.

In the~$\sqrt{g}=1$ gauge, the metric appearing in~\eqref{eq:CKS1},~\eqref{eq:CKS2} is given by
\be
\label{eq:metric2}
ds^2 = \sinh^2\eta \,  d\tau^2   + d\eta^2 + d\psi^2 + \sin^2\psi \, d\phi^2 \, ,
\ee
where we have changed the radial variable from~\eqref{eq:full-BPS-metric} to~$r=\cosh\eta$. The~$T$-tensor is as in~\eqref{eq:TTensor} with~$w=4$ in the gauge-fixed theory.

We now observe that a set of solutions to the conformal Killing spinor equations can be found simply by setting
\be
\eta^i = 0 \, ,
\ee
taking the spinor parameter~$\epsilon$ to be a solution of
\be
\label{eq:KS}
2\,\mathcal{D}_\mu(P_\pm\epsilon^i) \pm \tfrac1{16}(T^+_{ab} + T^-_{ab})\gamma^{ab}\gamma_\mu P_\mp\epsilon^i  = 0 \, ,
\ee
with Euclidean~$AdS_2 \times S^2$ boundary conditions, and with the field~$D$ being zero. Note that in~\eqref{eq:KS}, the covariant derivative only contains the spin-connection in our gauge-fixed theory since~$b_\mu$ and the R-symmetry gauge fields have been set to zero, \textit{i.e.}~$\mathcal{D}_\mu\epsilon^i = \partial_\mu\epsilon^i - \tfrac{1}{4}\omega_\mu^{\;\;ab}\gamma_{ab}\epsilon^i$. 

For the spinor~$\epsilon^i{}_{(\mathrm{D})} = P_+\epsilon^i + P_-\epsilon^i$,~\eqref{eq:KS} reduces to 
\be 
\label{eq:KSfull}
\mathcal{D}_\mu\epsilon^i{}_{(\mathrm{D})} - \tfrac{1}{32}(T^+_{ab} + T^-_{ab})\gamma^{ab}\gamma_\mu\gamma^5\epsilon^i{}_{(\mathrm{D})} = 0 \, .
\ee
The solutions to Killing spinor equations of the form~\eqref{eq:KSfull} have been obtained for general~$AdS_n \times S^m$ geometries of both Minkowski and Euclidean signature in~\cite{Lu:1998nu}. We obtain the following four complex, linearly independent solutions of~\eqref{eq:KSfull}:
\begin{align}
\epsilon_{(\mathrm{D})}^{(1)} =&\, \tfrac{e^{\tfrac{\mathrm{i}}{2}(\phi+\tau)}}{\sqrt{2}}\begin{pmatrix}[1.0] \sinh\tfrac{\eta}{2}\cos\tfrac{\psi}{2} \\ i\,\cosh\tfrac{\eta}{2}\cos\tfrac{\psi}{2} \\ -\sinh\tfrac{\eta}{2}\sin\tfrac{\psi}{2} \\ i\,\cosh\tfrac{\eta}{2}\sin\tfrac{\psi}{2} \end{pmatrix}\, , \quad
\epsilon_{(\mathrm{D})}^{(2)} = \tfrac{e^{-\tfrac{\mathrm{i}}{2}(\phi+\tau)}}{\sqrt{2}}\begin{pmatrix}[1.0] \cosh\tfrac{\eta}{2}\sin\tfrac{\psi}{2} \\ i\,\sinh\tfrac{\eta}{2}\sin\tfrac{\psi}{2} \\ \cosh\tfrac{\eta}{2}\cos\tfrac{\psi}{2} \\ -i\,\sinh\tfrac{\eta}{2}\cos\tfrac{\psi}{2} \end{pmatrix} \, , \cr
\epsilon_{(\mathrm{D})}^{(3)} =&\, \tfrac{e^{\tfrac{i}{2}(\phi-\tau)}}{\sqrt{2}}\begin{pmatrix}[1.0] \cosh\tfrac{\eta}{2}\cos\tfrac{\psi}{2} \\ i\,\sinh\tfrac{\eta}{2}\cos\tfrac{\psi}{2} \\ -\cosh\tfrac{\eta}{2}\sin\tfrac{\psi}{2} \\ i\,\sinh\tfrac{\eta}{2}\sin\tfrac{\psi}{2} \end{pmatrix} \, , \quad
\epsilon_{(\mathrm{D})}^{(4)} = \tfrac{e^{-\tfrac{i}{2}(\phi-\tau)}}{\sqrt{2}}\begin{pmatrix}[1.0] \sinh\tfrac{\eta}{2}\sin\tfrac{\psi}{2} \\ i\,\cosh\tfrac{\eta}{2}\sin\tfrac{\psi}{2} \\ \sinh\tfrac{\eta}{2}\cos\tfrac{\psi}{2} \\ -i\,\cosh\tfrac{\eta}{2}\cos\tfrac{\psi}{2} \end{pmatrix} \, . 
\end{align}
From these four complex spinor solutions to~\eqref{eq:KSfull}, one may build eight linearly independent, symplectic Majorana-Weyl solutions to~\eqref{eq:CKS1},~\eqref{eq:CKS2} upon imposing the chiral projections and reality conditions~\eqref{eq:weyl-proj},~\eqref{eq:sympl-Majo-4D}. Note that these solutions all have~$\eta^i = 0$.

We are interested in the supercharge~\eqref{eq:specificQ} which squares to~$(L_0 - J_0)$. This supercharge is parametrized explicitly by the following symplectic Majorana-Weyl spinor~$\widehat{\epsilon}$, where we display the Weyl-projected~$\mathrm{SU}(2)$ components:
\begin{align}
\label{eq:killingspinor}
P_+\widehat{\epsilon}^{\,1} =&\, \frac{e^{-\tfrac{i}{2}(\tau + \phi)}}{\sqrt{2}}\begin{pmatrix}[1.0] \cosh\tfrac{\eta}{2}\sin\tfrac{\psi}{2} \\ 0 \\ 0 \\ -i\,\sinh\tfrac{\eta}{2}\cos\tfrac{\psi}{2} \end{pmatrix}\, , \quad
P_+\widehat{\epsilon}^{\,2} = \frac{e^{\tfrac{i}{2}(\tau + \phi)}}{\sqrt{2}}\begin{pmatrix}[1.0] \sinh\tfrac{\eta}{2}\cos\tfrac{\psi}{2} \\ 0 \\ 0 \\ i\,\cosh\tfrac{\eta}{2}\sin\tfrac{\psi}{2} \end{pmatrix}\, , \nonumber \\
P_-\widehat{\epsilon}^{\,1} =&\, \frac{e^{-\tfrac{i}{2}(\tau + \phi)}}{\sqrt{2}}\begin{pmatrix}[1.0] 0 \\ i\,\sinh\tfrac{\eta}{2}\sin\tfrac{\psi}{2} \\ \cosh\tfrac{\eta}{2}\cos\tfrac{\psi}{2} \\ 0 \end{pmatrix} \, , \quad \quad
P_-\widehat{\epsilon}^{\,2} = \frac{e^{\tfrac{i}{2}(\tau + \phi)}}{\sqrt{2}}\begin{pmatrix}[1.0] 0 \\ i\,\cosh\tfrac{\eta}{2}\cos\tfrac{\psi}{2} \\ -\sinh\tfrac{\eta}{2}\sin\tfrac{\psi}{2} \\ 0 \end{pmatrix}\, .
\end{align}

\section{Off-shell supersymmetry algebra}
\label{subsec:offshellsusy}

Now that we have obtained the explicit spinor parameter for the supercharge~\eqref{eq:specificQ} used in the localization procedure, we proceed to the supersymmetry transformations of the fluctuations around the localizing manifold. 

Hereafter, the spinor~$\epsilon^i$ is taken to be specifically the one given in~\eqref{eq:killingspinor} and we omit the hat for clarity. Moreover, we will be interested in writing the action of the supercharge on the various fields in a cohomological form (\textit{i.e.} as an operator squared rather than an anti-commutator). To this end, we will use \emph{commuting} spinor parameters in this Chapter. This is achieved by extracting a Grassmann number in the expressions for the spinor parameters and the supercharge. We use this convention in order to stay as close as possible to what is usually used in the literature~\cite{Pestun:2007rz, Hama:2012bg}.

\noindent \textbf{Vector multiplets:} 
The supersymmetry transformation rules for the vector multiplet on our localizing background are, using~\eqref{eq:vector-good-vars}:
\begin{align} 
\label{eq:SUSYvect}
Q X^I =&\, \bar{\epsilon}_i P_+\lambda^{I\,i} \, , \cr 
Q \bar{X}^I =&\, \bar{\epsilon}_i P_-\lambda^{I\,i} \, ,\cr
Q W_\mu^I =&\, \bar{\epsilon}_i\gamma_\mu P_+\lambda^{I\,i} - \bar{\epsilon}_i \gamma_\mu P_- \lambda^{I\,i} \, , \\
Q(P_+\lambda^{I\,i}) =&\, \tfrac12\mathcal{F}_{ab}^{I\,-}\gamma^{ab}P_+\epsilon^i + 2\Slash{\partial}X^I\,P_-\epsilon^i - Y^{I\,i}{}_j\,P_+\epsilon^j \, , \cr
Q(P_-\lambda^{I\,i}) =&\, \tfrac12\mathcal{F}_{ab}^{I\,+}\gamma^{ab}P_-\epsilon^i - 2\Slash{\partial}\bar{X}^I\,P_+\epsilon^i - Y^{I\,i}{}_j\,P_-\epsilon^j \, , \nonumber \\
Q Y^{I\,i}{}_j =&\,  -2\,\bar{\epsilon}_j\gamma^5\Slash{\mathcal{D}}\lambda^{I\,i} + \delta^i{}_j\bar{\epsilon}_k\gamma^5\Slash{\mathcal{D}}\lambda^{I\,k} \, , \nonumber
\end{align}
where
\be
\mathcal{F}^{I\,-}_{ab} := F_{ab}^{I\,-} - \frac{1}{4}\bar{X}^I T_{ab}^- \, , \qquad \mathcal{F}^{I\,+}_{ab} := F_{ab}^{I\,+} - \frac{1}{4}X^I T_{ab}^+ \, , \nonumber 
\ee
and~$F^{I\,\pm}_{ab}$ is the (anti)self-dual part of the vector field strength. The covariant derivative on spinors is given by~$\mathcal{D}_\mu = \partial_\mu - \tfrac{1}{4}\omega_\mu^{\;\;ab}\gamma_{ab}$ in the gauge-fixed theory.
 
The square of the supersymmetry transformations can be obtained either by acting twice with~\eqref{eq:SUSYvect}, or using the general algebra derived in Section~\ref{sec:sugra-alg} and evaluating it on the full-BPS~$AdS_2 \times S^2$ background for commuting spinor parameters. We find:
\begin{align} 
\label{eq:Qsquarevect}
Q^2X^I =\,& v^\mu\partial_\mu X^I \, , \cr
Q^2\bar{X}^I =&\, v^\mu\partial_\mu \bar{X}^I \, , \cr
Q^2W^I_\mu =\,& v^\nu F^I_{\nu\mu} + \partial_\mu\bigl(2K_- \bar{X}^I + 2K_+ X^I\bigr) \, , \\
Q^2(P_+\lambda^{I\,i}) =\,& v^\mu \mathcal{D}_\mu P_+\lambda^{I\,i} + \tfrac14\,D_av_b\gamma^{ab} P_+\lambda^{I\,i} \, , \cr
Q^2(P_-\lambda^{I\,i}) =\,& v^\mu \mathcal{D}_\mu P_-\lambda^{I\,i} + \tfrac14\,D_av_b\gamma^{ab} P_-\lambda^{I\,i} \, , \cr
Q^2Y^i{}_j =\,& v^\mu \partial_\mu Y^i{}_j \nonumber \, .
\end{align}
The transformation parameters in~\eqref{eq:Qsquarevect} are given by
\be 
\label{eq:Killingcontract}
v^\mu = 2\,\bar{\epsilon}_i\gamma^\mu P_-\epsilon^i \, , \quad K_+ = \bar{\epsilon}_i P_-\epsilon^i \, , \quad K_- = \bar{\epsilon}_i P_+\epsilon^i \, . 
\ee
In the right-hand side of \eqref{eq:Qsquarevect}, we have used the following useful identities 
\be 
\label{eq:useid}
D_{[a}v_{b]} = -\frac{1}{4}K_-T^+_{ab} - \frac{1}{4}K_+T^-_{ab} \, ,
\ee
and 
\be
\partial_\mu K_\pm = \frac{1}{8}v^\nu T_{\mu\nu}^\pm \, , 
\ee 
which can be derived directly from the definition of the Killing vector and the conformal Killing spinor equations~\eqref{eq:CKS1} with~$\eta^i = 0$. 

Using the explicit form of the Killing spinor~\eqref{eq:killingspinor}, we find that 
\be
\label{eq:vmu}
v^\mu \=  \bigl(i \quad  0 \quad  0 \;\, -i\bigr)^{T} \, , 
\ee
and 
\be 
\label{eq:kpm}
K_\pm \= \frac{1}{2}\left(\cosh\eta \pm \cos\psi\right) \, ,
\ee
which we will use in the next section. 

\vspace{0.4cm}

\noindent \textbf{Hypermultiplets:}
We consider a set of~$n_h$ hypermultiplets where the scalars are denoted by~$A_i{}^{\alpha}$ with~$\alpha=1\ldots 2\,n_h$. The index~$i$ is a doublet under the~$\mathrm{SU}(2)$ R-symmetry, so that we have total of~$4\,n_{h}$ real scalars. The~$4\,n_h$ fermions are the~$2\,n_h$ positive-chirality spinors~$\zeta_\alpha$ and the~$2\,n_h$ negative-chirality spinors~$\zeta^\alpha$. We take the hypermultiplet fields to be neutral under the~$\mathrm{U}(1)$ gauge symmetry of the vector multiplet, as this is consistent with the classical attractor solution in asymptotically flat space. The scalars~$A_i{}^{\alpha}$ span a quaternionic-K\"{a}hler manifold and we will assume that the target-space of the hypermultiplet sigma model is flat~\cite{deWit:1984px}.

Hypermultiplets do not participate in the classical attractor black hole background -- they take zero or constant values as shown in~\eqref{eq:full-BPS-metric}, and as a consequence, they do not contribute to the classical action. Their \emph{quantum fluctuations}, however, are relevant for our discussion, and we will need an off-shell supersymmetry algebra to treat these fluctuations within our approach. For vector multiplets, we were able to directly use the formalism of off-shell conformal supergravity. For hypermultiplets, however, there is no known off-shell formalism for the full~$\CN=2$ supersymmetry algebra with a finite number of auxiliary fields. 

There is, however, a formalism for the off-shell closure of the algebra of one supercharge for vector and hyper multiplets with a finite number of auxiliary fields~\cite{Berkovits:1993hx}. This formalism was used in localization problems in four-dimensional field theory as in~\cite{Pestun:2007rz, Hama:2012bg}. This algebra acting on vector multiplets is exactly the one given by the conformal~$\CN=2$ supergravity formalism that we used above. As was emphasized many times now, the localization solutions~\eqref{eq:attractor-scalars-metric} are universal, in the sense that they do not depend on the physical action of the theory and continue to hold even in the presence of other matter fields (which are all constant as in the classical background~\eqref{eq:full-BPS-metric}). We can therefore use the formalism of~\cite{Berkovits:1993hx} and~\cite{Pestun:2007rz, Hama:2012bg} for hypermultiplets in black hole backgrounds.

The Q-supersymmetry transformation rules are given by a modification of~\eqref{eq:hyper-good-vars}:
\begin{align} 
\label{eq:SUSYhyp}
QA_i{}^\alpha =&\, 2\mathrm{i}\,\bar{\epsilon}_i P_+\zeta^\alpha - 2\mathrm{i}\,\bar{\epsilon}_i P_-\zeta^\alpha \, , \cr
Q(P_+\zeta^\alpha) =&\, -\mathrm{i}\,\Slash{\partial}A_i{}^\alpha P_-\epsilon^i - 2\,H_i{}^\alpha P_+\breve{\epsilon}^{\,i} \, , \cr
Q(P_-\zeta^\alpha) =&\, -\mathrm{i}\,\Slash{\partial}A_i{}^\alpha P_+\epsilon^i - 2\,H_i{}^\alpha P_-\breve{\epsilon}^{\,i} \, , \\
QH_i{}^\alpha =&\, \bar{\breve{\epsilon}}_i \Slash{\mathcal{D}}(P_-\zeta^\alpha) - \bar{\breve{\epsilon}}_i \Slash{\mathcal{D}}(P_+\zeta^\alpha) \, , \nonumber
\end{align}
where the action of the covariant derivative on the spinors is exactly as in the vector multiplet. Here,~$H_i{}^\alpha$ are~$4\,n_h$ scalar auxiliary fields. Indeed, upon setting~$H_i{}^\alpha=0$, one recovers the on-shell transformation rules of Chapter~\ref{chap:sugra}. 

In the off-shell transformations~\eqref{eq:SUSYhyp}, the parameters~$\breve{\epsilon}^{\,i}$,~$\breve{\epsilon}_i$ are built to satisfy:
\begin{align} 
\label{eq:constrainedparam}
\bar{\epsilon}_i P_- \breve{\epsilon}^{\,j} - \bar{\epsilon}_i P_+ \breve{\epsilon}^{\,j} =&\, 0 \, , \cr
\bar{\breve{\epsilon}}_i P_\mp \breve{\epsilon}^{\,i} - \bar{\epsilon}_i P_\pm \epsilon^i =&\, 0 \, , \\
\bar{\breve{\epsilon}}_i\gamma^\mu P_- \breve{\epsilon}^{\,i} - \bar{\epsilon}_i\gamma^\mu P_- \epsilon^i =&\, 0 \, . \nonumber
\end{align}
In these equations, the spinors~$\epsilon^i$,~$\epsilon_i$ are given by~\eqref{eq:killingspinor} as before. As mentioned in~\cite{Hama:2012bg}, the constraints~\eqref{eq:constrainedparam} do admit non-trivial solutions to~$\breve{\epsilon}$. An explicit solution is given by
\be 
\label{eq:xicheck}
P_+\breve{\epsilon}^{\,i} = \left(\frac{\cosh\eta-\cos\psi}{\cosh\eta+\cos\psi}\right)^{-1/2}P_+\epsilon^i \, , \quad P_-\breve{\epsilon}^i = \left(\frac{\cosh\eta-\cos\psi}{\cosh\eta+\cos\psi}\right)^{1/2}P_-\epsilon^i \, .
\ee
With these constraints, the Q-supersymmetry transformations close off-shell:
\begin{align} 
\label{eq:Qsquarehyp}
Q^2A_i{}^\alpha =&\, v^\mu \partial_\mu A_i{}^\alpha \, , \cr
Q^2(P_+\zeta^\alpha) =&\, v^\mu \mathcal{D}_\mu P_+\zeta^\alpha + \tfrac{1}{4}\,D_a v_b\gamma^{ab} P_+\zeta^\alpha \, , \\
Q^2(P_-\zeta^\alpha) =&\, v^\mu \mathcal{D}_\mu P_-\zeta^\alpha + \tfrac{1}{4}\,D_a v_b\gamma^{ab} P_-\zeta^\alpha  \, , \cr
Q^2H_i{}^\alpha =&\, v^\mu \partial_\mu H_i{}^\alpha \nonumber \, . 
\end{align}
For the localization analysis, we set all the fermion variations in~\eqref{eq:SUSYhyp} to zero. It is clear that the configuration where the auxiliary field~$H_i{}^\alpha=0$ and the hypermultiplet scalars~$A_i{}^{\a} = \textnormal{constant}$ is a solution to the above BPS equations. In order to find an exhaustive list of all solutions, one needs to do an analysis as in~\cite{Gupta:2012cy} by separating the different tensor structures on the right-hand side. For now, we proceed with the trivial solutions.

\noindent \textbf{Supersymmetry algebra of $Q$:} Inspection of~\eqref{eq:Qsquarevect} and~\eqref{eq:Qsquarehyp} shows that supersymmetry algebra of~$Q$ acting on all fields of the vector and hypermultiplets in the Euclidean~$AdS_{2}\times S^{2}$ background is:
\be 
\label{eq:Qsquarefullcov} 
Q^2 = \delta_\text{cgct}(v) + \delta_M\left(L_{ab}\right) + \delta_{\text{gauge}}(\theta^I) \, , 
\ee
where the quantities on the right-hand side are as follows. 

The operator~$\delta_\text{cgct}(v)$ is the covariant general coordinate transformation, which is the variation under all gauge symmetries of the conformal supergravity theory (including regular general coordinate transformations, but also \textit{e.g.}~the~$\mathrm{U}(1)$ gauge symmetry of the vector multiplets), with the gauge parameters determined by the vector~$v^\mu$ (given by~\eqref{eq:Killingcontract} for our background). In our case, it is equal to the sum of the Lie derivative along the vector~$v^\mu$ and the~$\mathrm{U}(1)$ gauge transformation parametrized by~$-v^{\mu} W^{I}_{\mu}$. 

The transformation~$\delta_M$ is a Lorentz transformation parametrized by (see \eqref{eq:useid})
\be
L_{ab} \, := \, -\frac{1}{4}\bigl(K_-T^+_{ab} + K_+T^-_{ab}\bigr) = D_{[a}v_{b]} \, ,
\ee
which, on our background solution, equals
\be  
\label{eq:ourlab}
L_{ab} \= \begin{pmatrix}[.7] 0 & -i\,\cosh \eta & 0 & 0 \\ i\,\cosh \eta & 0 & 0 & 0 \\ 0 & 0 & 0 & -i\,\cos \psi \\ 0 & 0 & i\,\cos \psi 
& 0 \end{pmatrix}\, . 
\ee
Lastly, the transformation~$\delta_{\text{gauge}}$ is a~$\mathrm{U}(1)$ gauge transformation parametrized by
\be
\theta^I := 2K_- \bar{X}^I + 2K_+ X^I \, .
\ee
In the following, we will combine the off-shell supersymmetry~$Q$ with the BRST symmetry encoding the~$\mathrm{U}(1)$ gauge symmetry of the vector multiplet. To do so, we isolate the~$\mathrm{U}(1)$ gauge connection term present in the covariant general coordinate transformation of~\eqref{eq:Qsquarefullcov} and combine it with the gauge transformation already present in the algebra of~$Q$. We thus rewrite the off-shell supersymmetry algebra in the form\footnote{A similar procedure can be used to combine the spin-connection term appearing in the covariant general coordinate transformation of fermions with the Lorentz transformation parameter~$L_{ab}$. In the gauge where $\omega_\tau^{\;12} = -\cosh\eta\, ,\omega_\phi^{\;34} = \cos\psi$, this yields $L_{ab} - v^\mu\omega_{\mu\,ab} = 0$, so that the supersymmetry algebra is simply $Q^2 = v^\mu\partial_\mu + \delta_{\text{gauge}}(\wh{\theta}^I)$. It will be enough to stay in a generic Lorentz gauge where such cancellations need not happen.}
\be 
\label{eq:Qsquare}
Q^2 \=  \mathcal{L}_v + \delta_M(L_{ab}) + \delta_{\text{gauge}}(\wh{\theta}^I) \, ,
\ee 
where~$\mathcal{L}_v$ is the Lie derivative along the vector~$v^\mu$, and
\be
\wh{\theta}^I \, := \, 2K_- \bar{X}^I + 2K_+ X^I - v^\mu W_\mu^I \, .
\ee
Using the values~\eqref{eq:full-BPS-metric} of the background gauge fields~$W_{\mu}^{I}$ on the localizing manifold, we obtain the explicit expression:
\be 
\label{eq:ourthhat}
\wh{\theta}^I = e^{I}_{*} + 2\,C^{I} = \phi^{I} \, . 
\ee
Note that, on the localizing manifold~$\mathcal{M}_Q$, the gauge parameters on the right-hand side of the supersymmetry algebra are precisely the coordinates on~$\mathcal{M}_Q$. 

We note here that the algebra~\eqref{eq:Qsquare} of the supercharge~$Q$ is similar in structure, but not quite the same, as the one appearing in~\cite{Pestun:2007rz,Hama:2012bg}. Before specifying the background manifold, the off-shell supersymmetry transformations~\eqref{eq:SUSYvect},~\eqref{eq:SUSYhyp} are the same as the corresponding ones in~\cite{Pestun:2007rz,Hama:2012bg}. The reason for the difference is simply that the background values of the supergravity fields are different. In particular, the right-hand side of the algebra~\eqref{eq:Qsquare} involves the~$\mathrm{SU}(2)$ R-symmetry of supergravity in the case of the sphere, while this term is absent in our case. Instead, the Euclidean~$AdS_{2} \times S^{2}$ algebra contains a Lorentz rotation which the sphere algebra does not have. This fact will play a role in our analysis of the index theorem in Section~\ref{sec:DetCalc}.

\section{Gauge-fixing and BRST ghosts}
\label{sec:ghosts}

We now turn to the issue of gauge-fixing the~$\mathrm{U}(1)$ symmetry in each vector multiplet. The main problem is that the action of fixing a gauge does not commute with the off-shell supersymmetry -- which is central to our localization methods. To treat this problem, we will need to extend the off-shell supersymmetry algebra of~$Q$ to include the effect of the gauge-fixing. We also saw a hint of this appearing in the fact that the supercharge~$Q$ squares to a compact bosonic generator \emph{only} modulo a gauge transformation in~\eqref{eq:Qsquare}.

It is natural to solve this problem by combining the conformal~$\CN=2$ supergravity formalism with the covariant BRST formalism\footnote{Another, more hands-on method is to choose a 
suitable gauge-fixed background and to compute the bosonic and fermionic eigenmodes around this background. The non-cancellation then happens because the naive~$Q$ operator, upon acting on a certain eigenmode, moves us out of the gauge-fixing condition and one therefore has to modify~$Q$ as in \textit{e.g.}~\cite{Alday:2013lba}.} by adding Fadeev-Popov ghosts to the theory. The technical task is to set up a BRST complex for the gauge symmetries of the theory, and combine it with the off-shell supersymmetry complex generated by~$Q$. This procedure builds a new supercharge~$\wh{Q}$ which, as we will demonstrate, is suitable for localization and encodes both the gauge symmetry and the supersymmetry of the action. Once this formalism has been set up, the approach turns out to be extremely compact, and we can use index theory to elegantly compute the required functional determinants as laid out some time ago in~\cite{Pestun:2007rz}.

To treat the~$\mathrm{U}(1)$ gauge symmetry of the vector multiplet, we introduce a standard BRST ghost system. A~$\mathrm{U}(1)$ gauge transformation acts on the vector fields as
\be
\delta_g W^I_\mu = \partial_\mu\xi^I
\ee
where~$\xi^I$ is the parameter of the transformation in each vector multiplet. To each of these transformations we associate a ghost~$c^I$ along with an anti-ghost~$b^I$ and a Lagrange multiplier~$B^I$. Notice that the operator~$\partial_{\mu}$ has normalizable zero modes on a compact space, namely any constant function. In order to treat these zero modes we need to introduce the so-called ghost-for-ghosts: the constant field~$c_0^I$, along with two BRST-trivial pairs~$(\bar{\eta}^I,B^I_1)$ and~$(\eta^I,\bar{B}^I_1)$. This is the required field content to properly fix the gauge in the QEF path-integral~\eqref{eq:QEF2}. This fact is most easily understood by making use of the Batalin-Vilkovisky formalism~\cite{Batalin:1981jr, Gomis:1994he} and noting that the gauge theory at hand is a \emph{first stage reducible} theory. 

The BRST transformation laws of the vector multiplet fields in the adjoint of the~$\mathrm{U}(1)$ gauge group are:
\be 
\label{eq:BRSTvect}
\delta_B W^I_\mu   =  \Lambda\,\partial_\mu c^I \, ,  \quad \delta_BX^I = 0 \, , \quad  \delta_B\bar{X}^I =0 \, , \quad \delta_B\lambda^{i\;I}_+  =  0 \, , \quad \delta_B\lambda^{i\;I}_- = 0 \, , \quad \delta_BY^I_{ij} = 0 \, ,  
\ee
with~$\Lambda$ a constant anti-commuting parameter parametrizing the BRST transformation. We also have the following transformations on the ghost fields:
\be
\delta_Bb^I = \Lambda B^I \, , \quad \delta_BB^I = 0 \, , \quad \delta_B\eta^I = \Lambda\bar{B}^I_1 \, , \quad \delta_B\bar{B}^I_1 = 0 \, , \quad \delta_B\bar{\eta}^I = \Lambda B_1^I \, , \quad \delta_BB_1^I = 0 \, ,
\ee 
and
\be
\delta_Bc^I = \Lambda c^I_0 \, , \quad \delta_B c^I_0 = 0.
\ee
The operator~$Q_B$ defined by~$\delta_B\phi := \Lambda\,Q_B\phi$ ($\phi$ being any field of the theory) is a nilpotent operator, due to the fact that the field~$c^I_0$ is constant.

We now add to the~$\CN=2$ supergravity Lagrangian a~$Q_B$-exact gauge-fixing term:
\be 
\label{eq:gflagrangian}
\CL_{\text{g.f.}} = Q_B\left[b^I\left(-\frac{B^I}{2\xi_W} + G^W(W^I_\mu)\right) + \bar{\eta}^I\left(-\frac{B^I_1}{2\xi_c} + G^c(c^I)\right) + \eta^I\left(-\frac{\bar{B}^I_1}{2\xi_b} + G^b(b^I)\right)\right] \, , 
\ee
where~$G^W, G^c$ and~$G^b$ are appropriate gauge-fixing functions for the vector field, the ghost and the anti-ghost, respectively, and~$\xi_W$,~$\xi_b$ and~$\xi_c$ are constant parameters. The gauge-fixed action
\be
S_{\text{gauge-fixed}} = S_0 + \int d^4x \,\CL_{\text{g.f.}} \, ,
\ee
where~$S_0$ is the action of vector and hypermultiplets coupled to conformal supergravity, is BRST invariant since~$\CL_{\text{g.f.}}$ is~$Q_B$-exact and~$Q_B$ is nilpotent. Expanding~\eqref{eq:gflagrangian} using the BRST transformation rules leads to the expression
\begin{align} 
\label{eq:gfaction}
S_{\text{g.f.}} =&\, \int d^4x \,\CL_{\text{g.f.}} \cr
=&\, \int d^4x \,\Big\{B^I\left(G^W(W^I_\mu) - \frac{B^I}{2\xi_W} - \eta^I\frac{\delta G^b}{\delta b^I}\right) - b^I\frac{\delta G^W(W^I_\mu)}{\delta W^J_\mu}\partial_\mu c^J \cr
&\,+\bar{B}^I_1\left(G^b(b^I) - \frac{\bar{B}^I_1}{2\xi_b}\right) + B_1^I\left(G^c(c^I) - \frac{B_1^I}{2\xi_c}\right) - c^I_0\bar{\eta}^J\frac{\delta G^c(c^I)}{\delta c^J}\Big\} \, . 
\end{align}
One can recognize in this action the field~$B^I$ as a Gaussian-weighted Lagrange multiplier for the gauge condition~$G^W(W^I_\mu) = \eta^J\frac{\delta G^b(b^I)}{\delta b^J}$, the field~$B_1^I$ as a Gaussian-weighted Lagrange multiplier for the gauge condition~$G^c(c^I) = 0$ and the field~$\bar{B}_1^I$ as a Gaussian-weighted Lagrange multiplier for the gauge condition~$G^b(b^I) = 0$. For the case at hand, these last two gauge-fixing functions are meant to freeze the freedom one has in shifting the ghost and anti-ghost by a constant function, and we can thus take them specifically to be~$G^c(c^I) = c^I$ and~$G^b(b^I) = b^I$. The~$B_1^I, \bar{B}_1^I$ Lagrange multipliers then impose the conditions that~$\int c^I = 0$ and~$\int b^I = 0$, respectively. The gauge-fixing function for the gauge field~$W_\mu^I$ is then fixed to~$G^W(W^I_\mu) = \eta^I$ through the equation of motion for the Lagrange multiplier~$B^I$. Note also that the partition function computed from this gauge-fixed action is independent of the~$\xi_W,\xi_c$ and~$\xi_b$ parameters~\cite{Pestun:2007rz}.

We pause here for a moment in order to make a technical comment on the ghost set up that was used in the original work of Pestun~\cite{Pestun:2007rz}. For non-Abelian gauge theories, like the one considered in~\cite{Pestun:2007rz}, constant functions like~$c_0$ are not zero modes of the operator~$D^a_\mu$ (where~$a$ is a color index). One could have tried to set up the ghost-for-ghost~$c_0$ to be a zero mode of the covariant derivative and thus take it to be a covariantly constant function -- indeed, this may seem natural from a certain point of view. Doing so, however, would render the integrations over the gauge field and the ghost-for-ghost inter-dependent inside the path-integral, which is difficult to implement in practice. The strategy for non-Abelian gauge fields considered in~\cite{Pestun:2007rz} was to keep~$c_0$ as a constant function, and use a BRST charge which is non-nilpotent. In our case the gauge symmetry is Abelian, so that we may use an honestly nilpotent BRST charge.

We now apply the above formalism to our problem of Abelian vector multiplets on Euclidean~$AdS_{2} \times S^{2}$. The non-compact nature of the space introduces some subtleties. 

Firstly, we need to specify boundary conditions on all the fields. For the physical fields, we choose boundary conditions as in~\cite{Castro:2008ms,Murthy:2009dq}. For the ghost fields, we impose Dirichlet boundary conditions on the fields~$b^I,\,c^I$. This implies that there is no normalizable zero modes for these fields, and therefore no ghost-for-ghosts. This is consistent with the boundary conditions used in~\cite{Sen:2011ba} for the gauge parameters. Using this formalism, we set all the ghost-for-ghost fields to zero hereafter.  

Secondly, there is the issue of boundary modes which are normalizable modes of the gauge fields~$W_\mu^I$ that are formally pure gauge, but with gauge parameters that do not vanish at infinity (these have been called ``discrete modes''~\cite{Sen:2011ba}). These modes are zero-modes of the Laplacian on the~$AdS_2\times S^2$ background. The four-dimensional bulk action depends only on gauge invariant quantities and therefore does not depend on these discrete modes -- thus naively giving a divergence in the path integral. These special modes have been treated carefully in~\cite{Sen:2011ba}, and the idea is to obtain their contribution separately using arguments of ultra-locality. This gives rise to a factor of~$\ell^{-2\beta}$ to the functional integral, where $\ell$ is the background length scale of the problem and $\beta$ depends on the field under consideration. The non-zero modes can be treated as usual, but since one needs a complete set of local fields in the computation, one should add and subtract one set of zero modes to the non-zero modes, thus obtaining the contribution of a complete local set of modes and a factor of~$\ell^{2}$. As a result, one needs to multiply the answer found by using a complete set of local field observables by a factor~$\ell^{2-2\beta}$. In the on-shell theory of~\cite{Sen:2011ba}, it was found that for the gauge fields~$\beta=1$, which effectively means that the discrete modes do not contribute. In the context we examine in this Chapter, the discrete modes are \emph{not} zero modes of the~$H$ operator introduced below in~\eqref{eq:Qalg} used for the calculation of the determinants. We believe that their contribution can nevertheless still be ignored. In order to justify this procedure more carefully in our localization computation, one needs to analyze the cut-off theory and carefully take an infinite-volume limit. This must be done in such a way as to keep the local superalgebra and the completeness of the basis intact. Another possible resolution of this subtlety is that boundary effects will lift these zero modes in the localization action, as consistent with the fact that~$H$ takes non-zero values on these modes. The boundary conditions introduced in the context of the AdS/CFT correspondence in~\cite{Henningson:1998cd} may be relevant to this discussion. Since additional work is required to treat these modes in the off-shell theory, we will for now proceed with the assumption that they can be ignored. In contrast, such modes are expected to play a role in the graviton determinant calculation since already in the on-shell theory, they have~$\beta \neq 1$.

\vspace{0.4cm} 

\noindent \textbf{The combined supercharge~$\wh{Q}$ and its algebra:} 
We now consider the combined transformation for the BRST symmetry and the off-shell supersymmetry, generated by~$\wh{Q} \equiv Q+Q_B$. We require this new supercharge to square to
\be 
\label{eq:Qalg}
\wh{Q}^2 = \mathcal{L}_v + \delta_M\left(L_{ab}\right) \, \equiv \, H \, , 
\ee
where~$\mathcal{L}_v$ and~$\delta_M$ are the Lie derivative and the Lorentz transformations defined around Equation~\eqref{eq:Qsquare}. Note that the vector multiplet gauge transformation is no longer present on the right-hand side of the algebra~\eqref{eq:Qalg} -- precisely because it is already encoded in the BRST symmetry. This algebra allows us to systematically derive the supersymmetry transformation rules on the ghost system.
 
Expanding~$\wh{Q}^2$, and using the algebra~\eqref{eq:Qsquare} for~$Q$ and the nilpotency of~$Q_{B}$, we obtain
\be
\wh{Q}^2 = Q^2 + Q_B^2 + \left\{Q,Q_B\right\} =  \mathcal{L}_v + \delta_M\left(L_{ab}\right) + \delta_{\text{gauge}}(\wh{\theta}^I) + \left\{Q,Q_B\right\} \, .
\ee
Comparing with~\eqref{eq:Qalg}, we deduce that the anti-commutator of a supersymmetry and a BRST transformation on the physical and auxiliary fields of the theory should compensate for the gauge transformation parametrized by the vector and scalar fields of the vector multiplet. Applying this observation to the various fields leads to the supersymmetry transformation rules for the ghost system. 

As an example, consider the vector field~$W_\mu^I$:
\be
\left\{Q,Q_B\right\}W_\mu^I \= Q\left(\partial_\mu c^I\right) \= -\partial_\mu(\wh{\theta}^I) \, , 
\ee
which immediately yields 
\be
Qc^I = -\wh{\theta}^I \, .
\ee
Applying~$\wh{Q}^2$ to the other fields of the theory, we obtain the remaining supersymmetry transformations\footnote{The same procedure can be applied to also determine the transformation rules for the ghost-for-ghost fields when they are present, \textit{e.g.}~as in~\cite{Pestun:2007rz}.}
\be
Qb^I = 0 \, , \quad QB^I = \mathcal{L}_v b^I \, . 
\ee
We can now write down the various anticommutators on all fields of the theory as
\begin{align}
Q^2\Phi^{(')} =&\, \Bigl(\mathcal{L}_v + \delta_M(L_{ab})+ \delta_{\text{gauge}}(\wh{\theta}^I)\Bigr)\Phi^{(')}\, , \qquad Q^2(\text{gh.}) = 0 \, , \cr
Q_B^2\Phi^{(')} =&\, 0 \, , \qquad \qquad \qquad \qquad \qquad \qquad \qquad \quad \, Q_B^2(\text{gh.}) = 0 \, , \\
\left\{Q,Q_B\right\}\Phi^{(')} =&\, -\delta_{\text{gauge}}(\wh{\theta}^I)\Phi^{(')} \, , \qquad \qquad \qquad \;\; \left\{Q,Q_B\right\}(\text{gh.}) = \mathcal{L}_v(\text{gh.})\, , \nonumber
\end{align}
where~$\Phi^{(')}$ stands for bosonic (fermionic) physical and auxiliary fields, and~$\text{gh.}$ stands for all the ghost field of the gauge-fixing complex. Using these transformation rules, we conclude that the complete set of fields (including the ghosts) now admits a symmetry~$\wh{Q}$ realized off-shell with algebra~\eqref{eq:Qalg}. This is the supercharge that we would like to use to perform localization, and the localizing arguments of Section~\ref{sec:susy-loc} need to be applied with this new operator.

The first observation to be made is that the complete gauge-fixed action is closed under~$\wh{Q}$,
\be
\wh{Q}\left(S_0 + S_{\text{g.f.}}\right) = 0 \, .
\ee
This is the case since the~$S_0$ action is gauge and supersymmetry invariant by definition, and as was established in~\cite{Pestun:2007rz}, one may replace~$Q_B$ in~\eqref{eq:gflagrangian} by~$\wh{Q}$ without changing the value of the path-integral under consideration. Thus, the gauge-fixed action we built by introducing the gauge-fixing complex is closed under the~$\wh{Q}$ operator, and this operator squares to a sum of bosonic symmetries. This is the correct setup for localization. 

We also need to revisit the conditions for the saddle point around which the localization is performed. This means we now look for solutions to the equation
\be
\wh{Q}\psi_\a = Q\psi_\a + Q_B\psi_\a = 0
\ee
for all physical fermions~$\psi_\a$ in the theory. For the gaugini in the adjoint representation of the gauge group, the added terms~$Q_B P_\pm\lambda^{I\,i}$ are zero and therefore do not modify the initial solution found for~$Q\lambda=0$ in~\cite{Dabholkar:2010uh}. A similar statement can be made for the fermions of the hypermultiplets.

Finally, we need to modify the deformation operator~$Q\CV$ used in localization to the operator~$\wh{Q}\wh{\CV}$ which now includes the gauge-fixing part of the action~\eqref{eq:gflagrangian}:
\be
\wh{\CV} \equiv \CV + \CV_{\text{g.f.}} =  \int d^{4} x \, \Bigl[\sum_{\a} \, \left(Q {\psi}_{\a}\, , \, \psi_{\a}\right) + b^I G^W(W^I_\mu)\Bigr] \, ,
\ee
where, following the discussion below Equation~\eqref{eq:gfaction}, we have discarded the ghost-for-ghost fields and taken the parameter~$\xi_W$ to infinity in the gauge-fixing action.

We now have the full formalism in place to compute the super-determinant of the~$\wh Q \wh \CV$ operator over the $\wh{Q}$-complex~\eqref{eq:SUSYvect},~\eqref{eq:SUSYhyp},~\eqref{eq:BRSTvect}.

\section{Calculation of the one-loop determinant}
\label{sec:DetCalc}

In this section we compute the one-loop determinant of the~$\wh{Q}\wh{\CV}$ operator using an index theorem. We follow the procedure as explained in~\cite{Pestun:2007rz,Hama:2012bg, LeeSJ, Hosomichi:2015jta}. We will first organize the various fields on which the~$\wh{Q}$ operator acts in bosonic and fermionic quantities as:
\be
X^a \xrightarrow{\wh{Q}} \wh{Q} X^a \, , \qquad \Psi^\alpha \xrightarrow{\wh{Q}} \wh{Q}\Psi^\alpha\, ,
\ee
where~$X^a$ and~$\Psi^\alpha$ stand for fundamental bosons and fermions, respectively. The full set of bosonic and fermionic fields of the theory are thus organized as: 
\be
\mathfrak{B} \, \equiv \, \{X^a\,,\,\wh{Q}\Psi^\alpha\} \; \, (\text{bosonic}) \, , \qquad  \mathfrak{F}  \, \equiv \, \{\Psi^\alpha\,, \,\wh{Q}X^a\}  \; \, (\text{fermionic})\, .
\ee
With this change of variables, the deformation operator~$\wh\CV = \CV + \CV_\text{gf}$ can be written, up to quadratic order in the fields, as follows:
\be 
\label{eq:deformationoperator}
\wh \CV|_{\text{quad.}} \= \left( \wh Q X \; \Psi \right) \; \left( \begin{matrix} D_{00} & D_{01} \\ D_{10} & D_{11} \end{matrix} \right) \left( \begin{matrix} X  \\ \wh Q \Psi \end{matrix} \right) \, . 
\ee
This implies the following form for~$\wh Q \wh \CV$:
\be  
\label{eq:QVLbLf}
\wh Q \wh \CV|_{\text{quad.}} \= \int \, d^{4}x \, \Bigl( \mathfrak{B}  \, K_{b} \, \mathfrak{B} \+  \mathfrak{F} \, K_{f} \, \mathfrak{F}  \, \Bigr) \equiv \, \CL_{b} \+ \CL_{f}\, ,   
\ee
\be 
\label{eq:Lb}
\CL_{b} \= \left(  X \; \wh Q \Psi \right) \; \left( \begin{matrix} H & 0 \\ 0 & 1 \end{matrix} \right) \left( \begin{matrix} D_{00} & D_{01} \\ D_{10} & D_{11} \end{matrix} \right) \left( \begin{matrix} X  \\ \wh Q \Psi \end{matrix} \right) \, , 
\ee
and
\be  
\label{eq:Lf}
\CL_{f} \= \left( \wh Q X \; \Psi \right) \; \left( \begin{matrix} D_{00} & D_{01} \\ D_{10} & D_{11} \end{matrix} \right) \left( \begin{matrix} 1 & 0 \\ 0 & H \end{matrix} \right) \left( \begin{matrix} \wh Q X  \\  \Psi \end{matrix} \right) \, ,
\ee
and where~$H = \wh Q^{2}$ as defined in~\eqref{eq:Qalg}. 

By definition, the one-loop determinant for the operator~$\wh Q \wh \CV$ is:
\be 
\label{eq:Z1loop}
Z_\text{1-loop} \= \left(\frac{\det K_{f}}{\det K_{b}} \right)^{\half} \, . 
\ee
Using equations~\eqref{eq:QVLbLf},~\eqref{eq:Lb} and~\eqref{eq:Lf}, we find
\be 
\label{eq:detratio}
\frac{\det K_{f}}{\det K_{b}} \= \frac{\det_{\Psi} H}{\det_{X} H} \=  \frac{\det_\text{Coker$D_{10}$} H}{\det_\text{Ker$D_{10}$} H} \, . 
\ee
The above ratio of determinants can be computed from the knowledge of the \emph{index}
\be 
\label{eq:indD10}
\text{ind}(D_{10})(t) := \Tr_\text{Ker$D_{10}$}  \, e^{-iHt} - \Tr_\text{Coker$D_{10}$}  \, e^{-iHt} \, . 
\ee
Indeed the expansion of the index 
\be
\text{ind}(D_{10})(t) \= \sum_{n} a(n) \, e^{-i \lambda_{n} t} \, , 
\ee 
encodes the eigenvalues~$\l_{n}$ of~$H$, as well as their indexed degeneracies~$a(n)$, and we can thus write the ratio of determinants in \eqref{eq:detratio} as:
\be 
\label{eq:detratio1}
\frac{\det_\text{Coker$D_{10}$} H}{\det_\text{Ker$D_{10}$} H} \= \prod_{n} \, \l_{n}^{-a(n)} \, . 
\ee
This infinite product is a formal expression, and we will discuss a suitable regulator in the following. 

From a mathematical point of view, the index~\eqref{eq:indD10} is an \emph{equivariant index} with respect to the action of~$H$. This operator acts on all the fields as~$H=\mathcal{L}_v + \delta_M(L_{ab})$ according to~\eqref{eq:Qalg}. The action of~$H$ on the spacetime manifold is simply through the Lie derivative, \textit{i.e.}~the $\mathrm{U}(1)$ action~$H= (i\,\partial_\tau - i\,\partial_\phi) = L_{0} - J_{0}$. A $\mathrm{U}(1)$-equivariant index of this type can be computed in an elegant manner using the \emph{Atiyah-Bott index theorem} for transversally elliptic operators~\cite{Atiyah:1974}, as was explained in detail in~\cite{Pestun:2007rz}. Here we will make use of this index theorem even though~$AdS_2$ is a non-compact space. We note in this context that the~$AdS$ space is effectively compact, in the sense that there is a gravitational potential well which localizes physical excitations around the fixed point of the~$\mathrm{U}(1)$ action. This suggests that continuous modes do not contribute to the index, which is what we will assume. We leave a detailed analysis of the boundary conditions and boundary action as an interesting problem to be analyzed in the future. 

We now summarize the ideas of the index theorem very briefly from a working point of view. The standard starting point for the considerations of index theory is that of an \emph{elliptic operator} on a manifold, which generalizes the notion of a Laplacian. If the operator is linear and of second order, we can write it in local coordinates~$x^{i}$ as 
\be
a^{ij} (x) \, \p_{i} \, \p_{j} + b^{i} (x) \p_{i} + c_{i} (x) \, . 
\ee
An elliptic operator is one for which the matrix~$a^{ij}$ is positive-definite\footnote{For technical reasons, the theory of elliptic operators often also assumes that the eigenvalues are bounded.}. This can be restated as follows: if we replace the derivatives by momenta, \textit{i.e.}~consider the Fourier transform of the linear operator, we obtain the \emph{symbol} of the operator. An operator is elliptic if the \emph{principal symbol}~$a^{ij} \, p_{i} \, p_{j}$ does not vanish for any non-zero~$p_{i}$. The operator~$D_{10}$ introduced above, however, is not elliptic -- but it can still be treated by index theory~\cite{Atiyah:1974}. The point is that we have a certain special~$\mathrm{U}(1)$ action (that of~$H$), and our operator~$D_{10}$ commutes with this action. In the directions transverse to the~$\mathrm{U}(1)$ orbits, the operator~$D_{10}$ \emph{is} elliptic -- such an operator is called \emph{transversally elliptic}, and there is a version of the index theorem which deals with such operators. In terms of the symbol, an operator is called transversally elliptic if its symbol does not vanish for any~$p_{i}$ that is transversal to the generator of the~$\mathrm{U}(1)$ action. This means that the matrix~$a_{ij}$ is allowed to degenerate, but only along the one-dimensional locus generated by the~$\mathrm{U}(1)$ action. With this definition, the operator~$D_{10}$ is transversally elliptic with respect to the~$\mathrm{U}(1)$ symmetry generated by~$H$, as shown in~\cite{Murthy:2015yfa}. The proof is rather technical so it will not be reproduced here.

The result of the theorem applied to our problem is that the index of the~$D_{10}$ operator~\eqref{eq:indD10} reduces to the fixed points of the manifold under the action of~$H$. Denoting this action by $x \mapsto \wt x = e^{-iHt} x$, we have:  
\be 
\label{eq:ASindthm}
\text{ind} (D_{10})(t) \= \sum_{\{x \mid \wt x = x\}} \frac{\Tr_{X,\Psi} \, (-1)^{F} \, e^{-iHt }}{\det (1- \p \wt x/\p x)} \, . 
\ee
In our case the action of $H$~on $AdS_{2} \times S^{2}$ decomposes into the separate actions of~$L_{0}$ and~$J_{0}$ on the~$AdS_{2}$ and~$S^{2}$ factors, respectively. There are two fixed points -- at the center $\eta=0$ of the~$AdS_{2}$ factor (fixed by the rotation~$L_{0}$), and at the two poles on the~$S^{2}$ factor (fixed by the rotation~$J_{0}$). To apply the index theorem, we further need to know the explicit field content of~$X$ and~$\Psi$, and the charges they carry under~$H$. Once we know the eigenvalues of all the fields under~$H$, we can compute the trace in the numerator of~\eqref{eq:ASindthm}. 

As we discussed in Section~\ref{subsec:offshellsusy}, the off-shell algebra that we use has the same structure as the one used in~\cite{Pestun:2007rz,Hama:2012bg}, insofar that the field content and the gauge invariances are the same. This allows us to use the splitting of fields into~$X$,$\,\Psi$ used by those authors. On the other hand, as was emphasized at the end of Section~\ref{subsec:offshellsusy}, the physical transformations on the right-hand side of the algebra as well as the background manifold are different, and we should use the algebra~\eqref{eq:Qalg} that is relevant to our problem. 

The action of the Lie derivative~$\CL_{v}$ on any field of the theory is composed of two parts: a local translation on the spacetime coordinates along the vector~$v^{\mu}$, and an action on the tensor indices of the field. At the fixed points of spacetime under~$H$, the former action vanishes by definition. Thus, in order to compute the action of~$H$, we only need to keep track of the latter action of the Lie derivative, as well as the action of the Lorentz rotation~$L_{ab}$. The vector~$v^{\mu}$~\eqref{eq:vmu} translates us along the angles~$\tau$ and~$\phi$ in the metric~\eqref{eq:metric2} and is therefore essentially a rotation around the fixed points. The operator~$L_{ab}$~\eqref{eq:ourlab} at the fixed points is also the same rotation (acting on the spin part of the fields). Therefore, we only need to compute the charges of all the fields under a rotation around the center of~$AdS_{2}$ combined with a rotation around the~$S^{2}$. 

The calculation is simplified by going to complex coordinates in which the Euclidean~$AdS_{2} \times S^{2}$ metric is
\begin{equation}
\label{eq:metriccomplex}
ds^2 \= \ell^{2} \biggl( \frac{4 dw d\bar w}{(1- w \bar w)^2} + \frac{4 dz d\bar z}{(1 + z \bar z)^2} \biggr) \, .
\end{equation}
Here~$\ell$ is the overall physical size of the~$AdS_{2} \times S^{2}$ metric, which is governed by the field-dependent physical metric~$e^{-\CK (X^{I})} g_{\mu\nu}$ which depends on the position in the~$AdS_{2}$ space. At the fixed points, \textit{i.e.}~the center of~$AdS_{2}$, this size is given by~$\ell^{2} = e^{-\CK (\phi^{I})}$ in the gauge $\sqrt{g}=1$.\footnote{Here and in the following, we write~$\CK (\phi^{I})$ to mean~$\CK((\phi^{I}+ip^{I})/2)$.} At the fixed points, we have~$w=0$, and~$z=0$ or~$1/z=0$. There, the action of the operator~$e^{-iHt}$ on the spacetime coordinates is~$(z,w) \to e^{- i t/\ell} (z,w)$. Therefore, the determinant factor in the denominator of~\eqref{eq:ASindthm} is~$(1-q)^{2} \, (1-q^{-1})^{2}$, with~$q := e^{- i t/\ell}$. 

Near the fixed points the space looks locally like~$\IR^{4}$ with an associated~$\mathrm{SO}(4)=\mathrm{SU}(2)_{+} \times \mathrm{SU}(2)_{-}$ rotation symmetry. The planes labelled by  the two complex coordinates~$(z,w)$ rotate in the same direction under the~$\mathrm{SU}(2)_{+}$, and in opposite directions under~$\mathrm{SU}(2)_{-}$. Comparing the two forms of the metric~\eqref{eq:metric2} and~\eqref{eq:metriccomplex} (noting the change in orientiation of the~$S^2$ part when going from one to another), and recalling that~$H = i\,\partial_\tau - i\,\partial_\phi$, we identify the action of~$H$ with the Cartan generator of~$\mathrm{SU}(2)_{+}$ at the North Pole, and with the Cartan of~$\mathrm{SU}(2)_-$ at the South Pole according to: 
\be
H = 2\,J_+ \quad \text{(NP)}\, , \qquad H = 2\,J_- \quad \text{(SP)}\, .
\ee
We now need to compute the charges of all the fields under this generator.

\vspace{0.4cm}

\noindent \textbf{Vector multiplets:} 
In the vector multiplet sector, the fields are separated into $X = \{ X^I-\bar{X}^I,W^I_\mu \}$ and~$\Psi=\{ \Xi^I_i{}^j\,, \,c^I,b^I \}$, along with their respective images under~$\wh{Q}$.
The fermions~$\Xi^I$ are defined as
\be 
\label{eq:defxiij}
\Xi^I_i{}^j \, := \,  2\,\bar{\epsilon}_i P_+\lambda^{I\,j} + 2\,\bar{\epsilon}_i P_-\lambda^{I\,j} \, . 
\ee
The scalars~$(X^I-\bar{X}^I)$, $c^{I}, b^{I}$ are in the~$(\bf{0},\bf{0})$ of~$\mathrm{SO}(4) = \mathrm{SU}(2)_+ \times \mathrm{SU}(2)_-$, and therefore are all uncharged under~$H$. The vector rotates with spin one, and therefore is in the~$(\bf{2},\bf{2})$ of the $\mathrm{SO}(4)$. There are two modes ($W_{z}$, $W_{w}$) with charges~$+1$ and two modes ($W_{\overline{z}}$, $W_{\overline{w}}$) with charges~$-1$ under~$H$.\footnote{Our convention is that a field~$\varphi$ of charge~$e$ transforms as~$\varphi \to e^{- i e H t} \, \varphi$.} To compute the charges of the spinor bilinears, we notice that the spinor~$P_+\epsilon^i$ given in~\eqref{eq:killingspinor} vanishes at the North Pole, and so the bilinear~$\Xi^{I}$ is in the~$(\bf{1},\bf{3})$ of the $\mathrm{SO}(4)$. The spinor bilinears~$\Xi^{I}$ thus carry charge~$0$ under~$H$. Similarly, at the South Pole, the spinor bilinears are in the~$(\bf{3},\bf{1})$, while~$H$ is the Cartan of the~$\mathrm{SU}(2)_{-}$. 

Putting all this together, we find that, at each of the poles of the~$S^2$, the contribution to the index is:
\be 
\left[\frac{2q}{(1-q)^{2}}\right] \, . 
\ee
We see that there is a pole in this expression when~$q=1$. This pole arises due to the fact that our operator is not elliptic but \emph{transversally} elliptic. At a hands-on level, the pole presents a problem in the interpretation of the index -- namely, how to compute the Fourier coefficients of this expression. Depending on whether we expand around~$q=0$ or~$q^{-1}=0$, we will obtain~$2\sum_{n \ge 1} n \, q^{n}$ or~$2\sum_{n \ge 1} n\,q^{-n}$, which clearly have different Fourier coefficients. This problem is resolved by giving a certain regularization defined by the behavior of the operator in the neighborhood of each fixed point~\cite{Pestun:2007rz}. Accordingly, we write: 
\be 
\label{eq:vecind}
\text{ind}_{\text{vec}} (D_{10}) \= \left[\frac{2q}{(1-q)^{2}}\right]_{\text{NP}} + \left[\frac{2q}{(1-q)^{2}}\right]_{\text{SP}}\, . 
\ee
Here we have indicated the North Pole and South Pole contributions. As we shall see, the effect of the different regulators in our final results for the determinant will only be in an additive constant which we ignore in the functional determinant. 

\vspace{0.4cm}

\noindent \textbf{Hyper multiplets:} 
We perform a similar analysis for the hypermultiplets. The fields are separated 
into~$X=\{ A_i{}^\alpha \}$ and~$\Psi=\{ \Xi_i{}^\alpha \}$, with
\be
\Xi_i^{\;\alpha} := 2\,\bar{\breve{\epsilon}}_i P_+\zeta^\alpha + 2\,\bar{\breve{\epsilon}}_i P_-\zeta^\alpha \, ,
\ee
again inspired by~\cite{Pestun:2007rz,Hama:2012bg}. 
 
The scalars~$A_i{}^\alpha$ do not transform under rotations. To compute the charges of the fermions, we note that, in contrast to the vector multiplet analysis, it is the spinor~$P_-\breve{\epsilon}^{\,i}$ which vanishes at the North Pole (as can be seen from its explicit expression~\eqref{eq:xicheck}), and therefore the spinor bilinear~$\Xi_i{}^\alpha$ is in the~$2\times(\bf{2},\bf{1})$ of~$\mathrm{SO}(4)$, where the factor of~$2$ counts both~$\alpha$ components of a given hypermultiplet. Similarly at the South Pole,~$P_+\breve{\epsilon}^{\,i}$ vanishes and therefore the bilinear is in the~$2\times(\bf{1},\bf{2})$ of~$\mathrm{SO}(4)$. 

Putting everything together, we obtain the index for one hyper multiplet:
\be 
\label{eq:hypind}
\text{ind}_{\text{hyp}} (D_{10}) \= \left[-\frac{2q}{(1-q)^{2}}\right]_{\text{NP}} + \left[-\frac{2q}{(1-q)^{2}}\right]_{\text{SP}} \, .
\ee

\noindent \textbf{Zeta function regularization:}
We now use the expressions~\eqref{eq:vecind} and~\eqref{eq:hypind} for the index of the vector and hyper multiplets to compute their one-loop determinants. Given the infinite product~\eqref{eq:detratio1}, we write a formal expression for the logarithm of the one-loop determinant as:
\be 
\label{eq:sdetH}
\log \frac{\det_{\Psi} H}{\det_\text{X} H} \= -\sum_{n \ge 1} a(n)\,\log\l_{n} \, . 
\ee
In order to regularize this infinite sum, we use the method of zeta functions\footnote{The zeta function regularization has been used with great success to compute the perturbative one-loop corrections to the physical quantum gravity path integral (see~\cite{Hawking:1976ja} and follow-ups). Here we use the technique for the exact computation using localization methods.}. We first construct the zeta function:
\be
\zeta_{H} (s) \= \sum_{n \ge 1} \, a(n) \, \l_{n}^{-s} \, . 
\ee
This converges for suitably large values of~$\text{Re}(s)$, and we then analytically continue it to the complex~$s$ plane. The superdeterminant~\eqref{eq:sdetH} is then defined as:
\be 
\label{eq:Hzeta}
\log  \frac{\det_{\Psi} H}{\det_\text{X} H}  \= \zeta_{H}'(s) \mid_{s=0} \, . 
\ee
One of the advantages of the zeta function method is that it easily yields the dependence of the determinant on the physical parameters of the problem. In our case, we have only one parameter in the background: the overall size of the metric~$\ell^{2}=e^{-\CK(\phi^{I})}$. The dependence on~$\ell$ is easily calculated using the scaling properties of the zeta function~\cite{Hawking:1976ja}. Note that this size is measured in Planck units, and thus we are implicitly assuming here that the UV cut-off regularizing the supergravity theory is of order unity \emph{in Planck units}.\footnote{We are indebted to A. Sen for comments on this point.}

We consider the contribution to the index at the North and South Poles separately. At the North Pole, we have an expression which is expanded around~$q=0$:
\be
\left[\frac{2q}{(1-q)^{2}}\right]_{\text{NP}} \= 2\sum_{n \ge 1}n\,q^n = \sum_{n \ge 1}2n\,e^{-it\tfrac{n}{\ell}} \, .
\ee
In the above language, this index has 
\be
a(n)\=2n \, , \quad \l_{n} \= \frac{n}{\ell} \, .
\ee
The zeta function for this piece of the determinant is 
\be
\zeta_{H}^\text{NP} (s) \=\sum_{n \ge 1} 2n \, \bigl(\frac{n}{\ell}\bigr)^{-s} \=  2 \, \ell^{s} \, \zeta_{R}(s-1) \, , 
\ee
where we have introduced the Riemann zeta function 
\be
\zeta_{R}(s) \= \sum_{n \ge 1} \, \frac{1}{n^{s}} \, .  
\ee
At the South Pole, where we expand in powers of~$q^{-1}$, we get a similar expression but the zeta function~$\zeta_{H}^\text{SP} (s) $ there differs from the north pole answer by a factor of~$(-1)^s$. We thus need to deal with expressions of the type~$\log(-n)$, for which we use the positive branch of the logarithm.  

Putting together the North and South Pole contributions, we obtain 
\begin{align} 
\label{eq:zetaHans}
\zeta_{H}'(s) \mid_{s=0} =&\, 4 \zeta'_{R}(-1) + 2\pi i \zeta_R(-1) + 4 \, \zeta_{R}(-1) \, \log \ell \nonumber \\
=&\, 4 \zeta'_{R}(-1) + 2\pi i \zeta_R(-1) + \frac{1}{6} \, \CK (\phi^{I}) \, . 
\end{align}
Since we are not keeping track of purely numerical overall constants, we drop the finite numbers~$4 \zeta'_{R}(-1)$ and~$2\pi i\zeta_R(-1)$ in further expressions. Putting together Equations~\eqref{eq:Z1loop}, \eqref{eq:detratio}, \eqref{eq:Hzeta}, and \eqref{eq:zetaHans}, we finally obtain:
\be 
\label{eq:1loopdetvec}
Z_\text{1-loop}^\text{vec} (\phi^{I}) \= \exp\bigl(\CK(\phi^{I})/12 \bigr) \, , 
\ee
with~$\CK(\phi^{I})$ the generalized K\"ahler potential defined in Equation~\eqref{eq:EminK}.

For the hypermultiplets, we use the same technique, and we find that the index is equal and opposite to that of the vector multiplet -- as can be seen directly from the expressions~\eqref{eq:vecind},~\eqref{eq:hypind}. Our final result is thus:
\be 
\label{eq:1loopdet}
Z_\text{1-loop}^\text{vec} (\phi^{I}) \= \left(Z_\text{1-loop}^\text{hyp} (\phi^{I})\right)^{-1} \= \exp\bigl(\CK(\phi^{I})/12 \bigr) \, . 
\ee
Although we have only worked out the details of the vector and hyper multiplets, it is clear that the above calculation will also go through essentially unchanged once we have fixed the off-shell complex of any multiplet. Since there is only one scale set by~$e^{-\CK}$ in the localization background, the functional determinant will have the symplectically invariant form~$e^{-a_{0} \CK(\phi^{I})}$. The number~$a_{0}$ receives contributions from each multiplet of the~$\CN=2$ supergravity theory:
\be 
\label{eq:defa0}
a_{0} \= a_{0}^\text{grav} \+ n_{3/2} \, a_{0}^{3/2} \+ (\nv + 1) \, a_{0}^\text{vec} \+  \nh \, a_{0}^\text{hyp} \, , 
\ee
where~$n_{3/2}$, $(\nv+1)$, $\nh$ are the number of gravitini, vectors and hypers in the off-shell theory, respectively. From our results in this section,~$a_{0}^\text{vec}=-a_{0}^\text{hyp}=-1/12$. We will see in the following section how we can use existing on-shell computations to check our formula~\eqref{eq:1loopdet} for the vectors and hypers, as well as to deduce the coefficients~$a_{0}$ for the other multiplets.

\section{Relation to previous results}
\label{sec:relations-previous-results}

The leading logarithmic corrections to the semi-classical black hole entropy have been obtained in~\cite{Banerjee:2010qc, Banerjee:2011jp, Sen:2011ba} by explicitly evaluating the one-loop determinant of the kinetic terms of all the quadratic fluctuations of the theory around the classical attractor background~\eqref{eq:metric2}. This is a very intricate computation which requires a diagonalization of the kinetic terms of all the fields of the theory, and it depends on the fact that the values of the metric, fluxes and scalars in the attractor solution are related by supersymmetry.
In contrast, the localization method involves the one-loop determinant of the localizing action~$Q\CV$, which does not depend on the equations of motion and the associated kinetic mixings. At a practical level, the on-shell computation of~\cite{Banerjee:2010qc, Banerjee:2011jp, Sen:2011ba} proceeds by solving for the spectrum of eigenvalues of the various Laplacians of the theory, and observing that there is a huge cancellation among them. The index theorem, on the other hand, reduces this problem to a very simple computation at the fixed points of a certain~$\mathrm{U}(1)$ action. 

The results of the on-shell and off-shell methods agree in the large-charge limit, as expected. 
To show this, we consider a limit in which all the charges~$(q_{I}, p^{I})$ scale uniformly by a large parameter~$\Lambda$, \textit{i.e.}~$(q_{I}, p^{I}) \to \Lambda (q_{I}, p^{I})$. In the leading $\Lambda \to \infty$ limit, one can evaluate the quantum entropy~\eqref{eq:master-integral} using saddle-point methods. If we ignore the determinant factor~$Z_\text{1-loop}$, the saddle-point equations are simply the extremization equations of the exponent in~\eqref{eq:master-integral}. These extremization equations are precisely the attractor equations~\eqref{eq:attractor-scalars-metric},~\eqref{eq:attractor-charges} (with~$w=4$ in the gauge-fixed theory), and the saddle-point values~$\phi^{I}_{*}= 2\,\textnormal{Re} X^{I}_{*}$ are the attractor value of the scalar fields. 

From the attractor equations~\eqref{eq:attractor-scalars-metric}, we see that the attractor values~$\phi^{I}_{*} \sim \Lambda$ for large~$\Lambda$, and the attractor entropy~\eqref{eq:semi-class-1/2bps} scales as~$\Lambda^{2}$. From Equation~\eqref{eq:EminK}, we see that the determinants~\eqref{eq:1loopdet} scale as~$\Lambda^{-2a_{0}}$ and therefore they will contribute to the entropy as~$\log{\Lambda}$, so that it is indeed justified to ignore them to leading order. The resulting semi-classical entropy is: 
\be 
\label{SclassAgain}
\mathcal{S}_{BHW} = -\pi q_{I} \,e^{I}_{*} + 4\pi \, \textnormal{Im} F^{(0)} ((e^{I}_{*}+ip^{I})/2) \, \approx \, \frac{A_{H}}{4} \, , 
\ee
where $F^{(0)}$ denotes the prepotential without any $\wh{A}$-dependence, corresponding to the two-derivative effective action which is consistent with the large-charge approximation. This entropy agrees with the attractor mechanism result~\eqref{eq:semi-class-1/2bps}.

The first corrections to the leading large-charge entropy are given by the first corrections to the saddle-point value~\eqref{SclassAgain} of~\eqref{eq:master-integral}. In the large-charge limit, we know that~$A_{H} \sim e^{-\CK} \sim \Lambda^{2}$. From Equation~\eqref{eq:1loopdet} we deduce that the quantum entropy goes like
\be 
\label{leadlog}
\mathcal{S}_{Q} = \frac{A_{H}}{4} + a_{0} \log A_{H} \+ \ldots \, ,
\ee
where the number~$a_{0}$ is precisely the coefficient defined in~\eqref{eq:defa0}. In Section~\ref{sec:DetCalc}, we saw that
\be 
\label{a0vechyp}
a_{0}^\text{vec} \= - a_{0}^\text{hyp} \= -\frac{1}{12}  \, , 
\ee
which indeed agrees with the corresponding on-shell computations of the log corrections to the black hole entropy~\cite{Sen:2011ba}, performed using the heat-kernel method.

We defined the number~$a_{0}$ as appearing in the off-shell one-loop determinant in Section~\ref{sec:DetCalc}, and we saw above that the same number is the coefficient of the logarithmic correction to the large-charge expansion of black hole entropy. We can actually use this consistency between on-shell and off-shell methods to deduce the value of~$a_{0}$ for the graviton and gravitini multiplets. The results of~\cite{Sen:2011ba} demand that~$a^{3/2}_{0}=-\frac{11}{12}$, and~$a^\text{grav}_{0}=2$ in the gauge $\sqrt{g}=1$.

\subsection*{Miraculous cancellations in~$\CN=2$ truncations of~$\CN=8$ and~$\CN=4$ supergravities}

Armed with the knowledge of the one-loop determinants, we can now come back to the second assumption that was used in the original calculation and agreement~\cite{Dabholkar:2010uh, Dabholkar:2011ec} for the entropy of 1/8-BPS black hole in $\CN=8$ theory in both the macroscopic and microscopic theories. 
As outlined at the end of Chapter~\ref{chap:sugra}, the physical low energy macroscopic field content is that of an~$\CN=8$ graviton multiplet which, in the $\CN=2$ language that we are considering here, consists of one $\CN=2$ graviton multiplet, $n_{3/2}=6$ gravitini multiplets, $\nv=15$ vector multiplets, and $\nh=10$ hyper multiplets. The macroscopic entropy was computed using localization in~\cite{Dabholkar:2011ec} in the \emph{truncated} theory first considered in~\cite{Shih:2005he}, where the physical spectrum consists only of the $\CN=2$ graviton multiplet coupled to $\nv^\text{trun}=7$ vector multiplets. 

In this truncated theory, only the measure for the zero modes of~$Q$ was taken into account in~\cite{Dabholkar:2011ec}, and it was computed to be~$Z_{0} =  e^{(\nv^\text{trun} +1)\CK/2} \times O(\Lambda^{0})$. As explained below~\eqref{eq:master-integral}, the localizing computation using only the contribution of these zero modes agreed precisely with the string theory prediction~\eqref{eq:rademsp}.

We now have an understanding of this agreement. Let us split the contribution of one vector multiplet into two parts as~$a_{0}^\text{vec}=-\frac{1}{12} = -\half+\frac{5}{12}$, where the~$-\frac{1}{2}$ is the contribution considered in~\cite{Dabholkar:2011ec}, and~$\frac{5}{12}$ is the rest. Then, using the values of~$a_{0}$ for the various multiplets written in the previous subsection, the contribution to~$a_{0}$ ignored in~\cite{Dabholkar:2011ec} is 
\be
\nonumber
\frac{5}{12}  (\nv^\text{trun}+1) -\frac{1}{12} (\nv - \nv^\text{trun}) + \frac{1}{12} \nh -\frac{11}{12} n_{3/2} + 2 \, . 
\ee
For the field content of the~$\CN=8$ theory and the~$\CN=2$ truncation given above, this indeed adds up to zero, thus explaining the miraculous cancellation in the full string theory seen in~\cite{Dabholkar:2011ec}. This cancellation can already be seen at the leading log level in the entropy from the results of~\cite{Sen:2011ba}. It is now clear from the comments in this section that this cancellation holds to all orders in perturbation theory.

We can also consider~$\CN=4$ string theories (as in the next Chapter), where the physical low energy macroscopic field content is an~$\CN=4$ graviton multiplet coupled to~$N_\text{v}$ $\CN=4$ vector multiplets. In terms of~$\CN=2$ multiplets, we have one graviton multiplet, $n_{3/2}=2$ gravitini multiplets, $\nv=N_\text{v}+1$ vector multiplets, 
and $\nh=N_\text{v}$ hyper multiplets. The total logarithmic correction according to~\eqref{eq:defa0} is given by~$a_{0} = 2-\frac{11}{12}\times 2 - \frac{1}{12} \times 2 =0$, 
which is consistent with the on-shell computations in the limit when all the charges are scaled to be equally large.

\section{Exact formulas for $\CN=2$ quantum black hole entropy and the relation to topological strings \label{ExFor}}

The true power of the localization method clearly lies in the fact that one can go beyond the perturbative large charge 
approximation to get an exact result for black hole entropy. In this Section we propose such an exact 
entropy formula for BPS black holes in~$\CN=2$ supergravity coupled to~$\nv$ vector multiplets and~$\nh$ 
hyper multiplets. We then make some comments relating our formula to the microscopic formula 
of~\cite{Denef:2007vg}, as well as on some relations with topological string theory.

We have seen that one-loop determinant of the fluctuations around the localization manifold 
takes the symplectically invariant form\footnote{In this section we assume~$a_{0}^\text{grav} = 2$ (as argued for above) 
in the gauge $\sqrt{g}=1$ which we used throughout this Chapter. It is important to derive this result from a proper analysis 
of the fluctuating Weyl multiplet and the corresponding gauge-fixing. This is currently under investigation by the author and collaborators.}:
\be 
\label{1loopfinal}
Z_\text{1-loop} \= \exp\Bigl(-\CK(\phi^{I}) \bigl(2 -\frac{\chi}{24} \bigr) \Bigr) \, , \qquad \chi \= 2(\nv+1-\nh) \, .
\ee
Recall that, beyond the renormalized action, there are two pieces which contribute to the integrand of the Master Formula~\eqref{eq:master-integral} -- the one-loop fluctuations~$Z_\text{1-loop}$, and the measure from the curvature of field space itself. Combining these elements, we obtain:
\be \label{Wpert}
\wh{W}(q, p) = \int_{\mathcal{M}_{Q}}  \, \prod_{I=0}^{\nv} [d\phi^{I}] \, 
e^{- \pi  \, q_I  \, \phi^I  + 4 \pi \, {\rm Im}{F ((\phi^I+ip^I)/2)}}
e^{-\CK(\phi^{I}) (2 - \chi/24) } \, . 
\ee
We now need to discuss the details of the prepotential function~$F(X^{I},\wh{A})$ entering this equation, which is a holomorphic homogeneous function of weight 2 in its variables under the scalings~$X^{I} \to \lambda X^{I}$, $\wh{A} \to \lambda^{2} X^{I}$. Generically, we have an expansion of the form:
\be \label{eq:FAexp}
F(X^{I},\wh{A}) \= \sum_{g=0}^{\infty} \, F^{(g)}(X^{I}) \, \wh{A}^{g} \, 
\ee
which enters the Wilsonian effective action of the on-shell supergravity theory. The function~$F^{(0)}(X^{I})$ controls the two-derivative interactions, 
and the coefficients~$F^{(g)}$, $g \ge 1$, describe higher derivative couplings of the form~$C^{2} \, T^{2g-2}$ and 
terms related by supersymmetry, where~$C$ is related to the Weyl tensor, and~$T$ is related to the 
graviphoton field strength.

At the two-derivative level, the prepotential has the form
\be \label{classprep}
F^{(0)}(X^{I}) \= -\half \sum_{i,j,k=1}^{\nv} C_{ijk} \, \frac{X^{i} X^{j} X^{k}}{X^{0}} \, , 
\ee
for a choice of symmetric~$C_{ijk}$. At this level, we can think of the measure of the scalars in a geometric manner, and 
compute it from the knowledge of the two-derivative kinetic term of the scalar sigma model. To be more thorough, 
we should take into account all the fields in the theory -- this can be done by using 
duality invariance as a criterion for the measure as in~\cite{LopesCardoso:2006bg}. 
Both these approaches give rise to the measure:
\be \label{indmsr0}
[d\phi^I] \= \bigl(\det \text{Im}\,F^{(0)}_{IJ}\bigr)^{\frac12}\,d\phi^I \, . 
\ee
For a prepotential of the form~\eqref{classprep}, and for\footnote{In the type IIA setting, 
this means absence of D6-branes in the charge configuration making up the black hole.}~$p^{0}=0, q_{0}\neq 0$, 
we can compute the various expressions entering the exact formula~\eqref{Wpert}.  We have:
\be \label{EminKatt}
e^{-\CK^{(0)}} \= \frac{C_{ijk} \, p^{i} \, p^{j} \, p^{k}}{\phi^{0}} \, ,
\ee
and~$\det \text{Im}\,F^{(0)}_{IJ} = A/(\phi^{0})^{(\nv+3)/2}$ where~$A$ does not depend on~$\phi^{I}$ (but does depend 
on~$C_{ijk}$ and~$p^{i}$). 
However, using these expressions in our integral expression~\eqref{Wpert}
leads to a formula which does not match the corresponding microscopic BPS state counting formulas beyond 
the leading logarithmic correction (see \textit{e.g.}~\cite{Dabholkar:2005dt, Pioline:2006ni, Denef:2007vg}). 

We believe that this discrepancy arises from our lack of complete understanding of the induced measure term~$[d\phi^I]$. 
The current best understanding of the measure in the supergravity field space comes from the 
work of~\cite{LopesCardoso:2006bg, Cardoso:2008fr}, whose main guiding principle is duality invariance. 
These authors have argued that imposing duality invariance leads to a non-holomorphic modification to 
the induced measure. At the two-derivative level, including these corrections, one has:
\be \label{indmsr}
[d\phi^I] \= \bigl( \phi_{0}^{-2} \, \exp\bigl[-\CK^{(0)}(\phi^{I})\bigr] \bigr)^{\frac{\chi}{24}-1}\,d\phi^I \, , 
\ee
We note that the precise context in which these modifications have been derived 
is different from the one considered in this Chapter. Notwithstanding this difference, if we 
combine the expression~\eqref{indmsr} and the one-loop factor~\eqref{1loopfinal} 
in our exact formula~\eqref{Wpert}, we obtain:
\be \label{integral2}
\wh{W}(q, p) = \int_{\mathcal{M}_{Q}}  \, \prod_{I=0}^{\nv} d\phi^{I} \, \exp\Bigl[- \pi  \, q_I  \, \phi^I 
 + 4 \pi \, \textnormal{Im}\,F^{(0)}\big((\phi^I+ip^I)/2 \big)\Bigr] \bigl(\phi^{0}\bigr)^{2-\frac{\chi}{12}} \, e^{-\CK^{(0)}(\phi)} \, .
\ee
The black hole entropy formula conjectured in~\cite{Denef:2007vg} based on  
consistency with the Rademacher expansion of the \emph{microscopic} black hole degeneracies in string theory
has exactly the same form as~\eqref{integral2}, with the two-derivative expressions~$F^{(0)}$, $\CK^{(0)}$ replaced
by the all-order expressions~$F$, $\CK$, respectively. 

To go beyond the two-derivative level in our formalism, we need a formula for the induced measure 
at all orders. The work of~\cite{LopesCardoso:2006bg, Cardoso:2008fr} provides a formalism to take into 
account all the holomorphic corrections to the supergravity measure. More work, however, needs to be 
done to fully understand the non-holomorphic effects in the induced measure as defined in our treatment.  
It is possible that the \emph{a priori} induced measure in the original supergravity 
path integral suffers from a holomorphic anomaly. 
Similar ideas have been proposed in~\cite{Verlinde:2004ck} in the context of the topological string theory. 
A computation of this measure from first principles would complete
the derivation of the exact quantum black hole entropy in the gravitational theory.

\subsection*{Comments on relations to topological string theory}

Consider type IIA string theory compactified on a Calabi-Yau 3-fold~$CY_{3}$. 
The A-model topological string partition function on~$CY_{3}$ has the expression:
\be \label{Ftop}
F_\text{top} \= -i \frac{(2\pi)^{3}}{6\lambda^{2}} \, C_{ijk} \, t^{i} \, t^{j} \, t^{k} - \frac{i \pi}{12} \, c_{2i} \, t^{i} 
+ F_{GW}(\lambda, t^{i}) \, , 
\ee
where~$\lambda$ is the topological string coupling, $t^{i}$ are the moduli fields (the complexified K\"ahler 
structure in the type IIA theory), $c_{2i}$ are the second Chern classes of the 4-cycles of the~$CY_{3}$, 
and~$F_{GW}$ is the generating function of the Gromov-Witten (GW) invariants of the~$CY_{3}$ that admits 
an expansion in powers of~$\lambda$. 
By comparing~\eqref{Ftop} to the corresponding Wilsonian expression~\eqref{eq:FAexp} in supergravity,
we obtain\footnote{There are important subtleties associated with the above identification, having to do with 
the action of duality (symplectic transformations) on the geometry of the Calabi-Yau surface and in 
supergravity~\cite{Cardoso:2008fr, Cardoso:2014kwa}. We do not add anything to this discussion here.}: 
\be
F_{\text{top}}\=\frac{i \pi}{2} F, \qquad   t^{i}\= \frac{X^{i}}{X^{0}} \, , 
\qquad \lambda^{2} \= \frac{\pi^{2}}{8} \frac{\wh A}{(X^{0})^{2}} \, . 
\ee

The value of the topological string coupling constant on the supergravity localization manifold 
analyzed in this Chapter is $|\lambda| = 2 \pi \sqrt{2}/\phi^{0}$ -- which is small for large values of the charges. 
The microscopic analysis of~\cite{Dabholkar:2005dt, Pioline:2006ni, Denef:2007vg} is based on large~$\lambda$. 
Using the relation of the GW invariants to the Gopakumar-Vafa invariants related to counting M2-branes in M-theory,
then making a precise prediction for the degenerate instanton contribution at large topological string coupling,
and a subsequent analytic continuation, the authors of~\cite{Dabholkar:2005dt, Pioline:2006ni, Denef:2007vg} 
claimed that the 
the topological string partition function at weak coupling must have an additional logarithmic term:
\be \label{Ftop2}
\wt F_\text{top} \= -i \frac{(2\pi)^{3}}{6\lambda^{2}} \, C_{ABC} \, t^{A} \, t^{B} \, t^{C} -  \frac{\chi}{24} \log\lambda 
- \frac{i \pi}{12} \, c_{2A} \, t^{A} + F_{GW}(\lambda, t^{A})  \, .
\ee
where~$\chi$ is the Euler characteristic of the Calabi-Yau three-fold.
The puzzle then is to interpret the logarithmic term in supergravity. 
Being a non-local contribution, it cannot arise at any order in perturbation theory in~$\wh{A}$. 

From our point of view, the logarithmic contribution in~$\lambda$ (or equivalently in~$\wh{A}$) 
appears as a quantum effect. If we interpret the formula~\eqref{integral2} as an OSV-type formula~\cite{Ooguri:2004zv}, 
then the imaginary part of the prepotential contains precisely the additional non-local logarithmic piece 
with coefficient~$\chi/24$ that is predicted by the analytic continuation of the microscopic theory. 
(We recall that in a string compactification on a~$CY_{3}$, the number~$\chi=2(\nv+1-\nh)$ is the Euler 
characteristic of the~$CY_{3}$.)
Our~$AdS_{2}$ functional integral incorporates the integration 
over massless modes, and although the Wilsonian action of supergravity does not contain the logarithmic term, 
the effective 1PI action appearing in the exponent of Equation~\eqref{integral2}  
does.\footnote{A deeper explanation of this phenomenon appears in~\cite{Dedushenko:2014nya}.}$\,$\footnote{There are similar $\log g_s$ terms in the couplings of the low energy effective action 
of string theory in flat space, e.g.~\cite{Kiritsis:1997em}, which can be explained by mixing between the local 
and non-local part of the 1PI action when rescaling from string frame to Einstein frame~\cite{Green:2010sp}.}
We mention that most of this interpretation can be reconstructed by combining the duality arguments 
of~\cite{Cardoso:2008fr, Cardoso:2014kwa} with the computation of the leading logarithmic effects 
of~\cite{Sen:2011ba}. The one point we add to this discussion is the direct calculation of the one-loop 
effects proportional to~$e^{-a_{0}\CK}$. 

Finally, we note that, in addition to being at different values of coupling constants, the values of the moduli 
in our analysis and that of~\cite{Denef:2007vg} are also different. The authors of~\cite{Denef:2007vg} work 
with moduli~$t_{\infty}$ in asymptotically flat space, while we choose attractor values of moduli to 
define the black hole degeneracy since we are only interested in the single-center black holes. 
Our results could be interpreted to mean that the relevant index does not suffer any wall-crossing 
on moving from one regime to the other. 

These results may also point to a new ``black hole index'' that is simply constant over all of moduli space. 
Indeed, an argument was made in~\cite{Sen:2009vz, Dabholkar:2010rm} that, when a black hole preserves at least four 
supercharges and consequently at least an~$\mathrm{SU}(2)_{R}$ symmetry at its horizon, its quantum entropy is equal 
to a supersymmetric index. Defining this index in the microscopic theory is not an easy problem, but one can 
do so in~$\CN=4$ string theories, as we will review in the next Chapter. In that case the black hole index is given by the coefficient of a \emph{mock modular form}, 
defined using the attractor value of moduli, and it is constant all over of moduli space~\cite{Dabholkar:2012nd}. 
A similar phenomenon in~$\CN=2$ string theories would point towards a larger symmetry underlying the BPS states 
of~$\CN=2$ theories as proposed in~\cite{Harvey:1995fq, Harvey:1996gc}.

\section{Summary of results and assumptions}

In this Chapter, we have refined the Master Formula~\eqref{eq:master-integral} for the quantum entropy of 1/2-BPS black holes in~$\CN=2$ supergravity by computing explicitly the one-loop determinants of vector and hyper multiplets. To reach this result, we made some assumptions along the way, which we gather here for convenience.

In setting up the~$Q$-complex on the hypermultiplet fields and examining the solutions to the localizing equations, we have assumed that only the trivial solution for which the sections~$A_i{}^\alpha$ are constant and the auxiliary scalars~$H_i{}^\alpha$ are zero contribute to the localizing manifold. While they are clearly part of the localizing manifold according to~\eqref{eq:SUSYhyp}, it has not been proven that these are the \emph{only} solutions. Nevertheless, the agreement between the one-loop determinants and the on-shell computations of~\cite{Sen:2011ba} discussed in Section~\ref{sec:relations-previous-results} seem to corroborate this assumption.

We have also assumed a trivial contribution coming from the discrete modes of the gauge fields~$W_\mu{}^I$ introduced above~\eqref{eq:Qalg}. As discussed there, while this assumption is valid in the on-shell calculations of~\cite{Sen:2011ba}, more work is required to examine their impact on the off-shell results presented in this Chapter. The author intends to examine this question more closely in the near future. A related assumption was made in using the Atiyah-Bott index theorem to compute the equivariant index~\eqref{eq:indD10} even though the background space is non-compact. We assumed that continuous modes do not contribute to the index, which should again be rigorously proven by first going to a cut-off theory where the size of~$AdS_2$ is finite and then taking the cut-off to infinity.

A more implicit assumption was used in taking the UV cut-off of the supergravity theory to be set by the Planck scale, as explained below~\eqref{eq:Hzeta}. This assumption explains why the one-loop determinants~\eqref{eq:1loopdet} explicitly depend on the K\"{a}hler potential of the theory, which is part of the data contained in the supergravity action and not solely in the supergravity algebra generated by~$Q$.\footnote{We are indebted to A.Sen for a discussion of this point.}

Lastly, we have used a comparison to the on-shell results derived for the logarithmic corrections to the Bekenstein-Hawking-Wald entropy of black holes in~\cite{Sen:2011ba} to argue that the graviton multiplet will contribute to the one-loop determinants with a factor of~$e^{-2\CK}$. As we mentioned already, it is crucial to derive this result by examining the action of  a combined supersymmetry/BRST complex on the quantum fluctuations of the Weyl multiplet fields around the localizing background and using the Atiyah-Bott index theorem. The author is currently investigating this point with collaborators.

Using these assumptions, the results derived in this Chapter is the exact expression for the one-loop determinant contribution of Weyl, vector and hyper multiplets:
\be
Z_\text{1-loop} \= \exp\Bigl(-\CK(\phi^{I}) \bigl(2 -\frac{\chi}{24} \bigr) \Bigr) \, , \qquad \chi \= 2(\nv+1-\nh) \, .
\ee
This can be used in combination with the Master Formula~\eqref{eq:master-integral} for the quantum entropy of 1/2-BPS black hole in~$\CN=2$ superconformal gravity coupled to~$\nv+1$ vector multiplets and~$\nh$ hypermultiplets. As explained in Section~\ref{ExFor}, one can combine these results with some explicit assumptions regarding the integration measure~$[d\phi^I]$ to reach an exact, closed formula. Having established the general form that the functional determinants of the supergravity take in the localization program, we now move on to a computation of quantum black hole entropy in~$\CN = 4$ supergravity, truncated down to~$\CN = 2$ pure supergravity coupled to~$\nv+1 = 24$ vector multiplets. This theory admits interesting black hole solutions, which are more subtle than their 1/8-BPS~$\CN = 8$ counterparts due to the existence of a phenomenon known as \emph{wall-crossing}. This is related to the mock modularity of the counting functions in the microscopic string theory, as alluded to above.

\chapter{Localization in~$\CN=4$ supergravity}
\label{chap:n4-loc}


Based on the analysis of the previous Chapters, we now have a complete understanding of the quantum entropy of 1/8-BPS black holes in maximally supersymmetric ($\CN=8$) theories. Localization reduces the full perturbative path-integral of the QEF in these theories to a one-dimensional integral, which is the integral representation of a modified $I$-Bessel function. Going further, one can also identify all non-perturbative saddle-points of the full path-integral~\cite{Banerjee:2008ky, Murthy:2009dq} and compute the contributions of fluctuations around them~\cite{Dabholkar:2014ema}. The exact non-perturbative expression for the black hole entropy is thus given by an infinite sum over different saddle-points yielding a corresponding infinite sum over $I$-Bessel functions with successively suppressed arguments, which add up to precisely the integer degeneracies of the microscopic ensemble computed in~\cite{Maldacena:1999bp} and given explicitly in~\eqref{eq:rademsp}. 

This remarkable manner in which continuum gravity arranges integer black hole degeneracies relies on the equally remarkable successive approximation of an integer in terms of complex analytic functions---eventually arriving at a convergent analytic series. This formula is well-known in analytic number theory as the Hardy-Ramanujan-Rademacher expansion. As explained in Chapter~\ref{chap:modern-BH-S} and Appendix~\ref{app:modular}, it is a consequence of the modular symmetry of the corresponding microscopic ensemble of the black hole constituents. This modular symmetry of the black hole ensemble is, however, special to~$\CN=8$ string theory. In theories with lower supersymmetry, there are gravitational configurations other than the black hole which contribute to the full entropy formula~\cite{Denef:2000nb, Denef:2007vg} (unlike the case for~$\CN=8$ string theories~\cite{Dabholkar:2009dq}), and isolating the microstates belonging to the black hole will, in general, destroy modularity. 

We have learned about many aspects of the modular behavior of the microscopic partition functions in the generic setting of~$\CN=2$ theories based on the modular nature of the effective strings when black holes descend from wrapped strings, and from the spacetime duality symmetries of the underlying theory~\cite{Maldacena:1996gb,Dijkgraaf:2000fq,Gaiotto:2006wm,deBoer:2006vg,Manschot:2007ha,Denef:2007vg,Manschot:2009ia,Alexandrov:2012au}. However, the counting function of microstates of a \emph{single} black hole is still not understood in general, and in particular, it is not clear to what extent the modular symmetry of the original counting function has any remnant in the single-center black holes. In this Chapter, we begin to address this problem from the point of view of the bulk gravitational theory.

The main point is that localization allows us to compute the perturbatively exact macroscopic formula for the black hole entropy. This formula is a very good analytic approximation to the microscopic degeneracies of the single-center black hole, and thus constrains the modular nature of their generating function. Under explicit assumptions about the prepotential and the functional integral measure in the language of effective supergravity, the exact macroscopic entropy has a structure similar to the Rademacher expansion of modular forms. As was already derived in~\cite{Dabholkar:2005dt,Denef:2007vg}, following the OSV formula~\cite{Ooguri:2004zv}, the leading approximation to the degeneracy is given by a Bessel function with argument equal to a quarter of the area of the black hole, in the two-derivative approximation to the Wilsonian effective action of supergravity. Here we go beyond the leading order and show that including the infinite series of instanton effects in the holomorphic prepotential leads to a finite series of sub-dominant Bessel functions. 

We illustrate this formula in the concrete setting of the~$\CN=4$ string theory obtained as a Type II compactification on~$\mathrm{K}3 \times T^{2}$. In this situation we have a complete knowledge of the non-perturbative prepotential in the supergravity theory, as well as that of the microscopic BPS counting function for 1/4-BPS states (see~\cite{Sen:2007qy}). Further, it is known~\cite{Dabholkar:2009dq} that the only configurations, apart from dyonic 1/4-BPS black holes, which contribute to the relevant supersymmetric index are two-centered black holes which are each 1/2-BPS. Subtracting this two-centered contribution leads, as expected, to a breaking of modular symmetry for the single-centered black hole degeneracies of interest. It was shown in~\cite{Dabholkar:2012nd} that this breaking of modular symmetry happens in a very special manner and the single-centered black hole degeneracies are coefficients of \emph{mock modular forms}~\cite{Zwegers:2008zna,Zagier:2007}. As a consequence, analytic number-theoretical expressions for the degeneracies can be resurrected---at the expense of some modifications to the formula due to the mock nature of the partition functions~\cite{Bringmann:2010sd}. 

We will show in this Chapter that the macroscopic answer in the~$\mathrm{K}3 \times T^{2}$ theory has the following structure. The prepotential of the theory is exact at one-loop order. The one-loop contribution to the prepotential depends only on a special modulus in the theory~$S=-iX^{1}/X^{0}$, and it can be expanded as an infinite series in powers of the type~$e^{-nS}$, where~$n$ is identified as the instanton number. The zero-instanton sector gives rise to the leading $I$-Bessel function in the Rademacher expansion of the microscopic theory. In addition, the contribution from each of the infinite instanton sectors has the right structure to be identified with an $I$-Bessel function -- seemingly leading to a badly divergent contribution to the answer. However, the choice of integration contour ensures that one gets sub-leading $I$-Bessel functions only until a certain value of the instanton number, beyond which one obtains exponentially suppressed terms.  

The supergravity partition function can thus be expressed as a sum of Bessel functions with successively sub-leading arguments, with exactly the same arguments of the Bessels as those which appear in the Rademacher expansion of a Jacobi form. Quite remarkably, we find that the coefficients of the Bessel functions also agree exactly for the first many Bessel functions -- and begin to deviate from the Rademacher expansion of a true Jacobi form exactly when we expect them to do so due to the mock modular nature of the counting functions. This shows that the supergravity answer is sensitive to the polar coefficients of the microscopic function \emph{including} the coefficients of the \emph{mock modular} part. This looks to be the beginning of the answer to the question ``How does the continuum supergravity know about the mock modular nature of the black hole partition function?'' when multi-centered configurations are present in the spectrum.


We will also point out a potential interest from a mathematical point of view -- namely that our results look like the beginning of a consistent large-charge expansion for the coefficients of meromorphic Siegel modular forms which, in contrast to the Rademacher expansion for (mock) modular and Jacobi forms, is not fully understood in the mathematics literature as of yet. In order to complete this analysis, we need to classify and consider the effect of all gravitational saddle-points with Euclidean~$AdS_{2}$ boundary conditions (as was done in~\cite{Dabholkar:2014ema} for the~$\CN=8$ theory). We leave this interesting problem for the future.

\section{Single-centered black hole degeneracies and (mock) Jacobi forms} 
\label{sec:deg}

We first introduce the microscopic degeneracy formula for the 1/4-BPS black holes in~$\CN=4$ string theory that we study in this Chapter. We then present some details of the automorphic symmetry properties of the corresponding generating function, which leads to an analytic formula for the degeneracies of a single-centered black hole. This is the analogue of the situation described in Chapter~\ref{chap:modern-BH-S} for 1/8-BPS black holes in~$\CN = 8$ string theory, where the degeneracies~\eqref{eq:rademsp} were given by the Fourier coefficients of a Jacobi form. In the present case, subtleties due to wall-crossing phenomena lead to the fact that the black hole degeneracies are coefficients of mock Jacobi forms. Here we will review the statements relevant to us and refer the reader interested in more details of these mock modular functions to~\cite{Dabholkar:2012nd}.

Consider Type II string theory compactified on~$\mathrm{K}3 \times T^{2}$ or, equivalently, heterotic string theory on~$T^{6}$. At low energies the effective description of the theory is given by~$\CN=4$ supergravity coupled to 28~$\CN=4$ gauge field multiplets specified by the compactification. The quarter-BPS black holes carry electric and magnetic charges~$(Q_{e}^{i}, Q_{m}^{i})$ ($i=1,\cdots, 28$), under these gauge fields, where~$i$ is a vector index under the T-duality group~$\mathrm{SO}(6,22)$, and~$(Q_{e},Q_{m})$ transform as a doublet under the S-duality group~$\mathrm{SL}(2,\IZ)$. The U-duality group of the theory is~$\mathrm{SL}(2,\IZ) \times \mathrm{SO}(6,22)$. 1/4-BPS dyonic states in the theory are completely labeled by the three continuous T-duality invariants:
\be 
\label{nlm}
(Q_{e}^{2}/2, Q_{e} \cdot Q_{m}, Q_{m}^{2}/2) \, \equiv \, (n,\ell,m) \, ,
\ee
and, in addition, some discrete charge invariants~\cite{Banerjee:2008ri}. As in the~$\CN=8$ case of Chapter~\ref{chap:modern-BH-S}, we write the compactification manifold as $\mathrm{K}3 \times S^{1} \times \wt S^{1}$, and we can choose a duality frame in which the black hole consists of
\begin{myitemize}
\item $Q_5$ D5-branes wrapped on~$K3 \times S^{1}$,
\item $Q_1$ D1-branes wrapped on~$S^1$,
\item $n$ units of momentum along~$S^1$,
\item $\ell$ units of momentum along~$\wt S^{1}$,
\item one unit of Kaluza-Klein monopole charge on~$\wt{S}^1$.
\end{myitemize}
The charge invariants are~($Q_e^2/2 = n$, $Q_e\cdot Q_m=\ell$, $Q_m^2/2 = Q_1 Q_{5}$). The exact microscopic counting formula for the index of a generic 1/4-BPS state has been worked out completely~\cite{Dijkgraaf:1996it,Shih:2005uc, David:2006yn, Banerjee:2008pu, Dabholkar:2008zy}. For charges where the discrete invariants are trivial, the BPS indexed partition function is given by
\be
Z^\text{BPS}(\t,z,\s) \= \frac{1}{\Phi_{10}(\t,z,\s)} \, ,
\ee
where we now have three chemical potentials that couple to the three T-duality invariants. 
The function~$\Phi_{10}$ is the Igusa cusp form, which is the unique Siegel cusp form of weight~10.\footnote{See Appendix~\ref{app:modular} for more details on Siegel modular forms.} 
The microscopic degeneracy is given by the so-called \emph{Dikgraaf-Verlinde-Verlinde} (DVV) formula~\cite{Dijkgraaf:1996it}:
\be
\label{eq:DVV}
d(n,\ell,m) = (-1)^{\ell + 1}\int_\mathcal{C} d\t dz d\s \, \frac{e^{-i\pi\left(\tau n + 2z \ell + \sigma m \right)}}{\Phi_{10}(\t,z,\s)} \, ,
\ee
with a contour~$\CC$ that was spelled out in~\cite{Cheng:2007ch}.

\subsection*{Mock Jacobi forms}

There is an important new physical phenomenon which arises in the~$\CN=4$ theory as compared to the~$\CN=8$ theory. 
While the microscopic index that counts 1/8-BPS
states  in the~$\CN=8$ theory only gets contributions from single-centered black holes, the corresponding index that 
counts 1/4-BPS states in the~$\CN=4$ theory gets contributions from single-centered black holes as well as 
two-centered black hole configurations, depending on the value of the moduli at infinity~\cite{Dabholkar:2009dq}. 
This ambiguity is captured in the DVV formula by the choice of contour in~\eqref{eq:DVV}, 
which depends on the moduli fields at infinity~\cite{Dabholkar:2007vk, Cheng:2007ch}. 
Choosing the moduli to be at the attractor point yields the pure single-centered black hole degeneracies. 
Doing so, however, destroys the modular symmetry. From a physical point of view this breaking is related to the fact that 
we are throwing away a part of the spectrum of the theory. From a mathematical point of view it is because the 
partition function~$1/\Phi_{10}$ is a \emph{meromorphic} function with poles in the bulk of the Siegel upper half plane. 

Without the powerful handle given by the modular symmetry, it looks at first sight like the program followed 
to interpret the microscopic degeneracies in supergravity will not work. 
In particular, we do not know how to write down an analytic expansion like~\eqref{eq:rademsp} for 
the~$\CN=8$ black hole case.
This problem was solved in~\cite{Dabholkar:2012nd}, as we now briefly summarize. (We give more details in 
Appendix~\ref{app:modular}.) We can perform one of 
the three Fourier expansions in~\eqref{eq:DVV} near~$\s\to i\infty$ to obtain:
\be\label{reciproigusa}
  \frac 1{\Phi_{10}(\t, z, \sigma)} \= \sum_{m\geq -1} \psi_m (\t,z) \, e^{2 \pi i m \s}  \, .
\ee
The functions~$\psi_{m}$ are Jacobi forms of weight~$-10$ and index~$m$ that are meromorphic (in~$z$). 
These contain the degeneracies of states with magnetic charge~$m$, including both single and two-centered 
black holes. 
The single-centered black hole degeneracies are found by subtracting the generating function of two-centered degeneracies 
(called $\psi_{m}^\text{P}$) from~$\psi_{m}$. The difference, called the \emph{finite or Fourier part} of $\psi_{m}$
\be \nonumber
\psi_{m}^\text{F} = \psi_{m} - \psi_{m}^\text{P} \, ,
\ee
is holomorphic in $z$, and has an unambiguous Fourier expansion:
\be \label{psiF}
\psi_{m}^\text{F}(\t,z) \= \sum_{n,\ell} \, c^\text{F}_{m}(n,\ell) \, q^{n} \, \zeta^{\ell} \, .
\ee
It was shown in~\cite{Dabholkar:2012nd} that: 
\begin{myenumerate}
\item The microscopic indexed degeneracies~$d(n,\ell,m)$ of the single-centered black holes (\textit{i.e.}~corresponding to  
the attractor contour) are precisely related to the Fourier coefficients of this function 
\be 
d(n,\ell,m) = (-1)^{\ell+1} c^\text{F}_{m}(n,\ell) \, , 
\ee
\item The function~$\psi^\text{F}_{m}(\t,z)$ is a \emph{mock Jacobi form}.
\end{myenumerate}

The meaning of the word \emph{mock} is that 
the transformation rule~\eqref{eq:modtransform} is modified. The functions~$\psi^\text{F}_{m}$ themselves are not modular,
but one can add a correction term (called the \emph{shadow}) to get completed functions~$\wh{\psi^\text{F}_{m}}$ which 
\emph{are} modular, \textit{i.e.}~they transform exactly with the rule~\eqref{eq:modtransform}. 
The shadow is a non-holomorphic function\footnote{See~\cite{Troost:2010ud,Eguchi:2010cb,Ashok:2011cy,Murthy:2013mya,Ashok:2013pya,Harvey:2014nha,Pioline:2015wza} 
for the physical origin of such non-holomorphic terms from the point of view of conformal field theory. 
Understanding the physical basis of the non-holomorphicity of the specific functions~$\wh{\psi^\text{F}_{m}}$ 
is an interesting open problem.} and leads to a holomorphic anomaly equation as in~\eqref{eq:ddtbarhphi}.
This resurrection of modular symmetry means, in particular, that we can again use the circle method to get a 
formula for the Fourier coefficients. 
This formula differs from that of the analogous formula for true Jacobi forms (the Rademacher expansion)
due to the effect of the shadow term (see \cite{Bringmann:2010sd, Bringmann:2012zr}). 
In order to make sharp estimates about how the asymptotic expansion of mock Jacobi forms differs from 
that of true Jacobi forms, we need to know the explicit expressions of the mock Jacobi forms in question. 
This is a fairly complicated question but it has been addressed and solved in~(\cite{Dabholkar:2012nd}, Chapters 9, 10). 
We provide some relevant details in Appendix~\ref{app:modular}, and here we illustrate the main points with some examples.

In order to present the results, we need to introduce two Jacobi forms 
\be
\label{phi2}
A(\tau, z)=\varphi_{-2,1}(\tau, z) :=  \frac{\vth_1^2(\t, z)}{\eta^6(\t)} \, ,
\ee
\be
\label{phi0}
B(\tau, z)=\varphi_{0, 1} (\t, z) := 4 \left( \frac{\vth_2^2(\t, z)}{\vth_2^2(\t)} + \frac{\vth_3^2(\t, z)}{\vth_3^2(\t)} +\frac{\vth_4^2(\t, z)}{\vth_4^2(\t)} \right) \, ,
\ee
where $\vartheta_{i}, i=1,\dots,4$ are the four classical Jacobi theta functions. These two Jacobi forms generate the ring of 
all weak Jacobi forms of even weight over the ring of modular forms~\cite{Eichler:1985ja}. 
The word ``weak'' here refers to a growth condition on the functions, and it means in particular that for 
large values of~$\Delta=4mn-\ell^{2}$, the coefficients grow as (see Appendix~\ref{app:modular}) 
\be \label{Jacgrowth}
c(n,\ell) \simeq \exp(\pi\sqrt{4mn-\ell^{2}}) \, . 
\ee
The functions~$\psi^\text{F}_{m}$ can be worked out explicitly (see~\cite{Bringmann:2012zr}) for a given value of~$m$. 
The first couple of cases are:
\bea \label{psiF12}
\psi_1^\text{F} &= & \frac1{\eta(\t)^{24}} (3 E_4A- 648\mathcal{H}_{1}) \, , \label{m1}\\
 \psi_2^\text{F}&= & \frac{1}{3\eta(\t)^{24}}\big(22E_4 AB-10 E_6A^2-9600 \mathcal{H}_{2} \big) \, .\label{m2}
\eea
Here the functions~$\CH_{1}$, $\CH_{2}$ are mock Jacobi forms whose coefficients 
are linear combinations of the so-called Hurwitz-Kronecker class numbers, whose Fourier coefficients have 
purely polynomial growth. This is representative of the general structure proved in~\cite{Dabholkar:2012nd}:
the mock Jacobi forms~$\psi_m^\text{F}$ can always be written as a sum of two pieces:~$\varphi^\text{true}_{2,m}(\t,z)/\eta(\t)^{24}$ 
and~$\varphi^\text{opt}_{2,m}(\t,z)/\eta(\t)^{24}$. The function~$\varphi^\text{true}_{2,m}(\t,z)$ is a true weak Jacobi form 
(in particular, we can apply the usual Rademacher expansion~\eqref{eq:radi} to it), and the second is
a mock Jacobi form of a very special kind insofar that its Fourier coefficients grow extremely slowly. 
In the two examples above, this growth is purely polynomial -- this is the case whenever~$m$ is a prime power.
In general, the growth of the coefficients of~$\varphi^\text{opt}_{2,m}(\t,z)$ goes as
\be \label{optgrowth0}
c^\text{opt}(n,\ell) \sim \exp\bigl(\frac{\pi}{m}\sqrt{4mn-\ell^{2}} \bigr)  \, .
\ee
which can be contrasted with~\eqref{Jacgrowth}. 
What we need is to estimate the growth of the ratios like~$\varphi^\text{opt}_{2,m}(\t,z)/\eta(\t)^{24}$ that enter our expressions.
Such functions are called \emph{mixed} mock Jacobi forms, and their Rademacher expansion already differs at 
leading order in the asymptotic expansion compared to a true Jacobi form of the same weight and index 
(see Comment 1 below Theorem (1.3) of~\cite{Bringmann:2010sd}). 

We are now ready to reap the benefits of this technical analysis. 
If we want to analyze the Rademacher expansion of the black hole degeneracies encoded in~$\psi_{m}^\text{F}$, we can use 
the usual Rademacher expansion~\eqref{eq:radi} of Jacobi forms \emph{as long as} the growth of Bessel functions in~\eqref{eq:radi}
are larger than the growth of the mixed mock Jacobi forms~$\varphi^\text{opt}_{2,m}(\t,z)/\eta(\t)^{24}$. 
From what we said above, it is clear that we always have the contribution of the (denoting polynomial prefactors by~$p_{i}$ for 
now)
\be \label{LeadBes}
\text{Leading Bessel:} \qquad p_{0} \, \wt I_{23/2}  \biggl( {2\pi} \sqrt{ (m+4) \Bigl( n -\frac{\ell^{2}}{4m} \Bigr) } \, \biggr) \, ,
\ee
where~$p_{0}=(m+2)\frac{4\pi}{\sqrt{m} }  \Bigl(\frac{m+4}{n - \frac{\ell^2}{4m}}\Bigr)^{23/4}$
as for a true Jacobi form for any~$m$. This is then followed by the sub-leading Bessel functions in 
the~$c=1$ series of~\eqref{eq:radi}:
\begin{align}
&\text{Sub-leading $c=1$ series:} \quad p_{1}  \, \wt I_{23/2}  \biggl( {2\pi} \sqrt{ \Bigl(\frac{(m-1)^{2}}{m}+4\Bigr) \Bigl( n -\frac{\ell^{2}}{4m} \Bigr) } \, \biggr) \, +  \\
&\qquad \qquad \qquad \qquad \qquad \quad \;\;\, p_{2}  \, \wt I_{23/2}  \biggl( {2\pi} \sqrt{ \Bigl(\frac{(m-2)^{2}}{m}+4\Bigr) \Bigl( n -\frac{\ell^{2}}{4m} \Bigr) } \, \biggr) + \ldots \nonumber
\end{align}
But we should stop trusting this series when one of two things happen: 
firstly the~$c=2$ term begins to contribute at the order 
\be \label{c2Bes} 
\text{$c=2$ series:} \qquad \wt I_{23/2}  \biggl( {2\pi} \sqrt{ \frac{(m+4)}{4} \Bigl( n -\frac{\ell^{2}}{4m} \Bigr) } \, \biggr) \, .
\ee
Secondly the mock modular terms begin to contribute according to the discussion above.
We need to use a modified Rademacher expansion for the mixed mock Jacobi forms 
as in~\cite{Bringmann:2010sd}. Working out the details of the latter is an interesting problem in analytic number theory 
which we leave for the future (and for the experts!).  

We will use the analysis presented here in Section \ref{microforms} to 
work out some details of when exactly the signature of the mock nature appears in the Rademacher expansion 
on a case-by-case basis for the first few values of~$m$. We now move on to a supergravity 
analysis of the single-center black hole partition function. 

\section{Localization in~$\CN =4$ supergravity in the zero-instanton approximation}
\label{locsugra}

We consider the particular case of 1/4-BPS black holes in~$\CN=4$ string theory coming from the compactification of Type II string theory on~$\mathrm{K}3 \times T^{2}$. In the two-derivative limit of supergravity, we show how the functional integral in the near-horizon~$AdS_{2}$ reduces to a single Bessel function. We then set the stage for the inclusion of instantons in the holomorphic prepotential of the supergravity, which we will treat in the next Section. 

The theory under consideration is described by~$\mathcal{N}=2$ superconformal gravity in four dimensions with the Weyl multiplet coupled to~$\nv +1$ vector multiplets. In this theory we consider a BPS black hole solution carrying electric and magnetic charges~$q_{I}$, $p^{I}$. 
The exact quantum entropy of the black hole is given by the Master Formula~\eqref{eq:master-integral}.
As already explained in and around~\eqref{eq:FAexp}, the prepotential function~$F(X^{I},\wh{A})$ entering the localization formula can be expanded as 
\be 
\label{FAexp}
F(X^{I},\wh{A}) \= \sum_{g=0}^{\infty} \, F^{(g)}(X^{I}) \, \wh{A}^{g} \, .
\ee
We stress again that the function~$F^{(0)}(X^{I})$ controls the two-derivative interactions, and the coefficients~$F^{(g)}$, $g \ge 1$, describe higher derivative couplings in the theory.

Now we consider specifically the~$\mathrm{K}3 \times T^{2}$ compactification of the Type II theory. As explained at the end of Chapter~\ref{chap:sugra}, writing this theory as an~$\CN=2$ supergravity yields a field content, in addition to the Weyl multiplet, of vector multiplets, hypermultiplets and gravitino multiplets. Following the ideas of~\cite{Shih:2005he} one can truncate this theory to an~$\CN=2$ supergravity with a Weyl multiplet and~$\nv=23$ vector multiplets. In this case the perturbative prepotential has the form:
\be 
\label{Ftree}
F^\text{tree}(X) = -\frac{X^1}{X^0} \, X^a C_{ab} X^b + \frac{X^1}{X^0} \, ,
\ee
where~$C_{ab}$ is the intersection matrix on the middle homology of~$\mathrm{K}3$ (the 2-cycles). In the full theory, this is modified due to the effects of worldsheet instantons, as we shall consider in the following. 

In this theory, one can solve the exact functional integral explicitly, as we now briefly recall. The charge configuration corresponding to the microscopic brane setup introduced in the previous Section below~\eqref{nlm} corresponds to a non-zero value of~$(q_0,q_2,p^1,p^2,p^3)$ as explained in~\cite{Dabholkar:2011ec}. It was argued in~\cite{Dabholkar:2010uh}, based on the structure of the classical metric of the moduli space, that the induced measure on the localizing manifold in the large-charge limit is:
\be \label{msr}
[d\phi^I] \= P_{1} \, \frac{1}{p^1\phi^0}\,d \phi^{I} \, , 
\ee
where the prefactor~$P_{1}$ is a function only of the charges and independent of the coordinates~$\phi^{I}$. One generically expects the measure factor to change when we go beyond the tree-level approximation. We shall discuss this in the next section.

In order to obtain the truncated~$\CN=2$ conformal supergravity theory from the starting~$\CN=4$ theory, we dropped two~$\CN=2$ gravitini multiplets and 22~$\CN=2$ hypermultiplets. Following the analysis of the previous Chapter, the one-loop determinant contributions from the dropped fields amounts to
\be
Z_{\textrm{1-loop}}^{\textrm{dropped}} = \exp\Bigl[-\CK\bigl(2\times(-\tfrac{11}{12}) + 22\times(\tfrac1{12})\bigr)\Bigr] = 1 \, ,
\ee
where~$\CK$ is the generalized K\"{a}hler potential of the theory defined in~\eqref{eq:EminK}. The one loop determinant contributions of the fields in the truncated theory (which contains one Weyl multiplet and~$\nv+1 = 24$ vector multiplets) is
\be
Z_{\textrm{1-loop}}^{\textrm{trunc.}} = \exp\Bigl[-\CK\bigl(2 + 24\times(-\tfrac{1}{12})\bigr)\Bigr] = 1 \, .
\ee
Therefore, the one-loop determinants are trivial in the truncated theory, and the fields we dropped in reaching this truncation of the original~$\CN=4$ theory also do not contribute to the localized QEF at one-loop. 

Putting all the ingredients together, the quantum entropy~\eqref{eq:master-integral} takes the form
\begin{align}
\!\!\!\!\!\!\!\!\!\!\wh{W}^\text{tree}(p,q) = P_{1} \int \frac{d\phi^0 \, d\phi^1}{\phi^0 p^1}& \, \exp\bigl[-\pi\phi^0q_0 \bigr] \cr
&\!\!\!\!\!\!\!\!\!\!\!\!\!\!\!\!\!\times\int  \prod_{a=2}^{\nv}d \phi^{a} \; \exp\Bigl[-\pi\phi^2q_2 + 4 \pi \, \text{Im} F^\text{tree}\Bigl(\frac{\phi^I+ip^I}{2}\Bigr)\Bigr] .  
\end{align}
From~\eqref{Ftree}, we see that the last $(\nv-1)$ integrals are Gaussian integrals:
\begin{align}
\!\!\!\!\!\!\!\!\!\!\!\!\!\!\!\!\!\!\!\int \prod_{a=2}^{\nv}&\,d \phi^{a} \, \exp\Bigl[-\pi\phi^2q_2 + 4 \pi \, \text{Im} F^\text{tree}\Bigl(\frac{\phi^I+ip^I}{2}\Bigr)\Bigr] = \cr
\;\;=&\; \left(\frac{\phi^0}{p^1}\right)^{(\nv-1)/2}\! \exp\Bigl[\pi\frac{\phi^1}{p^1}p^1 q_2\Bigr] \, \exp\Bigl[{\pi\frac{\phi^1}{\phi^0} \Bigl(\frac{\phi^1}{p^1}+\frac{p^1}{\phi^1}\Bigr)p^aC_{ab}p^b+4\pi\frac{p^1}{\phi^0}}\Bigr] . 
\end{align}
The change of variables
$\tau_1 = \phi^1/\phi^0$, $\tau_2 =  p^1/\phi^0$
yields
\be 
\label{Wtree}
\wh{W}^\text{tree}(p,q) = P_{1}\displaystyle{\int} \frac{d\t_{1} d\t_{2}}{\tau_2^{(\nv+3)/2}} \,
\exp\Bigl[\frac{\pi}{\tau_2}\left(-p^1q_0+p^1q_2\tau_1+p^aC_{ab}p^b\tau_1^2+(p^aC_{ab}p^b+4)\tau_2^2\right) \Bigr] \, . 
\ee 
Upon identifying the four-dimensional electric and magnetic charge invariants as
\be
Q_{e}^{2}/2 := -q_0p^1 \, , \qquad  Q_m^{2}/2 := p^aC_{ab}p^b \, , \qquad  
Q_{e} \cdot Q_{m} := -q_{2} p^{1} \, , 
\ee
and with the identification~$(Q_{e}^{2}/2, Q_{e} \cdot Q_{m}, Q_{m}^{2}/2) := (n,\ell,m)$ 
this integral takes the form,
\be  
\wh{W}^\text{tree} (n,\ell,m) \= P_{1} \displaystyle{\int} \frac{d^2\tau}{\tau_2^{(\nv+3)/2}} 
\; \exp\Bigl[\frac{\pi}{\tau_2}\left(n -\ell\tau_1 + m\tau_1^2+(m+4)\tau_2^2\right) \Bigr] \, .
\ee
The $\tau_1$ integral is Gaussian and can be evaluated in a straightforward manner. 
The remaining integral over~$\t_{2}$ can be evaluated
using the contour integral representation of the Bessel function~\eqref{eq:intrep}, 
\be 
\label{Whatfin}
\wh{W}^\text{tree}(n,\ell,m) \= P_{1} \frac{2\pi}{\sqrt{m} }  \biggl(\frac{m+4}{n - \frac{\ell^2}{4m}}\biggr)^{23/4} 
I_{23/2}\left(2\pi\sqrt{(m+4) \biggl(n - \frac{\ell^{2}}{4m}\biggr)}\right)\, .
\ee
It has been argued recently in~\cite{Gomes:2015xcf} that the prefactor $P_{1}=2m+4$. 
The function~\eqref{Whatfin} then agrees precisely with the leading Bessel function in the Rademacher expansion 
of the microscopic theory~\eqref{eq:radi} with the right weight, argument, and prefactor, see~\eqref{LeadBes}.

We now move to the instanton contributions. 
Note that we kept only the perturbative prepotential to first sub-leading order while in general we 
have instanton sums that generate an infinite series of corrections to the prepotential~\eqref{Ftree}.
In general the instantons contribute to all the couplings~$F^{(g)}$. 
In the Type II theory on~$\mathrm{K}3 \times T^{2}$ the holomorphic prepotential is one-loop exact:
\be 
\label{Finst}
F(X) \= -\frac{X^1 X^a C_{ab} X^b}{X^0} +  \frac{1}{2\pi i} \CF^{(1)}_{K3\times T^{2}} (X^{1}/X^{0}) \, .
\ee
The one-loop contribution to the prepotential is:
\be \label{F1K3}
\CF^{(1)}_{K3\times T^{2}} (X^{1}/X^{0}) \=  \log\Bigl(\eta^{24} \bigl(X^{1}/X^{0}\bigr) \Bigr) \, ,
\ee
and has the expansion 
\be 
\CF^{(1)}_{K3\times T^{2}} (X^{1}/X^{0}) \= 2\pi i \, \frac{X^1}{X^0}  + \wt \CF^\text{inst} (X^{1}/X^{0}) \, .
\ee
Here the function~$\wt \CF^\text{inst}$ encodes the contributions of worldsheet instantons in the Type II theory 
to the prepotential:
\be \label{Ftlinst}
\wt \CF^\text{inst} (\t) \= -\log \prod_{n=1}^{\infty} (1-e^{2\pi i n \t})^{-24} \, .
\ee
A natural question is how to properly take these corrections coming from the instantons into account. 
The instantons can affect the exact answer~\eqref{eq:master-integral} in two ways -- by the explicit change of the prepotential in the 
renormalized action, and by an implicit effect on the measure of the integral (which was also computed above in the 
zero-instanton sector). This is what we turn to in the following Section.

\section{Including instantons in the localization formula}
\label{ContourPres}

In this Section, we work out the corrections to~\eqref{Whatfin} due to instantons. 
We write:
\be 
\label{WhatInst}
\wh{W} (n,\ell,m) =  \int_{\gamma} \frac{d^{2} \tau}{\tau_2^{(\nv+3)/2}} \; 
e^{\frac{\pi}{\tau_2}\left(n-\ell\tau_1+m\tau_1^2+m\tau_2^2\right)} \;
M(\t, \overline{\t}) \, e^{-\CF^{(1)}(\tau) - \CF^{(1)}(-\bar{\tau})} \, .
\ee 
Here we have taken into account the explicit effect on the prepotential function:
\be \label{Ffull}
F(X) \= -\frac{X^1 X^a C_{ab} X^b}{X^0} + \frac{1}{2i\pi}\CF^{(1)}\Bigl(\frac{X^1}{X^0}\Bigr) \, ,
\ee
with~$\CF^{(1)}$ given in~\eqref{F1K3} being the one-loop effect (it is exact in this case), 
which contains contributions from an infinite set of worldsheet instantons wrapping the torus~$T^2$.
Naively the inclusion of all these instantons leads to an infinite series of~$I$-Bessel functions. 
In this section we show that with an appropriate choice of contour~$\gamma$ in~\eqref{WhatInst}
most of these are in fact exponentially suppressed, leading to a \emph{finite} number of Bessel 
functions that contribute to the quantum entropy. This finite sum has precisely the same 
structure as the leading~$c=1$ term in the expansion~\eqref{eq:radi} for Jacobi forms.

We preface the calculation in this section with some remarks on the measure in Equation~\eqref{WhatInst}. 
We have parametrized the effect of instanton corrections on the measure of the functional integral 
by the function~$M(\t, \bar \t)$. 
In Section~\ref{locsugra} we did not take into account the full quantum effects on the measure in the localization computation. 
Indeed one needs to compute the 
induced measure from the supergravity field space taking the instanton corrections into account. 
This has been addressed in various papers~\cite{Cardoso:2008fr, LopesCardoso:2006bg, Gomes:2015xcf} although we think it is fair to say that a full satisfactory first-principles derivation of this measure has not been reached yet. We do not attempt to solve this problem in the present work. Instead we will use the fact that one knows the exact measure factor based on a saddle-point approximation of the Dijkgraaf-Verlinde-Verlinde formula discussed in Section \ref{sec:deg}~\cite{Shih:2005he, David:2006yn, LopesCardoso:2004xf}, as we shall now present. 
Note that this is a different expansion compared to~\cite{Dabholkar:2012nd} that is used to compute
the exact single-centered degeneracies. In particular, the expansion of~\cite{Dabholkar:2012nd} explicitly 
subtracts the two-centered black hole contributions from the Siegel modular form that is the 
full 1/4-BPS partition function, and then for each magnetic charge invariant, produces a mock 
Jacobi form (whose coefficients can then be again expanded in a Rademacher-type series).  
The formulas in this section, as we shall see below, follow from keeping only the residue at the leading divisor of the 
Siegel modular form. These two expansions are not a priori related, and so the results of this and the 
next section are non-trivial. They seem to point to a Rademacher-type expansion of the Siegel modular form, which was anticipated 
in~\cite{Murthy:2009dq}.  

We begin with the DVV formula~\eqref{eq:DVV}, which is a three-dimensional contour integral\footnote{For the 
next few lines we will use the variables~$(\s,v,\rho)$ instead of~$(\t,z,\s)$ to avoid confusion.}:
\be
d(n,\ell,m) = (-1)^{\ell + 1}\int_{\mathcal{C}} d\s dv d\r \, \frac{e^{-i\pi\left(\s n + 2v \ell + \rho m \right)}}{\Phi_{10}(\r,v,\s)} \, .
\ee
We can perform an exact contour integral in the~$v$-variable which reduces to 
picking up residues at the divisors of $1/\Phi_{10}$ in the Siegel upper-half plane, leaving a two-dimensional integral 
over~$\s,\r$ which are reexpressed as~$\s=\t_{1}+i\t_{2}$, $\r=-\t_{1}+i\t_{2}$. 
The result is~\cite{David:2006yn}:   
\be \label{dmicint}
d(n,\ell,m) \simeq \int_\gamma \frac{d\t_{1} d\t_{2}}{\tau_2^2}\;e^{-F(\tau_1,\tau_2)} \, ,
\ee
where $\,\simeq\,$ implies equality up to exponentially suppressed contributions coming from additional poles, which we shall discard from now on. 
The function~$F(\tau_1,\tau_2)$ is given by:
\begin{align}
\!\!\!\!\!\!F(\tau_1,\tau_2) = -&\frac{\pi}{\tau_2}\bigl(n-\ell\t_{1}+m(\tau_1^2 + \tau_2^2)\bigr) + \ln \eta^{24}(\tau_1 + i\tau_2) + \ln \eta^{24}(-\tau_1+i\tau_2) \cr 
+& 12\ln(2\tau_2)\!-\!\ln\Bigl[\frac{1}{4\pi}\Bigl\{26+\frac{2\pi}{\tau_2}\bigl(n-\ell \t_{1}+m(\tau_1^2 + \tau_2^2)\bigr)\Bigr\}\Bigr],
\end{align}
and the contour of integration $\gamma$ is required to pass through the saddle-point of $F(\tau_1,\tau_2)$. 
We rewrite this formula by adding the following total derivative to the integrand of~\eqref{dmicint},
\be
\label{eq:totdiv}
\displaystyle{\frac{d}{d\tau_2}}\Bigl(\frac{1}{\tau_2^{13}}\exp\Bigl[\frac{\pi}{\tau_2}(n-\ell\t_{1}+m\tau_1^2+m\tau_2^2) - \ln\eta^{24}(\t_{1}+i\t_{2}) - \ln\eta^{24}(-\t_{1}+i\t_{2})\Bigr]\Bigr) \, , 
\ee
which yields (with~$\t=\t_{1}+i\t_{2}$):
\be
\label{eq:davidsen}
d(n,\ell,m) = \frac{1}{2^{12}}\int_\gamma \frac{d^2\tau}{\tau_2^{13}}(m+E_2(\tau)+E_2(-\bar{\tau}))(\eta^{24}(\tau)\eta^{24}(-\bar{\tau}))^{-1}e^{\frac{\pi}{\tau_2}(n - \ell\tau_1+m\tau_1^2+m\tau_2^2)} \, , 
\ee
where $E_2$ is the Eisenstein series of weight 2. It is related to the Dedekind eta function as:
\be \label{E2etarel}
E_2(\t) \= \frac{1}{2\pi i} \,\frac{d}{d\tau}\log\eta^{24}(\t) \, .
\ee
Comparing this to our parametrization~\eqref{WhatInst}, we obtain:
\be \label{ME2}
M(\t,\overline{\t}) \= \frac1{2^{12}}(m + E_2(\t) + E_2(-\overline{\t})) \, . 
\ee
We note that~$M$ can be written, as anticipated in~\cite{Dabholkar:2005dt,Shih:2005he}, in terms of the generalized K\"{a}hler potential defined in~\eqref{eq:EminK},
which for the prepotential~$F$ given in~\eqref{Ffull} takes the form:
\be
e^{-\CK(X^{I})} \= \frac{2p^1}{\phi^0}\bigl(m + E_2(\tau) + E_2(-\overline{\t})\bigr) \, . 
\ee
This yields the relation  
\be
M(\t,\overline{\t}) \= \frac1{2^{13}}\frac{\phi^0}{p^1}\;e^{-\CK(\phi^{I})} \, .
\ee

The function~$\CF^{(1)}$ has a Fourier expansion in powers of~$q=e^{2 \pi i \t}$: 
\be 
\label{defdp}
e^{-\CF^{(1)}(\t)} = \sum_{p=-1}^\infty\,d(p)\,q^p \, ,
\ee
with~$d(p)$ for positive~$p$ being the number of instantons with charge~$p$. 
Combining the measure factor~\eqref{E2etarel},~\eqref{ME2}, we have (with~$N_{0}=2^{-12}$):
\begin{align}
\label{expM}
M(\t,\bar{\t})\,e^{-\mathcal{F}^{(1)}(\t) - \mathcal{F}^{(1)}(-\overline{\t})} =&\, N_{0} \sum_{p,\bar{p}\,=\,-1}^{\infty}\,(m - p -\bar{p})\,d(p)\,d(\bar{p})\,q^p\,\bar{q}^{\bar{p}} \, , \\
=&\, N_{0} \sum_{p,\bar{p}\,=\,-1}^{\infty}\,(m - p -\bar{p})\,d(p)\,d(\bar{p})\,e^{2\pi i(p-\bar{p})\t_1}\,e^{-2\pi(p+\bar{p})\t_2} \, . \nonumber
\end{align}
We now plug in the expansion~\eqref{expM} in the quantum entropy integral~\eqref{WhatInst}. 
For each term in this series, we can complete the square in~$\t_{1}$ to get a quadratic Gaussian integrand. 
If we perform the~$\t_{1}$ integral naively over the real line, each term in the above series would lead to 
an integral over~$\t_{2}$ of the form~\eqref{eq:intrep}. 
It would seem that we get an infinite series of $I$-Bessel functions for~$\wh{W}(n,\ell,m)$. 
We remind the reader that it is not surprising to find an infinite series of Bessel functions -- indeed the discussion 
of Section \ref{sec:deg} shows that the microscopic degeneracy has the same structure with the Bessel functions 
having successively sub-leading arguments. We find, however, that the arguments of the Bessel functions here 
decrease (as we expect) up to a point, but then increase indefinitely, thus showing that this 
sum is not convergent! 

A solution to this puzzle was presented recently in~\cite{Gomes:2015xcf} by making a choice of 
contour~$\gamma$ in~\eqref{WhatInst} and analyzing the contributions to the degeneracies from 
each term in the Fourier expansion. 
With this choice of contour, almost all of the infinite number of Bessel functions turn out to 
be highly suppressed, and one is left with a finite number of~$I$-Bessel functions, 
consistent with the structure of the leading~$c=1$ term of the Rademacher expansion~\eqref{eq:radi}. 
We now review this analysis, and use the contour prescription of~\cite{Gomes:2015xcf} to 
make a detailed comparison between the expansion of the integral~\eqref{eq:davidsen} and 
the~$c=1$ term of the Rademacher expansion~\eqref{eq:radi}.
We find, at the end of our analysis, that the two expansions actually agree in great detail, 
in the appropriate regime of validity, including the integer coefficients of the Bessel functions.
At first sight this observation may seem to be a pleasant surprise about this particular~$\CN=4$ string theory, but  
as we sketched in the introduction to this Chapter, it can be understood as a reflection of the deeper and broader ideas 
of~\cite{Ooguri:2004zv, Denef:2007vg}, namely that worldsheet instanton degeneracies 
encode the microscopic black hole degeneracies in a very precise manner.

By using the expansion~\eqref{expM} in the expression~\eqref{WhatInst}, splitting the contour~$\gamma$ into 
two contours~$\gamma_1$,~$\gamma_2$ for the~$\t_1$ and~$\t_2$ integrals, respectively, and 
completing the square in each term, we obtain:
\begin{align}
\label{eq:perfectsquare}
\wh{W}(n,\ell,m) =&\, N_0\sum_{p,\bar{p}\,\geq\,-1}(m - p - \bar{p})d(p)d(\bar{p})\,e^{i\pi(p-\bar{p})\frac{\ell}{m}} \, \times \cr
&\,\times \; \int_{\gamma_2}\frac{d\tau_2}{\tau_2^{(\nv+3)/2}} \exp\Bigl[-\pi\t_2 \frac{\Delta(p,\bar{p})}{m} + \frac{\pi}{\t_2}\bigl(n - \frac{\ell^2}{4m}\bigr)\Bigr] \, \times \\
&\,\times \; \int_{\gamma_1} d\tau_1 \exp\Bigl[\frac{\pi m}{\tau_2}\Bigl(\t_1 + i(p-\bar{p})\frac{\t_2}{m}-\frac{\ell}{2m}\Bigr)^2\,\Bigr] \, , \nonumber
\end{align}
where we have defined
\be 
\label{defK}
\Delta(p,\bar{p}) := 4m\bar{p} - (m-(p-\bar{p}))^2 \, .
\ee
We will see in the following that the function~$\Delta$ becomes precisely the polar discriminants entering the Rademacher expansion~\eqref{eq:radi}.
We now define the contours~$\gamma_1$,~$\gamma_2$ pertaining to the~$\t_1$ and~$\t_2$ integrals as~\cite{Gomes:2015xcf}
\begin{align} 
\label{contours}
\tau_1 = i\,\tau_2\,u\;\, :&\,  \;\; -1+\delta \leq u \leq 1-\delta \, , \cr
\tau_2 \;\, :&\, \;\; \epsilon - i\infty < \tau_2 < \epsilon + i\infty \, ,
\end{align}
with~$\delta$ small and positive and~$\epsilon$ positive. This choice ensures that~$|q|<1$ and $|\bar{q}|<1$ so that the Fourier expansion~\eqref{expM} is convergent. As we will see below, it also brings~$\wh{W}(n,\ell,m)$ to a form which is exactly of the same type as the~$c=1$ term in~\eqref{eq:radi}, namely coming from a generating
function that has the elliptic transformation property of a Jacobi form of index~$m$. We now define
\be 
\label{Iu}
I_u(p,\bar{p}) := \int_{\gamma_1} d\tau_1 \exp\Bigl[\frac{\pi m}{\tau_2}\Bigl(\t_1 + i(p-\bar{p})\frac{\t_2}{m}-\frac{\ell}{2m}\Bigr)^2\,\Bigr] \, .
\ee
Following the idea of~\cite{Gomes:2015xcf}, we can evaluate this integral. 
The analysis is somewhat technical and won't be reproduced here. The interested reader is referred to the original publication~\cite{Murthy:2015zzy} and its Appendix B in particular. Here, we simply quote the result.
 
Defining $\alpha := (p-\bar{p})/m$, there are two types of contributions to~$\wh{W}(n,\ell,m)$ depending on 
whether~$|\alpha| \leq 1-\delta$ or~$|\alpha| > 1-\delta$. The leading contributions to the sum~\eqref{eq:perfectsquare} 
are for~$|\alpha| \leq 1-\delta$, and the terms for which~$|\alpha| > 1-\delta$ are exponentially suppressed. 
We then need to take a~$\delta \rightarrow 0$ limit in the contour~$\gamma_1$. This limit is rather subtle, but it can be shown 
that once we take it, the leading contributions to the sum~\eqref{eq:perfectsquare} are the ones for which~$|\alpha| \leq 1$ 
(modulo what we call ``edge-effects'', see again Appendix B of~\cite{Murthy:2015zzy}).

Focusing on these contributions to the quantum entropy, we may evaluate the~$\tau_1$ integral in~\eqref{eq:perfectsquare} and we 
are left with the~$\tau_2$ integral. The latter will yield exponentially growing~$I$-Bessel functions~\eqref{eq:intrep} as long as~$\Delta(p,\bar{p}) < 0$.
Therefore, we now have two conditions,~$|\alpha| \leq 1$ and~$\Delta < 0$, which can be used to bound the sums 
over~$(p,\bar{p})$. Putting these facts together leads to the following expression for~$\wh{W}$:
\begin{align}
\wh{W}(n,\ell,m) \simeq&\, N_0\sum_{p,\bar{p}\,\geq\,-1} \; \sum_{\substack{-m \, \leq \, p-\bar{p} \, \leq \, m \\ \Delta(p,\bar{p}) \, < \, 0}}(m - p - \bar{p})\,d(p)\,d(\bar{p})\,e^{i\pi(p-\bar{p})\frac{\ell}{m}} \; \times \cr
&\,\times \; \frac{i}{\sqrt{m}}\int_{\gamma_2} \frac{d\tau_2}{\tau_2^{(\nv+2)/2}} \exp\Bigl[-\pi\t_2 \frac{\Delta}{m} + \frac{\pi}{\t_2}\bigl(n-\frac{\ell^2}{4m}\bigr)\Bigr] \, .
\end{align}
Here the~$\,\simeq\,$ sign means that we have thrown away exponentially suppressed contributions to the complete answer for~$\wh{W}$. We can now evaluate the remaining integral on the contour~$\gamma_2$, which yields a Bessel function:
\begin{align}
\label{Whatfinal}
\wh{W}(n,\ell,m) \simeq&\, N_0\sum_{p,\bar{p}\,\geq\,-1} \; \sum_{\substack{-m \, \leq \, p - \bar{p} \, \leq \, m \\ \Delta(p,\bar{p}) \, < \, 0}}(m - p - \bar{p})\,d(p)\,d(\bar{p})\,e^{i\pi(p-\bar{p})\frac{\ell}{m}} \, \times \\
&\,\times \;\frac{2\pi}{\sqrt{m}}\Biggl(\frac{-\Delta(p,\bar{p})/m}{n-\frac{\ell^2}{4m}}\Biggr)^{\nv/4}\,I_{\nv/2}\Biggl(2\pi\sqrt{-\frac{\Delta(p,\bar{p})}{m}\Bigl(n-\frac{\ell^2}{4m}\Bigr)}\Biggr) \, . \nonumber
\end{align}
The symmetry~$\Delta(p,\bar{p}) = \Delta(\bar{p},p)$ implies that one can write the above expression as 
a sum over~$p-\overline{p}$ from~$0$ to~$m$, with the replacement of the phase~$e^{i\pi(p-\bar{p})\frac{\ell}{m}}$
by~$\cos \bigl(\pi(p-\bar{p})\frac{\ell}{m} \bigr)$. 

To proceed further, we make the following change of variables:
\be
\ell' := m-(p-\bar{p}) \, , \qquad n' := \bar{p} \, .
\ee
In these variables, we have~$\Delta(n',\ell') = 4mn'-\ell'^{2}$ as anticipated, and~\eqref{Whatfinal} takes the form
\begin{align} 
\label{WhatRadvars}
\wh{W}(n,\ell,m) \simeq&\, 2N_0\sum_{\substack{0\,\leq\,\ell'\,\leq\,m \\ n'\,\geq\,-1}} \: \sum_{4n' - \frac{\ell'^2}{m} \, < \, 0}(\ell'-2n')\,d(m+n'-\ell')\,d(n')\cos\bigl(\pi(m-\ell')\frac{\ell}{m}\bigr) \cr 
&\times\frac{2\pi}{\sqrt{m}}\Biggl(\frac{\bigl|4n' - \frac{\ell'^2}{m}\bigr|}{n-\frac{\ell^2}{4m}}\Biggr)^{\nv/4}\,I_{\nv/2}\Biggl(2\pi\sqrt{\Bigl|4n' - \frac{\ell'^2}{m}\Bigr|\Bigl(n - \frac{\ell^2}{4m}\Bigr)}\Biggr) \, .
\end{align}
In this form,~$\wh{W}$ can readily be compared to the leading Rademacher expansion for a Jacobi form of index~$m$ and weight~$(3-\nv)/2$. Indeed for such a Jacobi form, the~$c=1$ term of the Rademacher 
expansion~\eqref{eq:radi},~\eqref{eq:Kloosc1} reads
\begin{align} 
\label{RadJac}
c(n,\ell) \simeq&\, \frac{1}{2^{(\nv-1)/2}}\sum_{0 \, \leq \, \ell' \, \leq \, m}\:\sum_{4n'- \tfrac{\ell'^2}{m} \, < \, 0}c(n',\ell')\,\cos\bigl(\pi(m-\ell')\frac{\ell}{m}\bigr) \cr
&\times\frac{2\pi}{\sqrt{m}} \Biggl(\frac{\Bigl|4n' - \frac{\ell'^2}{m}\Bigr|}{n-\frac{\ell^2}{4m}}\Biggr)^{\nv/4} I_{\nv/2}\Biggl(2\pi\sqrt{\Bigl|4n' - \frac{\ell'^2}{m}\Bigr|\Bigl(n - \frac{\ell^2}{4m}\Bigr)}\Biggr) \, .
\end{align}

We see that~\eqref{WhatRadvars} 
has exactly the same form as~\eqref{RadJac} if we make the identification:
\be  \label{cdrel}
c(n,\ell) = (\ell-2n)\,d(m+n-\ell)\,d(n) \, , \qquad 4mn-\ell^{2} <0, \; n \geq -1, \; 0 \leq \ell \leq m \, .
\ee
We interpret this formula as an explicit prediction for the left-hand side, which are the polar coefficients~$c^\text{F}(n,\ell)$ 
of the mock Jacobi forms~\eqref{psiF} that control the single-centered black hole degeneracies. 
The coefficients~$d(p)$ of the right hand side are the instanton degeneracies captured by the 
function~$\CF^{(1)}$~\eqref{defdp}
\be
\frac{1}{\eta(\t)^{24}} = \sum_{n\ge -1} d(n) \, q^{n} = q^{-1} \+ 24  \+ 324q \+ 3200q^2 \+ 25650q^3 \+ 176256q^4 \+ \ldots
\ee 
The fact that the instanton degeneracies~$d(n)$ vanish for~$n<-1$ is reflected in the fact that 
the single centered degneracies~$c^\text{F}(n,\ell)$ also vanish for~$n<-1$ as we explained briefly in Section~\ref{sec:deg}. 
In the next Section, we will show that the expansion~\eqref{WhatRadvars} agrees very precisely with the Rademacher-like 
expansion for the Fourier coefficients~$c^\text{F}(n,\ell)$ -- up to an order where the latter starts to deviate from the 
form~\eqref{RadJac} due to its mock modular nature. 

\section{Polar terms in 1/4-BPS black holes in~$\CN=4$ string and supergravity theory}
\label{microforms} 

In this Section, we verify the relation~\eqref{cdrel}
for the first few values of~$m$. There are three sources of approximations in our derivation which impose
a regime of validity for the comparison of the macroscopic and the microscopic formulas. 
The first source is that we have only 
kept the first~($c=1$) series in the microscopic Rademacher expansion while we should really keep all the terms 
from~$c=1,2,3,\ldots$ The second source, as we explained in Section~\ref{sec:deg}, is that the effects of the shadow of 
the mock modular forms (although small to leading order) can become relevant at a certain sub-leading order.
The third is what we call ``edge-effects''~\cite{Murthy:2015zzy} in the evaluation of the 
two-dimensional integral which is the result of the localized supergravity path integral. The first source can be
controlled in a fairly straightforward manner but typically this is the smallest effect. 
The second source is an interesting problem in analytic number theory, and the third is a problem for us to better 
define our contour prescription in supergravity. We leave these two problems for the future. 
We now analyze these three effects in specific examples.

We begin with~$m=1$.  We have:
\be
\psi_1^\text{F}(\tau,z) =  \frac{1}{\eta(\tau)^{24}}(3 E_4(\tau)A(\tau,z)- 648\mathcal{H}_{1}(\tau,z)) \, , \label{m1}\\
\ee
whose Fourier expansion begins as:
\begin{align}
\psi_1^\text{F}&(\tau,z) = \cr
&\,(3\z \+ 48\+ 3\z^{-1})q^{-1} \cr
+&\, (48\z^2 \+ 600\z - 648 \+ 600\z^{-1} \+ 48\z^{-2}) \cr
+&\,(3\z^3 - 648\z^2 \+ 25353\z - 50064 \+ 25353\z^{-1} - 648\z^{-2} \+ 3\z^{-3}) \, q \+ \cr
+&\,(600\z^3 - 50064\z^2 \+ 561576\z - 1127472 \+ 561576\z^{-1} - 50064\z^{-2} \+ 600\z^{-3})q^2 \cr
+&\, \ldots 
\end{align}
The polar terms are~$(n,\ell)=(-1,1)$, $(-1,0)$, and~$(0,1)$ or equivalently in terms of~$\Delta$, $(\Delta,\ell)=(-5,1)$, $(-4,0)$, $(-1,1)$.  
The corresponding coefficients~$c_{1}^\text{F}(n,\ell)$ are\footnote{We note that there is a textual error in 
the Appendix of~\cite{Bringmann:2012zr}. In the first paragraph, it says that the 
coefficients~$c_{m}^{F}(n,\ell)$ of the mock Jacobi forms are presented for the first four values of~$m$, 
while what is really presented is~$d(n,\ell,m)=(-1)^{\ell} c_{m}^{F}(n,\ell)$ to emphasize the positivity 
of those numbers. In particular, the polar coefficients~$c_{m}^{F}(n,\ell)$ (\textit{i.e.}~with~$4mn-\ell^{2}<0$) 
are strictly positive.}~\cite{Bringmann:2012zr}:
\be 
\label{polarm1}
c_{1}^\text{F}(-1, 1) = 3 \, , \qquad c_{1}^\text{F}(-1, 0) = 48 \, , \qquad c_{1}^\text{F}(0, 1) = 600 \, . 
\ee
The corresponding combinations of the~$(\ell-2n)\,d(m+n-\ell)\,d(n)$ are:
 \be
(n,\ell)=(-1,1): \; 3 \, , \qquad (n,\ell)=(-1,0): \; 48 \, , \qquad (n,\ell)=(0,1): \; 576 \, . 
\ee
We see that the first two coefficients agree, and the third does not. This is exactly what we expect, as we explained at the end of Section~\ref{sec:deg}. 
Indeed we have made an approximation in the Rademacher expansion keeping only the leading~$c=1$ term, 
and we have\footnote{Here and below we do the comparisons at~$\ell = 0$ for simplicity.} 
\begin{align}
\label{Wm1}
\frac{\wh{W}(n,0,1)}{4\pi N_0} =&\, 3\,\Bigl(\frac{5}{n}\Bigr)^{23/4}I_{23/2}\bigl(2\pi\sqrt{5n}\bigr) + 48\,\Bigl(\frac{4}{n}\Bigr)^{23/4}I_{23/2}\bigl(2\pi\sqrt{4n}\bigr) \cr
&\,+ 576\,\Bigl(\frac{1}{n}\Bigr)^{23/4}I_{23/2}\bigl(2\pi\sqrt{n}\bigr) \, ,
\end{align}
while the~$c=1$ term of the Rademacher expansion of a Jacobi form with the polar coefficients~\eqref{polarm1} is:
\begin{align}
\label{m1c1}
\frac{c^\text{F}_{1}(n,0)}{4\pi N_0} =&\, 3\,\Bigl(\frac{5}{n}\Bigr)^{23/4}I_{23/2}\bigl(2\pi\sqrt{5n}\bigr) + 48\,\Bigl(\frac{4}{n}\Bigr)^{23/4}I_{23/2}\bigl(2\pi\sqrt{4n}\bigr) \cr
&\,+ 600\,\Bigl(\frac{1}{n}\Bigr)^{23/4}I_{23/2}\bigl(2\pi\sqrt{n}\bigr)\, ,
\end{align}
with~$N_0 = 2^{-12}$. The~$c=2$ series in the expansion~\eqref{eq:radi} starts with~$I_{23/2}\bigl(2\pi\sqrt{5n/4}\bigr)$  
which is larger than the last term in~\eqref{m1c1}, and therefore we do not expect an agreement
at this order for the last coefficients in~\eqref{Wm1} and~\eqref{m1c1}.
This is one of the issues that we need to be careful about in our comparison. 

Secondly, we need to be careful about the interference of the mock nature of the functions~$\psi_{m}^\text{F}$. The first time\footnote{We find experimentally that 
for~$m=1,2$ the two expansions agree even including the mock piece, but we believe this is an accident, which will be explained if we work out the asymptotic expansion of the corresponding mock Jacobi form in detail.} we see this interference is for~$m=3$, where we have:
\begin{align} 
\label{Wm3}
\frac{\sqrt{3}}{4\pi N_0}\wh W(n,0,3)&\, = 5\left(\frac{7}{n}\right)^{23/4}I_{23/2}\left(2\pi\sqrt{7n}\right)+ 96 \left(\frac{16}{3n}\right)^{23/4}I_{23/2}\left(2\pi\sqrt{\frac{16}{3}n}\right) \nonumber \\
&\,+ 972\left(\frac{13}{3n}\right)^{23/4}I_{23/2}\left(2\pi\sqrt{\frac{13}{3}n}\right) + 6400\left(\frac{4}{n}\right)^{23/4}I_{23/2}\left(2\pi\sqrt{4n}\right) \nonumber \\
&\,+1728\left(\frac{3}{n}\right)^{23/4}I_{23/2}\left(2\pi\sqrt{3n}\right) + 15552\left(\frac{4}{3n}\right)^{23/4}I_{23/2}\left(2\pi\sqrt{\frac{4}{3}n}\right) \nonumber \\
&\,+76800\left(\frac{1}{3n}\right)^{23/4}I_{23/2}\left(2\pi\sqrt{\frac{1}{3}n}\right)\, .
\end{align}
Correspondingly, the $c=1$ term of the Rademacher expansion~\eqref{eq:radi} for $m=3$ is:
\begin{align} 
\label{m3c1}
\frac{\sqrt{3}}{4\pi N_0}c_3^\text{F}(n,0)&\, = 5\left(\frac{7}{n}\right)^{23/4}I_{23/2}\left(2\pi\sqrt{7n}\right) + 96 \left(\frac{16}{3n}\right)^{23/4}I_{23/2}\left(2\pi\sqrt{\frac{16}{3}n}\right) \nonumber \\
&\,+ 972\left(\frac{13}{3n}\right)^{23/4}I_{23/2}\left(2\pi\sqrt{\frac{13}{3}n}\right) + 6404\left(\frac{4}{n}\right)^{23/4}I_{23/2}\left(2\pi\sqrt{4n}\right) \nonumber \\
&\,+1728\left(\frac{3}{n}\right)^{23/4}I_{23/2}\left(2\pi\sqrt{3n}\right) + 15600\left(\frac{4}{3n}\right)^{23/4}I_{23/2}\left(2\pi\sqrt{\frac{4}{3}n}\right) \nonumber \\
&\,+85176\left(\frac{1}{3n}\right)^{23/4}I_{23/2}\left(2\pi\sqrt{\frac{1}{3}n}\right)\, .
\end{align}
The~$c=2$ term of the Rademacher expansion starts with~$I_{23/2}\bigl(2\pi\sqrt{7n/4}\bigr)$, and we should ignore terms of that order, \textit{i.e.} the last two Bessels in~\eqref{m3c1}. However, we still see a disagreement for the Bessel~$I_{23/2}\left(2\pi\sqrt{4n}\right)$. This is precisely the interference
from the mixed mock Jacobi form~$\varphi^\text{opt}_{2,3}(\t,z)/\eta(\t)^{24}$. Therefore we 
should only expect agreement up to the Bessel functions~$I_{23/2}(2\pi\sqrt{4n})$. In the 
expressions~\eqref{Wm3},~\eqref{m3c1}, this means that we should not expect a matching of the 
coefficients for the fourth terms,~$6400$ vs.~$6404$.

Thirdly, in deriving our Rademacher-like expression from the supergravity path integral, we made a choice of contour in~\eqref{contours}. 
As explained in Appendix B of~\cite{Murthy:2015zzy}, there are ``edge-effects'' in this contour that we have not taken into account properly here. 
These may go towards explaining the boxed discrepancies in the tables we present below for the larger values of~$m=5$ and~$m=7$. We believe a more detailed analysis of 
the integral~$I_u(p,\bar{p})$ in~\eqref{Iu} would resolve these discrepancies.

We checked up to~$m=7$ that this kind of an agreement holds exactly after taking into account these three effects. We present the data below. 

\textbf{Legend for tables}:
The pair~$(n,\ell)$ satisifies the conditions in~\eqref{cdrel}, that is $n\geq -1$,~$0\leq\ell\leq m$ and~$(4mn-\ell^2) = \Delta < 0$.
The third column is the coefficient~$c^\text{F}(n,\ell)$ of the mock Jacobi forms~$\psi^\text{F}_{m}$ \eqref{psiF}. Recall that the
black hole exists for positive values of~$\Delta$ and the degneracy~$c_{m}(n,\ell)$ is controlled by the polar coefficients 
through an expansion of the type~\eqref{RadJac}. 
(Essentially a polar term labelled by~$\Delta$ enters the analytic formula for the degeneracy~$c_{m}(n,\ell)$ for~$4mn-\ell^{2}>0$ 
at an order~$\exp[2\pi |\Delta| (4n-\ell^{2})]$.) 
The coefficients below the horizontal line have deviations from their true values because 
we have only included the~$c=1$ series of the Rademacher expansion, while 
at these orders we should necessarily start including the~$c\ge 2$ series. 
We indicate in bold face when the Rademacher expansion cannot be trusted because we have treated a 
mock Jacobi form as a true Jacobi form. (For~$m=1,2$ the coefficient still agree -- which we indicate by a~$\bf{}^{*}$.)
As we see clearly in the tables, the deviations for the bold-faced coefficients are small and should be resolved 
by including the effects of the shadow. The boxed values indicate possible edge-effects~\cite{Murthy:2015zzy} in the contour prescription. 

\vspace{1cm}

{$\bf m=1$:}

\begin{center}
\begin{tabular}{|c|c|c|c|}
\hline
$\Delta$ & $(n,\ell)$ & $c_1(n,\ell)$ & $(\ell-2n)\,d(1+n-\ell)\,d(n)$ \\
\hline
$-5$ & $(-1,1)$ & 3 & 3 \\
$-4$ & $(-1,0)$ & \bf{48}* & 48 \\
\hline
$-1$ & $(0,1)$ &  {600} & 576 \\
\hline
\end{tabular}
\end{center}

{$\bf m=2$:}

\begin{center}
\begin{tabular}{|c|c|c|c|}
\hline
$\Delta$ & $(n,\ell)$ & $c_2(n,\ell)$ & $(\ell-2n)\,d(2+n-\ell)\,d(n)$ \\
\hline
$-12$ & $(-1,2)$ & 4 & 4 \\
$-9$ & $(-1,1)$ & 72 & 72 \\
$-8$ & $(-1,0)$ & \bf{648}* & 648 \\
$-4$ & $(0,2)$ & 1152 & 1152 \\
\hline
$-1$ & $(0,1)$ &  {8376} &  7776 \\
\hline
\end{tabular}
\end{center}

{$\bf m=3$:}

\begin{center}
\begin{tabular}{|c|c|c|c|}
\hline
$\Delta$ & $(n,\ell)$ & $c_3(n,\ell)$ & $(\ell-2n)\,d(3+n-\ell)\,d(n)$ \\
\hline
$-21$ & $(-1,3)$ & 5 & 5 \\
$-16$ & $(-1,2)$ & 96 & 96 \\
$-13$ & $(-1,1)$ & 972 &  972 \\
$-12$ & $(-1,0)$ & \bf{6404} & 6400 \\
$-9$ & $(0,3)$ & 1728 & 1728 \\
\hline
$-4$ & $(0,2)$ &  {15600} & 15552 \\
$-1$ & $(0,1)$ &  {85176} &  76800 \\
\hline
\end{tabular}
\end{center}

\vspace*{\fill}

\newpage

\vspace*{\fill}

{$\bf m=4$:}

\begin{center}
\begin{tabular}{|c|c|c|c|}
\hline
$\Delta$ & $(n,\ell)$ & $c_4(n,\ell)$ & $(\ell-2n)\,d(4+n-\ell)\,d(n)$ \\
\hline
$-32$ & $(-1,4)$ & 6 & 6 \\
$-25$ & $(-1,3)$ & 120 & 120 \\
$-20$ & $(-1,2)$ & 1296 &  1296 \\
$-17$ & $(-1,1)$ & 9600 & 9600 \\
$-16$ & $(0,4)$ & 2304 & 2304 \\
$-16$ & $(-1,0)$ & \bf{51396} & 51300 \\
$-9$ & $(0,3)$ & 23328 & 23328 \\
\hline
$-4$ & $(0,2)$ &  {154752} & 153600 \\
$-1$ & $(0,1)$ &  {700776} &  615600 \\
\hline
\end{tabular}
\end{center}

\vspace{1cm}

{$\bf m=5$:}

\begin{center}
\begin{tabular}{|c|c|c|c|}
\hline
$\Delta$ & $(n,\ell)$ & $c_5(n,\ell)$ & $(\ell-2n)\,d(5+n-\ell)\,d(n)$ \\
\hline
$-45$ & $(-1,5)$ & 7 & 7 \\
$-36$ & $(-1,4)$ & 144 & 144 \\
$-29$ & $(-1,3)$ & 1620 &  1620 \\
$-25$ & $(0,5)$ & 2880 & 2880 \\
$-24$ & $(-1,2)$ & 12800 & 12800 \\
$-21$ & $(-1,1)$ & 76955 & \boxed{76950}  \\
$-20$ & $(-1,0)$ & \bf{353808} & 352512 \\
$-16$ & $(0,4)$ & 31104 & 31104 \\
\hline
$-9$ & $(0,3)$ &  {230472} & 230400 \\
$-5$ & $(1,5)$ &  {315255} & 314928 \\
$-4$ & $(0,2)$ &  {1246800} & 1231200 \\
$-1$ & $(0,1)$ &  {4930920} & 4230144 \\
\hline
\end{tabular}
\end{center}

\vspace*{\fill}

\newpage

{$\bf m=6$:}

\renewcommand{\arraystretch}{1}
\begin{center}
\begin{tabular}{|c|c|c|c|}
\hline
$\Delta$ & $(n,\ell)$ & $c_6(n,\ell)$ & $(\ell-2n)\,d(6+n-\ell)\,d(n)$ \\
\hline
$-60$ & $(-1,6)$ & 8 & 8 \\
$-49$ & $(-1,5)$ & 168 & 168 \\
$-40$ & $(-1,4)$ & 1944 &  1944 \\
$-36$ & $(0,6)$ & 3456 & 3456 \\
$-33$ & $(-1,3)$ & 16000 & 16000 \\
$-28$ & $(-1,2)$ & 102600 & 102600 \\
$-25$ & $(0,5)$ & 38880 & 38880 \\
$-25$ & $(-1,1)$ & \bf{528888} & 528768 \\
$-24$ & $(-1,0)$ & \bf{2160240} & 2147440 \\
$-16$ & $(0,4)$ & 307200 & 307200 \\
\hline
$-12$ & $(1,6)$ &  {419904} &  419904 \\
$-9$ & $(0,3)$ &  {1848528} & 1846800 \\
$-4$ & $(0,2)$ &  {8615040} & 8460288 \\
$-1$ & $(0,1)$ &  {30700200} & 25769280 \\
$-1$ & $(1,5)$ &  {3118776} & 3110400 \\
\hline
\end{tabular}
\end{center}

{$\bf m=7$:}

\begin{center}
\begin{tabular}{|c|c|c|c|}
\hline
$\Delta$ & $(n,\ell)$ & $c_7(n,\ell)$ & $(\ell-2n)\,d(7+n-\ell)\,d(n)$ \\
\hline
$-77$ & $(-1,7)$ & 9 & 9 \\
$-64$ & $(-1,6)$ & 192 & 192 \\
$-53$ & $(-1,5)$ & 2268 &  2268 \\
$-49$ & $(0,7)$ & 4032 & 4032 \\
$-44$ & $(-1,4)$ & 19200 & 19200 \\
$-37$ & $(-1,3)$ & 128250 &  128250 \\
$-36$ & $(0,6)$ & 46656 & 46656 \\
$-32$ & $(-1,2)$ & 705030 & \boxed{705024} \\
$-29$ & $(-1,1)$ & 3222780 & \boxed{3221160} \\
$-28$ & $(-1,0)$ & \bf{11963592} & 11860992 \\
$-25$ & $(0,5)$ & 384000 & 384000 \\
$-21$ & $(1,7)$ & 524880 & 524880 \\
\hline
$-16$ & $(0,4)$ &  {2462496} & 2462400 \\
$-9$ & $(0,3)$ &  {12713760} & 12690432 \\
$-8$ & $(1,6)$ &  {4147848} & 4147200 \\
$-4$ & $(0,2)$ &  {52785360} & 51538560 \\
$-1$ & $(0,1)$ &  {173032104} & 142331904 \\
\hline
\end{tabular}
\end{center}

\chapter{Conclusion and open questions}
\label{chap:conclusion}
In this work, we have shown how the supersymmetric localization technique could be successfully applied to certain black holes in order to compute their quantum entropy exactly. The two cases investigated here are~$\mathcal{N}=8$ four-dimensional superconformal gravity and~$\mathcal{N}=4$ four-dimensional superconformal gravity. When considering both of these theories as truncated~$\mathcal{N}=2$ four-dimensional theories, it is possible to exactly evaluate the quantum entropy function introduced in Chapter~\ref{chap:intro} for 1/8-BPS and 1/4-BPS black holes, respectively. Armed with the knowledge of the partition functions for the corresponding 1/8-BPS states and 1/4-BPS states in the full string theory, it is possible to compare the macroscopic (supergravity) and microscopic (string theory) answers.

In the case of 1/8-BPS states in four-dimensional~$\mathcal{N}=8$ string theory, the degeneracies of states are given by the Fourier coefficient of the~$\varphi_{-2,1}(\tau,z)$ Jacobi form, as explained in Chapter~\ref{chap:modern-BH-S}, and powerful techniques of number theory (the Rademacher expansion) allow for the complete evaluation of these degeneracies. They are given by a sum over Bessel functions~\eqref{eq:rademsp}. Correspondingly, the supergravity result for the exact quantum entropy of 1/8-BPS black holes can be obtained using supersymmetric localization, which leads to the Master Formula~\eqref{eq:master-integral} (after confirming that full-superspace integrals do not contribute to the entropy, the purpose of the second half of Chapter~\ref{chap:n8-loc}, and computing the one-loop determinant exactly, which was done in Chapter~\ref{chap:n2-loc}), where the prepotential is simply the cubic one given in~\eqref{eq:prepot-n8}. Localization around the leading saddle-point configuration of the supergravity partition function (the~$AdS_2 \times S^2$ configuration) yields precisely the~$c=1$ term in~\eqref{eq:rademsp}, and inclusion of smooth orbifold configurations~$AdS_2 / \mathbb{Z}_c$ provides the full sum over~$c$ in~\eqref{eq:rademsp}, as was shown in~\cite{Dabholkar:2014ema}.  Thus, it is fair to say that there is an exact correspondence between the macroscopic and microscopic pictures in this specific case. Nevertheless, some clarification would still be welcome, specifically regarding the point of the truncation down to an~$\mathcal{N}=2$ supergravity theory where the localization is performed. The complete~$\mathcal{N}=8$ supergravity theory contains more fields than the truncated one, and it is still somehow mysterious why these fields can be dropped from the theory altogether, and how one can still obtain the complete, correct answer in the truncated theory. A careful analysis of the consistency of such truncation would be of interest in order to provide a first-principle explanation of why the truncated theory encodes all the low-energy dynamic of the full~$\mathcal{N}=8$ four-dimensional string theory.

Moving on the case of 1/4-BPS states in four-dimensional~$\mathcal{N}=4$ string theory, Chapter~\ref{chap:n4-loc} explained in some detail how one can write down the partition function for such states, and how it is possible to obtain an approximation to its Fourier coefficients. This approximation stems from the fact that, contrary to the previous case, the counting functions are now mock modular forms, and thus an expansion akin to the Rademacher one is not known exactly for their Fourier coefficients. One can still obtain an estimate for these degeneracies by relying on the fact that mock modular forms are \emph{almost} modular, as explained in Chapter~\ref{chap:n4-loc}. In this setting, it is in fact the~$\CN=2$ supergravity localization computation which is under better control. Indeed, there it is possible to evaluate the quantum entropy function exactly as in~\eqref{WhatRadvars}. If the matching with string theory is to hold, one can interpret this formula as a \emph{prediction} for the degeneracies of mock modular forms. It would still be worthwhile to derive the Fourier coefficients of mock modular forms independently using number theory techniques, and then conduct the comparison with the supergravity answer. Also, again in this case, the issue of truncating the full~$\mathcal{N}=4$ supergravity theory down to an~$\mathcal{N}=2$ raises the same questions as in the previous case, and it would again be very interesting to examine this truncation in more details.

Lastly, when examining the more general case of 1/2-BPS states and black holes in~$\mathcal{N}=2$ four-dimensional string theory and supergravity, a number of questions remain open. In light of the work contained herein, an important aspect which should be explored is how to derive the complete one-loop determinant factor entering the Master Formula~\eqref{eq:master-integral}, not only for the vector and hyper multiplets but also for the Weyl multiplet. In Chapter~\ref{chap:n2-loc}, we have used a comparison to an on-shell calculation to fix this contribution. But as we explained there, it should be possible to derive this contribution directly by analyzing the interplay between supersymmetry and gauge symmetries acting on the fields of the Weyl multiplet. Since for the Weyl multiplet, supersymmetry itself is gauged, one should be able to construct a single BRST charge encoding all gauge symmetries (including supersymmetry), and use this charge for a computation of the index theorem~\eqref{eq:ASindthm}. This is currently under investigation by the present author and collaborators.
Another issue concerns the exact evaluation of the Master Formula for a generic~$\mathcal{N}=2$ string theory compactification on a Calabi-Yau three-fold. In such a compactification to four-dimensions, the prepotential of the low-energy supergravity theory is not know at all orders in the~$\widehat{A}$ expansion, as explained in Chapter~\ref{chap:n2-loc}. This makes the exact knowledge of the one-loop determinants, the localization measure and the renormalized action itself difficult. The microscopic string theory suffers from similar shortcomings, and it is the belief of the author that advancing this computation on either front would shed light on the correspondence between thermodynamical and statistical entropy of 1/2-BPS black holes.

In closing, although much work remains to be done to examine other, more intricate examples of the matching between the microscopic (string-theoretic) and the macroscopic (supergravity) calculations of the quantum entropy of black holes, it is encouraging to have found a few non-trivial examples in which such correspondence can be verified explicitly and exactly, at all orders in perturbation theory.  Furthering our understanding of the microstates counting in string theory, of the mathematical theory of mock modular forms and their possible generalizations and of supersymmetric localization in quantum field theories will certainly allow us to find more successful examples of a statistical interpretation for the thermodynamical entropy of black holes, including all possible quantum effects. 


\begin{appendix}

\chapter{Conventions}
\label{app:conv}
\section*{Space-time and tensor conventions}

Space-time (curved) indices are denoted by Greek letters $\mu,\nu,\ldots$ while tangent space (flat) indices are denoted with Roman letters~$a,b,\ldots$
(Anti-)symmetrization of indices is always done with weight one.

The dual of a rank-2 tensor in four dimensions and Euclidean signature is defined as
\be
\wt{T}_{ab} = \frac12\,\varepsilon_{abcd}\,T^{cd} \, , \quad \text{with} \quad \varepsilon_{1234} = \varepsilon^{1234} = -1 \, .
\ee
Note that the dual is an involution in Euclidean signature,~$\wt{\wt{T}} = T$. The (anti) self-dual part of a tensor is defined as
\be
T^\pm_{ab} = \frac12(T_{ab} \pm \wt{T}_{ab}) \, .
\ee

\section*{Spinor and Clifford conventions}

In Euclidean signature and in four dimensions, the Clifford algebra is
\be
\{\gamma^a,\,\gamma^b\} = 2\,\delta^{ab} \, .
\ee
In Chapter~\ref{chap:n2-loc}, we use an explicit Hermitian representation of the Clifford algebra given by
\be 
\label{cliffordrep}
\gamma_1 = \sigma_1 \otimes 1\, , \quad \gamma_2 = \sigma_2 \otimes 1\, , \quad \gamma_3 = \sigma_3 \otimes \sigma_1\, , \quad \gamma_4 = \sigma_3 \otimes \sigma_2\, ,
\ee
where~$\sigma_i$,~$i=1,2,3$ are the Pauli matrices. We also define the usual combination~$\gamma^{ab} = \tfrac{1}{2}[\gamma^a,\gamma^b]$ and similarly for higher-rank~$\gamma$ matrices. In addition,
\be
\gamma_5 = -\gamma_1\gamma_2\gamma_3\gamma_4 \, . 
\ee
These matrices obey the following useful identities in four dimensions:
\begin{align}
\gamma_{ab} =&\, -\tfrac12\,\varepsilon_{abcd}\gamma^{cd}\gamma_5 \, , \qquad \qquad \qquad \gamma^b\gamma_a\gamma_b = -2\gamma_a\, , \cr
\gamma^{ab}\gamma_{ab} =&\, -12 \, , \qquad \qquad \qquad \qquad \;\;\, \gamma^{cd}\gamma_{ab}\gamma_{cd} = 4\gamma_{ab}\, , \cr
\gamma^c\gamma_{ab}\gamma_c =&\, 0 \, , \qquad \qquad \qquad \qquad \qquad \;\;\; \gamma^{ab}\gamma_c\gamma_{ab} = 0\, , \\
\bigl[\gamma^c,\,\gamma_{ab}\bigr] =&\, 4\,\delta_{[a}{}^c\gamma_{b]} \, , \qquad \qquad \qquad \quad \;\; \bigl\{\gamma^c,\,\gamma_{ab}\bigr\} = 2\,\varepsilon_{ab}{}^{cd}\gamma_5\gamma_d \, , \cr
\bigl[\gamma_{ab},\,\gamma^{cd}\bigr] =&\, -8\,\delta_{[a}{}^{[c}\gamma_{b]}{}^{d]} \, , \qquad \qquad \quad \: \bigl\{\gamma_{ab},\,\gamma^{cd}\bigr\} = -4\,\delta_{[a}{}^c\delta_{b]}{}^d + 2\,\varepsilon_{ab}{}^{cd}\gamma_5 \, . \nonumber
\end{align}

In this work, we deal mainly with four-dimensional symplectic Majorana spinors transforming under an~$\mathrm{SU}(2)_R$ symmetry. The summation convention for~$\mathrm{SU}(2)$ indices is NW-SE and (anti)symmetrization of indices is done with weight one. The antisymmetric tensor of~$\mathrm{SU}(2)$ is such that
\be
\varepsilon^{ij}\varepsilon_{jk} = -\delta^i_k \quad \text{and} \quad \varepsilon^{ij}\varepsilon_{ij} = 2\, .
\ee
We define the Dirac conjugate of a spinor in four dimensions and Euclidean signature as
\be
\bar{\psi}_i := (\psi^i)^\dagger \, .
\ee
The symplectic Majorana reality condition reads
\be
\label{eq:symplMajo}
C^{-1}\bar{\psi}_i{}^\mathrm{T} = \varepsilon_{ij}\,\psi^j \, ,
\ee
where the charge conjugation matrix~$C$ is such that
\be
C\,\gamma_a\,C^{-1} = -\gamma_a{}^T \, , \quad C\,\gamma^5\,C^{-1} = \gamma^5 \, , \quad C^{-1} = C^\dagger \, , \quad C^T = -C \, .
\ee
In the explicit representation~\eqref{cliffordrep}, we take
\be
C\gamma^5 = \sigma_1 \otimes \sigma_2 \, .
\ee
In Euclidean signature and four dimensions, the symplectic Majorana condition is compatible with the Weyl projection onto positive and negative chirality spinors. That is, if we define the chiral parts of a spinor as
\be
\psi_\pm^i := \frac{1 \pm \gamma_5}{2}\,\psi^i \, ,
\ee
each of the chiral projections enjoys the property~\eqref{eq:symplMajo}:
\be
\label{eq:symplMajoWeyl}
C^{-1}\bar{\psi}_{\pm\,i}{}^\mathrm{T} = \pm\,\varepsilon_{ij}\,\psi_{\pm}{}^j \, ,
\ee
%
A useful property of spinors and antisymmetric tensors is that when~$T_{ab}\gamma^{ab}$ acts on a spinor of (positive) negative chirality, it is projected onto its (anti)self-dual part:
\be
T_{ab}\gamma^{ab}\epsilon^i_+ = T^-_{ab}\gamma^{ab}\epsilon^i_+ \quad \text{and} \quad T_{ab}\gamma^{ab}\epsilon^i_- = T^+_{ab}\gamma^{ab}\epsilon^i_-\, .
\ee
The Fierz rearrangement formula for two four-dimensional anti-commuting spinors~$\psi$ and~$\chi$ reads
\be
\chi\,\bar{\psi} = -\tfrac14(\bar{\psi}\chi)\oneone - \tfrac14(\bar{\psi}\gamma^5\chi)\gamma^5 - \tfrac14(\bar{\psi}\gamma^a\chi)\gamma_a + \tfrac14(\bar{\psi}\gamma^a\gamma^5\chi)\gamma_a\gamma^5 + \tfrac18(\bar{\psi}\gamma^{ab}\chi)\gamma_{ab} \, .
\ee 
\chapter{Modular, Jacobi and Siegel forms}
\label{app:modular}
\section*{Modular forms}

Let~$\mathbb{H}$ denote the upper half-plane, which is the set of complex numbers~$\tau$ whose imaginary part is positive. Let~$\mathrm{SL}(2;\mathbb{Z})$ be the group of $2\times 2$ matrices with integer entries and unit determinant. A \emph{modular form}~$f(\tau)$ of weight~$w$ on~$\mathrm{SL}(2;\mathbb{Z})$ is a holomorphic function on~$\mathbb{H}$ which transforms as
\be
f\Bigl(\frac{a\tau+b}{c\tau+d}\Bigr) = (c\tau+d)^w \, f(\tau)  \qquad \forall \quad \Bigl(\begin{array}{cc} a&b\\ c&d \end{array} \Bigr) \in SL(2; \IZ) \, ,
\ee
for an integer~$w$. It follows from this definition that~$f(\tau)$ is periodic under~$\tau \rightarrow \tau+1$, and thus admits a Fourier expansion
\be
f(\tau) = \sum_{n=-\infty}^{+\infty} a(n)\,q^n \, , \qquad q:=e^{2\pi i \tau} \, .
\ee
If~$a(0) = 0$, then the modular form vanishes at infinity and is called a cusp form. Weakening the growth condition at infinity to~$f(\tau) = O(q^{-N})$ (rather than~$O(1)$) for some~$N\geq0$, then the Fourier coefficients satisfy~$a(n)=0$ for~$n<-N$. Such a function is called a weakly holomorphic modular form. An important example of a modular form is the \emph{discriminant function}~$\Delta(\tau)$ introduced in Chapter~\ref{chap:modern-BH-S}:
\be
\Delta(\tau) = q\prod_{n=1}^{\infty}(1-q^n)^{24} = q - 24q^2 + 252q^3 + \ldots \, .
\ee

\section*{Jacobi forms}

A \emph{Jacobi form}~$\varphi(\tau,z)$ of weight~$w$ and index~$m$ is a holomorphic function from~$\mathbb{H}\times\mathbb{C}$ to~$\mathbb{C}$ whose defining properties are the following two transformations. It is ``modular in~$\t$'',~\textit{i.e.} it transforms under the modular group~$SL(2;\IZ)$ as
\be
\label{eq:modtransform}  
\varphi\Bigl(\frac{a\t+b}{c\t+d},\frac{z}{c\t+d}\Bigr) \= (c\t+d)^w\,e^{\frac{2\pi imc z^2}{c\t+d}}\,\varphi(\t,z)  \qquad \forall \quad \Bigl(\begin{array}{cc} a&b\\ c&d \end{array} \Bigr) \in SL(2; \IZ) \, ,
\ee
and it is ``elliptic in~$z$'',~\textit{i.e.} it transforms under the translations of $z$ by $\mathbb{Z} \tau + \mathbb{Z}$ as
\be
\label{eq:elliptic}
\varphi(\t, z+\l\tau+\mu)\= e^{-2\pi i m(\l^2 \t + 2 \l z)} \varphi(\t, z) \qquad \forall \quad \l,\,\m \in \IZ \, . 
\ee
These symmetry properties are very powerful, in particular when investigating the Fourier coefficients of Jacobi forms:
\be
\label{eq:fourierjacobi}
\varphi(\t,z) \= \sum_{n, \ell \, \in \, \IZ} c(n, \ell)\,q^n\,y^\ell \, , \quad q := e^{2\pi \mathrm{i}\tau} \, , \;\; y := e^{2\pi \mathrm{i} z} \, .
\ee
As a simple example, the elliptic transformation property~\eqref{eq:elliptic} implies that the Fourier coefficients of a Jacobi form of index~$m$ obey the property 
\be
\label{eq:cnrprop}
c(n, \ell) \= C_{\ell}(4 n m - \ell^2) \ , \quad \mbox{where} \; C_{\ell}(\Delta) \; \mbox{depends only on} \; \ell \, \text{mod}\, 2m \ . 
\ee
The same property also implies that~$\varphi(\tau,z)$ has a ``theta decomposition''
\be
\varphi(\tau,z) = \sum_{\ell \in \mathbb{Z}/2m\mathbb{Z}} h_\ell(\tau)\,\vartheta_{m,\ell}(\tau,z) \, .
\ee
Here,~$\vartheta_{m,\ell}(\tau,z)$ denotes the standard index~$m$, weight 1/2 theta function, 
\be
\vartheta_{m,\ell}(\tau,z) := \sum_{n\in\mathbb{Z}}q^{(\ell+2mn)^2/4m}\,y^{\ell+2mn} \, ,
\ee
and~$h_\ell(\tau)$ are vector-valued modular forms of weight~$w-1/2$.  In terms of the coefficients~$C_{\ell}(\Delta)$ in~\eqref{eq:cnrprop},
\be
h_\ell(\tau) = \sum_\Delta C_\ell(\Delta)q^{\Delta/4m} \qquad \textnormal{with} \;\; \ell \in \mathbb{Z}/2m\mathbb{Z} \, .
\ee
The precise mathematical definition of Jacobi forms~\cite{Eichler:1985ja} includes some technical conditions on the growth of the Fourier coefficients, in addition to the transformation formulas~\eqref{eq:modtransform},~\eqref{eq:elliptic}. Two types of Jacobi forms will be relevant in the course of this work.
 
The first is a \emph{weakly holomorphic Jacobi form}, which means that the Fourier expansion in~\eqref{eq:fourierjacobi} obeys~$n \ge -n_{0}$ for a fixed positive~$n_{0}$. Note that this implies that there are only a finite number of terms with non-zero Fourier coefficients for negative values of~$\Delta = 4mn-\ell^2$. These coefficients are called the \emph{polar coefficients} in the Fourier expansion of the Jacobi form. 
The second type is a \emph{weak Jacobi form}, which means that~$n_{0}=0$ above. We refer the reader to~\cite{Eichler:1985ja} for a detailed theory of these functions. 

The modular transformation property~\eqref{eq:modtransform} is so constraining that one has an analytic formula for all the Fourier coefficients of a Jacobi form in terms of its polar coefficients. This formula, called the \emph{Hardy-Ramanujan-Rademacher} expansion, takes the form of an infinite convergent sum of Bessel functions and is established by the so-called circle method in analytic number theory. The formula for the coefficients~$C_{\ell}(\Delta)$ of a Jacobi form of weight~$w+1/2$ and index~$m$, with~$\Delta=4mn-\ell^{2}$, has the following form:
\begin{align}
\label{eq:radi}
C_\ell (\Delta) =&\, (2\pi)^{2-w} \sum_{c=1}^\infty c^{w-2} \; \times \nonumber \\
&\times \; \sum_{\wt\ell \in \IZ/2m \IZ} \, \sum_{\wt \Delta < 0} C_{\wt\ell}(\wt \Delta) \,  K\ell(\Delta,\ell,\wt \Delta,\wt\ell;c) \left| \frac{\wt \Delta}{4m} \right|^{1-w} \, \wt I_{1-w} \Bigl( {\pi\over m c} \sqrt{| \wt \Delta| \Delta}\Bigr) \, ,
\end{align}
where 
\be
\label{eq:intrep}
\wt{I}_{\rho}(z)=\frac{1}{2\pi i}\int_{\epsilon-i\infty}^{\epsilon+i\infty} \, \frac{d\s}{\s^{\r +1}}\exp \Bigl( {\s+\frac{z^2}{4\s}} \Bigr) \, ,
\ee
is called the modified Bessel function of index $\r$, and is related to the standard Bessel function of the first kind $I_{\rho}(z)$ by
\be
\wt I_{\rho}(z) = \Bigl(\frac{z}{2} \Bigr)^{-\rho} I_{\rho}(z) \, .
\ee
The latter function has an asymptotic expansion for large arguments:
\be
\label{eq:Bessel-Exp}
I_{\rho}(z) \underset{z \rightarrow \infty}{\sim} \frac{e^z}{\sqrt{2\pi z}}\Bigl(1 - \frac{\mu - 1}{8z} + \frac{(\mu - 1)(\mu - 3^2)}{2!(8z)^3} - \frac{(\mu - 1)(\mu - 3^2)(\mu - 5^2)}{3!(8z)^5} + \ldots\Bigr) \, ,
\ee
with~$\mu = 4\rho^2$. The coefficients $K\ell(\Delta,\ell,\wt \Delta,\wt\ell;c)$ in~\eqref{eq:radi} are the so-called generalized Kloosterman sums~\cite{Dijkgraaf:2000fq}, and they consist essentially in sums of phases. For~$c=1$, they are given by:
\be 
\label{eq:Kloosc1}
K\ell(\Delta,\ell,\wt \Delta,\wt\ell;c=1) \= \sqrt{\frac{2}{m}}\,e^{i\pi(m-\ell')\frac{\ell}{m}} \, . 
\ee

The remarkable thing about~\eqref{eq:radi} is that the coefficients~$C_{\ell}(\Delta)$ for~$\Delta>0$ are completely determined by the coefficients~$C_{\wt\ell}(\wt \Delta)$ associated to the polar terms~$q^{\wt \Delta}$ with~$\wt \Delta<0$, which are finite in number. The asymptotic formula of the Bessel function~$I_\r(z) \sim e^{z}$ for large~$z$ shows that the terms with~$c>1$ are exponentially suppressed compared to the leading~$c=1$ terms.

\section*{Siegel forms and mock Jacobi forms}

There exists a generalization of Jacobi forms called \emph{Siegel modular forms}. They are holomorphic functions of three variables~$(\tau,z, \sigma)$, which are arranged in a matrix
\be
\Omega = \begin{pmatrix}[.9] \tau & z \\ z& \sigma \end{pmatrix} \, ,
\ee
satisfying
\be
\textnormal{Im}\,\tau > 0\, , \quad \textnormal{Im}\,\sigma > 0 \, , \quad \textnormal{det}(\textnormal{Im}\,\Omega) > 0 \, .
\ee
This defines the Siegel upper half-plane, where the Siegel forms are well-defined. A Siegel form~$F(\Omega)$ of weight~$w$ satisfies a property analogous to~\eqref{eq:modtransform}, namely
\be
F\bigl((A\Omega + B)(C\Omega + D)^{-1}\bigr) = \textnormal{det}(C\Omega + D)^w\,F(\Omega) \, ,
\ee
where the matrices~$A,\,B,\,C$ and~$D$ are~$2\times 2$ matrices with integer entries satisfying
\be
AB^T = BA^T \, , \quad CD^T = DC^T\, , \quad AD^T - BC^T = 1 \, .
\ee
Just like the Jacobi forms, Siegel forms have a Fourier expansion
\be
F(\Omega) = \sum_{\substack{n,\,r,\,m \in \mathbb{Z} \\ r^2 \leq 4mn}} a(n,r,m) q^n \, y^r \, p^m \, ,
\ee
again with the standard notation~$q := e^{2\pi \mathrm{i}\tau}, \, y := e^{2\pi \mathrm{i} z} , \, p := e^{2\pi \mathrm{i}\sigma}$. If one now writes this Fourier expansion as
\be
F(\Omega) = \sum_{m=0}^\infty \varphi_m(\tau,z) p^m \, ,
\ee
then each~$\varphi(\tau,z)$ is a Jacobi form of weight~$k$ and index~$m$. We will refrain from giving a complete characterization of Siegel forms and their properties here, and we now briefly review some facts from~\cite{Dabholkar:2012nd} which are relevant for the present work. 

As reviewed in Chapter~\ref{chap:n4-loc}, Siegel forms naturally appear in the counting problem of 1/4-BPS dyons in~$\mathcal{N}=4$ string theory. The partition function for such states is given by the inverse of the so-called Igusa cusp form:
\be
Z^{\textnormal{BPS}}(\tau,z,\sigma) = \frac{1}{\Phi_{10}(\tau,z,\sigma)} \, .
\ee
The Igusa cusp form has double zeroes at~$z=0$ (and its~$\mathrm{Sp}(2;\mathbb{Z})$ images), so that the partition function is a \emph{meromorphic} Siegel form of weight -10. The first step in~\cite{Dabholkar:2012nd} to analyze its Fourier coefficients is to expand the 
microscopic partition function in $e^{2 \pi i \s}$:
\be
\label{eq:recipro-igusa}
\frac 1{\Phi_{10}(\t, z, \sigma)} \= \sum_{m\geq -1} \psi_m (\t,z) \, e^{2 \pi i m \s}  \, .
\ee
One then defines the \emph{polar part} of $\psi_{m}$
\be
\label{eq:Tm}
\psi^{\text{P}}_m (\t, z) := \; \frac{p_{24}(m+1) }{\eta^{24}(\t)} \, \sum_{s\in\IZ} \, \frac{q^{ms^2 +s}y^{2ms+1}}{(1 -yq^s )^2} \, , 
\ee
where $p_{24}(n)$ counts the number of  partitions of an integer $n$ with $24$ colors. The function~$\psi^{\text{P}}_{m}$ is the average over the lattice $ \mathbb{Z} \t + \IZ$ of the leading behavior of the function near the pole $z=0$
\be 
\label{eq:simplewall}  
\frac{p_{24}(m+1)}{\eta(\t)^{24}} \frac{y}{(1-y)^2}\,. 
\ee
The function $\psi_{m}^{\text{P}}$ is an example of an Appell-Lerch sum, and it encodes the physics of all the wall-crossings due to the decay of two-centered black holes presented in Chapter~\ref{chap:n4-loc}.

The two functions $\psi_{m}$ and $\psi_{m}^{\text{P}}$ have, by construction, the same poles and residues, so the difference 
\be  
\psi_{m}^\text{F} := \psi_{m} - \psi_{m}^\text{P} \, ,
\ee
called the \emph{finite or Fourier part} of $\psi_{m}$, is holomorphic in $z$, and has an unambiguous Fourier expansion:
\be 
\label{eq:psiF2}
\psi_{m}^\text{F}(\t,z) \= \sum_{n,\ell} \, c^{\text{F}}_{m}(n,\ell) \, q^{n} \, y^{\ell} \, .
\ee
The indexed degeneracies of the single-centered black hole with magnetic charge invariant~$Q_{m}^{2}/2=m$, as defined by the attractor mechanism, are related to the Fourier coefficients of the function~$\psi_{m}^{\text{F}}$ as 
\be
d(n,\ell,m) = (-1)^{\ell+1} c^{\text{F}}_{m}(n,\ell)\, ,
\ee
the overall sign coming from an analysis of the fermion zero modes described in~\cite{Dabholkar:2010rm}. The statement of the main theorem of~(\cite{Dabholkar:2012nd}, Chapter 8) is that the single-center black hole partition function~$\psi^\text{F}_{m}(\t,z)$ is a \emph{mock Jacobi form}.

What this means is that~$\psi_m^\text{F}$ has the same elliptic transformation property~\eqref{eq:elliptic} as a regular Jacobi form governed by the index~$m$. Its modular transformation property~\eqref{eq:modtransform}, however, is modified. The lack of modularity is governed by an explicit non-holomorphic function called the~\emph{shadow}:
\be
\psi_m^\text{S} (\t,z) \=  \frac{1}{\eta(\t)^{24}} \sum_{\ell\in \IZ/2m\IZ} \vth^{*}_{m,\ell}(\t,0) \, \vth_{m,\ell}(\t,z) \, ,
\ee
where
 the operation~$*$ is defined such that a modular form~$g$ of weight~$w$ obeys 
\be
\label{eq:ddtbarh}   
(4\pi\t_2)^w\,\,\frac{\partial g^{*}(\t)}{\partial \overline{\tau}} \= -2\pi i\;\overline{g(\tau)}\, .  
\ee
The function
\be
\widehat{\psi_m^\text{F}}(\t,z) = \psi_m^\text{F} (\t,z) + \psi_m^{\text{S}} (\t,z) \, , 
\ee 
called the \emph{completion} of~$\psi_m^\text{F}$, transforms as a Jacobi form of weight~$-10$ and index~$m$, but it is not holomorphic. It obeys the holomorphic anomaly equation:
\be
\label{eq:ddtbarhphi}   
(4\pi\t_2)^{1/2} \,\,\frac{\partial \widehat{\psi_m^\text{F}} (\t,z)}{\partial \overline{\tau}} \= -2\pi i\; \frac{1}{\eta(\t)^{24}} \sum_{\ell\in \IZ/2m\IZ} \overline{\vth_{m,\ell}(\t,0)} \, \vth_{m,\ell}(\t,z) \, . 
\ee

We now briefly present some relevant facts about the growth of the coefficients of the mock Jacobi forms~$\psi^\text{F}_{m}$. By multiplying~$\psi^\text{F}_{m}$ by the function~$\eta(\t)^{24}$, we get a function~$\varphi_{2,m}^\text{mock} = \eta^{24} \psi^\text{F}_{m}$ which is a mock Jacobi form of weight~2 and index~$m$. It was shown in~(\cite{Dabholkar:2012nd}, Chapters~9,~10) that~$\varphi_{2,m}$ can be written\footnote{Recall that the definition of a mock Jacobi form only holds modulo the addition of a true Jacobi form.} as a linear combination of a (true) weak Jacobi form and a mock Jacobi form 
\be
\varphi^\text{mock}_{2,m}(\t,z) \= \varphi^\text{true}_{2,m}(\t,z) + \varphi^\text{opt}_{2,m}(\t,z) \, , 
\ee
such that the mock Jacobi form~$\varphi^\text{opt}_{2,m}$ has \emph{optimal growth}. This means that the Fourier-Jacobi coefficients of~$\varphi^\text{opt}_{2,m}(\t,z)$ grow at most as
\be 
\label{eq:optgrowth}
c^\text{opt}(n,\ell) \sim \exp\bigl(\frac{\pi}{m}\sqrt{4mn-\ell^{2}} \bigr)  \, .
\ee
If we look at the Rademacher expansion~\eqref{eq:radi}, the growth~\eqref{eq:optgrowth} is the smallest possible one, governed by the value of~$|\wt\Delta|=1$. In fact, for~$m$ a prime power, the coefficients of the optimal mock Jacobi form has only polynomial growth.

\chapter{Building Euclidean~$\CN=2$ conformal supergravity}
\label{app:sugra}
Starting from the superconformal transformations for $5D$ supermultiplets presented in \eqref{eq:Weyl-susy-var}, \eqref{eq:sc-vector-multiplet} and \eqref{eq:hypertransf}, we perform a reduction on the time coordinate and obtain the corresponding results for the $4D$ Euclidean superconformal transformations as well as the relevant supermultiplets. The Weyl multiplet contains the gauge fields
associated with the superconformal transformations as well as
additional supercovariant fields, which act as a background for all
other supermultiplets.  Therefore this multiplet must be considered
first. Here a subtle complication is that the Weyl multiplet becomes
reducible upon the reduction. In $D=5$ it comprises $32+32$ bosonic
and fermionic degrees of freedom, which, in the reduction to $D=4$
dimensions decomposes into the Weyl multiplet comprising $24+24$
degrees of freedom, and a vector multiplet comprising $8+8$ degrees of
freedom.

In Section \ref{sec:euclid-superg-from-dim-red} we also described the
Kaluza-Klein decomposition of the metric and the dilatational gauge
field that ensure that the $4D$ fields transform covariantly under the
$4D$ diffeomorphisms. Since these decompositions involve gauge choices
on the vielbein and the dilatational gauge field, compensating Lorentz
and special conformal transformations must be included when deriving
the $4D$ Q-supersymmetry transformations to ensure that these gauge
conditions are preserved. Here the parameter of the compensating Lorentz
transformation is most relevant. It is equal to
\begin{equation}
  \label{eq:comp-Lor}
  \epsilon^{a5} = -\epsilon^{5a} = \mathrm{i} \phi\,
  \bar\epsilon_i\gamma^a\psi^i \; \Longleftrightarrow \;
  \epsilon^{a0} = -\epsilon^{0a} =  \phi\,
  \bar\epsilon_i\gamma^a\psi^i \,,
\end{equation}
where we assumed the standard Kaluza-Klein decomposition on the
gravitino fields, 
\begin{equation}
  \label{eq:gravitino-KK}
  \psi_M{}^i =  \begin{pmatrix}\psi_\mu{}^i+ B_\mu \psi^i\\[4mm]
    -\mathrm{i}\psi^i \end{pmatrix}\;, 
\end{equation}
which ensures that $\psi_\mu{}^i$ on the right-hand side transforms as
a $4D$ vector. Owing to the factor of~$\mathrm{i}$ in this decomposition, both $\psi_\mu{\!}^i$ and
$\psi^i$ are symplectic Majorana spinors. Upon including this extra
term, one can write down the Q- and S-supersymmetry transformations on
the $4D$ fields defined above. As a result of this, the $4D$ and $5D$
supersymmetry transformation will be different. For instance, the
supersymmetry transformations of the $4D$ fields $e_\mu{}^a$, $\phi$
and $B_\mu$ read,
\begin{align}
  \label{eq:susy-e-B-phi}
  \delta e_\mu{}^a =&\,  \bar\epsilon_i\gamma^a\psi_\mu{}^i
  \,, \nonumber\\[.2ex]
  \delta\phi =&\, \mathrm{i} \phi^2\,\bar\epsilon_i\gamma^5\psi^i\,,
  \nonumber \\[.2ex]
  \delta B_\mu=&\,  -\phi^2 \,\bar\epsilon_i\gamma_\mu\psi^i -\mathrm{i}
  \phi \,\bar\epsilon_i\gamma^5\psi_\mu{}^i \,,
\end{align}
where the first term in $\delta B_\mu$ originates from the
compensating transformation \eqref{eq:comp-Lor}. Consequently the
supercovariant field strength of $B_\mu$ contains a term that is not
contained in the supercovariant five-dimensional curvature
$R(P)_{MN}{}^A$.  Therefore the $5D$ spin-connection components are
not supercovariant with respect to $4D$ supersymmetry, as is reflected
in the second formula below,
\begin{align}
  \label{eq:spin-connection}
  \omega_M{}^{ab} =&\, \begin{pmatrix} \omega_\mu{}^{ab} \\[4mm] 
    0 \end{pmatrix} - \tfrac12  \phi^{-2} \hat F(B)^{ab} \, 
  \begin{pmatrix} B_\mu \\[4mm] -\mathrm{i} \end{pmatrix} \;, \nonumber \\[.8ex]
  \omega_M{}^{a5} =&\, -\tfrac12 \mathrm{i}\begin{pmatrix} \phi^{-1}
    \hat F(B)_\mu{}^a - \phi\,\bar\psi_{\mu i}\gamma^a\psi^i \\[4mm]
    0 \end{pmatrix} -\mathrm{i} \phi^{-2} D^{a}\phi  \,    
  \begin{pmatrix} B_\mu \\[4mm] -\mathrm{i}\end{pmatrix} \;.
\end{align}
Here we introduced the supercovariant field strength and derivative
(with respect to $4D$ supersymmetry), 
\begin{align}
  \label{eq:supercov-FB-Dphi}
  \hat F(B)_{\mu\nu} =&\, 2\,\partial_{[\mu} B_{\nu]} + \phi^2\,
  \bar\psi_{[\mu i}\gamma_{\nu]} \psi^i +\tfrac12 \mathrm{i}
  \phi\,\bar\psi_{\mu i} \gamma_5 \psi_{\nu}{}^i \,,\nonumber\\
  D_\mu\phi =&\, (\partial_\mu -b_\mu) \phi -\tfrac12\mathrm{i} \phi^2
  \,\bar\psi_{\mu i}\gamma_5 \psi^i \,.
\end{align}
We should mention that the dilatational gauge field (as
well as the composite gauge fields, such as $\omega_\mu{}^{ab}$, that
depend on it) will not necessarily acquire the form that is familiar
from $4D$. This may require to include an  additional compensating
conformal boost transformation. 

Subsequently one writes corresponding Kaluza-Klein decompositions for
some of the other fields of the Weyl multiplet, which do not require
special gauge choices,
\begin{equation}
  \label{eq:1Weyl-KK}
    {V}_{M i}{}^j=
  \begin{pmatrix}{V}_{\mu i}{}^j+ B_\mu
    {V}_{i}{}^j\\[4mm] 
    -\mathrm{i} {V}_{i}{}^j \end{pmatrix}\;,\qquad
  \phi_M{}^i =  \begin{pmatrix}\phi_\mu{}^i+ B_\mu \phi^i\\[4mm]
    -\mathrm{i}\phi^i \end{pmatrix}\;\qquad 
   T_{AB} = \begin{pmatrix} T_{ab} \\[4mm]  T_{a5}\equiv \tfrac16
     \mathrm{i} A_a \end{pmatrix}  \,. 
\end{equation}

Hence we are now ready to consider the Q- and S-supersymmetry
transformations of the spinor fields originating from the $5D$
gravitino fields. Up to possible higher-order spinor terms, one
derives the following results from \eqref{eq:Weyl-susy-var},
\begin{align}
  \label{eq:susy-W-gravitino}
  \delta(\phi^2\,\psi^i) =&\, -\tfrac12
   \big[-\hat F(B)_{ab} + \gamma_5 \phi (
  3\,T_{ab} +\tfrac14  \phi^{-1} \hat F(B)_{ab}\gamma_5
  )\big]   \gamma^{ab}\epsilon^i \nonumber\\
  &\,
  +\mathrm{i}\big[ \Slash{D} \phi \,\gamma^5 - 
  \Slash{A}\phi\big] \epsilon^i - \phi^2 
  {V}^i{}_j \,\epsilon^j \nonumber\\[.2ex]
  &\,
    +\gamma_5 \phi \big[ \eta^i
    -\tfrac13\mathrm{i} \Slash{A}\gamma_5\epsilon^i  -\tfrac1{8} 
  \gamma_5\phi^{-1}(\hat F(B)_{ab}-4\phi
   T_{ab}\gamma_5)\gamma^{ab} \epsilon^i\big]  \,, \nonumber\\[.2ex] 
  \delta\psi_\mu{}^i =&\,2\,\big(\partial_\mu
  -\tfrac14\omega_\mu{}^{ab}\gamma_{ab}+\tfrac12 b_\mu
  +\tfrac12 e_\mu{}^a \,A_a \gamma_5 \big)\epsilon^i
  +  {V}_{\mu j}{}^i \epsilon^j \\
  &\, + \tfrac12 \mathrm{i}\big[3\, T_{ab} +\tfrac14 
  \phi^{-1} \hat F(B)_{ab}\gamma_5  \big] \gamma_{ab} \gamma_\mu
  \epsilon^i  \nonumber\\
  &\,
  - \mathrm{i} \gamma_\mu \big[\eta^i
   -\tfrac13\mathrm{i}\Slash{A}\gamma_5 \epsilon^i  -\tfrac1{8}
   \gamma_5\phi^{-1}(\hat F(B)_{ab}-4\phi
   T_{ab}\gamma_5)\gamma^{ab} \epsilon^i \big]\,.  \nonumber
\end{align}
Clearly, the fields $e_\mu{}^a$ and $\psi_\mu{}^i$ must belong to the
Weyl multiplet, whereas $\phi$, $B_\mu$ and $\phi^2\psi^i$ correspond
to the Kaluza-Klein vector multiplet, as the transformations shown in
\eqref{eq:susy-e-B-phi} and \eqref{eq:susy-W-gravitino} have many
features in common with the expected $4D$ transformations of these
supermultiplets. Note that we have multiplied $\psi^i$ with a factor
$\phi^2$ to give it the expected Weyl weight ${w=\tfrac32}$. At this
stage we have only identified one of the two $w=1$ scalars that must reside
in a $4D$ vector multiplet.  The field $A_a$ seems to play
the role of an R-symmetry connection because it appears to
covariantize the derivatives on $\phi$ and $\epsilon^i$ in
\eqref{eq:susy-W-gravitino}.  Furthermore, a particular
linear combination of the $5D$ tensor components $T_{ab}$ and the
(dual) supercovariant field strength $\hat F(B)_{ab}$ appears in the
transformations \eqref{eq:susy-W-gravitino} in precisely the same form
as the $4D$ auxiliary tensor $T_{ab}$, so that the latter is not just
proportional to the original $5D$ tensor field. The same combination
will also appear in other transformation rules, as we shall see in,
for instance, Section \ref{sec:shell-dimens-reduct-matter}. Finally,
S-supersymmetry transformations are accompanied by extra contributions
characterized by a field-dependent parameter proportional to
$\epsilon^i$.  

However, the result \eqref{eq:susy-W-gravitino} is not yet complete as we have
suppressed the variations quadratic in the spinor fields. First of all
we did not include the non-covariant term in
\eqref{eq:spin-connection} and we ignored the compensating Lorentz
transformation \eqref{eq:comp-Lor}. Secondly we ignored the variation
of the field $B_\mu$ in the decomposition of the $4D$ gravitino
\eqref{eq:gravitino-KK}, and thirdly the multiplication of $\psi^i$ with
$\phi^2$ will also generate a variation quadratic in $\psi^i$ . Since
these terms will play an important role we summarize them below,
\begin{align}
  \label{eq:susy-W-gravitino-nonlin}
  \delta(\phi^2\psi^i)\big\vert_\mathrm{non-linear} =&\, \tfrac12
  \mathrm{i} \phi^3\,
  \bar\epsilon_j\gamma^a\psi^j\, \gamma_a\gamma_5\psi^i   +
  2\mathrm{i}\, \phi^3 \,\bar\epsilon_j\gamma^5\psi^j\, \psi^i\,,
  \nonumber\\[.2ex]  
  \delta\psi_\mu{}^i\big\vert_\mathrm{non-linear} =&\, -\tfrac12
  \mathrm{i} \phi \,\bar\psi_{\mu j} \gamma^a\psi^j \, \gamma_a\gamma^5
  \epsilon^i 
  + \tfrac12 \mathrm{i} \phi\, \bar\epsilon_j\gamma^a\psi^j\,
  \gamma_a\gamma_5\psi_\mu{\!}^i \\
  &\,+\big(\phi^2
  \,\bar\epsilon_j\gamma_\mu\psi^j +\mathrm{i} \phi
  \,\bar\epsilon_j\gamma^5\psi_\mu{}^j \big) \psi^i \,.  \nonumber
\end{align}

The systematic pattern already noticed in \cite{Banerjee:2011ts} for
the space-like reduction is that  the $5D$ supersymmetry
transformations can uniformally  be decomposed in terms of the $4D$
supersymmetry transformation and field-dependent S-supersymmetry, and
$\mathrm{SU}(2)$ R-symmetry transformations with field-dependent
parameters. Since the derivation is
identical to what was carried out in \cite{Banerjee:2011ts}, we just
present the universal formula for $5D$ Q-supersymmetry transformations
of fields $\Phi$ that transform covariantly in the $4D$ setting,
\begin{equation}
  \label{eq:D5-D4-decomp}
  \delta_\mathrm{Q}(\epsilon)\big|^\mathrm{reduced}_{5D} \Phi =
  \delta_\mathrm{Q}(\epsilon)\big|_{4D} \Phi + 
  \delta_\mathrm{S}(\tilde\eta)\big|_{4D} \Phi +
  \delta_{\mathrm{SU}(2)}(\tilde\Lambda)\big|_{4D}\Phi  +
  \delta^\prime(\tilde\Lambda^0) \Phi\,.  
\end{equation}
Here the first term on the right-hand side defines the $4D$
supersymmetry transformation, while $\tilde\eta$ and $\tilde\Lambda$
denote the (universal) field-dependent parameters of accompanying
S-supersymmetry and $\mathrm{SU}(2)$ R-symmetry transformations. The
last variation denoted by $\delta^\prime(\tilde\Lambda^0)$ is a linear
transformation on the fields $\Phi$ that signals the emergence of an
extra component in the $4D$ Euclidean R-symmetry group. Note that
$\tilde\eta$, $\tilde\Lambda$ and $\tilde\Lambda^0$ are all linearly
proportional to the supersymmetry parameter $\epsilon^i$. The explicit
form of these field-dependent parameters is as follows, 
\begin{align}
  \label{eq:Q-susy decom-par}
  \tilde{\eta}^i =&\,-\tfrac13\mathrm{i} \Slash{A}\gamma_5\epsilon^i
  -\tfrac1{8}\gamma_5\phi^{-1}\big(\hat F(B)_{ab}-4\phi
  T_{ab}\gamma_5\big)\gamma^{ab} \epsilon^i \nonumber\\
  &-\tfrac14\mathrm{i}\,\phi^2\big(\bar{\psi}_j\gamma^5\psi^i\gamma_5
    -\bar{\psi}_j\psi^i + \bar{\psi}_j\gamma^a\psi^i\gamma_a
    +\tfrac12\bar{\psi}_k\gamma^5\gamma^a\psi^k\gamma_5
    \gamma_a\delta_j{}^i\big)\epsilon^j\, , \nonumber\\
  \tilde{\Lambda}_j{}^i =&\,- \mathrm{i}\,\phi\big(\bar{\epsilon}_j\gamma^5\psi^i
    -\tfrac12\delta_j{}^i\bar{\epsilon}_k\gamma^5\psi^k\big)\, ,\\
  \tilde{\Lambda}^0 =&\,\mathrm{i}\,\phi\,\bar{\epsilon}_k\psi^k \,. \nonumber
\end{align}
After verifying that the decomposition is universally realized, these
extra symmetries with field-dependent coefficients can be dropped
provided they define local symmetries of the $4D$ theory.

Evaluating the terms of higher order in the fermions is subtle; here we
can only partly rely on the results of \cite{Banerjee:2011ts} because
the phases of the spinor bilinears cannot always be converted directly
from $5D$ (as noted just below equation \eqref{eq:bilinear}). It leads
to the following redefinitions of the various bosonic fields,
\begin{align}
  \label{eq:field-redef}
   \hat A_\mu =&\, A_a \,e_\mu{}^a -\tfrac12\mathrm{i}\, \phi\,
  \bar\psi_j\psi_\mu{}^j-\tfrac1{4} \phi^2\,
  \bar\psi_j\gamma^5\gamma_\mu\psi^j \,, \nonumber \\[.2ex]
  \hat{T}_{ab}=&\, 24\,T_{ab} - \phi^{-1}\,\varepsilon_{abcd}\,
  \hat F(B)^{cd} 
   +\mathrm{i} \phi^{2}\, \bar\psi_i\gamma_{ab} \psi^i
   \,,\nonumber \\[.2ex] 
   \hat V_j{}^i=&\, \phi^2\, V_j{}^i + \tfrac32\mathrm{i} \phi^3\, 
   \bar\psi_j\,\gamma^5\psi^i \,, \\ 
   \hat V_\mu{}_j{}^i =&\, {V}_\mu{}_j{}^i
  +\mathrm{i}\phi \big( \bar\psi_{\mu j} \gamma^5 \psi^i
   - \tfrac12 \delta_j{}^i\,
  \bar\psi_{\mu k} \gamma^5 \psi^k  \big)
  +\tfrac12  \phi^2\,\bar\psi_{j} \gamma_\mu \psi^i \,. \nonumber
\end{align}
Note that in the last two equations possible contributions
proportional to $\bar\psi_k\gamma^5\psi^k$ and
$\bar\psi_k\gamma_\mu\psi^k$ do not appear as they vanish owing to the
Majorana condition. 

The modifications given in \eqref{eq:field-redef} lead to important
changes in the supersymmetry transformations. For instance, the
S-supersymmetry transformations are given by
\begin{align}
  \label{eq:S-ATV}
   \delta\hat A_\mu=&\,\tfrac12\mathrm{i}\bar\psi_{\mu j}\gamma^5 \eta^j\,,
   \nonumber \\[.2ex] 
   \delta\hat T_{ab} =&\, 0\,, \nonumber \\[.2ex]
   \delta \hat V_j{}^i=&\, 0\,, \\[.2ex]
   \delta\hat V_{\mu j}{}^i=&\,  -2\mathrm{i}\big(\bar\psi_{\mu j}\,\eta^i
   -\tfrac1{2}\delta_j{}^i\,\bar\psi_{\mu k}\,\eta^k \big) \,. \nonumber
\end{align}
In particular, note that the factor in the variation of $\hat V_{\mu
  i}{}^j$ has now changed as compared to the corresponding $5D$
S-variation given in \eqref{eq:Weyl-susy-var}. Furthermore, $\hat
A_\mu$ is not supercovariant because its Q-supersymmetry variation
contains a term proportional to the derivative of the supersymmetry
parameter. This suggest that $\hat A_\mu$ will be related to a gauge
field associated with an extra $4D$ R-symmetry, which will indeed
be consistent with the fact that $\hat A_\mu$ transforms into the
gravitino fields under S-supersymmetry.

Let us now present the supersymmetry transformations for the redefined
fields, suppressing the field-dependent S-supersymmetry and
$\mathrm{SU}(2)$ transformations indicated in
\eqref{eq:D5-D4-decomp}. For the vierbein and gravitini, we find
\begin{align}
  \label{eq:susy-weyl1}
  \delta e_\mu{}^a =&\,  \bar\epsilon_i\gamma^a\psi_\mu{}^i
  \,, \\[.2ex]
    \delta\psi_\mu{}^i =&\,2\,\big(\partial_\mu
  -\tfrac14\omega_\mu{}^{ab}\gamma_{ab}+\tfrac12 b_\mu
  +\tfrac12  \hat A_\mu \gamma_5 \big)\epsilon^i
  +  {\hat V}_{\mu j}{}^i\, \epsilon^j + \tfrac1{16} \mathrm{i} \hat T_{ab} \gamma^{ab} \gamma_\mu
  \epsilon^i 
  -\mathrm{i} \gamma_\mu \eta^i \,. \nonumber
\end{align}
For the scalar $\phi$, the spinor $\hat\psi^i\equiv \phi^2 \psi^i$ and
the Kaluza-Klein photon field $B_\mu$ we have the following Q- and
S-supersymmetry transformations,
\begin{align}
  \label{eq:susy-KK}
    \delta\phi =&\, \mathrm{i}\,\bar\epsilon_i\gamma^5\hat\psi^i\,,
  \nonumber \\[.2ex]
  \delta B_\mu=&\, -\bar\epsilon_i\gamma_\mu\hat \psi^i - \mathrm{i}
  \phi \,\bar\epsilon_i\gamma^5\psi_\mu{}^i \,,\nonumber
  \\[.2ex]  
  \delta\hat\psi^i  =&\, \tfrac12 \big[\hat
  F(B)_{ab}-\tfrac18 \phi\, \hat T_{ab}\gamma_5 \big]
  \gamma^{ab}\epsilon^i  \\
  &\, -\mathrm{i} \gamma^5 \gamma^\mu \big[\mathcal{D}_\mu\phi
  -\tfrac12\mathrm{i}(\bar\psi_{\mu j} 
  \gamma^5 \hat\psi^j -\bar\psi_{\mu j}\hat \psi^j \gamma^5) - \hat
    A_\mu \phi\,\gamma^5 \big] \epsilon^i + {\hat V}_j{}^i \,\epsilon^j +\phi \gamma^5 \eta^i\,,\nonumber
\end{align}
where the derivative $\mathcal{D}_\mu$ is covariant with respect
to $4D$ local Lorentz, dilatation and $\mathrm{SU}(2)$ transformations.

At this point we make a number of important comments. First of all, we
have suppressed the chiral transformations proportional to the
field-dependent parameter $\tilde\Lambda^0$, 
\begin{equation}
  \label{eq:}
  \delta\psi_\mu{}^i =  - \tfrac12 \tilde\Lambda^0 \,\gamma^5
  \psi_\mu{}^i\,, \quad  \delta\hat\psi^i  =  -\tfrac12
  \tilde\Lambda^0 \, \gamma^5\hat\psi^i \,.
\end{equation}
Note, however, that we were not allowed to do this as these transformations are
at this stage not realized as local transformation of the $4D$
theory. Furthermore, the variations of $\hat\psi^i$ that are
proportional to $\bar\psi_{\mu j} \hat \psi^j$ are not part of a
supercovariant derivative of the field $\phi$. And finally the field
$\hat A_\mu$ is not a gauge field associated with the chiral
transformations (although it appears in a suggestive way). However, it
is not a proper matter field either as it does not transform
supercovariantly. We will address these issues momentarily. 

Rather than resolving these issues now, we prefer to first
continue. Therefore it is convenient to first define a
composite fermionic gauge field $\hat\phi_\mu{}^i$ which serves as a
$4D$ connection for S-supersymmetry. It is the solution of the
equation (in the ensuing analysis we will not exhibit terms quadratic
in the spinor fields)
\begin{equation}
  \label{eq:intermediate-4d-connection}
  \gamma^\mu \Big[\big(\mathcal{D}_{[\mu} + \tfrac12 \hat A_{[\mu} \gamma^5
  \big)\psi_{\nu]}{\!}^i  
  -\tfrac12 \mathrm{i} \,\gamma_{[\mu} \,\hat\phi_{\nu]}{}^i   
  +\tfrac1{32} \mathrm{i}\,\hat T_{ab} \gamma^{ab} \,\gamma_{[\mu}
  \,\psi_{\nu]}{}^i \Big]  =0\,,
\end{equation} 
and transforms under S- and Q-supersymmetry as 
\begin{align}
  \label{eq:delta-hat-phi-mu}
  \delta\hat \phi_\mu{}^i =&\, 2\,\big(\mathcal{D}_\mu -\tfrac12 \hat
  A_\mu \gamma_5 \big)\eta^i + \tfrac{1}{48}\mathrm{i}\gamma_\mu
  \hat{T}_{ab}\gamma^{ab}\eta^i \nonumber\\
  &\, + 2\mathrm{i} \,\hat f_\mu{}^a\gamma_a\epsilon^i
  +\tfrac1{16}\bigl(\gamma^\nu \gamma^{ab}\gamma_\mu -
  \tfrac13\gamma_\mu\gamma^{ab}\gamma^\nu\bigr)D_\nu\hat
  T_{ab}\epsilon^i \\
  &\, -\tfrac14\mathrm{i}\bigl(\gamma^{ab}\gamma_\mu -
  \tfrac13\gamma_\mu\gamma^{ab}\bigr)R(\hat V)_{abj}{}^i\epsilon^j
  - \tfrac12\mathrm{i}\bigl(\gamma^{ab}\gamma_\mu +
  \tfrac13\gamma_\mu\gamma^{ab}\bigr)R(\hat A)_{ab}\gamma^5\epsilon^i \,, \nonumber
\end{align}
where $\hat f_\mu{}^a$ reads
\begin{equation}
  \label{eq:def-hat-f}
  \hat f_\mu{}^a =\tfrac12  R(\omega,e)_\mu{}^a - \tfrac1{12} R(\omega,e)
  \,e_\mu{}^a -\tfrac12\,\tilde R(\hat A)_\mu{}^a
  -\tfrac1{128}  (\hat T- \tilde{\hat T})_{\mu b} \,(\tilde T+\tilde{\hat T})^{ba} \,, 
\end{equation}
where $R(\omega,e)_\mu{}^a = R(\omega)_{\mu\nu}{\!}^{ab}\,e_b{}^\nu$
is the generalized (non-symmetric) Ricci tensor. Its anti-symmetric
part is equal to $R(\omega,e)_{[\mu\nu]} =R(b)_{\mu\nu}= \partial_\mu b_\nu
-\partial_\nu b_\mu$.  This follows from the identity
$R(\omega)_{[ab,c]}{}^d = -R(b)_{[ab}\,\delta_{c]}{}^d$, which
reflects the fact that the spin connection $\omega_\mu{\!}^{ab}$ depends on
the dilatational gauge field $b_\mu$. As a result
the generalized Riemann tensor $R(\omega)_{\mu\nu}{\!}^{ab}$ is not
symmetric under pair-exchange,
\begin{equation}
    \label{eq:pair-exchange}
    R(\omega)_{ab,cd}-R(\omega)_{cd,ab}  =
    -2\,\eta_{[a[c}\,R(\omega,e)_{d]b]} +2\,\eta_{[c[a}\,R(\omega,e)_{b]d]}\,.    
\end{equation}
Finally, $\tilde R(\hat A)_{\mu\nu}$ denotes the dual of
$R(\hat A)_{\mu\nu}= \partial_\mu \hat A_\nu -\partial_\nu\hat A_\mu$.  

The $5D$ S-supersymmetry gauge field $\phi_M{}^i$ follows from the
fermionic conventional constraint given in
\eqref{eq:conv-constraints-5} and can be decomposed as follows under
the $4D$ reduction, 
\begin{align}
  \label{eq:S-susy-gauge fields}
  \phi_\mu{}^i \vert_{5D}  &- 
  \tfrac16\mathrm{i}\, \Slash{\hat{A}}\gamma_5\psi_\mu{}^i +
  \tfrac1{96}\hat{T}_{ab}\gamma^{ab}\psi_\mu{}^i
  -\tfrac1{12}\phi^{-1}\hat{F}_{ab}\gamma^{ab}\gamma_5\psi_\mu{}^i \nonumber\\
   &= \tfrac12\hat \phi_\mu{}^i
   +\tfrac13\phi^{-1}\gamma_5 D_\mu\hat{\psi}^i +
  \tfrac1{12}\phi^{-1}\gamma_\mu \gamma_5\Slash{D}\hat{\psi}^i
  - \tfrac23\phi^{-1}\hat{A}_\mu\hat{\psi}^i
  +\tfrac16\phi^{-1}\gamma_\mu\Slash{\hat{A}}\hat{\psi}^i \nonumber\\
  &\,\quad +\tfrac1{3}\mathrm{i}\phi^{-2}\gamma^\nu\hat{F}_{\mu\nu}\hat{\psi}^i
  - \tfrac1{24}\mathrm{i}
  \phi^{-2}\gamma_\mu\hat{F}_{ab}\gamma^{ab}\hat{\psi}^i -
  \tfrac1{96}\mathrm{i}\phi^{-1}
  \hat{T}_{ab}\gamma^{ab}\gamma_\mu\gamma_5\hat{\psi}^i \\
  &\,\quad-\tfrac23\phi^{-2}\gamma_5\big(\mathcal{D}_\mu\phi - \hat{A}_\mu\phi\gamma_5\big)\hat{\psi}^i -
  \tfrac16\phi^{-2}\gamma_\mu\gamma_5
  \big(\Slash{\mathcal{D}}\phi-\Slash{\hat{A}} \phi\gamma_5\big)\hat{\psi}^i \, . \nonumber
\end{align}
The right-hand side of this equation contains only supercovariant $4D$
expressions, with the exception of the field $\hat \phi_\mu{}^i$ which
is a gauge field. For instance $D_\mu\hat\psi^i$ is the $4D$ fully
supercovariant derivative given by (at linear order in the spinor
fields)
\begin{align}
  D_\mu\hat{\psi}^i =&\, \big(\mathcal{D}_\mu
  +\tfrac12\hat{A}_\mu\gamma^5\big)\hat{\psi}^i
  -\tfrac12\phi\,\gamma^5\hat\phi_\mu{}^i -\tfrac14 \big[\hat
  F(B)_{ab}-\tfrac18 \phi\, \hat T_{ab}\gamma_5 \big]
  \gamma^{ab}\psi_\mu{}^i  \nonumber \\
  &\, +\tfrac12\mathrm{i} \gamma^5 \gamma^\nu
  \big[\mathcal{D}_\nu\phi - \hat A_\nu \phi\,\gamma^5 \big]
  \psi_\mu{}^i - \tfrac12{\hat V}_j{}^i \,\psi_\mu{}^j\, ,
\end{align}
which also contains the S-supersymmetry gauge field
$\hat\phi_\mu{}^i$. The terms on the left-hand side of
\eqref{eq:S-susy-gauge fields} that depend explicitly on
$\psi_\mu{}^i$ seem to affect the covariance under
Q-supersymmetry. However, they are to be expected because, according
to \eqref{eq:D5-D4-decomp}, the $5D$ Q-supersymmetry differs from the
$4D$ one by a field dependent S-supersymmetry transformation
parametrized by $\tilde \eta^i$ given in \eqref{eq:Q-susy
  decom-par}. 

The correctness of this result can be verified by considering the Q 
and S transformations of the $4D$ $\mathrm{SU}(2)$ gauge
fields $\hat{V}_{\mu\,i}{}^j$. After taking into account the
Kaluza-Klein decomposition, one has to correct for the field-dependent
S-supersymmetry transformation indicated in \eqref{eq:D5-D4-decomp},
which precisely cancels against the terms in \eqref{eq:S-susy-gauge
  fields} that depend explicitly on $\psi_\mu{\!}^i$. Furthermore one
has to take into account the redefinitions in \eqref{eq:field-redef}
and the field-dependent $\mathrm{SU}(2)$ transformation in
\eqref{eq:D5-D4-decomp}. There combined effect will only lead to terms such
as
\begin{equation}
  \label{eq:3}
  \mathrm{i}(\delta \bar \psi_{\mu i} -2\,\mathcal{D}_\mu \epsilon_i)
  \,\gamma^5 \phi^{-1} \hat\psi{}^j\,,\quad
  -2\mathrm{i}\,\bar\epsilon_i \,[\gamma^5 D_\mu (\phi^{-1} \hat\psi^j)
  +\tfrac12 \hat \phi_\mu{}^j] \,, \quad
  \phi^{-1} \bar{\hat\psi}_i \,\delta(\phi^{-1} \hat\psi^j) \,,  
\end{equation}
where the derivative $D_\mu$ is supercovariant. Combining this with the
result of the Kaluza-Klein decomposition and with
\eqref{eq:S-susy-gauge fields}, one obtains
\begin{equation}
  \label{eq:delta-V-hat-su2}
      \delta\hat{V}_{\mu\,i}{}^j =
      2\,\mathrm{i}\,\bar{\epsilon}_i\hat{\phi}_\mu{}^j  
      - 2\, \bar{\epsilon}_i\gamma_\mu\hat{\chi}^j - 2\mathrm{i}\,\bar\eta_i
      \psi_\mu{}^j- \tfrac12\delta_i{}^j\bigl(2\mathrm{i} \,
      \bar{\epsilon}_k\hat{\phi}_\mu{}^k - 2\,
      \bar{\epsilon}_k\gamma_\mu\hat{\chi}^k - 2\mathrm{i}\,\bar\eta_k
      \psi_\mu{}^k \bigr) \, , 
\end{equation}
where $\hat{\chi}^i$ is a supercovariant spinor field equal to 
\be
  \label{eq:hat-chi}
  \hat{\chi}^i = 8\,\chi^i\big\vert_{5D} -
  \tfrac1{4}\mathrm{i}\phi^{-1}\gamma^5\Slash{D}\hat{\psi}^i 
  - \tfrac12\phi^{-2}\hat{V}_k{}^i\hat{\psi}^k + \tfrac1{8}\phi^{-2}\bigl[\hat{F}_{ab} 
  - \tfrac14\phi\hat{T}_{ab}\gamma^5\bigr]\gamma^{ab}\hat{\psi}^i -\tfrac12
  \mathrm{i}\phi^{-1}\Slash{\hat{A}}\hat{\psi}^i \, . 
\ee

Let us subsequently turn to the Q- and S-supersymmetry transformations
of the field $\hat\chi^i$, which contains the remaining independent
fermion field $\chi^i\vert_\mathrm{5D}$ of the $5D$ Weyl multiplet
according to the equation above. When writing its variation in terms
of the $4D$ quantities, we naturally obtain terms that depend
exclusively on the $4D$ Weyl multiplet components and others that will
involve both the Weyl multiplet and the Kaluza-Klein vector
multiplet. The latter terms should then cancel by the variations of
the additional terms in \eqref{eq:hat-chi}, because $\hat\chi^i$ must
vary exclusively into the components of the $4D$ Weyl multiplet. Here
one should again compensate for the composite S-supersymmetry
variation parametrized in terms of $\tilde\eta^i$. This leads to the
following expression,
\begin{align}
  \label{eq:QS-transfo-chi5}
  \delta\hat\chi^i =&\,8\,\delta\chi^i\big\vert_{5D} - \tfrac3{2}T_{AB}\,\gamma^{AB}
  \tilde\eta^i   -\tfrac1{4}\mathrm{i} \, \delta\big[\phi^{-1} \gamma^5
  \Slash{D} \hat\psi^i\big] \nonumber\\ 
  &\,-\tfrac18\,  \delta \big[ 
  4\, \phi^{-2} \hat V_j{}^i \,\hat\psi^j
  - \phi^{-2} [ \hat F(B)^{ab} -\tfrac1{4} \phi\,
  \hat T^{ab} \gamma^5]\gamma_{ab}   \,\hat\psi^i  +4\mathrm{i}\,
  \phi^{-1} \Slash{\hat A}\, \hat\psi^i \big] \,,
\end{align}
where we use the definition \eqref{eq:delta-hat-phi-mu} for the
supercovariant derivative of $\hat\psi^i$ based on the S-supersymmetry
gauge field $\hat\phi_\mu{\!}^i$. Eventually we will make another,
more suitable, choice for this composite gauge field, but for the
moment we adopt this definition.

Restricting ourselves to terms linearly proportional to fermion fields, the
variation of $\hat\chi^i$ takes the following form, 
\begin{align}
  \label{eq:delta-hat-chi}
  \delta\hat\chi^i=&\, \tfrac1{24}\hat T_{ab} \gamma^{ab} \eta^i  +
  \tfrac1{6}R(V)_{abj}{}^i \,\gamma^{ab} \epsilon^j + \tfrac1{24}
  \mathrm{i} \,\gamma^{ab} \, \Slash{\mathcal{D}}\hat T_{ab}\epsilon^i
  -\tfrac13 R(A)_{ab} \,\gamma^{ab}\gamma^5 \epsilon^i  + \hat D\, \epsilon^i \,,
\end{align}
where $\hat D$ is defined as (up to terms quadratic in spinor fields)
\begin{align}
  \label{eq:def-hat-D}
  \hat{D} =&\, 4\,D\big\vert_{5D} - \tfrac1{4} \phi^{-2} \,\hat
  V_j{}^k\, \hat V_k{}^j + \tfrac14 \phi^{-1}\,\bigl[(\mathcal{D}_a)^2
  + \tfrac16 R(\omega,e)\bigr] \phi
  - \tfrac1{12} (\hat A_a)^2  \nonumber \\
  &\, - \tfrac1{12}\phi^{-2}\hat{F}^{ab}\hat{F}_{ab} +\tfrac1{192}
  \phi^{-1}\epsilon_{abcd}\,\hat{T}^{ab}\hat{F}^{cd}
  +\tfrac1{384}\hat{T}^{ab}\,\hat{T}_{ab} \, ,
\end{align}
where we have made use of~\eqref{eq:def-hat-f}.  Note that all 
bosonic terms in \eqref{eq:delta-hat-chi} have been included. 

We conclude this part
of the analysis by giving the Q- and S-supersymmetry transformations
for the remaining fields, where we give also some further details about terms
quadratic in the spinor fields:
\begin{align}
 \label{eq:susy-weyl2}
  \delta b_\mu =&\, \tfrac12 \mathrm{i}\, \bar\epsilon_i\,\hat
  \phi_\mu{}^i -\tfrac12 \epsilon_i\gamma_\mu\hat\chi^i
  +\tfrac12 \mathrm{i}\,  \bar\eta_i \,\psi_\mu{\!}^i  \,, \nonumber\\
  \delta\hat{T}_{ab} =&\,
  -8\mathrm{i}\,\bar{\epsilon}_i\,R(Q)_{ab}{}^i
  + 4\mathrm{i}\,\bar{\epsilon}_i\gamma_{ab}\hat{\chi}^i +\tfrac12
  \varepsilon_{abcd}\, \hat T^{cd} \, \tilde\Lambda^0 \,,\nonumber \\
  \delta\hat{A}_\mu =&\;
  \bar{\epsilon}_i\gamma_\mu\gamma^5\hat{\chi}^i
  -\tfrac12\mathrm{i}\,\bar{\epsilon}_i\gamma^5\hat{\phi}_\mu{}^i -
  \tfrac12\mathrm{i} \,\bar\eta_i \gamma^5 \psi_\mu{}^i + \partial_\mu
  \tilde \Lambda^0   \,, \\
  \delta \hat V_j{}^i=&\, 2\,\bar\epsilon_j (\Slash{D}\hat{\psi}^i
  -\mathrm{i} \gamma^5 \phi\,\hat \chi^i) 
  - \delta_j{}^i\,  \bar\epsilon_k (\Slash{D}\hat{\psi}^k  -\mathrm{i}
  \gamma^5 \phi\,\hat \chi^k)  \,,  \nonumber\\
  \delta\hat{D} =&\, \bar{\epsilon}_i \,\Slash{D}\hat{\chi}^i +
  \ldots\,. \nonumber
\end{align}
In the derivation of the first result for $\delta b_\mu$ we note that the same
phenomenon takes place as when deriving the transformation rules for
$\hat V_{\mu\,i}{}^j$ in \eqref{eq:delta-V-hat-su2}. Namely, the
S-supersymmetry transformation with field-dependent parameter $\tilde\eta^i$ in
\eqref{eq:D5-D4-decomp} cancels against the terms in
\eqref{eq:S-susy-gauge fields} that depend explicitly on
$\psi_\mu{\!}^i$. After that we use the definition of $\hat\chi^i$ in
\eqref{eq:hat-chi}, and the remaining terms are absorbed into the
$4D$ conformal boost transformation. Since $b_\mu$ is the only field
that transforms under conformal boosts, this will only affect the explicit
form of the supersymmetry algebra. The transformation rules of
$\hat T_{ab}$, $\hat A_\mu$ and $\hat V_j{}^i$ do not involve further
subtleties, except that $\hat A_\mu$ does not seem to transform
supercovariantly. The transformation rule of $\hat D$, however, cannot
be realiably calculated at this stage, because we have not yet determined
the contributions quadratic in the spinor fields in its definition
\eqref{eq:def-hat-D}. In view of
the fact that the original $5D$ theory as well as its reduced~$4D$ version are consistent, there is no doubt that the present calculation can be completed to all orders.

We have thus shown in sufficient detail how the $5D$ Weyl multiplet
reduces to the $4D$ Euclidean Weyl multipet and a Kaluza-Klein vector
supermultiplet. However, the latter multiplet involves only seven
bosonic and eight fermionic degrees of freedom, so that one bosonic
field seems to be missing in the Kaluza-Klein vector multiplet. A
similar counting for the Weyl multiplet reveals that the Weyl
multiplet has twenty-five bosonic and twenty-four fermionic degrees of
freedom (in this off-shell counting one always corrects for the number of gauge
invariances, so that for instance each gravitino corresponds to only
eight fermionic degrees of freedom).

The reason for the mismatch is well known; under dimensional reduction
one obtains the lower-dimensional theory in a partially gauge-fixed
form. The R-symmetry is extended to
$\mathrm{SU}(2)\times\mathrm{SO}(1,1)$, where the non-compact
$\mathrm{SO}(1,1)$ factor acts by a chiral transformations on the
fermions (it will also act on some of the bosonic fields). At this
point the $\mathrm{SO}(1,1)$ group is, however, not realized as a
local invariance. Although the vector field $\hat A_\mu$ seems to play
the role of an $\mathrm{SO}(1,1)$ gauge field, it is not transforming
under a corresponding gauge symmetry and represents four bosonic
dergrees of freedom. This is the underlying reason why the combined
Weyl and Kaluza-Klein supermultiplets are not yet fully irreducible.

Full irreducibility can be obtained by introducing a compensating scalar
field $\varphi$ and writing
\begin{equation}
  \label{eq:def-A-gauge}
  \hat A_\mu= A_\mu - \partial_\mu \varphi \,,
\end{equation}
where $A_\mu$ and $\varphi$ transform under {\it local}~$\mathrm{SO}(1,1)$ gauge
transformations as 
\begin{equation}
  \label{eq:so-11-gauge transf}
  A_\mu \to A_\mu + \partial_\mu\Lambda^0 \,, \qquad
  \varphi \to \varphi + \Lambda^0\,,
\end{equation}
so that $\hat A_\mu$ remains invariant. Under supersymmetry we assume
that $\varphi$ changes according to 
\begin{equation}
  \label{eq:transf-varphi}
  \delta\varphi = - \tilde\Lambda^0 = -\mathrm{i}
  \phi^{-1}\bar\epsilon_i\,\hat\psi^i\,. 
\end{equation}
Subsequently one uniformly redefines all fields and parameters with a
$\varphi$-dependent $\mathrm{SO}(1,1)$ transformation, which will
remove all explicit terms in the transformation rules proportional to
$\tilde\Lambda^0$. When re-imposing the gauge condition $\varphi=0$,
then all the $\tilde\Lambda^0$-terms will re-emerge in the form of
compensating gauge transformations.

We now summarize all the $\varphi$-dependent field redefinitions. The
R-covariant spinors, transforming under local
$\mathrm{SU}(2)\times\mathrm{SO}(1,1)$ R-symmetry transformations, are
as follows, 
\begin{equation}
  \label{eq:compensating-chiral-tr}
  \begin{array}{rcl}
  \epsilon^i\vert^\mathrm{Rcov} &\!\!=\!\!& 
  \exp[-\tfrac12\varphi\,\gamma^5]\, 
  \epsilon^i\,,\\[2mm] 
  \eta^i\vert^\mathrm{Rcov} &\!\!=\!\!&
  \exp[\tfrac12\varphi\,\gamma^5]\,\eta^i \,,\\[2mm] 
  \chi^i\vert^\mathrm{Rcov} &\!\!=\!\!&
  \exp[-\tfrac12\varphi\,\gamma^5]\, \hat\chi^i \,,\\
   \end{array}
  \qquad
  \begin{array}{rcl}
  \psi_\mu{}^i\vert^\mathrm{Rcov} &\!\!=\!\!&
  \exp[-\tfrac12\varphi\,\gamma^5]\,\psi_\mu{}^i\,,\\[2mm] 
  \phi_\mu{\!}^i\vert^\mathrm{Rcov} &\!\!=\!\!&
  \exp[\tfrac12\varphi\,\gamma^5]\, \hat\phi_\mu^{\,i} \,,\\[2mm]
  \psi^i\vert^\mathrm{Rcov} &\!\!=\!\!&
  \exp[-\tfrac12\varphi\,\gamma^5]\, \hat\psi^i \,,\\
  \end{array}
\end{equation}
Also some of the bosons will have to be redefined so that they
transform covariantly under $\mathrm{SO}(1,1)$. First of all the
tensor fields $\hat T_{ab}$, when decomposed into the self-dual and
anti-selfdual (real) components, take the form 
\begin{equation}
  \label{eq:cov-T}
  T_{ab}{\!}^{\pm\mathrm{Rcov}} = \exp[\pm\varphi]\, \hat
    T_{ab}{\!}^\pm\,.  
\end{equation}
Furthermore the scalars $\phi$ and $\varphi$ are combined into
\begin{equation}
  \label{eq:cov-phi-varphi}
  \phi^\mathrm{Rcov} = \exp[-\varphi]\phi \, , \qquad \bar\phi^\mathrm{Rcov} = \exp[\varphi]\phi \, .
\end{equation}
After these redefinitions the Weyl multiplet is now irreducible. It includes the $\mathrm{SO}(1,1)$ gauge field~$A_\mu$ and comprises 24+24 off-shell degrees of freedom. The compensator $\varphi$ belongs to the
Kaluza-Klein vector multiplet, defined in a background made up of the Weyl multiplet and comprising 8+8 degrees of freedom.

At this stage we will make some further field redefinitions to bring the
results in closer contact with the Minkowski version of $\CN=2$
conformal supergravity. First of all we will redefine the S-supersymmetry gauge field
according to 
\begin{equation}
    \label{eq:S-convetional constraint}
  \phi_\mu{}^i = \phi_\mu{}^i\vert_{\mathrm{old}} - \tfrac12\mathrm{i}\,\gamma_\mu\hat{\chi}^i \,.
\end{equation}
This will correspond to a different conventional constraint (the
previous one was given by
\eqref{eq:intermediate-4d-connection}) which is S-supersymmetric. At the same time, we make use of the R-covariant fields defined above. As a result, the transformation rules of the various fields will acquire a simpler form. For instance, because of the redefinition~\eqref{eq:S-convetional constraint}, the explicit expressions for the dependent gauge fields~$\phi_\mu{}^i$ and~$f_\mu{}^a$ become
\begin{align}
&\phi_\mu{}^i = -\tfrac12\mathrm{i}\left( \gamma^{\rho \sigma} \gamma_\mu - \tfrac{1}{3} \gamma_\mu \gamma^{\rho \sigma} \right) \left(\mathcal{D}_\rho \psi_{\sigma}{}^i + \tfrac{1}{32}\mathrm{i}\,(T_{ab}^+ + T_{ab}^-) \gamma^{ab} \gamma_\rho\psi_{\sigma}{}^i + \tfrac{1}{4} \gamma_{\rho \sigma} \chi^i \right) \, , \cr
&f_\mu{}^a = \tfrac12\,R(\omega,e)_\mu{}^a - \tfrac14\,\bigl(D+\tfrac13 R(\omega,e)\bigr) e_\mu{}^a - \tfrac12\,\wt{R}(A)_\mu{}^a - \tfrac1{32}\,T_{\mu b}^-\,T^{+\,ba} \, .
\end{align}
where we restrict ourselves to bosonic terms in the last expression, and we have dropped the caret on the field~$D$ (which we will consistently do from now on). Furthermore, the derivative~$\mathcal{D}_\mu$ is now covariant with respect to local Lorentz, dilatations and the~\emph{full} R-symmetry~$\mathrm{SU}(2)\times\mathrm{SO}(1,1)$ transformations. All the field strengths are also supercovariant. The presence of the fully supersymmetric covariant quantities has not been verified in every possible detail, but these covariantizations are implied by the supersymmetry algebra.
 
Another change concerns the spinors. In view of the fact that we are now dealing with a Euclidean theory, we prefer to change the definition of the Dirac conjugate fields accordingly, so that~$\bar{\chi}_i := (\chi^i)^\dagger$.This requires us to replace all the barred spinors~$\bar{\chi}_i$ in the previous equations by~$\bar{\chi}_i\gamma^5$. The corresponding symplectic Majorana condition on the spinors can still be written in the same form as the five-dimensional, Minkowski one~\eqref{eq:sympl-Majo-5D}:
\be
\label{eq:sympl-Majo-4D}
\wt{C}^{-1}\bar{\chi}_i{}^\mathrm{T} = \varepsilon_{ij}\chi^j \, ,
\ee
albeit with a new charge conjugation matrix
\be
\wt{C} = C\,\gamma^5 \, .
\ee
Therefore, the Hermitian gamma matrices~$\gamma_a$ now satisfy\footnote{This should be contrasted with the properties given in footnote~\ref{foot:charge-conj-5D}.}
\be
\wt{C}\gamma_a\wt{C}^{-1} = -\gamma_a{}^\mathrm{T} \qquad (a = 1\ldots4) \, .
\ee
We still use the convention according to which raising or lowering~$\mathrm{SU}(2)$ indices is effected by complex conjugation. For four-dimensional fermionic bilinears, with spinor fields~$\psi^i$ and~$\phi^i$ and a spinor matrix~$\Gamma$ built out of products of gamma matrices, we note the following result:
\be
\label{eq:bilinear-4D}
(\bar\varphi_j\Gamma^{\dagger}\psi^i)^\dagger = \bar\psi_i\,\Gamma\,\varphi^j= - \delta_i{}^j \,\bar\varphi_k\,\wt{C}^{-1}\, \Gamma^{\rm T}\,\wt{C}\, \psi^k   + \bar\varphi_i\,\wt{C}^{-1}\, \Gamma^{\rm T}\,\wt{C}\, \psi^j \, .
\ee 

With these new conventions, and the further field redefinition
\begin{equation}
\label{eq:redefs}
\mathcal{V}_\mu{}^i{}_j := \hat{V}_{\mu\,j}{}^i \, ,
\end{equation}
the transformation rules for the independent Weyl multiplet fields ($e_\mu{}^a,\,\psi_\mu^i,\,b_\mu$, $\,A_\mu,\,\mathcal{V}_\mu{}^i{}_j,\,T_{ab}^\pm,\,\chi^i,\,D)$ displayed in~\eqref{eq:susy-weyl1}, \eqref{eq:delta-V-hat-su2}, \eqref{eq:delta-hat-chi} and \eqref{eq:susy-weyl2} take the form given in \eqref{eq:weyl-multiplet}. We refrain from displaying the transformation rules of the Kaluza-Klein vector multiplet, since we display the dimensional reduction of generic matter vector multiplets in Section~\ref{sec:shell-dimens-reduct-matter}.

\end{appendix}


\chapter*{English summary}
\addcontentsline{toc}{chapter}{English summary}
\pagestyle{fancy}
\fancyhf{}
\rhead{English summary}
\lhead{\thepage}

Physicists strive to describe our entire reality as a single, unified theory. To this day, two main pillars of physics have been identified: Quantum Theory and General Relativity. The former describes the interaction of fundamental particles and applies to microscopic scales. The latter is Einstein's theory of gravity, and describes the interactions of large-scale objects, such as planets, stars and galaxies. It has been a long-standing problem to try and unify these two foundations of our reality into a single unified description. One possible avenue into this reconciliation is to examine black holes in detail. Black holes are predictions of General Relativity, which describes what happens when a large mass (at least a few solar masses) is concentrated into a tiny area, for instance once a sufficiently massive star has burnt out all of its fuel for fusion and collapses unto itself. This results in a singularity surrounded by an imaginary spherical surface, the so-called event horizon. Any observers or light particles falling inside the black hole past this horizon is unable to escape due to the extreme gravitational forces at play. What makes black holes fascinating is that, within relatively small distances around the horizon, quantum phenomena are believed to be relevant, which means that one has to deal with both gravitational and quantum effects to arrive at a correct description. Their detailed examination can thus teach us more about the unified theory of quantum gravity. Such examination is undertaken in the present theoretical work, where we focus on the so-called ``quantum entropy'' of certain specific black holes. This research involves the use of new mathematical techniques which have recently become available and allow for highly detailed predictions.

Entropy is a quantity known from classical physics in the context of thermodynamics. There, one deals with physical systems containing a large number of constituents and their associated degrees of freedom, such as a gas of atoms or molecules in a box. When studying the behavior of such a gas under changes in the total energy, temperature, volume, pressure, or density (proportional to the number of molecules or atoms), $19^\mathrm{th}$ century physicists were able to derive certain 
\pagestyle{fancy}
\fancyhf{}
\lhead{English summary}
\rhead{\thepage}
relations between the various quantities used to describe the system. One such quantity is the entropy function, which depends on the extensive quantities of the gas: its energy, its volume, and the number of its constituents. According to the so-called second law of thermodynamics, the classical evolution of the system always takes place in the direction of increasing or constant entropy.

At the tail-end of the $19^\mathrm{th}$ century, Ludwig Boltzmann understood that a deep connection exists between the aforementioned macroscopic (thermodynamic) description of a classical system, and its microscopic description. He realized that the thermodynamic properties of a gas could be obtained from the microscopic behavior of its atomic constituents. The latter can be described using methods of statistical mechanics, and upon averaging over the behavior of a large number of atoms or molecules, it is possible to recover the thermodynamic properties of the gas. As such, entropy can be explained in a Boltzmann interpretation as being the logarithm of the number of degrees of freedom accessible to the atoms or molecules making up the gas. 

It was recognized in the 1970s that a similar situation arises for black holes in general relativity. Bardeen, Carter and Hawking showed that black hole evolution is governed by a set of laws which they dubbed the ``laws of black hole mechanics''. One such law states that the surface area of the horizon of a black hole never decreases when undergoing a physical process. For instance, when two black holes collide, they will merge into a single black hole whose surface area is necessarily greater or equal to the sum of the surface areas of the horizons of the initial black holes. Furthermore, the surface area of the horizon behaves, under changes of the other parameters entering the description of the black hole (its mass, electric-magnetic charges and angular momentum), in a way akin to the behavior of the thermodynamic entropy under changes of energy, volume or density. On this basis, Bekenstein and Hawking proposed to formally identify the thermodynamic entropy of a black hole with the surface area of its horizon. Associated to this thermodynamic entropy, it is also possible to formally define a notion of temperature for the black holes. This might \textit{a priori} seem in contradiction with the classical statement that nothing can escape the horizon of a black hole. To understand how this is possible, one needs to adopt a semi-classical picture, where the black hole itself is still classical and described within the framework of general relativity but is interacting with an intrinsically quantum field outside its horizon. This field undergoes quantum vacuum fluctuations arbitrarily close to the horizon of the black hole, which leads to creations of pairs of particles and anti-particles. A member of one such pair can then fall inside the horizon of the black hole, while 
\pagestyle{fancy}
\fancyhf{}
\rhead{English summary}
\lhead{\thepage}
the other member escapes away from the black hole. The net result of such a process is therefore the emission of a so-called ``Hawking radiation''. The spectrum of this radiation is almost exactly thermal, with a given temperature and entropy.

The discovery of black hole entropy then opens the way for a natural and ultimately deep question: is there a Boltzmann interpretation of their thermodynamic entropy? In other words, we should ask what are the corresponding microscopic degrees of freedom, or ``gravitational atoms'', making up the black hole. This is where a tentative description of quantum gravity comes into play, since these microscopic constituents should be sensitive to both gravitational and quantum-mechanical effects. Are there existing theories of quantum gravity which could provide such a description? A straightforward way of obtaining such a theory would be to try and directly ``quantize'' Einstein's theory of general relativity. Unfortunately, this procedure is riddled with technical complications and does not provide sensible predictions. There is, however, an extension of general relativity, which combines Einstein's theory with supersymmetry. This symmetry relates bosons and fermions (particles with different quantum statistics), and imposes additional constraints on the theory which imply a better conceptual and computational control of its behavior at extremely short distances. The combination of general relativity and supersymmetry is known as supergravity. Even though, as of yet, supersymmetry has not been confirmed experimentally as being a fundamental symmetry of Nature, supergravity theories should be thought of as a convenient theoretical framework, allowing us to start gathering clues regarding the behavior of quantum gravity, albeit in a slightly idealized context.

Another foray into the quantum gravity regime is provided by string theory. String theory differs from general relativity or supergravity in the sense that the fundamental objects in the theory are not fields defined at every point in space-time and describing point-particles, but extended objects: tiny (typically of a size close to the Planck length, $10^{-33}$ cm) vibrating strings, whose excitation spectrum generates what we observe in our macroscopic world as particle manifestations. Some of these particles correspond to the elementary particles which have been observed experimentally, while other are as of yet inaccessible to our current detection methods. At large distances, string theory effectively reduces to a supergravity theory. On the other hand, it also contains its own extended objects, called branes, and it is possible to provide a microscopic description of black holes in terms of these branes. An invaluable insight provided by Strominger and Vafa in 1996 showed that it is indeed possible to give a microscopic description of the entropy of black 
\pagestyle{fancy}
\fancyhf{}
\lhead{English summary}
\rhead{\thepage}
holes in the context of string theory by describing interacting branes and examining their available degrees of freedom. Upon averaging over a large number of branes, they were able to recover, in a certain limit, the thermodynamic entropy of Bekenstein and Hawking for the black hole. This was the first encouraging hint that a Boltzmann interpretation of the Bekenstein-Hawking entropy could be achieved using string theory methods.

Since this discovery, the relationship between a macroscopic and microscopic description of black holes has been investigated in various ways, and in increasing level of details. Both the predictions from string theory and supergravity have been clarified and generalized. In this respect, a more general definition of the entropy of a black hole is often used, so that the special limit which Strominger and Vafa used to recover the Bekenstein-Hawking entropy is no longer needed. Hence, one often considers the Bekenstein-Hawking-Wald entropy when investigating quantum mechanical corrections to the original result of Bekenstein and Hawking. These corrections may be best incorporated by making use of the so-called ``quantum entropy function'', which was defined by Sen in 2008, in the context of the AdS/CFT correspondence. This definition makes use of a path-integral which is an integral over the infinite-dimensional space of all possible field configurations in the supergravity theory. At first glance, the exact computation of such a quantity might seem rather hopeless. We show, however, that there exist mathematical techniques which make the exact calculation possible in certain situations. These are known as localization techniques, and they reduce the path-integral to a standard, well-defined integral which can be evaluated using traditional methods. As shown in this thesis, it is possible to use localization techniques to compute the quantum entropy function of certain black holes exactly in supergravity, at all orders in perturbation theory (and also including some non-perturbative effects). The result can then be compared to the microscopic predictions of string theory made on the basis of the brane description of the same black holes. They are found to be in agreement, which indicates that there are indeed two different ways of determining the entropy of certain black holes, in accordance with the interpretation of Boltzmann.

The present work begins with laying a solid foundation for the evaluation of the quantum entropy function by carefully defining the four-dimensional supergravity theory under consideration. Within this theory, a first examination of highly supersymmetric black holes and their quantum entropy is conducted. Subsequently, the main ingredients of the localization method are derived in some generality, before being applied to specific black holes possessing less supersymmetry. In each 
\pagestyle{fancy}
\fancyhf{}
\rhead{English summary}
\lhead{\thepage}
case investigated, an agreement with known string theory predictions is found. In the less supersymmetric cases, such an agreement is in fact quite non-trivial due to the presence of so-called ``multi-center'' solutions in the spectrum of both the supergravity and string theory. This eventually relates in an interesting fashion to the mathematical theory of so-called modular, mock-modular and Jacobi forms.

\pagestyle{fancy}
\fancyhf{}
\lhead{English summary}
\rhead{\thepage}

\chapter*{Nederlandse samenvatting}
\addcontentsline{toc}{chapter}{Nederlandse samenvatting}
\pagestyle{fancy}
\fancyhf{}
\rhead{Nederlandse samenvatting}
\lhead{\thepage}

Natuurkundigen streven ernaar om verschillende materiële verschijnselen te beschrijven vanuit een allesomvattende theorie. De belangrijkste theorieën in dit verband zijn de quantum-mechanica en de algemene relativiteitstheorie. De eerste beschrijft de werkelijkheid op microscopische afstandschalen. De tweede is Einsteins theorie van de zwaartekracht, die van toepassing is voor grote massa’s zoals sterren en melkwegstelsels. Tot dusver bestaat er geen experimenteel getoetste theorie die de uitgangspunten van beide theorieën in zich verenigt. De problemen kunnen nader onderzocht worden in de context van zwarte gaten, die lang geleden werden voorspeld door de algemene relativiteitstheorie en die inmiddels in ons heelal worden waargenomen. Zwarte gaten ontstaan als een grote massa (groter dan de zon) in een klein volume wordt geconcentreerd, zoals bijvoorbeeld gebeurt als een zware ster is opgebrand en implodeert. Dit resulteert in een singulariteit omgeven door een denkbeeldig boloppervlak, de zogenaamde ``horizon''. Materie en lichtsignalen die de ster benaderen tot binnen die horizon zijn niet meer in staat om te ontsnappen ten gevolge van de extreem sterke zwaartekracht. Op relatief kleine afstanden rond de horizon worden quantum-mechanische verschijnselen relevant en dat maakt dat we gelijktijdig te maken hebben met de effecten van zowel de zwaartekracht als de quantum-mechanica. Het theoretisch onderzoek aan zwarte gaten kan daarom leiden tot nieuwe inzichten over de wederzijdse relatie tussen de twee theorieën en op termijn tot een consistente theorie voor quantum-gravitatie. In dit proefschrift wordt de zogenaamde ``quantum-entropie'' onderzocht voor zeer specifieke zwarte gaten. Hierbij wordt gebruik gemaakt van nieuwe wiskundige technieken die recent beschikbaar zijn gekomen en die zeer gedetailleerde voorspellingen mogelijk maken. 
 
Entropie is een begrip dat bekend is van de klassieke natuurkunde in de context van de thermodynamica. In de thermodynamica onderzoekt men systemen bestaande uit een zeer groot aantal bestanddelen, zoals bijvoorbeeld een gas van moleculen of atomen.  Door het gedrag van het gas te bestuderen onder veranderingen van de totale energie, de temperatuur, het volume, de druk, of de dichtheid van het gas (evenredig met het aantal moleculen) bepaalde men in de $19^\mathrm{de}$ eeuw relaties tussen de verschillende grootheden die gebruikt kunnen worden om het systeem te beschrijven. Een daarvan was de zogenaamde entropie-functie die afhangt van de extensieve grootheden  van het gas, zoals 
\pagestyle{fancy}
\fancyhf{}
\lhead{Nederlandse samenvatting}
\rhead{\thepage}
de totale energie, het volume, en het aantal moleculen.  Volgens de zogenaamde tweede hoofdwet van de thermodynamica kan een systeem alleen maar zodanig veranderen dat de entropie toeneemt of eventueel  constant blijft. 

Op het eind van de $19^\mathrm{de}$ eeuw begreep Ludwig Boltzmann dat er een diep verband bestond tussen de macroscopische (thermodynamische) beschrijving van het klassieke systeem en de microscopische beschrijving  in termen van moleculen of atomen. Hij realiseerde zich dat de thermodynamische eigenschappen van een gas bepaald kunnen worden uitgaand van het microscopisch gedrag van de bestanddelen. Dat laatste kan worden beschreven met de statistische mechanica door te middelen over het gedrag van een zeer groot aantal moleculen was het mogelijk om de thermodynamische eigenschappen van het gas te reproduceren.  In Boltzmanns interpretatie kon worden bewezen dat de entropie evenredig is met de logaritme van het aantal vrijheidsgraden dat beschikbaar is voor de moleculen van het gas. 

Rond 1970 realiseerde men zich dat er een soortgelijke situatie bestond voor zwarte gaten in de algemene relativiteitstheorie. Bardeen, Carter en Hawking toonden aan dat zwarte gaten voldoen aan de zogenaamde ``wetten van de mechanica van zwarte gaten''. Een van die wetten geeft aan dat de grootte van het oppervlak van de horizon van een zwart gat nooit afneemt als gevolg van een fysisch proces. Bijvoorbeeld, twee zwarte gaten die botsen kunnen een nieuw zwart gat vormen waarvan het horizon-oppervlak gelijk is aan of groter is dan de som van de horizon-oppervlakken van de oorspronkelijke zwarte gaten.  Voorts verandert de grootte van het horizon-oppervlak van een zwart gat door veranderingen van andere grootheden die het zwarte gat bepalen, en wel op een soortgelijke manier als waarop de thermodynamische entropie veranderd volgens de tweede hoofdwet van de thermodynamica. Vandaar dat Bekenstein en Hawking voorstelden om aan het zwart gat een thermodynamische entropie toe te kennen gelijk aan de grootte van het horizon-oppervlak. Volgens deze analogie is het ook mogelijk een temperatuur toe te kennen aan het zwarte gat. Dit lijkt \textit{a priori} in strijd met het feit dat een zwart gat geen straling kan uitzenden zoals alle lichamen met een eindige temperatuur doen. Om te begrijpen hoe dit mogelijk is moeten we gebruikmaken van een semi-klassieke benadering waarin het zwarte gat wordt voorgesteld als een klassiek zwart gat in interactie met een quantum-veld in de buurt van de horizon. 
\pagestyle{fancy}
\fancyhf{}
\rhead{Nederlandse samenvatting}
\lhead{\thepage}
Zo’n veld kan als gevolg van quantum fluctuaties een deeltje en een anti-deeltje produceren. Een daarvan kan in het zwarte gat verdwijnen en de andere kan dan ontsnappen als zogenaamde Hawking-straling. Het spectrum van die straling is thermisch en gekarakteriseerd door een bepaalde temperatuur en entropie. 

De ontdekking van entropie voor zwarte gaten leidt tot een voor de hand liggende vraag, namelijk of er ook een mogelijke interpretatie van deze entropie bestaat  analoog aan die van Boltzmann voor de thermodynamica van gassen. Met andere woorden, bestaan er ook elementaire microscopische ``bestanddelen'' hier die een statische verklaring kunnen geven van het bestaan van de entropie van zwarte gaten. Deze microtoestanden zouden onderhevig moeten zijn aan zowel de quantum-mechanica en de zwaartekracht. Bestaat er een theorie van quantum-gravitatie die dit zou kunnen verklaren?  Een antwoord op deze vraag zou kunnen worden gegeven door bijvoorbeeld Einsteins gravitatie-theorie te ``quantiseren'', maar helaas heeft deze theorie teveel technische complicaties en het is ook niet duidelijk hoe hier de gewenste microtoestanden geïdentificeerd kunnen worden. Er bestaat een uitbreiding van de relativiteitstheorie die Einsteins theorie combineert met supersymmetrie. Deze symmetrie relateert fermionen en bosonen (deeltjes met een verschillende quantum-statistiek)  hetgeen extra restricties impliceert voor de theorie die aanleiding geven tot een beter gedrag op extreem korte afstanden. Supersymmetrie is niet experimenteel aangetoond in de natuur, maar supergravitatie is desalniettemin een geschikt theoretische model om een verklaring te zoeken voor het bestaan van entropie van zwarte gaten.
  
Een ander idee is gebruik te maken van de snaartheorie. Snaartheorie verschilt van de algemene relativiteitstheorie of van supergravitatie in die zin dat de fundamentele objecten in de theorie geen velden zijn, gedefinieerd op elk punt in de ruimte-tijd, en geen puntdeeltjes beschrijven, maar kleine (de grootte orde is ongeveer Planck lengte, $10^{-33}$ cm.) trillende snaren, waarvan de eigentrillingen corresponderen met deeltjes. Sommige van die deeltjes corresponderen met de elementaire deeltjes die we experimenteel waarnemen, maar anderen zijn vooralsnog niet waarneembaar met de huidige detectiemethoden. Op grote afstanden neemt de snaartheorie de vorm aan van supergravitatie. Maar de snaartheorie kent ook andere uitgebreide objecten, de zogenaamde branen, en het is mogelijk om een microscopische beschrijving van zwarte gaten te geven in termen van deze branen. Een belangrijke aanwijzing werd in 1996 gegeven door Strominger en Vafa, die erin slaagden een microscopische beschrijving van de entropie van zwarte gaten te geven door  deze interagerende branen te beschrijven en hun beschikbare vrijheidsgraden te onderzoeken. Het resultaat hiervan werd vergeleken met de entropie 
\pagestyle{fancy}
\fancyhf{}
\lhead{Nederlandse samenvatting}
\rhead{\thepage}
van zwarte gaten die voorkomen in supergravitatie, gebruikmakend van de entropie zoals gedefinieerd door Bekenstein en Hawking. In een bepaalde limiet werd bewezen dat identieke resultaten konden worden verkregen, hetgeen suggereerde dat Boltzmanns interpretatie ook van toepassing kon zijn op de Bekenstein-Hawking entropie binnen het kader van de snaartheorie. 

Inmiddels is deze relatie op verschillende manieren en in meer detail onderzocht. Zowel de voorspellingen gebaseerd op snaartheorie als die gebaseerd op supergravitatie theorieën zijn gepreciseerd en gegeneraliseerd. In dit verband wordt vaak een wat algemenere definitie gebruikt voor de entropie in het kader van de algemene relativiteitstheorie, die soms wordt aangeduid als de Bekenstein-Hawking-Wald-entropie. In dat geval kunnen ook al quantum-correcties worden toegevoegd en is de limiet van Strominger en Vafa niet langer nodig om vergelijkbare resultaten te verkrijgen. De quantum-mechanische correcties kunnen echter nog vollediger  geïncorporeerd worden door gebruik te maken  van de zogenaamde ``quantum entropie'', die werd gedefinieerd door Sen in 2008 op basis van de AdS-CFT correspondentie. Deze beschrijving, die wordt gebruikt in dit proefschrift,  leidt tot een ``pad-integraal'': een integraal over de oneindig-dimensionale ruimte van alle mogelijke veldconfiguraties in de supergravitatie theorie. Op het eerste gezicht is de exacte berekening van zo’n integraal onmogelijk, maar er bestaan wiskundige technieken die een exacte berekening mogelijk maken onder bepaalde omstandigheden. Met dergelijke ``lokalisatie'' technieken reduceert het antwoord tot een standaard integraal over een eindig aantal variabelen, die vervolgens kan worden uitgerekend. Zoals aangetoond in dit  proefschrift kunnen we met behulp van lokalisatie de quantum-entropie van bepaalde zwarte gaten exact berekenen in supergravitatie. Het resultaat kan vervolgens worden vergeleken met de microscopische voorspellingen van snaartheorie op basis van de beschrijving in termen van branen. Het feit dat de resultaten onderling in overeenstemming zijn, geeft aan dat er inderdaad twee verschillende manieren zijn om de entropie van zwarte gaten te bepalen in overeenstemming met de interpretatie van Boltzmann.

Dit proefschrift begint met een uitgebreide onderbouwing van de vier-dimensionale supergravitatie theorieën die nodig zijn voor de bepaling van de quantum-entropie-functie. Binnen deze theorieën worden supersymmetrische zwarte gaten bestudeerd en vervolgens worden de belangrijkste ingrediënten van de lokalisatie-methode besproken alvorens deze toe te passen op specifieke oplossing van zwarte gaten. In alle gevallen die onderzocht worden is er overeenstemming met resultaten die zijn 
\pagestyle{fancy}
\fancyhf{}
\rhead{Nederlandse samenvatting}
\lhead{\thepage}
verkregen binnen het kader van de snaartheorie.  In situaties met minder supersymmetrie is een dergelijke overeenstemming minder vanzelfsprekend omdat oplossingen met meerdere zwarte gaten kunnen bijdragen aan de entropie. Deze bijdragen spelen een rol in zowel de supergravitatie als in de snaar-theoretische beschrijving. Dit leidt uiteindelijk tot een interessante relatie met de wiskundige theorie van zogenaamde mock-modulaire en Jacobi vormen.

\pagestyle{fancy}
\fancyhf{}
\lhead{Nederlandse samenvatting}
\rhead{\thepage}


\chapter*{Acknowledgments}
\addcontentsline{toc}{chapter}{Acknowledgments}
\pagestyle{fancy}
\fancyhf{}
\rhead{Acknowledgments}
\lhead{\thepage}
It is my privilege and pleasure to acknowledge my supervisor and co-supervisor, Bernard de Wit and Sameer Murthy, whose invaluable guidance and understanding throughout the four years we've worked together have made this work not only possible, but also very much enjoyable. When I set out on this path, my goal was to broaden my horizons and further my understanding of physics, if not of the world in general, and it rapidly became clear to me that they were the perfectly-suited mentors who could help me achieve this. I hold both of them in the highest regard, and will always be grateful to have had the opportunity to work with such great scientists and human beings. I can only hope to be worthy of the knowledge they have imparted upon me, and will strive to adopt the same positive attitude towards our field that they have displayed so consistently.\\
My thanks go to my former advisor Laurent Baulieu, for introducing me to supergravity theories and various interesting aspects of field theory in general.\\
I would like to take this opportunity to thank Ashoke Sen and Gabriel Cardoso for invaluable comments and excellent advice regarding my work in general, and the writing of this manuscript in particular.\\
I am also grateful to have been working in close contact with a wonderful group of people at Nikhef and elsewhere during my Ph.D. To my colleague, peer, but above all my friend, Ivano, I wish to communicate my sincerest thanks for having made every day working together an amazing and thoroughly fun experience, as well as for the (sometimes fatherly) advice and perspective he offered on all aspects of life. For the sake of brevity (and quite possibly decency), these advice will not be repeated here, but I'm sure you know how much they mean to me.\\
To my friend and roommate Franz, heartfelt thanks for sharing the daily life of a Ph.D. student with me, especially during the unavoidable low points. I have gained a lot from listening to your outlook and perspective, and will be sure to always remember our conversations, scientific or otherwise. Moreover, it was extremely enjoyable to be able to share the music I hold dear with someone so eager and enthusiastic in this respect.\\
To my close colleagues Adolfo, Dan and Gianluca, thank you for your helpful and invaluable guidance, for all the interesting exchanges we have had, for the regular Brouwerij and Polder sessions, and for allowing me to delve back into Dungeons\&Dragons, if only for a little while.\\
To all the people of the Nikhef theory group, new and old, and to our neighbors across the street at Universiteit van Amsterdam, Andrea, Diego, Domenico, Elisa, Francesca, Franz, Giuseppe, Jacopo, Jan, Lisa, Michael, Robbert, Sabrina, Tomas, and Alex, Lisa and Wilco, you have all contributed to a great work and social environment. I will be hard-pressed to find a group of co-workers and friends like the one you provided.\\
I would also like to communicate my thanks to all the staff members of Utrecht University, where I have had stimulating discussions and a great time exchanging ideas. Each of my stays there has been very rewarding.\\
To all my friends in Utrecht, this cute and crazy Amsterdam suburb, for all the wonderful times together, the chats, the parties, the beers, and of course the hungover market sessions. Thank you Andrea, Davide, the lads at the Croatian embassy Dra\v{z}en, Sandro and Marko (won't be forgetting your annual party anytime soon), Jonas, Tania, Benedetta, Dewie, Alexis and Nava. Of course, my thanks also go to Luisa, Marta, Diane, Iannis and all the wonderful people I have had the chance to meet during my time spent in St. Gotthard. It has been a great pleasure and a humbling experience to get to know all of you over the years!\\
A tous les parisiens, Andr\'{e}a, Benoit, Chatoune, Claire, Cl\'{e}mence, Cubi, Dominika, Fanny et Fanny, Guillaume, Jeanne, Julie, Justine, Laetitia, L\'{e}o, Marguerite, Mussard, Pierre-Yves, Tibo, Marta, Nil et Willy (m\^{e}me exil\'{e}s a Londres) et Za~\ldots~you all know what you did. So I'll just say this: I'm so very glad I met all of you. Your friendship and support means the world to me.\\
Maybe somewhat unusually, I would like to pass my thanks and gratitude to Kiran Sande of the Blackest Ever Black label and Oscar Powell of the Diagonal Records label, for making what I consider to be the two best radio shows currently on the air, on Berlin Community Radio and NTS Radio. They have provided me with an amazing soundtrack on many a days spent working at my desk, and anyone who knows me a little will know how important this aspect can be to me.\\
Lastly, but quite certainly most importantly, I want to express my deepest gratitude to my family for their love and support, and in particular my parents Vincent and Claudine, and my brother and sister Nicolas and Philippine. I love you all very much, I'm sure you already know that.\\
Now, thanks for reading if you've made it this far, but honestly, just put down this book. Put on a record, go have a drink, and remember: everything is important.


\newpage

\bibNote{}
\renewcommand\bibname{Bibliography}

\providecommand{\href}[2]{#2}\begingroup\raggedright\endgroup
\addcontentsline{toc}{chapter}{Bibliography}

\end{document}